\documentclass[aps,prd,showpacs,eqsecnum,twocolumn,superscriptaddress]{revtex4-1}
\usepackage{amsmath,amssymb,graphicx,color,ulem,multirow}
\usepackage{hyperref,ulem}
\hypersetup{colorlinks=true}

\begin{document}

\title{Sub-radian-accuracy gravitational waves from coalescing binary neutron stars in numerical relativity II: Systematic study on equation of state, binary mass, and mass ratio}

\author{Kenta Kiuchi}
\affiliation{Max Planck Institute for Gravitational Physics (Albert Einstein Institute), Am M\"{u}hlenberg, Potsdam-Golm, 14476, Germany}
\affiliation{Center for Gravitational Physics, Yukawa Institute for Theoretical Physics, Kyoto University, Kyoto, 606-8502, Japan} 

\author{Kyohei Kawaguchi}
\affiliation{Institute for Cosmic Ray Research, The University of Tokyo, 5-1-5, Kashiwanoha, Kashiwa, Chiba, 277-8582, Japan}
\affiliation{Center for Gravitational Physics, Yukawa Institute for Theoretical Physics, Kyoto University, Kyoto, 606-8502, Japan}
\affiliation{Max Planck Institute for Gravitational Physics (Albert Einstein Institute), Am M\"{u}hlenberg, Potsdam-Golm, 14476, Germany}

\author{Koutarou Kyutoku}
\affiliation{Department of Physics, Kyoto University, Kyoto 606-8502, Japan}
\affiliation{Center for Gravitational Physics, Yukawa Institute for Theoretical Physics, Kyoto University, Kyoto, 606-8502, Japan} 
\affiliation{Theory Center, Institute of Particle and Nuclear Studies, KEK, Tsukuba 305-0801, Japan}
\affiliation{Interdisciplinary Theoretical and Mathematical Sciences Program (iTHEMS), RIKEN, Wako, Saitama 351-0198, Japan}

\author{Yuichiro Sekiguchi}
\affiliation{Department of Physics, Toho University, Funabashi, Chiba, 274-8510, Japan}
\affiliation{Center for Gravitational Physics, Yukawa Institute for Theoretical Physics, Kyoto University, Kyoto, 606-8502, Japan} 

\author{Masaru Shibata}
\affiliation{Max Planck Institute for Gravitational Physics (Albert Einstein Institute), Am M\"{u}hlenberg, Potsdam-Golm, 14476, Germany}
\affiliation{Center for Gravitational Physics, Yukawa Institute for Theoretical Physics, Kyoto University, Kyoto, 606-8502, Japan}

\date{\today}

\begin{abstract}
We report results of numerical relativity simulations for {\it new} 26 non-spinning binary neutron star systems with 6 grid resolutions using an adaptive mesh refinement numerical relativity code {\tt SACRA-MPI}. The finest grid spacing is $\approx 64$--$85$ m, depending on the systems. 
First, we derive long-term high-precision inspiral gravitational waveforms and show that the accumulated gravitational-wave phase
error due to the finite grid resolution is less than $0.5$ rad during more than $200$ rad phase evolution irrespective of the systems.
We also find that the gravitational-wave phase error for a binary system with a tabulated equation of state (EOS) is comparable to that for a piecewise polytropic EOS. 
Then we validate the SACRA inspiral gravitational waveform template, which will be used to extract tidal deformability from gravitational wave observation, and find that accuracy of our waveform modeling is $\lesssim 0.1$ rad in the gravitational-wave phase and $\lesssim 20 \%$ in the gravitational-wave amplitude up to the gravitational-wave frequency $1000$ Hz. Finally, we calibrate the proposed universal relations between a post-merger gravitational wave signal and tidal deformability/neutron star radius in the literature and show that they suffer from systematics and many relations proposed as universal are not very universal. Improved fitting formulae are also proposed. 
\end{abstract}


\maketitle


\section{Introduction}

On August 17, 2017, advanced LIGO~\cite{TheLIGOScientific:2014jea} and advanced Virgo~\cite{TheVirgo:2014hva} detected gravitational waves from a binary neutron star (BNS) merger, GW170817, for the first time~\cite{TheLIGOScientific:2017qsa}.
In this event, not only gravitational waves but also the electromagnetic signals in the gamma-ray~\cite{Monitor:2017mdv, Goldstein:2017mmi,Savchenko:2017ffs}, ultraviolet-optical-infrared~\cite{Evans:2017mmy,Drout:2017ijr,Kilpatrick:2017mhz,Kasliwal:2017ngb,Nicholl:2017ahq,Utsumi:2017cti,Tominaga:2017cgo,Chornock:2017sdf,Arcavi:2017vbi,Diaz:2017uch,Shappee:2017zly,Coulter:2017wya,Soares-Santos:2017lru,Valenti:2017ngx,Pian:2017gtc,Smartt:2017fuw}, X-ray~\cite{Haggard:2017qne,Margutti:2017cjl,Troja:2017nqp}, and radio bands~\cite{Alexander:2017aly,Hallinan:2017woc,Margutti:2018xqd,Dobie:2018zno,Mooley:2017enz,Mooley:2018dlz} were detected.
This monumental event GW170817, GRB170817A, and AT2017gfo heralded the opening of the multi-messenger astrophysics. Furthermore, advanced LIGO and advanced Virgo have started a new observation run, O3, from April 2019 and 
a new BNS merger event, GW190425, was reported~\cite{Abbott:2020uma} and 7 candidates of a BNS merger as of Feb. 17, 2020, have been detected~\cite{GCN}.

One noteworthy finding in GW170817 is that tidal deformability of the neutron star (NS) was constrained for the first time. 
Due to a tidal field generated by a companion, NSs in a binary system could be deformed significantly in the late inspiral stage~\cite{Flanagan:2007ix}. 
 The response to the tidal field, the tidal deformability, is imprinted as a phase shift in gravitational waves and its measurement gives a constraint on the equation of state (EOS) of NSs because the tidal
deformability depends on EOSs. GW170817 constrained the binary tidal deformability in the range of $100 \lesssim \tilde{\Lambda} \lesssim 800$ with the binary total mass of $2.73^{+0.04}_{-0.01}M_\odot$~\cite{TheLIGOScientific:2017qsa,Abbott:2018exr,De:2018uhw,Abbott:2018wiz} where the precise value depends on the analysis methods.

To extract information of the tidal deformability from observed gravitational wave data, a high precision template for gravitational waveforms plays an essential role.
Numerical relativity simulation is the unique tool to derive high-precision gravitational waveforms in the late inspiral stage during which the gravitational-wave phase shift due to the tidal deformation becomes prominent. During this stage, any analytic techniques break down.
Dietrich and his collaborators constructed a gravitational wave template for the inspiral stage based on the numerical relativity simulations in a series of papers~\cite{Dietrich:2015pxa,Dietrich:2017feu,Dietrich:2017aum,Dietrich:2018uni,Dietrich:2018phi,Dietrich:2019kaq} and their template was used in gravitational wave data analysis by LIGO Scientific and Virgo Collaborations to infer the tidal deformability from GW170817~\cite{Abbott:2018exr}.
  However, the residual phase error caused mainly by the finite grid resolution in their simulations is $\approx 0.5$--$2.3$ rad~\cite{Dietrich:2019kaq}.  
  The phase error of $O(1)$ rad could be an obstacle to construct a high-quality inspiral gravitational waveform template (see also Refs.~\cite{Haas:2016cop,Foucart:2018lhe}).

In Ref.~\cite{Kiuchi:2017pte},
we tackled this problem by using our numerical relativity code {\tt SACRA-MPI} and performed long-term simulations with the highest grid resolution to date 
(see also Refs.~\cite{Hotokezaka:2013mm,Hotokezaka:2015xka,Hotokezaka:2016bzh,Shibata:2005xz} for our effort in the early stage of this project). 
In our numerical results, the gravitational-wave phase error caused by the finite grid resolution is less than $0.5$ rad for $31$--$32$ inspiral gravitational wave cycles. On the basis of these high-precision gravitational waveforms,
Ref.~\cite{Kawaguchi:2018gvj} presented a waveform template, the SACRA inspiral gravitational waveform template, of BNS mergers.
Specifically, we multiply the tidal-part phase of the $2.5$ Post-Newtonian (PN) order derived in Ref.~\cite{Damour:2012yf} by a correction term composed of the PN parameter and the binary tidal deformability.
Then, we validated it by confirming that it reproduces the high-precision gravitational waveforms derived in Ref.~\cite{Kiuchi:2017pte}. 
We also validated a correction term in the tidal-part amplitude of the $1$ PN order derived in Refs.~\cite{Damour:2012yf,Vines:2011ud}. 

In Refs.~\cite{Kiuchi:2017pte,Kawaguchi:2018gvj}, we performed simulations for a limited class of BNS systems, i.e., two equal-mass and two unequal-mass systems.
Thus, the applicable range of the SACRA inspiral gravitational waveform template has not quantified precisely yet.
In this paper, we derive a number of gravitational waveforms from BNS mergers by performing numerical-relativity simulations in a wider parameter space for EOSs, binary total mass, and mass ratio than that in the previous papers~\cite{Kiuchi:2017pte,Kawaguchi:2018gvj}. For each binary parameter, we perform an in-depth
resolution study to assess the accuracy of our waveforms. On the basis of newly derived high-precision gravitational waveforms, we validate the template. 

In addition, we analyze post-merger gravitational wave signals derived in this paper. 
The post-merger signal in GW170817 has not been detected~\cite{Abbott:2017dke}, but a post-merger signal
could be detected in near future for the nearby events or in the third generation detectors such as Einstein Telescope or Cosmic Explorer~\cite{Punturo:2010zz,Evans:2016mbw}.
The signal could bring us information of the EOS complementary to that imprinted in the late inspiral signal.
To extract such information, we should explore a heuristic relation between post-merger signals and the tidal deformability/NS radius in numerical relativity simulations.
In several previous papers, such an attempt has been made~\cite{Rezzolla:2016nxn,Read:2013zra,Zappa:2017xba,Bauswein:2011tp,Bauswein:2012ya,Bernuzzi:2014owa,Bernuzzi:2015rla}.
However, systematics contained in these relations are unclear because of the lack of resolution study, the approximate treatment of relativistic gravity, the lack of the estimation for the systematics with the uncertainty of the NS EOS, and
the narrow range of the BNS parameter space explored.
In this paper, we assess to what extent the proposed universal relations between the post-merger gravitational wave signal and tidal deformability/NS radius~\cite{Rezzolla:2016nxn,Read:2013zra,Zappa:2017xba,Bauswein:2011tp,Bauswein:2012ya,Bernuzzi:2014owa,Bernuzzi:2015rla} hold. 

To stimulate an independent attempt by other researchers for constructing a gravitational waveform template based on the numerical relativity simulations and/or to stimulate a comparison to numerical relativity waveforms derived by other groups, we release our simulation data on a website  
\href{https://www2.yukawa.kyoto-u.ac.jp/~nr_kyoto/SACRA_PUB/catalog.html}{SACRA Gravitational Waveform Data Bank}~\cite{DB}.

This paper is organized as follows. Section~\ref{sec:model} describes our method, grid setup, and initial condition of the simulations. Section \ref{sec:result} is devoted to describing the accuracy of inspiral gravitational waveforms. 
Section~\ref{sec:WFmodel} presents validation of the SACRA inspiral gravitational waveform template. 
Section~\ref{sec:universal-relation} describes the assessment of the universal relations of the post-merger signals. This section also presents the energy and angular momentum carried by gravitational waves. 
We summarize this paper in Sec.~\ref{sec:summary}.
Throughout this paper, we employ the geometrical unis of $c=G=1$ where $c$ and $G$ are the speed of light and the gravitational constant, respectively. 


\section{Method, grid setup, and initial models}\label{sec:model}

\subsection{Method and grid setup}

We use our numerical relativity code, {\tt SACRA-MPI}~\cite{Yamamoto:2008js,Kiuchi:2017pte}, 
to simulate a long-term inspiral stage of BNS up to early post-merger. {\tt SACRA-MPI} implements
the Baumgarte-Shapiro-Shibata-Nakamura-puncture formulation~\cite{SN,BS,Capaneli,Baker}, {\it locally}
incorporating a Z4c-type constraint propagation prescription~\cite{Hilditch:2012fp}, to solve Einstein's equation.
We discretize the field equation with the 4th-order accuracy in both the space and time.
We also apply the 4th-order lop-sided finite difference scheme for the advection term~\cite{Bruegmann:2006at}.

In {\tt SACRA-MPI}, a conservation form of general relativistic hydrodynamics equations is employed and we implement a high-resolution shock capturing scheme proposed by
Kurganov and Tadmor~\cite{Kurganov} together with the 3rd-order accurate cell reconstruction~\cite{Colella:1982ee}.

We also implement the Berger-Oliger type adaptive mesh refinement (AMR) algorithm~\cite{BergerOliger} to enlarge a simulation domain to a local wave zone of gravitational waves
while guaranteeing a high spatial grid resolution around NSs. A simulation domain consists of two sets of the 4 Cartesian AMR domains which follow orbital motion of each of NSs
and the 6 Cartesian AMR domains whose
center is fixed to the coordinate origin throughout all the simulations.
The grid spacing of a coarser refinement level is twice as large as that of its finer refinement level. Thus, the grid spacing of a refinement level $l$
is given by $\Delta x_l = L/(2^{l}N)$ with $l=0,1,\cdots 9$. $L$ denotes the distance from the coordinate origin to the outer boundary along each coordinates axis. $N$ is an even number and each of the AMR domains possesses the grid point $(2N+1,2N+1,N+1)$ in the $(x,y,z)$ directions where we assumed the orbital plane symmetry.

In this work, we performed simulations with $N=182,150,130,110,102,$ and $90$ for all the systems to check the convergence of gravitational waveforms with respect to the
grid resolution. The values of $L$ and $\Delta x_9$ are summarized in Table~\ref{tb:model}.

\subsection{Binary system parameters and gravitational wave extraction}

Table~\ref{tb:model} shows the list of the binary systems as well as the grid setup for the simulations.

\subsubsection{Equation of state}

Following the previous papers~\cite{Kiuchi:2017pte,Kawaguchi:2018gvj},
we employ a parameterized piecewise polytropic EOS to describe the NS matter~\cite{rlof2009}. Specifically, we assume that the pressure and
specific internal energy consist of two segments with respect to the rest-mass density:
\begin{align}
  &P_\text{cold}(\rho) = \kappa_i \rho^{\Gamma_i},\nonumber\\
  &\epsilon_\text{cold}(\rho) = \frac{\kappa_i}{\Gamma_i-1}\rho^{\Gamma_i-1} + \Delta \epsilon_i~(\rho_i \le \rho < \rho_{i+1}), \nonumber
\end{align}
with $i=0,1$, $\rho_0=0{\rm~g~cm^{-3}}$, and $\rho_2 = \infty$. $\rho_1$ is the rest-mass density which divides the pressure and specific internal energy into the two segments. 
Given the adiabatic indices $\Gamma_0,\Gamma_1$ and one of the polytropic constants $\kappa_0$, the other polytropic constant $\kappa_1$ is calculated from the
continuity of the pressure at $\rho=\rho_1$ by $\kappa_0\rho_1^{\Gamma_0}=\kappa_1\rho_1^{\Gamma_1}$. $\Delta \epsilon_1$
is also calculated from the continuity of the specific internal energy at $\rho=\rho_1$ by $\kappa_0\rho_1^{\Gamma_0-1}/(\Gamma_0-1)=\kappa_1\rho_1^{\Gamma_1-1}/(\Gamma_1-1)+\Delta \epsilon_1$. Note that $\Delta \epsilon_0=0$. Following Ref.~\cite{rlof2009}, we fix $\Gamma_0=1.3562395$, $\Gamma_1=3$, and $\kappa_0=3.594\times 10^{13}$ in cgs units.
By varying the remaining parameter $\rho_1$ for a wide range as shown in Table~\ref{tb:pwp}, we can derive plausible NS models with a variety of the radii and tidal deformability (see Table~\ref{tb:eos_model}).

In addition to the piecewise polytropic EOS, we employ one tabulated EOS, SFHo~\cite{Steiner:2012rk}. 
To model an EOS for cold NS, we simply set $T=0.1$ MeV which is the minimum temperature in the table of SFHo EOS. We also impose the neutrinoless low-temperature $\beta$-equilibrium condition to set the value of $Y_e$. 
Then, the original tabulated EOS is reduced to a one dimensional SFHo (tabulated) EOS, i.e., $P_\text{cold}(\rho)$ and $\epsilon_\text{cold}(\rho)$ (see also Table~\ref{tb:eos_model} for the NS radius and
tidal deformability).

During simulations (in particular for the post-merger stage), we employ a hybrid EOS to capture the shock heating effect. Specifically, we assume that the pressure consists of the cold and thermal parts:
\begin{align}
  P = P_\text{cold}(\rho) + ( \Gamma_\text{th} - 1 )\rho( \epsilon - \epsilon_\text{cold} (\rho) ), \label{eq:pres}
\end{align}
where $\epsilon$ is the specific internal energy and we assumed that the thermal part is described by the $\Gamma$-law EOS with the index $\Gamma_\text{th}$.
Following Refs.~\cite{Kiuchi:2017pte,Kawaguchi:2018gvj}, we fix $\Gamma_\text{th}=1.8$.
We note that gravitational waveforms for the post-merger stage depend on the value of $\Gamma_{\rm th}$~\cite{Shibata:2005ss}, although inspiraling waveforms do not. Since the major purpose of the present paper is to derive the accurate inspiraling waveforms,
the choice of $\Gamma_{\rm th}$ does not have any essential importance. On the other hand, it has been long known that the post-merger waveform depends strongly on this value (see, e.g., Ref.~\cite{Shibata:2005ss}). Thus, we have to keep in mind that the systematics exist due to the
uncertainty of this value~\cite{Carbone:2019pkr}. 

\subsubsection{Binary systems}

In this paper, we consider 6 irrotational binary systems assuming that NSs have no spin before merger. We fix a chirp mass, ${\cal M}_c$, and symmetric mass ratio, $\eta$, to be
$({\cal M}_c,\eta)=(1.1752M_\odot,0.2500)$, $(1.1752M_\odot,0.2485)$, $(1.1752M_\odot,0.2455)$, $(1.1752M_\odot,0.2450)$, $(1.0882M_\odot,0.2470)$, and $(1.0882M_\odot,0.2440)$.
With this setting, gravitational masses of a less massive and massive components for the infinite orbital separation is $(m_1,m_2)=(1.35M_\odot,1.35M_\odot)$, $(1.25M_\odot,1.46M_\odot)$, $(1.18M_\odot,1.55M_\odot)$, $(1.17M_\odot,1.56M_\odot)$, $(1.12M_\odot,1.40M_\odot)$, and $(1.07M_\odot,1.46M_\odot)$ (see Table~\ref{tb:model}). For the SFHo (tabulated) EOS, we only consider the equal-mass binary system with $m_1=1.35M_\odot$ and $m_2=1.35M_\odot$. 

Table~\ref{tb:model} also shows the binary tidal deformability for all the binary systems~\cite{Wade:2014vqa,Favata:2013rwa}:
\begin{align}
  \tilde{\Lambda} &= \frac{8}{13}\Big[(1+7\eta-31\eta^2)(\Lambda_1 + \Lambda_2) \nonumber \\
                  &- \sqrt{1-4\eta}(1+9\eta-11\eta^2)(\Lambda_1-\Lambda_2)\Big],
\end{align}
where $\Lambda_1(\Lambda_2)$ is the tidal deformability of the less massive (massive) component. The value of the tidal deformability in this paper covers a wide
range of $\approx 300$--$1800$.

\begin{figure}
         \hspace{-13.6mm}
 	 \includegraphics[width=1.15\linewidth]{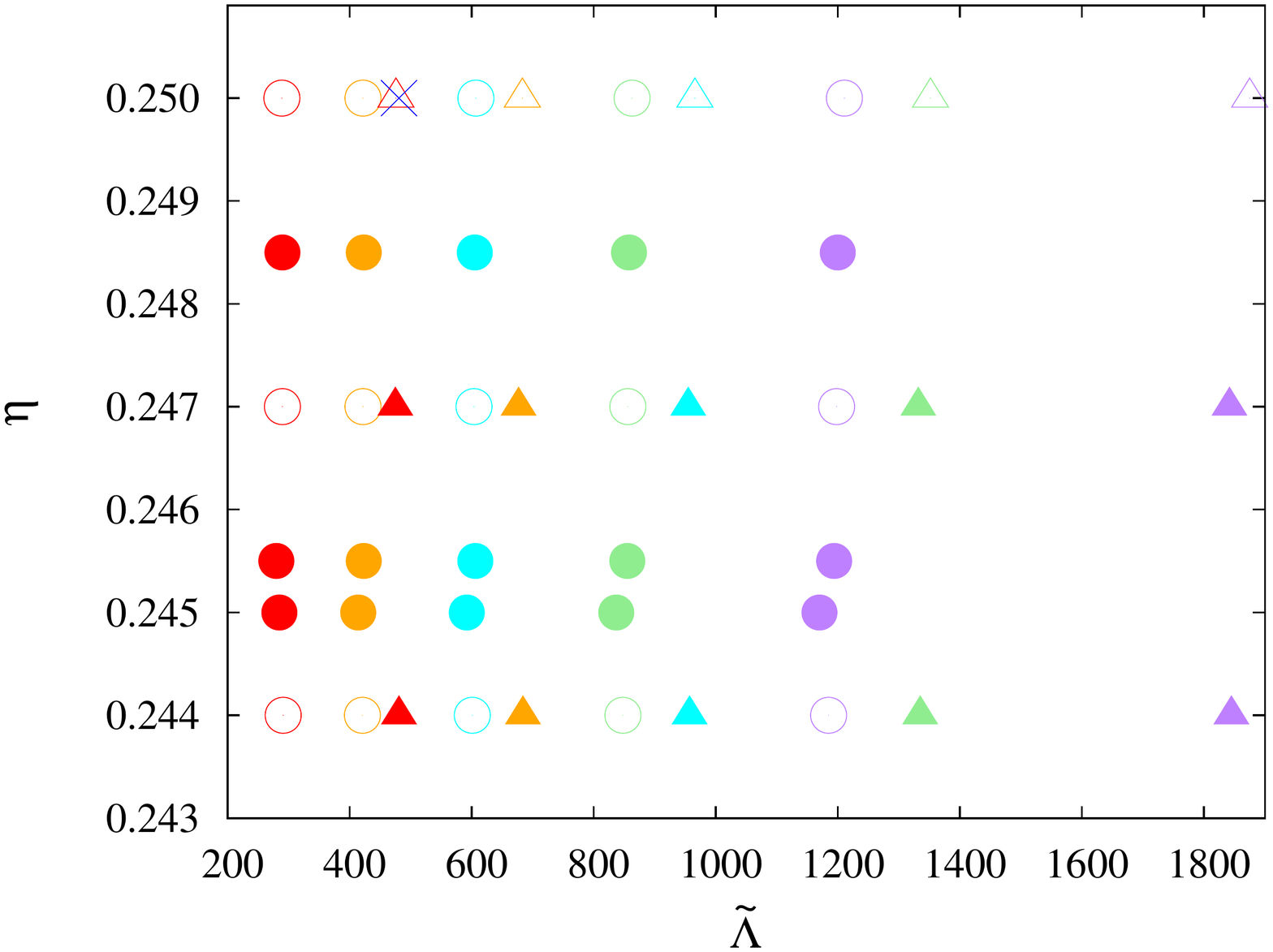}
 	 \caption{Symmetric mass ratio, $\eta$, and binary tidal deformability, $\tilde{\Lambda}$, of all the models simulated for long durations by our group. The circle and triangle symbols denote BNS systems with ${\cal M}_c =  1.1752 M_\odot$ and with ${\cal M}_c =  1.0882 M_\odot$, respectively. The open symbols denote the systems reported in Refs.~\cite{Kiuchi:2017pte,Kawaguchi:2018gvj}. The filled symbols are the systems newly simulated in this study. The purple, green, cyan, orange, and red colors are for the systems with EOS 15H, 125H, H, HB, and B, respectively.
           The blue cross symbol is for SFHo135-135.
         }\label{fig:model}
\end{figure}

Figure~\ref{fig:model} plots the BNS systems simulated for long durations by our group to date. For the SFHo (tabulated) EOS case, an interpolation of the thermodynamic variables is necessary in the simulations.
Because we implement the linear interpolation scheme for this purpose, the associated truncation error can be a non-negligible error source for generating high-precision gravitational waveforms.
This system is used to assess the error budget possibly caused by employing tabulated EOS (see also Ref.~\cite{Foucart:2019yzo} for the gravitational-wave phase error stemming from different analytical descriptions of the EOSs). 

We name all the systems according to the EOS, the mass of the less massive component, and that of the massive component.
For example, 15H125-146 refers to the system with 15H EOS, $m_1= 1.25M_\odot$, and $m_2= 1.46 M_\odot$. We set the initial orbital angular velocity to
be $m_0\Omega_0 = 0.0150$--$0.0155$ with $m_0=m_1+m_2$. With this, the BNSs experience $15$--$16$ orbits before the onset of merger for all the systems.

To generate a high-precision inspiral waveform from BNS inspirals by a numerical relativity simulation, initial data with low orbital eccentricity are necessary
because the orbital motion of a BNS in the late inspiral stage is circularized due to the gravitational-wave emission. 
We numerically obtain quasi-equilibrium sequences of the BNSs by a spectral-method library, LORENE~\cite{LORENE,Taniguchi}. Then, we reduce orbital eccentricity by using the prescription in Ref.~\cite{Kyutoku:2014yba}. With this method, 
  we confirm that the initial orbital eccentricity is reduced typically to $\approx 10^{-3}$ which is low enough to generate a high-precision inspiral waveform
  (see also Appendix in Refs.~\cite{Kiuchi:2017pte,Kawaguchi:2018gvj}).

\subsection{Gravitational wave extraction}

We calculate a complex Weyl scalar $\Psi_4$ from simulation data to derive gravitational waveforms~\cite{Yamamoto:2008js}.
Given an extraction radius $r_0$, the Weyl scalar
$\Psi_4$ is decomposed into $(l,m)$ modes with the spin-weighted spherical harmonics by
\begin{align}
  \Psi_4(t_\text{ret},r_0,\theta,\phi) = \sum_{l,m} \Psi_4^{l,m}(t_\text{ret},r_0) _{-2}Y_{lm}(\theta,\phi),
\end{align}
where $t_\text{ret}$ is a retarded time defined by
\begin{align}
  t_\text{ret} \equiv t - \left[ D + 2 m_0 \ln\left(\frac{D}{2m_0}-1\right) \right], \label{eq:tret}
\end{align}
with $D=\sqrt{A/4\pi}$. $A$ is a proper area of the extraction sphere. We apply the Nakano's method~\cite{Nakano:2015} to extrapolate $\Psi_4^{l,m}$ to
infinity by
\begin{align}
  D \Psi_4^{l,m,\infty}(t_\text{ret}) & \equiv C(r_0)\Big[D \Psi_4^{l,m}(t_\text{ret},r_0) \nonumber\\
    &  - \frac{(l-1)(l+2)}{2}\int^{t_\text{ret}} \Psi_4^{l,m}(t',r_0) dt' \Big],  
\end{align}
where $C(r_0)$ is a function of $r_0$.  Following Ref.~\cite{Kiuchi:2017pte}, we choose $D \approx r_0[1+m_0/(2r_0)]^2$ and $C(r_0)=1-2m_0/D$ because
our coordinates are similar to isotropic coordinates of non-rotating black holes in the wave zone.

Gravitational waves of each harmonic mode are calculated by integrating $\Psi_4^{l,m,\infty}$ twice in time:
\begin{align}
  h^{l,m,\infty}(t_\text{ret}) &= h^{l,m,\infty}_+ (t_\text{ret}) - i h^{l,m,\infty}_\times(t_\text{ret}) \nonumber\\
  &= - \int^{t_\text{ret}} dt'\int^{t'} \Psi_4^{l,m,\infty}(t'')dt''.
\end{align}
For the time integration, we employ the fixed frequency method~\cite{Reisswig:2010di} by
\begin{align}
  h^{l,m,\infty}(t_\text{ret}) = \int df' \frac{\tilde{\Psi}_4^{l,m,\infty}(f')}{(2\pi\max[f',f_\text{cut}])^2} \exp(2\pi i f' t_\text{ret}),
\end{align}
where $\tilde{\Psi}_4^{l,m,\infty}(f)$ is the Fourier component of $\Psi_4^{l,m,\infty}(t)$ and $f_\text{cut}$ is set to be $0.8m\Omega_0/(2\pi)$. 

To check the convergence with respect to the extraction radius $r_0$, we repeat this analysis for $r_0$ $=244\,m_0$, $199 \,m_0,$ and $155 \,m_0$ for ${\cal M}_c= 1.1752M_\odot$
and $r_0=262\,m_0$, $213\,m_0$, and $156\,m_0$ for
${\cal M}_c= 1.0882M_\odot$ (see Table~\ref{tb:model}).

In general, gravitational waves for each $(l,m)$ mode are decomposed into the amplitude and phase as
\begin{align}
  h^{l,m,\infty}(t_\text{ret}) = A^{l,m,\infty}(t_\text{ret}) e^{-i\Phi^{l,m}(t_\text{ret})}, \label{eq:freqGW}
\end{align}
and instantaneous gravitational-wave frequency is defined by $d\Phi^{l,m}/dt_\text{ret}$.
In Sec.~\ref{sec:result}, we explore the accuracy of the gravitational-wave phase of the $(l,m)=(2,2)$ mode, and simply refer to $\Phi^{2,2}$ as the gravitational-wave phase. 
With Eq.~(\ref{eq:freqGW}), the instantaneous frequency of the $(l,m)=(2,2)$ mode is calculated by 
\begin{align}
  f_\text{GW} = \frac{1}{2\pi}{\rm Im} \left(\frac{h^{*2,2,\infty}\dot{h}^{2,2,\infty}}{|h^{2,2,\infty}|^2}\right),
  \label{eq:GWfreq}
\end{align}
where the asterisk symbol denotes the complex conjugate of $h^{2,2,\infty}$.

We also calculate the energy and angular momentum flux due to gravitational-wave emission by~\cite{Shibata:text}
\begin{align}
  \frac{dE_\text{GW}^{l,m}}{dt} &= \lim_{r \to \infty} \frac{r^2}{16\pi} \left| \int^t \Psi_4^{l,m,\infty}(t')dt'\right|^2, \label{eq:EGW}\\
  \frac{dJ_\text{GW}^{l,m}}{dt} &= - \lim_{r \to \infty} \frac{r^2}{16\pi} \text{Im} \Big[m \left(\int^t \Psi_4^{l,m,\infty}(t')dt' \right)^* \nonumber\\
    &\times \int^t dt' \int^{t'}dt'' \Psi_4^{l,m,\infty}(t'')\Big]. \label{eq:JGW}
\end{align}
Thus, the energy and angular momentum carried by gravitational waves are calculated by
\begin{align}
  E_\text{GW}^{l,m} &= \int^{t_\text{sim}} \frac{dE_\text{GW}^{l,m}}{dt}dt, \label{eq:EGW2}\\
  J_\text{GW}^{l,m} &= \int^{t_\text{sim}} \frac{dJ_\text{GW}^{l,m}}{dt}dt, \label{eq:JGW2}
\end{align}
where $t_\text{sim}$ denotes the time we terminate the simulations. 

\begin{table*}[t]
\centering
\caption{List of the systems for which we performed new simulations. The names of the systems are given in the 1st column. The 2nd and 3rd columns show gravitational mass of less massive NS, $m_1$, and massive NS, $m_2$, respectively.
  The 4th column shows EOS. 
  Dimensionless initial orbital angular velocity, $m_0\Omega_0$, with the total gravitational mass of the binary systems, $m_0=m_1+m_2$, is given in the 5th column. The 6th, 7th, and 8th columns show chirp mass, ${\cal M}_c=(m_1m_2)^{3/5}(m_1+m_2)^{-1/5}$, symmetric mass ratio, $\eta=m_1m_2(m_1+m_2)^{-2}$, and binary tidal deformability, $\tilde{\Lambda}$, respectively.
  Location of outer boundary in a computational domain, $L$,  and grid spacing of a finest AMR level, $\Delta x_9$,
  are given in the 9th and 10th columns, respectively. The grid spacing with $N=182,150,130,110,102,$ and $90$ is shown in the parenthesis in the 10th column. The final column shows the extraction radii of gravitational waves.
  }
\begin{tabular}{c|cccc|ccc|ccc}
\hline\hline
System         & $m_1~[M_\odot]$  & $m_2~[M_\odot]$   & EOS   & $m_0\Omega_0$ & ${\cal M}_{\rm c}$ & $\eta$ & ${\tilde \Lambda} $& $L\,[{\rm km}]$ & $\Delta x_9\,[{\rm m}]$ & $r_0/m_0$ \\\hline
15H125-146    & 1.25            & 1.46            & 15H   & 0.0155            &  1.1752            & 0.2485  & 1200               & 7823            & (84,102,117,138,149,169) & (244,199,155) \\
125H125-146   & 1.25            & 1.46            & 125H  & 0.0155            &  1.1752            & 0.2485  & 858                & 7323            & (78,95,110,129,140,158)  & (244,199,155) \\
H125-146      & 1.25            & 1.46            & H     & 0.0155            &  1.1752            & 0.2485  & 605                & 6824            & (73,89,102,121,130,147)  & (244,199,155) \\
HB125-146     & 1.25            & 1.46            & HB    & 0.0155            &  1.1752            & 0.2485  & 423                & 6491            & (69,84,97,115,124,140)   & (244,199,155) \\
B125-146      & 1.25            & 1.46            & B     & 0.0155            &  1.1752            & 0.2485  & 290                & 5992            & (64,78,90,106,114,129)   & (244,199,155) \\
15H118-155    & 1.18            & 1.55            & 15H   & 0.0155            &  1.1752            & 0.2455  & 1194               & 7889            & (84,102,118,139,150,170) & (242,198,154) \\
125H118-155   & 1.18            & 1.55            & 125H  & 0.0155            &  1.1752            & 0.2455  & 855                & 7390            & (79,96,111,131,141,159)  & (242,198,154) \\
H118-155      & 1.18            & 1.55            & H     & 0.0155            &  1.1752            & 0.2455  & 606                & 6990            & (75,91,105,124,133,151)  & (242,198,154) \\
HB118-155     & 1.18            & 1.55            & HB    & 0.0155            &  1.1752            & 0.2455  & 423                & 6491            & (69,84,97,115,124,140)   & (242,198,154) \\
B118-155      & 1.18            & 1.55            & B     & 0.0155            &  1.1752            & 0.2455  & 292                & 5992            & (64,78,90,106,114,129)   & (242,198,154) \\
15H117-156    & 1.17            & 1.56            & 15H   & 0.0155            &  1.1752            & 0.2450  & 1170               & 7889            & (84,102,118,139,150,170) & (242,198,154) \\
125H117-156   & 1.17            & 1.56            & 125H  & 0.0155            &  1.1752            & 0.2450  & 837                & 7323            & (78,95,110,129,140,158)  & (242,198,154) \\ 
H117-156      & 1.17            & 1.56            & H     & 0.0155            &  1.1752            & 0.2450  & 592                & 6990            & (75,91,105,124,133,151)  & (242,198,154) \\
HB117-156     & 1.17            & 1.56            & HB    & 0.0155            &  1.1752            & 0.2450  & 414                & 6491            & (69,84,97,115,124,141)   & (242,198,154) \\
B117-156      & 1.17            & 1.56            & B     & 0.0155            &  1.1752            & 0.2450  & 285                & 6058            & (65,79,91,107,115,131)   & (242,198,154) \\
15H112-140    & 1.12            & 1.40            & 15H   & 0.0150            &  1.0882            & 0.2470  & 1842               & 7989            & (85,104,120,141,152,172) & (262,214,167) \\
125H112-140   & 1.12            & 1.40            & 125H  & 0.0150            &  1.0882            & 0.2470  & 1332               & 7490            & (80,97,112,132,143,162)  & (262,214,167) \\
H112-140      & 1.12            & 1.40            & H     & 0.0150            &  1.0882            & 0.2470  & 955                & 6990            & (75,91,105,124,133,151)  & (262,214,167) \\
HB112-140     & 1.12            & 1.40            & HB    & 0.0150            &  1.0882            & 0.2470  & 677                & 6491            & (69,84,97,115,124,140)   & (262,214,167) \\
B112-140      & 1.12            & 1.40            & B     & 0.0150            &  1.0882            & 0.2470  & 475                & 6092            & (65,79,91,108,116,131)   & (262,214,167) \\
15H107-146    & 1.07            & 1.46            & 15H   & 0.0150            &  1.0882            & 0.2440  & 1845               & 7989            & (85,104,120,141,152,172) & (261,213,166) \\
125H107-146   & 1.07            & 1.46            & 125H  & 0.0150            &  1.0882            & 0.2440  & 1335               & 7490            & (80,97,112,132,143,162)  & (261,213,166) \\
H107-146      & 1.07            & 1.46            & H     & 0.0150            &  1.0882            & 0.2440  & 957                & 6990            & (75,91,105,124,133,151)  & (261,213,166) \\
HB107-146     & 1.07            & 1.46            & HB    & 0.0150            &  1.0882            & 0.2440  & 684                & 6591            & (71,86,99,117,126,142)   & (261,213,166) \\
B107-146      & 1.07            & 1.46            & B     & 0.0150            &  1.0882            & 0.2440  & 481                & 6091            & (65,79,91,108,116,131)   & (261,213,166) \\
SFHo135-135   & 1.35            & 1.35            & SFHo  & 0.0155            &  1.1752            & 0.2500  & 460                & 6491            & (69,84,97,115,124,140)   & (244,200,156) \\
\hline\hline
\end{tabular}\label{tb:model}
\end{table*}

\begin{table}
\centering
\caption{List of $\rho_1$ in two-piecewise polytropic EOSs.}
\begin{tabular}{c|c}\hline\hline
EOS             & $\rho_1[\rm g~cm^{-3}]$ \\\hline
15H 		& $9.3108 \times 10^{13}$\\
125H            & $1.0711 \times 10^{14}$\\
H 		& $1.2323 \times 10^{14}$\\
HB 		& $1.4177 \times 10^{14}$\\
B 		& $1.6309 \times 10^{14}$\\
\hline
\end{tabular}\label{tb:pwp}
\end{table}

\begin{table*}
\centering
\caption{The radius, $R_M$, and the dimensionless tidal deformability, $\Lambda_M$, for spherical NSs with gravitational mass $M=1.07$, $1.12$, $1.17$, $1.18$, $1.25$, $1.35$, $1.40$, $1.46$, $1.55$ and $1.56\,M_\odot$ for the given EOS.
  $R_M$ is listed in units of ${\rm km}$. For SFHo (tabulated) EOS, the quantities for the spherical star with $M=1.35\,M_\odot$ are listed. The last column in the upper table shows the maximum mass of the spherical NS in units of $M_\odot$.}
\begin{tabular}{c|ccccccccccccccccccccc}\hline\hline
EOS             & $R_{1.07}$ & $R_{1.12}$ & $R_{1.17}$& $R_{1.18}$ & $R_{1.25}$ & $R_{1.35}$ & $R_{1.40}$ &$~R_{1.46}$& $~R_{1.55}$& $~R_{1.56}$ & $M_\text{max}$\\\hline
15H 		& 13.54 & 13.58 & 13.61 & 13.62 & 13.65 & 13.69 & 13.71 & 13.72 & 13.74 & 13.74 & 2.53\\
125H            & 12.86 & 12.89 & 12.91 & 12.92 & 12.94 & 12.97 & 12.98 & 12.99 & 12.98 & 12.98 & 2.38\\
H 		& 12.22 & 12.23 & 12.24 & 12.24 & 12.26 & 12.27 & 12.28 & 12.18 & 12.26 & 12.25 & 2.25\\
HB 		& 11.60 & 11.59 & 11.60 & 11.60 & 11.61 & 11.61 & 11.60 & 11.59 & 11.55 & 11.55 & 2.12\\
B 		& 10.97 & 10.97 & 10.98 & 10.98 & 10.98 & 10.96 & 10.95 & 10.92 & 10.87 & 10.86 & 2.00\\
SFHo            & --    & --    & --    & --    & --    & 11.91 & --    & --    & --    & --    & 2.06\\
\hline\hline
EOS             &$~\Lambda_{1.07}$&$~\Lambda_{1.12}$&$~\Lambda_{1.17}$&$~\Lambda_{1.18}$&$~\Lambda_{1.25}$&$~\Lambda_{1.35}$&$~\Lambda_{1.40}$&$~\Lambda_{1.46}$ & $~\Lambda_{1.55}$ & $~\Lambda_{1.56}$\\\hline
15H 		& 4361 & 3411 & 2692 & 2575 & 1871 & 1211 & 975 & 760 & 530 & 509 \\
125H            & 3196 & 2490 & 1963 & 1875 & 1351 & 863  & 693 & 535 & 366 & 350 \\
H 		& 2329 & 1812 & 1415 & 1354 & 966  & 607  & 484 & 369 & 249 & 238 \\
HB 		& 1695 & 1304 & 1013 & 966  & 684  & 422  & 333 & 252 & 165 & 157 \\
B 		& 1216 & 933  & 719  & 681  & 477  & 289  & 225 & 168 & 107 & 101 \\
SFHo            & --   & --   & --   & --   & --   & 460  & --  & --  & --  & --  \\\hline
\end{tabular}\label{tb:eos_model}
\end{table*}

\begin{figure*}[t]
  	 \includegraphics[width=.75\linewidth]{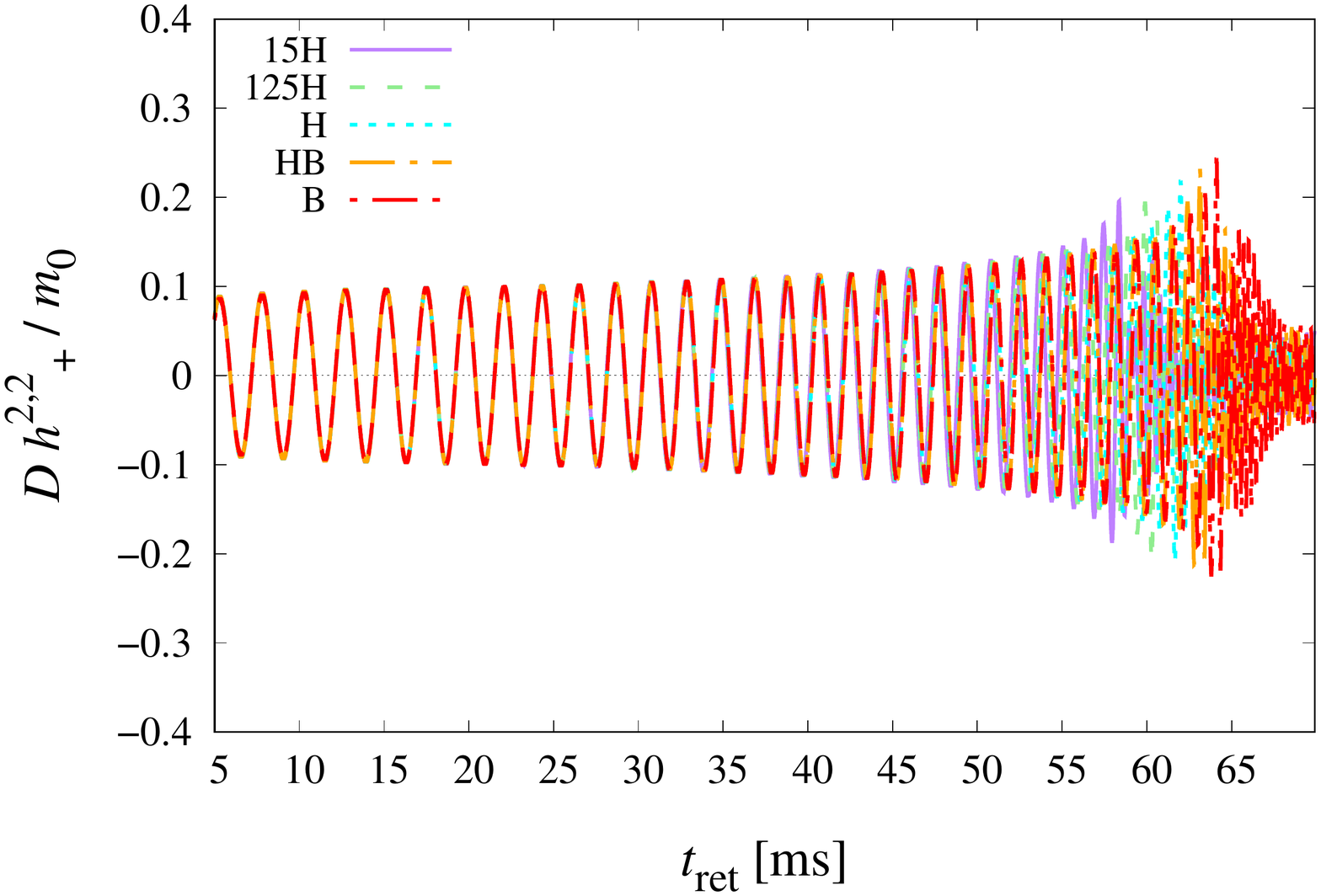}
 	 \includegraphics[width=.75\linewidth]{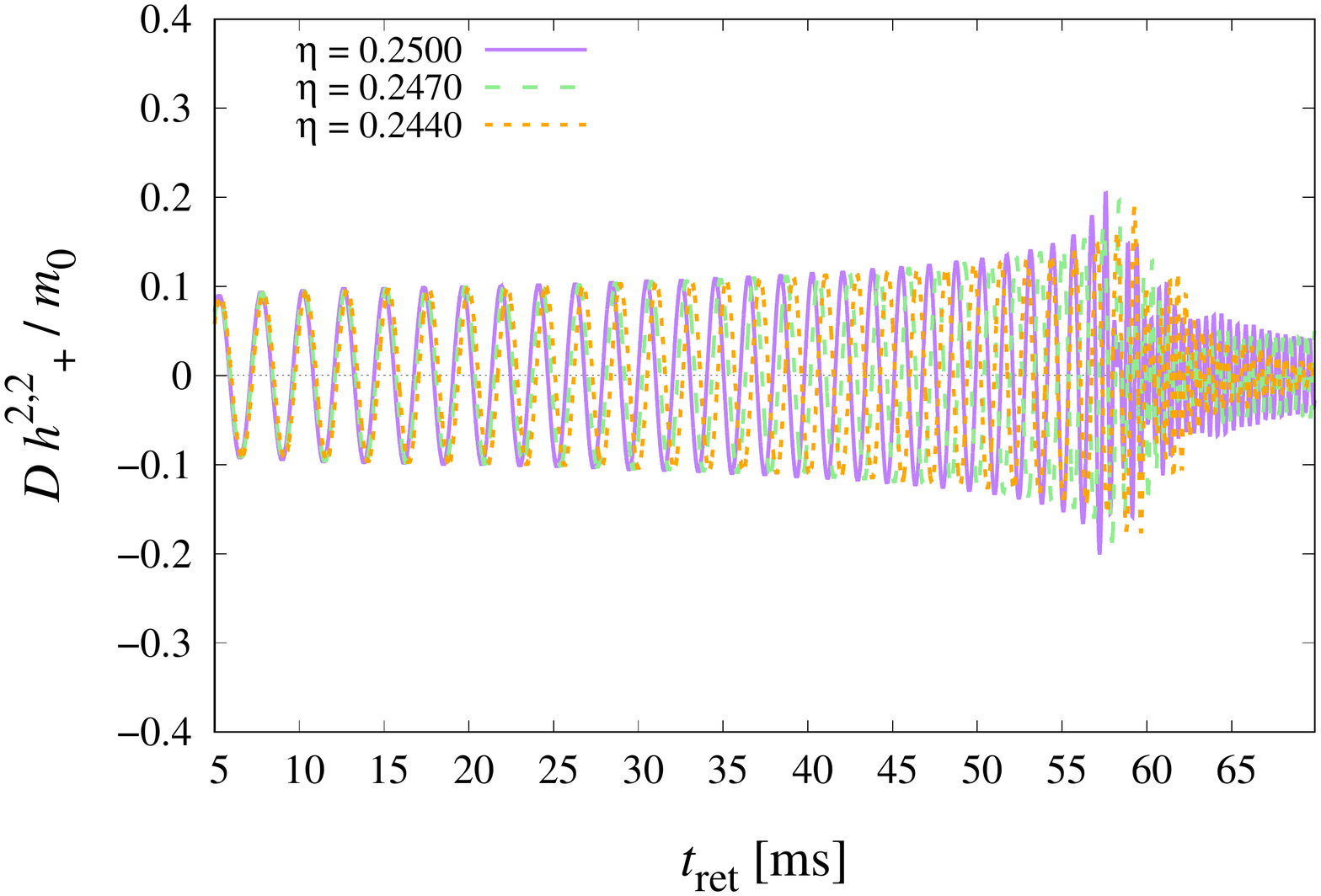}
         \caption{\label{fig:GW3}
           (Top)  $h_+$ for $(l,m)=(2,2)$ mode of the gravitational waveforms for binary systems with $m_1=1.12M_\odot$ and $m_2=1.40M_\odot$. 
           (Bottom) The same as the top panel, but for 15H125-125, 15H112-140, and 15H107-146. In both panels, the grid resolution is $N=182$. 
}
\end{figure*}

\begin{figure*}[t]
  \includegraphics[width=.5\linewidth]{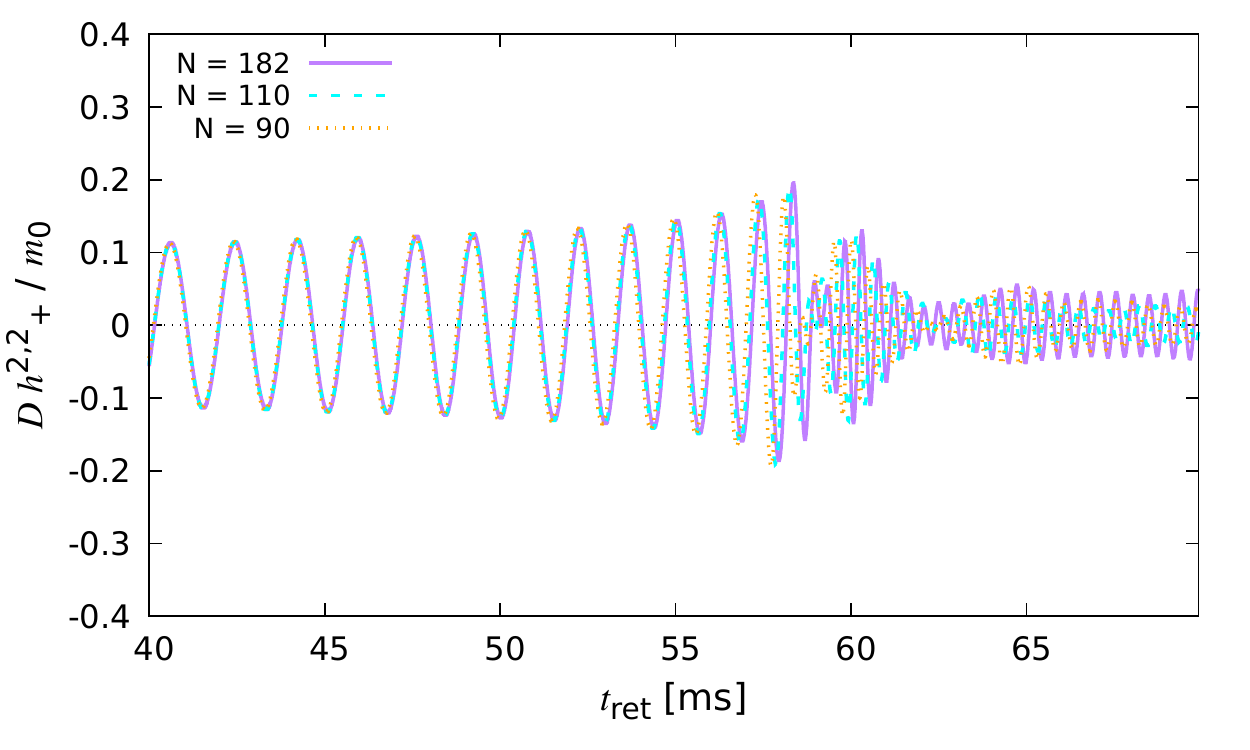}
  \includegraphics[width=.5\linewidth]{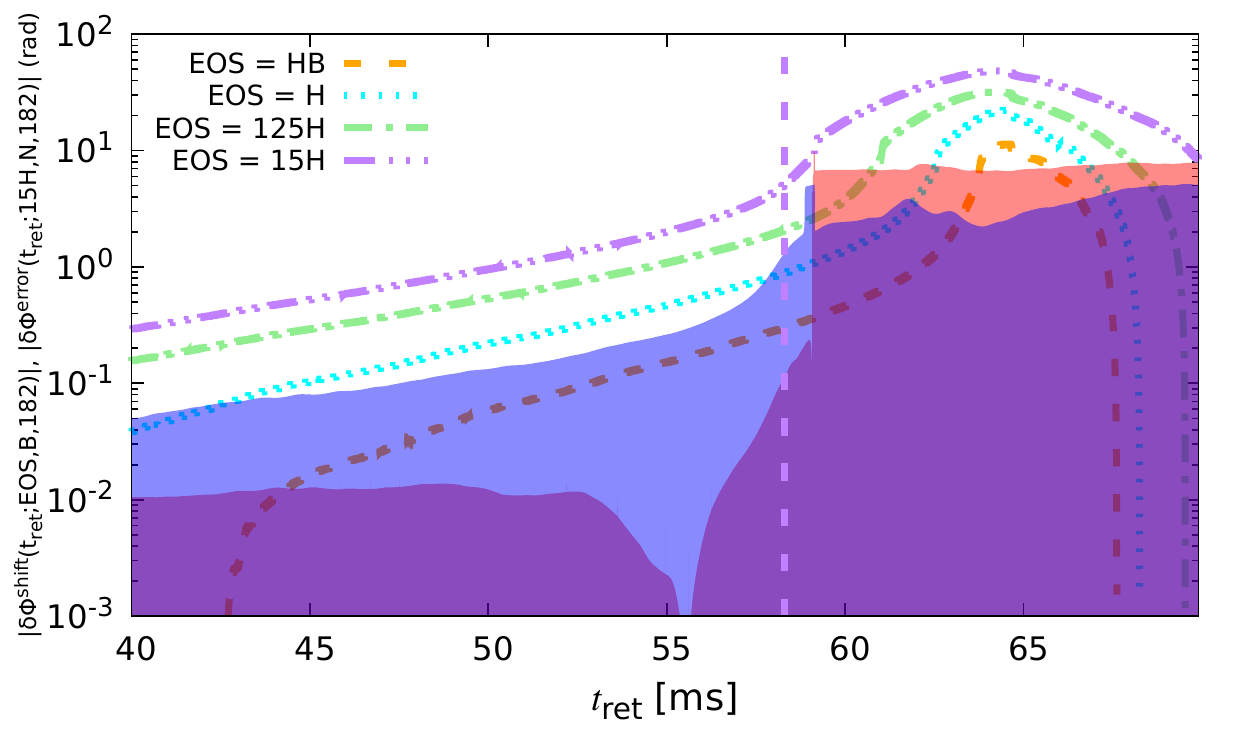}
         \caption{\label{fig:GW4}
           (Top) The same as Fig.~\ref{fig:GW3}, but for 15H112-140 with $N=182,110,$ and $90$. 
           (Bottom) The gravitational-wave phase shift, $\delta\Phi^\text{shift}(t_\text{ret};{\rm EOS},{\rm B},182)$, for the binary systems with $m_1=1.12M_\odot$, $*m_2=1.40M_\odot$, and $\text{EOS}={\rm 15H},{\rm 125H},{\rm H},{\rm HB}$. 
           The shaded region shows $\delta \Phi^\text{error}(t_\text{ret};15{\rm H},150,182)$ (red) and $\delta \Phi^\text{error}(t_\text{ret};15{\rm H},90,182)$ (blue), respectively, for 15H112-140. The overlapped region has purple color. The vertical dashed line denotes the peak time of the gravitational-wave amplitude for 15H112-140 with $N=182$ (see the text for details). 
}
\end{figure*}

\begin{figure*}
           \includegraphics[width=.47\linewidth]{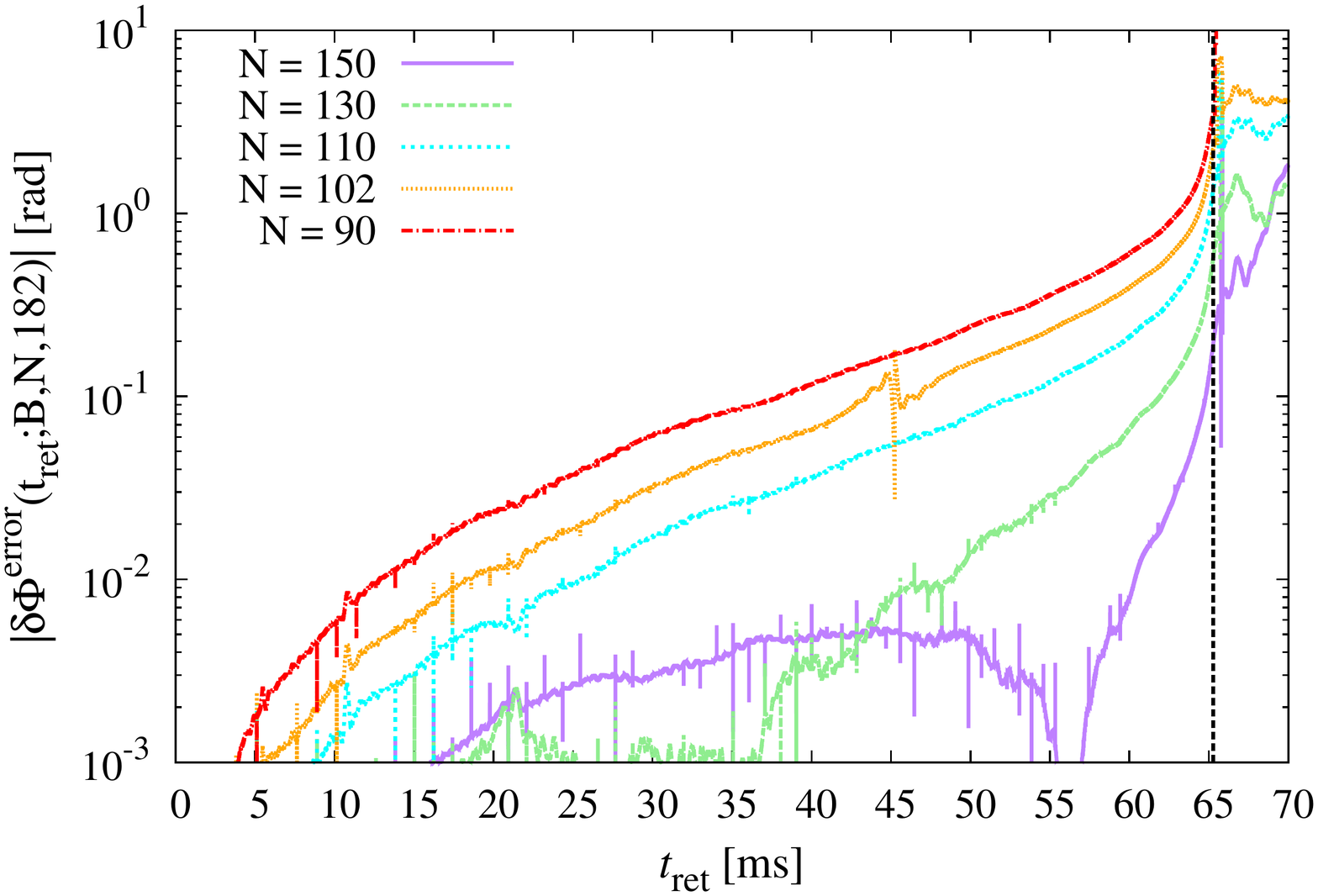}
           \includegraphics[width=.47\linewidth]{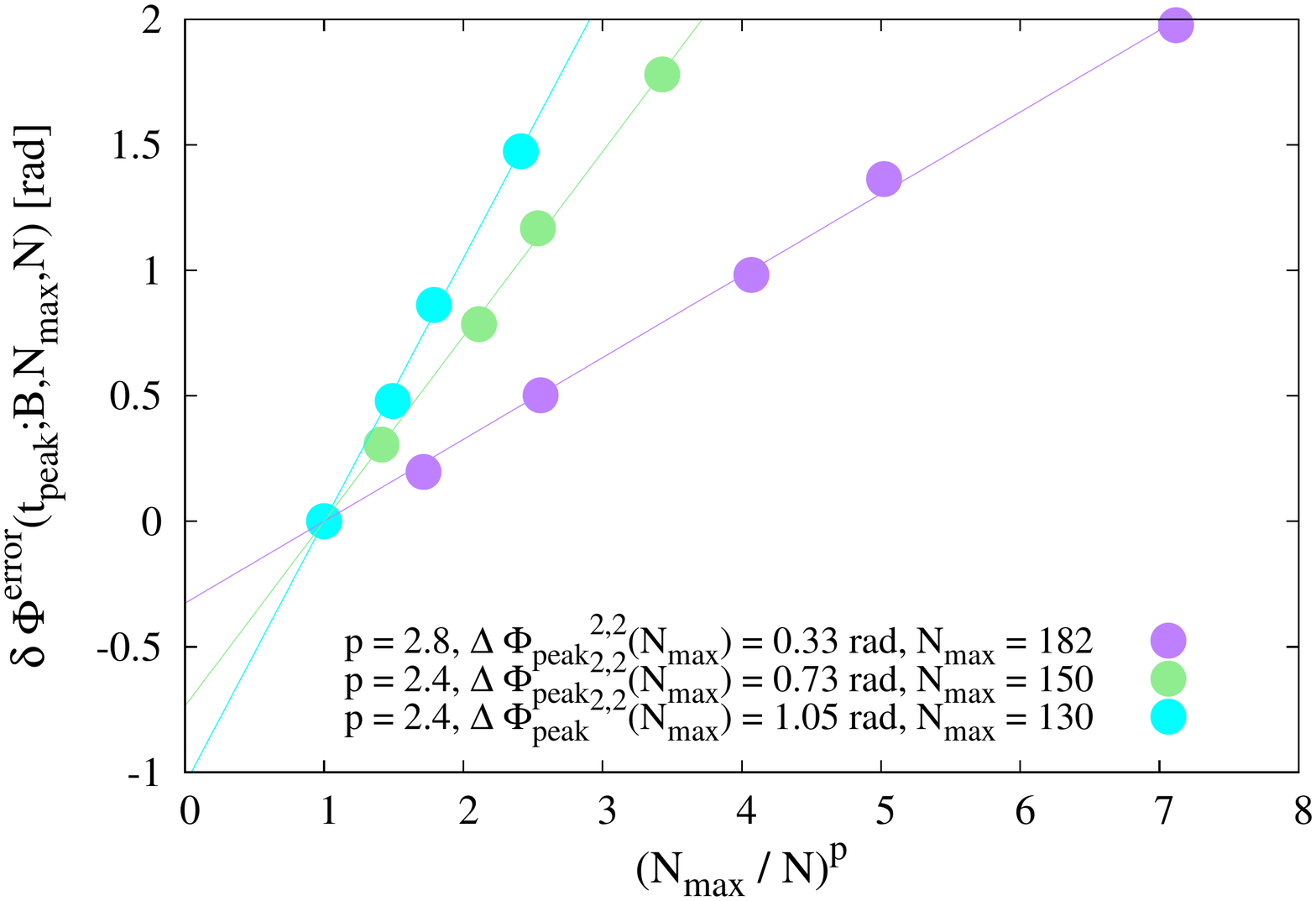}
           \caption{
             (Left) Gravitational-wave phase error, $\delta\Phi^\text{error}(t_\text{ret};{\rm B},N,182)$, with $N=150,130,110,102,$ and $90$ for B107-146. 
           The vertical dashed line denotes the peak time of the gravitational-wave amplitude for $N=182$. 
           (Right) Gravitational-wave phase error at the peak time, $\delta\Phi^\text{error}(t_\text{peak};{\rm B},N_\text{max},N)$, with a reference grid resolution denoted by $N_\text{max}$ and $N=90,102,\cdots,N_\text{max}$. The purple circles show the phase error with $N_\text{max}=182$
           and $N=150,130,110,102,90$. The light-green color is for $N_\text{max}=150$ and
           $N=130,110,102,90$, and the cyan color is for $N_\text{max}=130$ and $N=110,102,90$. 
           The fitting parameters $p$ and $\Delta \Phi^{2,2}_\text{peak}(N_\text{max})$ are listed in the legend. The purple, light-green, and cyan lines denote 
           $\Delta \Phi^{2,2}_\text{peak} (N_\text{max}) [(N_\text{max}/N)^p-1]$ with these fitting parameters for $N_\text{max}=182$, 150, and 130, respectively.
         }\label{fig:dephase}
\end{figure*}

\section{Accuracy of waveforms}\label{sec:result}

To date, we have simulated for long durations 46 binary systems with 6 grid resolutions for each model. 26 binary systems are newly reported in this paper and 20 binary systems have been reported in Refs.~\cite{Kiuchi:2017pte,Kawaguchi:2018gvj}. 
Our waveform data are publicly available on the website:\\

\href{https://www2.yukawa.kyoto-u.ac.jp/~nr_kyoto/SACRA_PUB/catalog.html}{SACRA Gravitational Waveform Data Bank}~\cite{DB}.\\

On the website, the waveform data are tabulated according to the system name, dimensionless initial orbital angular velocity, and grid resolution. For example, 15H\_135\_135\_00155\_182 refers to the employed EOS as 15H, $m_1= 1.35M_\odot$, $m_2= 1.35M_\odot$, $m_0 \Omega_0 = 0.0155$, and $N=182$ (see also Table~\ref{tb:model}).
A user can download the data for $\Psi^{2,2}_4(t_\text{ret},r_0)$ extracted at several values of $r_0$ and $h^{2,2,\infty}_{+,\times}(t_\text{ret})$ from the link on the system name. 

\subsection{Overview of physical and numerical phase shifts}\label{subsec:overview}
  First we briefly illustrate that the waveforms depend on EOSs and each mass of binary systems. 
  The top panel of Fig.~\ref{fig:GW3} shows the dependence of the gravitational waveforms on the EOSs for the binary systems with $m_1=1.12M_\odot$, $m_2=1.40M_\odot$ and $N=182$.
  It shows that the systems with the larger values of $\tilde{\Lambda}$ merge earlier than those with the smaller values of $\tilde{\Lambda}$ because the tidal force due to its companion induces the quadrupole moment and the resultant attractive force accelerates the orbital shrinkage. 
  The bottom panel of Fig.~\ref{fig:GW3} shows the dependence of the gravitational waveforms on the symmetric mass ratio for the binary systems with 15H125-125, 15H112-140, and 15H107-146 with $N=182$.
  It shows that the systems with the larger values of $\eta$ merge earlier than those with the smaller values of $\eta$ because the emissivity of gravitational waves decreases as the symmetric mass ratio decreases~\cite{Blanchet:2013haa}.

  The top panel of Fig.~\ref{fig:GW4} shows the dependence of the gravitational waveforms on the grid resolutions for 15H112-140 with $N=182,110$, and $N=90$. 
  Errors in the amplitude and phase caused by the finite grid resolution become prominent for the late inspiral and post-merger stages. 
  The bottom panel of Fig.~\ref{fig:GW4} plots the phase shift among the systems of different EOSs for $m_1=1.12M_\odot$, $m_2=1.40M_\odot$, and $N=182$. 
  The phase shift is defined by
\begin{align}
  &\delta \Phi^\text{shift}(t_\text{ret};\text{EOS1},\text{EOS2},N) \nonumber\\ 
  &= \Phi^{2,2}(t_\text{ret};{\rm EOS1},N) - \Phi^{2,2}(t_\text{ret};{\rm EOS2},N)
\end{align}
where $\Phi^{2,2}(t_\text{ret};\text{EOS},N)$ is the gravitational-wave phase for $l=|m|=2$ mode derived from a simulation with employing EOS and the grid number $N$.
Because we compare the phase among models with common masses of components, we omit the masses from the argument. 
The shaded region shows the evolution of the phase error defined by 
\begin{align}
  & \delta \Phi^\text{error} (t_\text{ret};\text{EOS},N_1,N_2) \nonumber\\
  & = \Phi^{2,2}(t_\text{ret};\text{EOS},N_1) - \Phi^{2,2}(t_\text{ret};\text{EOS},N_2),
\end{align}
where $N_1$ and $N_2$ denote the employed grid numbers. The red shaded region shows $\delta \Phi^\text{error}(t_\text{ret};15{\rm H},150,182)$ and the blue shaded region shows $\delta \Phi^\text{error}(t_\text{ret};15{\rm H},90,182)$, respectively, for 15H112-140. 
The overlapped region has purple color. The vertical dashed line denotes the peak time, $t_\text{peak}$, at which the gravitational-wave amplitude becomes maximal for 15H112-140 with $N=182$. 
Just after the peak time, burst-type gravitational waves are emitted for a short time as shown in the upper panel of Fig.~\ref{fig:GW4}, i.e., for $58~{\rm ms}\lesssim t_\text{ret} \lesssim 59~{\rm ms}$. 
    These waves cause very rapid increase in phase during this short-term interval and consequently the phase shift shows very rapid increase. This feature can be also seen in the phase error and the very rapid increase appears later in $\delta \Phi^\text{error}(t_\text{ret};15{\rm H},150,182)$ than in $\delta \Phi^\text{error}(t_\text{ret};15{\rm H},90,182)$  because the peak time becomes later with improving the grid resolution. 

    The phase shift and the phase error up to the peak time are comparable, in particular, for the case with the coarser grid resolution.  Therefore, unless a convergence study is sufficiently carried out, a capability of inspiral waveform models to measure the tidal deformability is unclear. This is also the case for the the post-merger stage. In particular, the phase evolution loses the convergence as found in the bottom panel of Fig.~\ref{fig:GW4}, i.e., $\delta \Phi^\text{error}(t_\text{ret};{\rm 15H},150,182)$ (red shaded region) is larger than $\delta \Phi^\text{error}(t_\text{ret};{\rm 15H},90,182)$ (blue shaded region). Therefore, time-domain post-merger gravitational waves derived in numerical-relativity simulations are not very reliable. Instead, we will discuss the post-merger signal
    in terms of the energy and angular momentum carried by gravitational waves and their spectrum amplitude. These quantities are calculated by a time integration of the gravitational waveforms and the convergence in the phase could be subdominant as discussed in Sec.~\ref{sec:universal-relation}.

\subsection{Estimation of the residual phase error in gravitational waves}\label{subsec:phase_error}

Following Refs.~\cite{Kiuchi:2017pte,Kawaguchi:2018gvj}, we estimate a residual gravitational-wave phase error at the peak time in the simulations. 
The left panel of Fig.~\ref{fig:dephase} plots evolution of the phase error, $\delta \Phi^\text{error}(t_\text{ret};{\rm B},N,182)$, with $N=150,130,110,102$, and $90$ for B107-146. 
The vertical dashed line denotes the peak time for B107-146 with $N=182$.
Although the phase error is accumulated with time, its value at the peak time decreases as improving the grid resolution.
We estimate the residual phase error by assuming that the gravitational-wave phase at the peak time obeys the following functional form,
\begin{align}
  &\Phi^{2,2}(t_\text{peak};\text{EOS},N) \nonumber\\
  &= \Phi_\text{peak}^{2,2,\infty}(N_\text{max}) - \Delta \Phi^{2,2}_\text{peak} (N_\text{max}) \left(\frac{N_\text{max}}{N}\right)^p, \label{eq:dephase}
\end{align}
where $\Phi_\text{peak}^{2,2,\infty}(N_\text{max})$ and $p$ denote the gravitational-wave phase at the peak time in the continuum limit of the finite difference $(N \to \infty)$ and an order of the convergence, respectively. 
$\Delta \Phi^{2,2}_\text{peak} (N_\text{max})$ should be recognized as the residual phase error for the simulation with $N=N_\text{max}$.
$N_\text{max}$ denotes a reference value of $N$ to estimate unknown quantities $\Phi^{2,2,\infty}_\text{peak}(N_\text{max})$, $\Delta \Phi^{2,2}_\text{peak} (N_\text{max})$, and $p$.
For example, with $N_\text{max}=182$, these unknowns are obtained by fitting the simulation results of $N=150,130,110,102,$ and $90$ with Eq.~(\ref{eq:dephase}) given an EOS, a chirp mass, and a symmetric mass ratio.

The right panel of Fig.~\ref{fig:dephase} plots the gravitational-wave phase error at the peak time, $\delta \Phi^\text{error}(t_\text{peak};\text{B},N_\text{max},N)$, as a function of $1/N^p$ with a reference grid number $N_\text{max}$ and $N=90,102,\cdots,N_\text{max}$.  
Assuming Eq.~(\ref{eq:dephase}), the phase error at the peak time in a binary system is given as
\begin{align}
  & \delta \Phi^\text{error}(t_\text{peak};\text{EOS},N_\text{max},N)\nonumber\\
  &= \Delta \Phi^{2,2}_\text{peak} (N_\text{max}) \left[ \left(\frac{N_\text{max}}{N}\right)^p - 1 \right]. \label{eq:dephase2}
\end{align}
The values of $\Delta \Phi^{2,2}_\text{peak} (N_\text{max})$ and $p$ are shown in the legend of this plot. It is clear that the order of the convergence $p$ is improved and
the residual gravitational-wave phase error is reduced as increasing $N_\text{max}$. 

Table~\ref{tb:residual_phase} summarizes the residual phase error and the order of the convergence of the gravitational-wave phase at the peak time for all the systems. 
We estimate the residual phase error with respect to three reference values of $N_\text{max}$ as $182,150,$ and $130$.  In some systems, the residual phase error and the order of the convergence show an irregular behavior. 
That is, the residual phase error (the order of convergence) for $N_\text{max}=130$ happens to be smaller (higher) than that for $N_\text{max}=150$.
Nonetheless, the residual phase error (the order of convergence) for $N_\text{max}=182$ is smaller (higher) than that for $N_\text{max}=150$ except for 125H125-146. 
Thus, we adopt the values for $N_\text{max}=182$ as the residual phase error in our waveforms and it is in the range of $\approx 0.1$--$0.5$ rad. 

For the SFHo (tabulated) EOS, we find that the residual phase error still remains within sub-radian accuracy. Because SFHo135-135 and HB135-135 have nearly identical values of $\tilde{\Lambda}$~\cite{Kiuchi:2017pte}, The phase error due to the tabulated EOS is estimated by comparing the results for them. 
For HB135-135, the residual phase error and the order of the convergence are $(\Delta \Phi^{2,2}_\text{peak}(182),p)=(0.17~\text{rad},3.6)$, $(\Delta \Phi^{2,2}_\text{peak}(150),p)=(0.48~\text{rad},3.2)$,
and $(\Delta \Phi^{2,2}_\text{peak}(130),p)=(2.0~\text{rad},1.7)$~\cite{Kiuchi:2017pte}. For SFHo135-135, the residual phase error and the order of the convergence are $(\Delta \Phi^{2,2}_\text{peak}(182),p)=(0.43~\text{rad},2.3)$, $(\Delta \Phi^{2,2}_\text{peak}(150),p)=(0.76~\text{rad},2.2)$, and $(\Delta \Phi^{2,2}_\text{peak}(130),p)=(0.33~\text{rad},4.2)$, respectively.
Thus, the system with the SFHo (tabulated) EOS has slightly larger residual phase error than with the piecewise polytropic EOS.
This indicates that the linear interpolation of the thermodynamic quantities could cause a phase error of $\approx 0.2$--$0.3$ rad. Nonetheless, it is encouraging that our waveforms have the sub-radian accuracy even for the SFHo (tabulated) EOS. 
For more detailed estimate of the error budget due to tabulated EOSs, we need to perform BNS simulations with a wide class of tabulated EOSs. In particular, we speculate that the phase error when using a tabulated EOS with a phase transition could be even larger. 

\begin{table*}[t]
\centering
\caption{Residual phase error (rad) and order of the convergence of the gravitational-wave phase at the peak time calculated by Eq.~(\ref{eq:dephase}) for $N_\text{max}=182,150$, and $130$.
  }
\begin{tabular}{c|ccc}
\hline\hline
System        & $(\Delta \Phi^{2,2}_\text{peak}(182),p)$ & $(\Delta \Phi^{2,2}_\text{peak}(150),p)$ & $(\Delta \Phi^{2,2}_\text{peak}(130),p)$ \\
\hline
15H125-146   & (0.11,~4.1) & (0.58,~2.7) & (5.44,~0.7) \\
125H125-146  & (0.31,~2.6) & (0.15,~4.5) & (0.45,~3.6) \\
H125-146     & (0.17,~3.4) & (0.78,~2.2) & (0.73,~2.8) \\
HB125-146    & (0.13,~3.7) & (1.10,~1.7) & (1.00,~2.2) \\
B125-146     & (0.12,~3.8) & (0.28,~3.7) & (0.45,~3.8) \\
15H118-155   & (0.22,~3.1) & (0.75,~2.2) & (0.47,~3.5) \\
125H118-155  & (0.26,~2.9) & (0.83,~2.1) & (1.44,~1.7) \\
H118-155     & (0.23,~3.1) & (0.48,~3.0) & (0.56,~3.4) \\
HB118-155    & (0.44,~2.3) & (1.21,~1.6) & (0.79,~2.5) \\
B118-155     & (0.29,~2.7) & (0.69,~2.2) & (0.47,~3.3) \\
15H117-156   & (0.26,~2.9) & (0.36,~3.2) & (0.39,~4.0) \\
125H117-156  & (0.28,~2.8) & (0.38,~2.8) & (0.92,~2.4) \\
H117-156     & (0.24,~3.0) & (0.31,~3.5) & (0.74,~2.9) \\
HB117-156    & (0.22,~3.0) & (0.84,~2.0) & (1.42,~1.7) \\
B117-156     & (0.42,~2.3) & (0.43,~2.8) & (0.23,~4.8) \\ 
15H112-140   & (0.19,~3.4) & (0.70,~2.5) & (0.66,~3.2) \\
125H112-140  & (0.21,~3.4) & (0.53,~3.0) & (0.66,~3.3) \\ 
H112-140     & (0.17,~3.5) & (0.92,~2.1) & (1.00,~2.4) \\
HB112-140    & (0.42,~2.5) & (0.48,~3.0) & (0.21,~5.5) \\ 
B112-140     & (0.19,~3.6) & (0.34,~3.7) & (39.59,~0.13) \\ 
15H107-146   & (0.38,~2.6) & (0.86,~2.2) & (0.43,~3.9) \\ 
125H107-146  & (0.54,~2.2) & (2.93,~1.0) & (0.61,~3.2) \\ 
H107-146     & (0.41,~2.4) & (0.60,~2.5) & (1.03,~2.3) \\ 
HB107-146    & (0.35,~2.8) & (0.44,~3.3) & (0.43,~4.2) \\ 
B107-146     & (0.33,~2.8) & (0.73,~2.4) & (1.05,~2.4) \\ 
SFHo135-135  & (0.43,~2.3) & (0.76,~2.2) & (0.33,~4.2) \\ 
\hline\hline
\end{tabular}\label{tb:residual_phase}
\end{table*}

\section{Inspiral gravitational waveform modeling} \label{sec:WFmodel}

\subsection{SACRA inspiral gravitational waveform template}
In the previous paper~\cite{Kawaguchi:2018gvj}, we developed a frequency-domain gravitational waveform model for inspiralling BNSs (with $l=|m|=2$) based on high-precision numerical-relativity data. In this section, we extend the examination of the inspiral waveform model to a parameter space wider than the previous papers~\cite{Kiuchi:2017pte,Kawaguchi:2018gvj} by employing new waveforms obtained in this paper.

Before moving on to the comparison, we briefly review our inspiral waveform model.
First we calculate the Fourier component for the quadrupole mode of gravitational waves for all the systems by
\begin{align}
   \tilde{h}_{+,\times}(f)=\int_{t_{\rm i}}^{t_{\rm f}}  h^{2,2,\infty}_{+,\times}(t)e^{-2\pi i f t}dt, \label{eq:deffreqdom}
\end{align}
where $t_{\rm i}$ and $t_{\rm f}$ are the initial and final time of the waveform data, respectively. 
Then, we decompose $\tilde{h}_+(f)$ in Eq.~(\ref{eq:deffreqdom}) into the frequency-domain amplitude, $A\left(f\right)$, and phase, $\Psi\left(f\right)$, (with an ambiguity in the origin of the phase) by 
\begin{align}
	{\tilde h}_+\left(f\right)=A\left(f\right) {\rm e}^{-i\Psi\left(f\right)}. 
\end{align}
We only use $h^{2,2,\infty}_+$ for modeling the inspiral gravitational waveforms because the difference between $h^{2,2,\infty}_+$ and $h^{2,2,\infty}_\times$ is approximately only the phase difference of $\pi/2$. 
We define the corrections due to the NS tidal deformation to the gravitational-wave amplitude and phase by
\begin{align}
	A^{\rm tidal}\left(f\right)=A\left(f\right)-A_{\rm BBH}\left(f\right)\label{eq:defAtidal}
\end{align}
and 
\begin{align}
	\Psi^{\rm tidal}\left(f\right)=\Psi\left(f\right)-\Psi_{\rm BBH}\left(f\right),\label{eq:defphitidal}
\end{align}
respectively. Here, $A_{\rm BBH}\left(f\right)$ and $\Psi_{\rm BBH}\left(f\right)$ are the gravitational-wave amplitude and phase of a binary black hole with the same mass as the BNS, respectively (hereafter referred to as the point-particle parts: see Ref.~\cite{Kawaguchi:2018gvj} for details).

Our numerical-relativity waveforms only contain the waveforms for the frequency higher than $\approx 400\,{\rm Hz}$. 
Thus, we employ the effective-one-body waveforms of Refs.~\cite{Hinderer:2016eia,Steinhoff:2016rfi,Lackey:2018zvw,Taracchini:2013rva} (SEOBNRv2T) to model the low-frequency part waveforms, in which the effect of dynamical tides is taken into account, and construct hybrid waveforms combining them with the numerical-relativity waveforms. The hybridization of the waveforms is performed in the time-domain by the procedure described in Refs.~\cite{Hotokezaka:2016bzh,Kawaguchi:2018gvj} and we set the matching region to be from $t_\text{ret} \approx 7.38$ ms to $14.78$ ms. 
After the hybridization, the waveforms are transformed into the frequency domain employing Eq.~\eqref{eq:deffreqdom}, and the tidal-part amplitude and phase are extracted by Eqs.~\eqref{eq:defAtidal} and~\eqref{eq:defphitidal}.

For modeling the tidal-part phase and amplitude, we employ the following functional forms motivated by the 2.5 PN order formula~\cite{Damour:2012yf}:
\begin{align}
&\Psi_{\rm model}^{\rm tidal}=\frac{3}{128\eta}\left[-\frac{39}{2}{\tilde \Lambda}\left(1+a\,{\tilde \Lambda}^{2/3} x^p \right)\right]x^{5/2}\nonumber
\\&\times\left(1+\frac{3115}{1248}x-\pi x^{3/2}+\frac{28024205}{3302208}x^2 -\frac{4283}{1092}\pi x^{5/2}\right)\label{eq:phimodel}
\end{align}
for the phase correction and
\begin{align}
	A_{\rm model}^{\rm tidal}&=\sqrt{\frac{5\pi\eta}{24}}\frac{m_0^2}{D_{\rm eff}} {\tilde \Lambda} x^{-7/4}\nonumber\\
	&\times \left(-\frac{27}{16}x^{5}-\frac{449}{64}x^{6}-b\,x^q\right)\label{eq:Amodel}
\end{align}
for the amplitude correction where $D_{\rm eff}$ is the effective distance to the binary~\cite{Hotokezaka:2016bzh} and $x\equiv (\pi m_0 f)^{2/3}$. $a$, $p$, $b$, and $q$ are the free parameters of the models. 
To focus on the inspiral waveform and to avoid the contamination from the post-merger waveforms of high frequency, which would have large uncertainties, we restrict the gravitational-wave frequency range in $10$--$1000\,{\rm Hz}$. The fitting parameters were determined by employing the hybrid waveforms of 15H125-125, which has the largest value of binary tidal deformability in the systems studied in the previous study~\cite{Kawaguchi:2018gvj}.  By performing the least square fit with respect to the phase shift and relative difference of the amplitude, we obtained $a=12.55$, $p=4.240$, $b=4251$, and $q=7.890$.

In Ref.~\cite{Kawaguchi:2018gvj}, the validity of the inspiral waveform model was examined employing hybrid waveforms which were not used for the parameter determination.
We should stress again that the parameters $a,p,b$, and $q$ in Eqs.~(\ref{eq:phimodel}) and (\ref{eq:Amodel}) were determined by the particular system 15H125-125. 
We found that the tidal-part waveform model always reproduced the tidal-part phase and amplitude of the hybrid waveforms within $\sim 0.1\,{\rm rad}$ and $15\%$, respectively,
for the equal-mass and unequal-mass cases with ${\cal M}_{\rm chirp}= 1.1752\,M_\odot$ and the equal-mass cases with ${\cal M}_{\rm chirp}= 1.0882\,M_\odot$, covering the parameter space of $0.244 \le \eta \le 0.250$ and $300\lesssim \tilde{\Lambda} \lesssim 1800$.

\subsection{Validation of SACRA inspiral gravitational waveform template}
  
While the validity of our inspiral waveform model was already examined in the most interesting part of the parameter space of BNSs~\cite{Kawaguchi:2018gvj}, there still remain some important cases which were not examined in the previous study~\cite{Kawaguchi:2018gvj}. First, the dependence of the error of the tidal correction on the mass ratio has to be checked for less massive BNSs. While unequal-mass cases with total mass of $\approx\,2.7\,M_\odot$ were checked in the previous study~\cite{Kawaguchi:2018gvj}, it is important to check whether our inspiral waveform models are also applicable to unequal-mass cases with smaller total mass, for which tidal effect is enhanced due to increase of tidal deformability. Second, the systematics due to simplification on the high-density part of the EOS should be checked. For the inspiral waveforms, we expect that the high-density part of the EOS has a minor effect, and thus, we employ simplified two-piecewise polytropic EOS models. However, we should confirm that this assumption is indeed valid. 

To check the points listed above, we compare our inspiral waveform model with hybrid waveforms employing the numerical-relativity waveforms obtained in this paper. Hybrid waveforms are constructed in the same manner as in the previous study~\cite{Kawaguchi:2018gvj} employing the SEOBNRv2T waveforms as the low-frequency part waveforms. In particular, we focus on the validity of the tidal correction model to the waveform, comparing it with the tidal-part phase and amplitude of the hybrid waveforms computed based on Eqs.~\eqref{eq:defAtidal} and~\eqref{eq:defphitidal} using the SEOBNRv2 waveforms with no-tides as the point-particle parts. 
  
\begin{figure*}[t]
\hspace{-25.0mm}
\begin{center}
\includegraphics[width=0.49\linewidth]{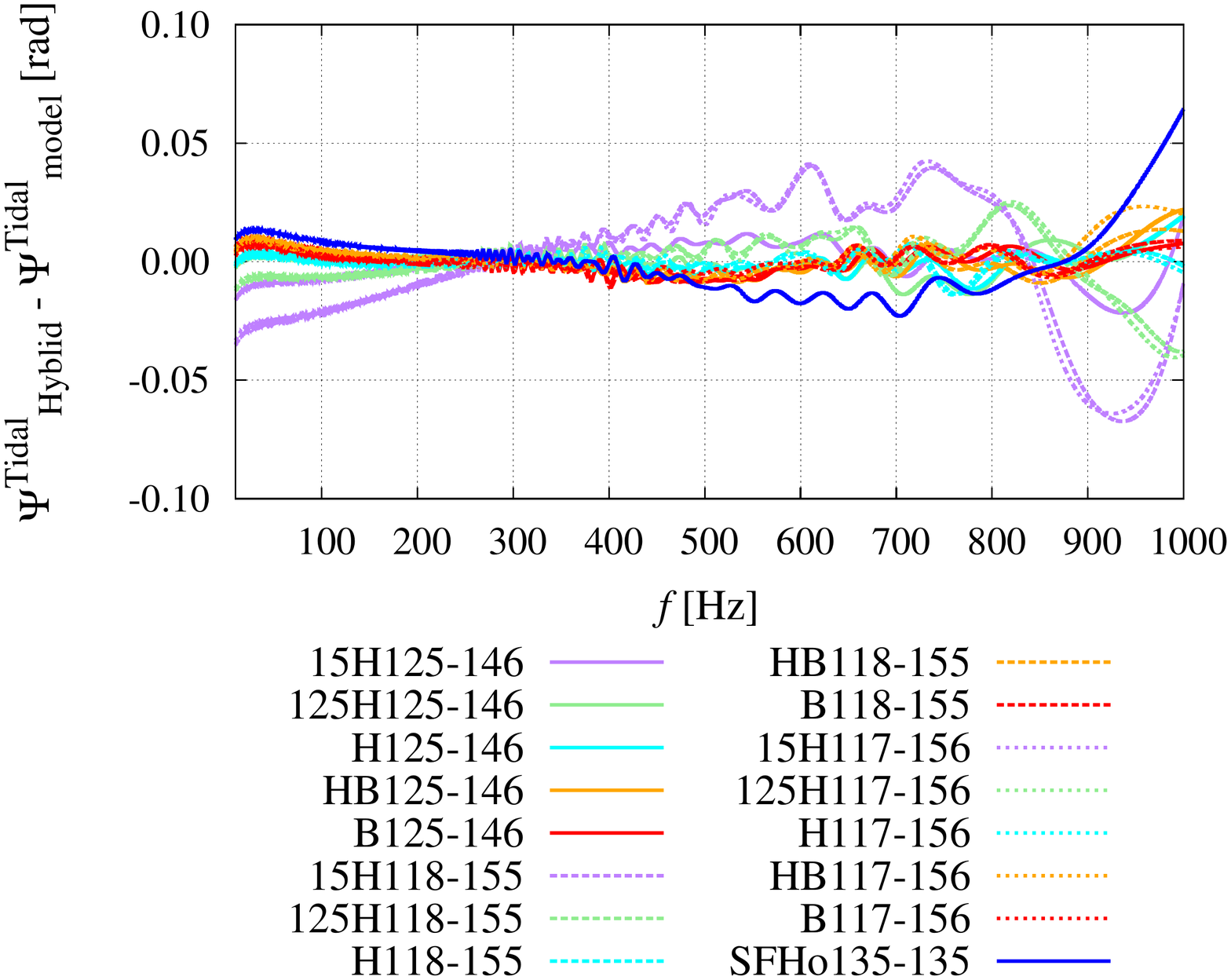}
\includegraphics[width=0.49\linewidth]{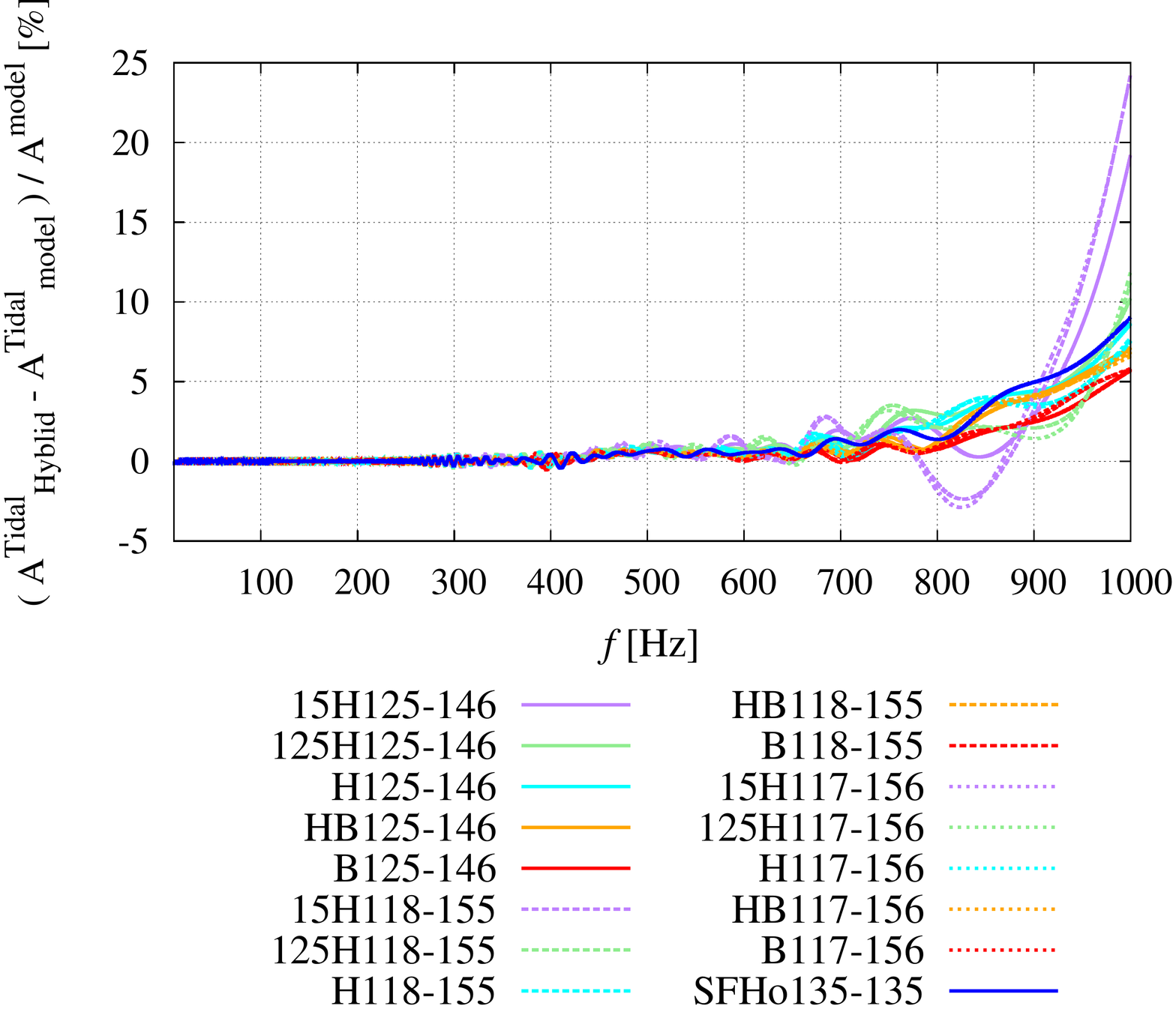}
\end{center}
\caption{(Left) Difference in the tidal-part phase between the hybrid waveforms and the model given by Eq.~\eqref{eq:dphi} for the binary systems with ${\cal M}_c=1.1752M_\odot$. Phase differences are plotted after the alignment in the frequency range of $10$--$1000,{\rm Hz}$. (Right) Relative difference of tidal-part amplitude between the hybrid waveforms and the model given by Eq.~\eqref{eq:reldA}.}\label{fig:model_comp_27}
\end{figure*}

\begin{figure*}[t]
\hspace{-25.0mm}
\begin{center}
  \includegraphics[width=.49\linewidth]{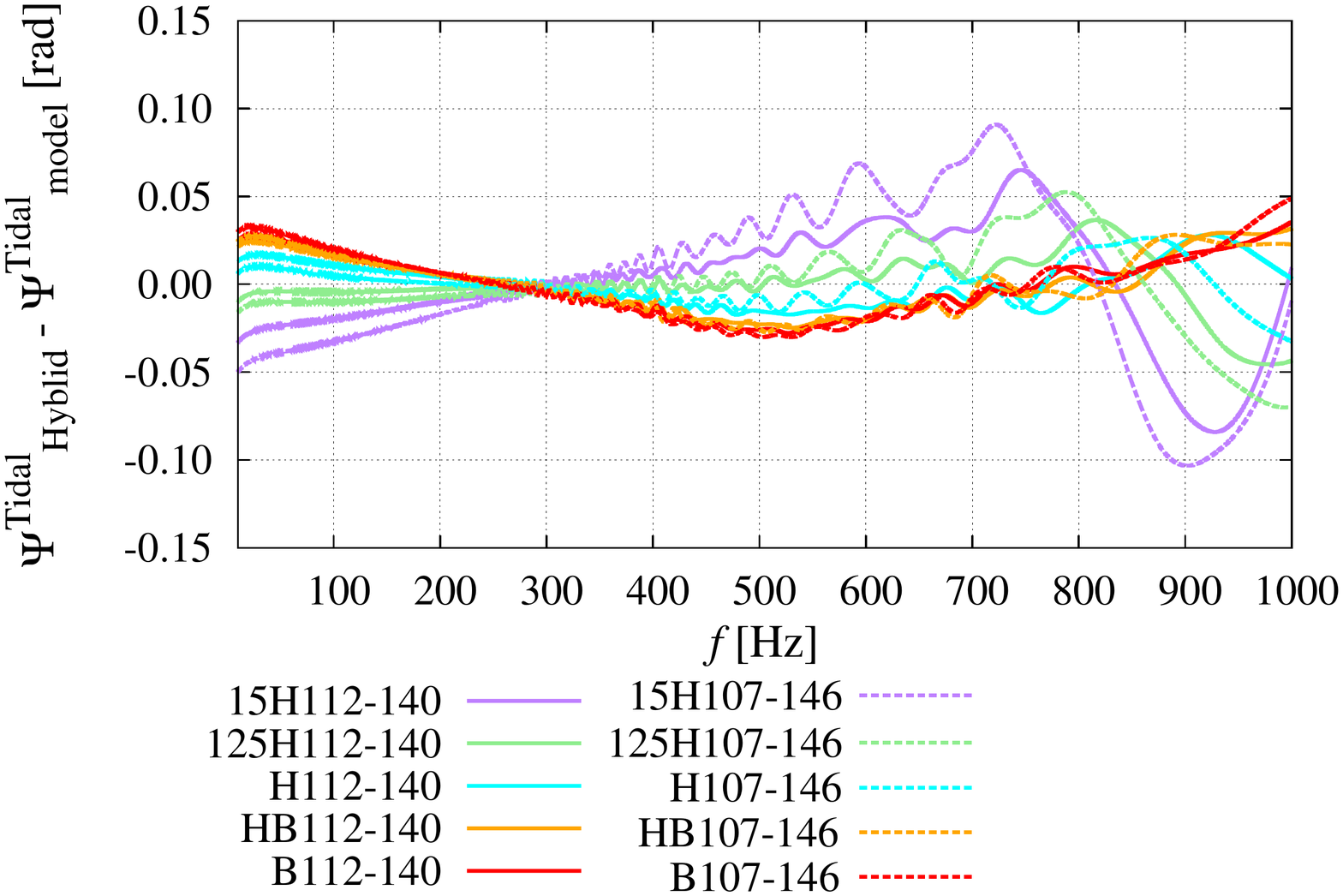}
\vspace{-15mm}
\includegraphics[width=.49\linewidth]{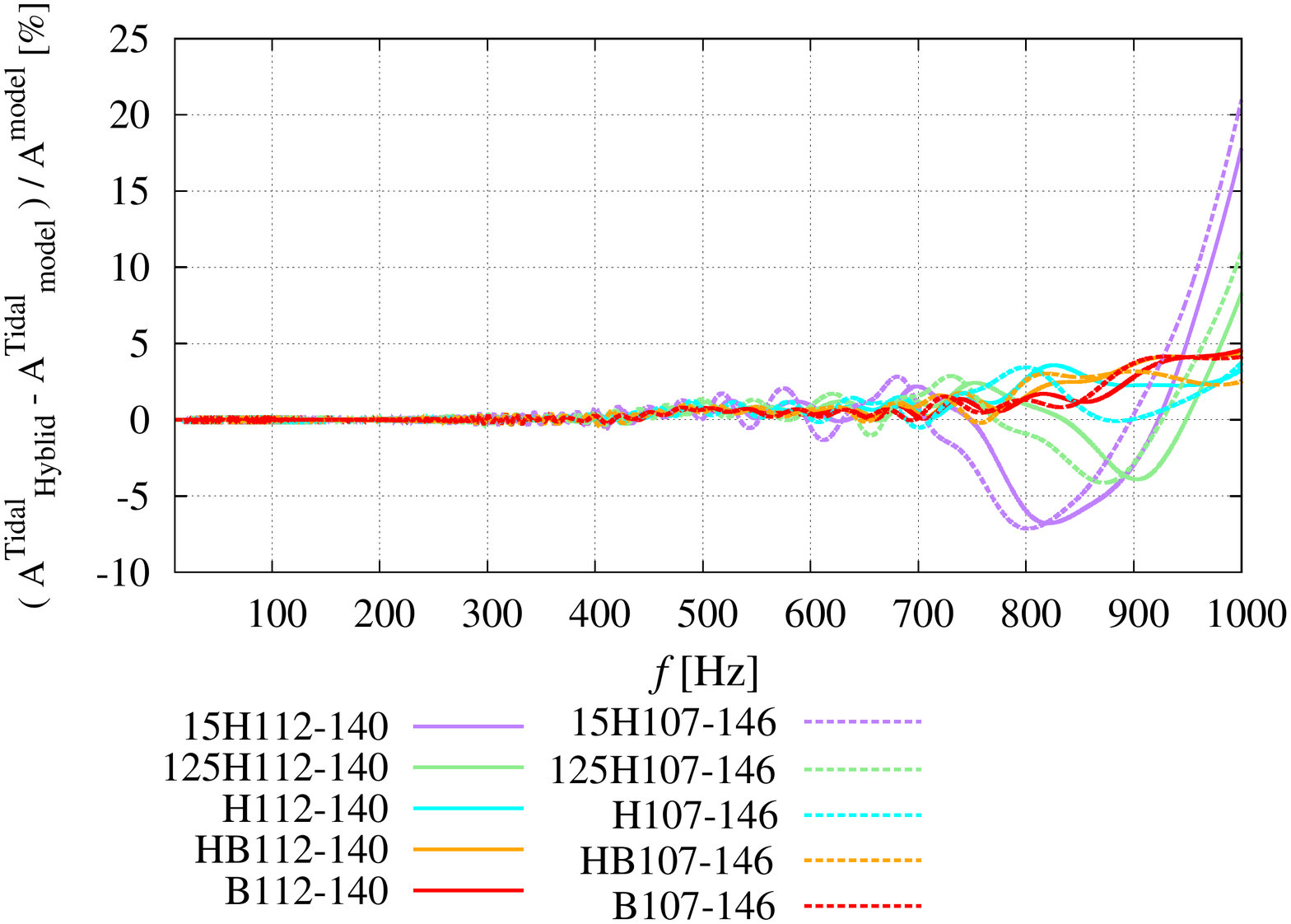}
\end{center}
\caption{The same as in Fig.~\ref{fig:model_comp_27} but for the models with ${\cal M}_c= 1.0882\,M_\odot$.}\label{fig:model_comp_25}
\end{figure*}

Figures~\ref{fig:model_comp_27} and \ref{fig:model_comp_25}  show the difference of the tidal-part phase and amplitude between our inspiral waveform model~\eqref{eq:phimodel} and \eqref{eq:Amodel} and the hybrid waveforms for the models with ${\cal M}_c= 1.1752\,M_\odot$ and  ${\cal M}_c= 1.0882\,M_\odot$. Here, the phase difference between the tidal-part phase of hybrid waveforms, $\Psi_{\rm Hybrid}^{\rm tidal}$, and that of our inspiral waveform model, $\Psi_{\rm model}^{\rm tidal}$, is computed by
\begin{align}
	\Delta\Psi(f)=\Psi_{\rm Hybrid}^{\rm tidal}(f)-\Psi_{\rm model}^{\rm tidal}(f)-2\pi f t_0+\phi_0,\label{eq:dphi}
\end{align}
where $t_0$ and $\phi_0$ are the free parameters which correspond to the degrees of freedom in choosing the origins of time and phase, respectively, and are determined by minimizing $\int |\Delta\Psi(f)|^2 df$ integrated in the range of $f=10$--$1000\,{\rm Hz}$. For the comparison of the tidal-part amplitude, relative difference of the amplitude, 
\begin{align}
\Delta A(f)/A(f)=(A_{\rm Hybrid}^{\rm tidal}(f)-A_{\rm model}^{\rm tidal}(f))/A_{\rm model}(f),\label{eq:reldA}
\end{align}
is computed, where $A_{\rm Hybrid}^{\rm tidal}$ and $A_{\rm model}=A_{\rm model}^{\rm tidal}+A_{\rm BBH}$ are the tidal-part amplitude of hybrid waveforms and the amplitude of the model waveforms including the point-particle part, respectively.
Again, we employ the amplitude of the SEOBNRv2 waveforms with no-tides for $A_{\rm BBH}$.

The systems of mass $1.25M_\odot$--$1.46M_\odot$, $1,18M_\odot$--$1.55M_\odot$, and $1.17M_\odot$--$1.56M_\odot$ are within the parameter space which we studied in the previous study~\cite{Kawaguchi:2018gvj}, and thus, we expect that those waveforms are well reproduced by our inspiral waveform model. Indeed Fig.~\ref{fig:model_comp_27} shows that differences in both phase and amplitude are within the error which we observed in the previous study~\cite{Kawaguchi:2018gvj}. Figure~\ref{fig:model_comp_27} also shows that tidal-part phase and amplitude for system SFHo135-135 are well reproduced by our inspiral waveform model. This confirms that, at least for the frequency range and $m_0$ we focus on, employing an EOS whose high-density part is simplified has only a minor effect on the systematics of the model. 
Figure~\ref{fig:model_comp_25}  shows the results in the unequal-mass cases with ${\cal M}_c= 1.0882\,M_\odot$. The difference in the tidal-part phase is larger than the cases with ${\cal M}_c= 1.1752\,M_\odot$. This is reasonable because we found that the error of tidal-part model becomes relatively large for a small mass ratio or a large value of tidal deformability in the previous study~\cite{Kawaguchi:2018gvj}. Nevertheless, the phase error is always smaller than $\approx0.1\,{\rm rad}$, which is smaller than the systematics in the waveforms stemming from the finite difference as shown in the previous
section. The deviation for the amplitude model is also the same level as for the models with ${\cal M}_c= 1.1752\,M_\odot$.

To quantify the deviation of our inspiral waveform model from the new sets of hybrid waveforms, we calculate the mismatch between those waveforms, ${\bar F}$, defined by
\begin{align}
	{\bar F}=1-\max_{\phi_0,t_0}\frac{\left({\tilde h}_1\middle|{\tilde h}_2{\rm e}^{2\pi i f  t_0 +i \phi_0}\right)}{||{\tilde h}_1||\,||{\tilde h}_2||},\label{eq:mismatch}
\end{align}
where $(\cdot|\cdot)$ and $||\cdot||$ are defined by
\begin{align}
	\left({\tilde h}_1\middle|{\tilde h}_2\right)=4{\rm Re}\left[\int_{f_{\rm min}}^{f_{\rm max}}
	\frac{{\tilde h}_1\left(f\right){\tilde h}^*_2\left(f\right)}{S_{\rm n}\left(f\right)}df\right],\label{eq:inp}
\end{align}
where $f_{\rm min}=10$~Hz and $f_{\rm max}=1000$~Hz and
\begin{align}
	||{\tilde h}||=\sqrt{\left({\tilde h}\middle|{\tilde h}\right)}.
\end{align} 
Here, $h_1$ and $h_2$ denote the hybrid waveforms and our inspiral waveform models, respectively.
The inspiral waveform model employs Eqs.~(\ref{eq:phimodel}) and (\ref{eq:Amodel}) as the tidal part and the SEOBNRv2 waveforms with no-tides as the point-particle baseline. 
$S_{\rm n}$ denotes the one-sided noise spectrum density of the detector, and we employ the noise spectrum density of the {\tt ZERO\_DETUNED\_HIGH\_POWER} configuration of advanced LIGO~\cite{aLIGOnoise} for it. 

\begin{table*}
\centering
\caption{Mismatch between the inspiral waveform model and hybrid waveforms.}
\begin{tabular}{c|c}\hline\hline
System&$~{\bar F}\,(\times10^{-5})$\\\hline
15H125-146	&	0.83\\
125H125-146	&	0.36\\
H125-146	&	0.29\\
HB125-146	&	0.28\\
B125-146	&	0.22\\

15H118-155	&	0.82\\
125H118-155	&	0.26\\
H118-155	&       0.30\\ 
HB118-155	&	0.32\\ 
B118-155	&	0.31\\

15H117-156	&	0.97\\
125H117-156	&	0.31\\
H117-156	&       0.25\\ 
HB117-156	&	0.30\\ 
B117-156	&	0.17\\

\hline
15H112-140	&	0.88\\ 
125H112-140	&	0.24\\ 
H112-140	&	0.37\\ 
HB112-140	&	0.71\\ 
B112-140	&	0.91\\

15H107-146	&	1.82\\ 
125H107-146	&	0.45\\ 
H107-146	&	0.30\\ 
HB107-146	&	0.79\\ 
B107-146	&	1.12\\ 
\hline
SFHo135-135	&	0.45\\
\hline
\end{tabular}\label{tb:mismatch}
\end{table*}

We summarize the values of mismatch between our inspiral waveform model and hybrid waveforms in Table~\ref{tb:mismatch}. For all the cases, the value of mismatch is smaller than $\approx 2\times10^{-5}$. According to our previous results~\cite{Kawaguchi:2018gvj}, these results indicate that the the signal to noise ratio of the difference between our inspiral waveform model and hybrid waveforms are as small as $1$ even for the case in which the total signal to noise ratio is as large as $200$.

\section{Assessment of universal relation for late inspiral and post-merger gravitational waves}\label{sec:universal-relation}
    
\subsection{frequency and amplitude}

Instantaneous gravitational-wave frequency defined by Eq.~(\ref{eq:GWfreq}) at some characteristic time in the late inspiral or post-merger stage is reported to be correlated with the tidal deformability or the tidal coupling constant~\cite{Read:2013zra,Rezzolla:2016nxn,Bernuzzi:2015rla,Bernuzzi:2014owa}. 
In addition, characteristic peak frequencies imprinted in the spectrum amplitude of post-merger gravitational waves are reported to be correlated with the tidal coupling constant or NS radius~\cite{Rezzolla:2016nxn,Shibata:2005xz,Hotokezaka:2013iia,Bauswein:2011tp}. 
We assess these proposed universal relations using our waveform data, for which the systematic study has been conducted in a wide range of the binary parameters with a wide range of the grid resolution of the simulations.
We also propose new relations in terms of the binary tidal deformability. 

\subsubsection{Peak frequency and binary tidal deformability relation}

\begin{figure*}
  	 \includegraphics[width=.41\linewidth]{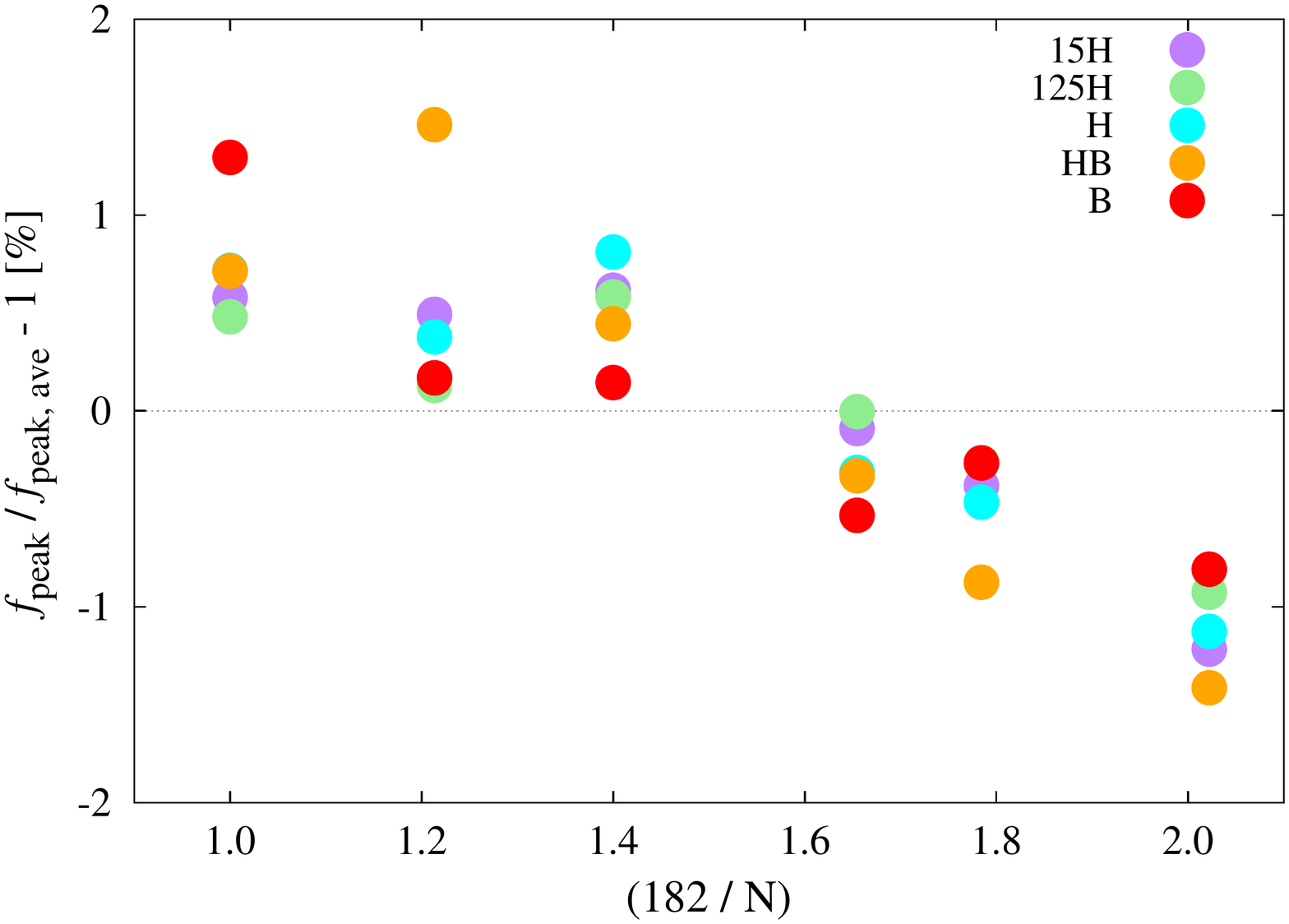} 
 	 \includegraphics[width=.45\linewidth]{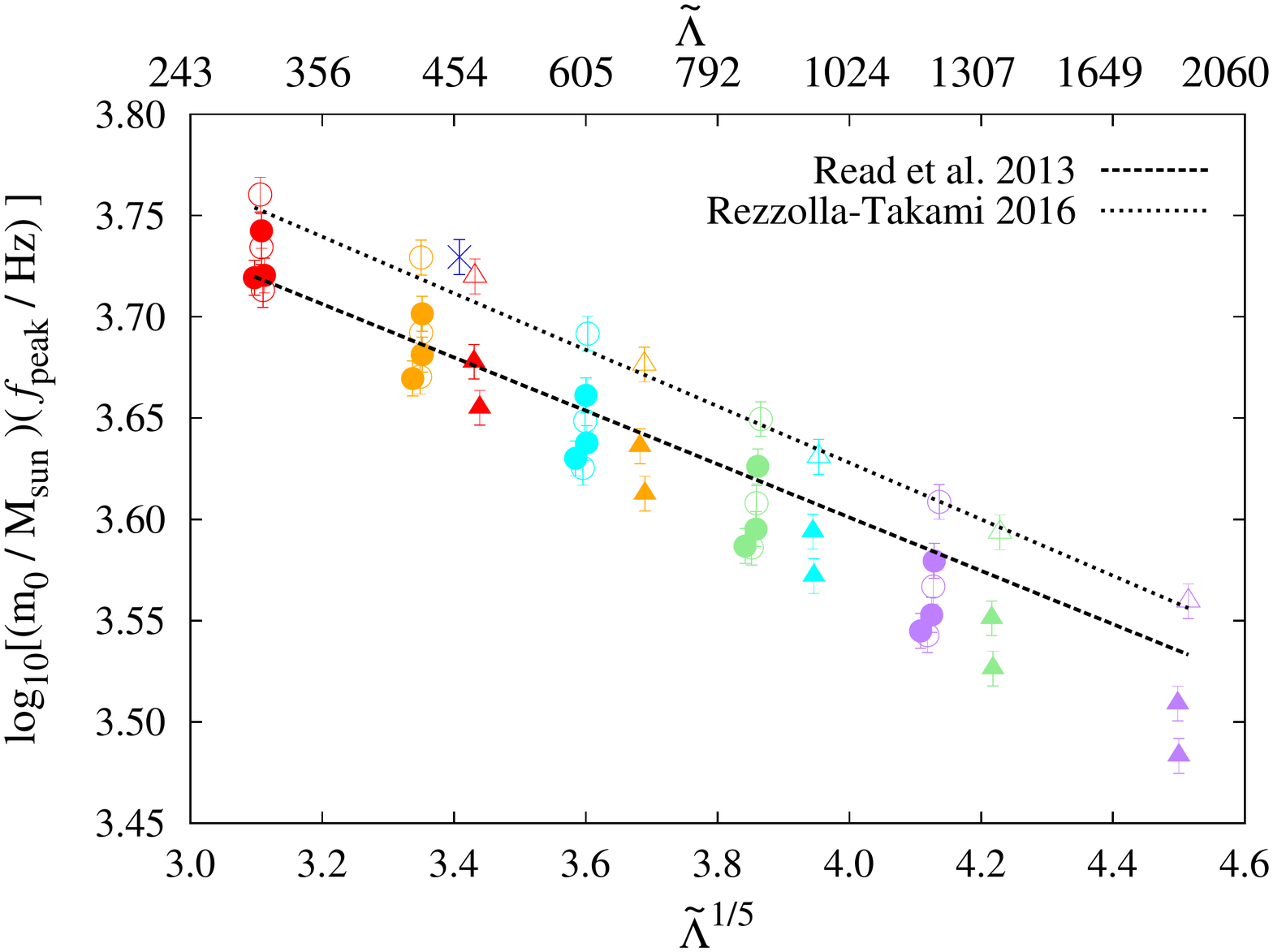}
 	 \caption{(Left) A deviation of instantaneous gravitational-wave frequency at the peak time $f_\text{peak}$ relative to $f_\text{peak,ave}$ as a function of $1/N$ for the binary systems with $m_1= 1.17 M_\odot$ and $m_2= 1.56 M_\odot$. $f_\text{peak,ave}$ is an average of $f_\text{peak}$ over the results with different grid resolutions. 
           (Right) $m_0f_\text{peak}$ as a function of $\tilde{\Lambda}^{1/5}$. Meaning of the color and symbols is the same as that in Fig.~\ref{fig:model}. 
           The error bar of $\pm 2\%$ comes from the systematics associated with the finite grid resolution in $f_\text{peak}$. The proposed universal relations in Refs.~\cite{Read:2013zra,Rezzolla:2016nxn} are shown. 
         }\label{fig:fpeak}
\end{figure*}

Reference~\cite{Read:2013zra} reported that the instantaneous gravitational-wave frequency (of $l=|m|=2$ mode) at the peak time $(t_\text{peak})$, $f_\text{peak}$, has a tight correlation with the binary tidal deformability $\tilde{\Lambda}$
(see also Refs.~\cite{Rezzolla:2016nxn,Bernuzzi:2015rla,Bernuzzi:2014owa} for the relation with the tidal coupling constant: In Ref.~\cite{Rezzolla:2016nxn}, they referred to it as $f_\text{max}$). Figure~\ref{fig:fpeak} plots the dependence of $f_\text{peak}$ on the grid resolution where $f_\text{peak,ave}$ is the average of $f_\text{peak}$ over the 
results with different grid resolutions.
$f_\text{peak}$ does not converge perfectly with respect to the grid resolution, but the fluctuation around the averaged value is less than 2$\%$ for a wide
range of the grid resolution. This is also the case for all the binary systems. Thus, we estimate a relative error due to the finite grid resolution in $f_\text{peak}$
to be 2$\%$ and tabulate the values of $f_\text{peak}$ in Table~\ref{tb:fpeak}.

The right panel of Fig.~\ref{fig:fpeak} plots $m_0 f_\text{peak}$ as a function of $\tilde{\Lambda}^{1/5}$. The error bar shows the systematics associated with the finite grid resolution in $f_\text{peak}$. 
We also plot the universal relations reported in Refs.~\cite{Read:2013zra} (black dashed line) and \cite{Rezzolla:2016nxn} (black dotted line).
We find that the universal relation in Ref.~\cite{Rezzolla:2016nxn} holds only for the symmetric binary systems with ${\cal M}_c= 1.1752M_\odot$ and ${\cal M}_c= 1.0882M_\odot$ (see also Table~\ref{tb:fpeak}). Given an EOS and a chirp mass, $f_\text{peak}$ shifts to a lower value as the symmetric mass ratio decreases.
This is attributed to following three facts. First, given the total mass $m_0$ and $f_\text{GW}$, $df_\text{GW}/dt$ decreases as the symmetric mass ratio decreases because 
the gravitational-wave luminosity is proportional to $\eta^2$~\cite{Blanchet:2013haa}. 
Second, the time at which the two NSs come into contact becomes earlier as the symmetric mass ratio decreases because the less massive companion is more subject to the tidal elongation 
and the resultant mass accretion on the massive component starts earlier than for the symmetric binary. Third, the difference between the peak time and the contact time becomes small as the symmetric mass ratio decreases because 
the peak time corresponds to the moment when a dumbbell-like density structure with double dense cores formed after the contact disappears as discussed in Ref.~\cite{Kiuchi:2017pte} 
and the dumbbell-like density structure becomes less prominent in the asymmetric binary systems. 
Due to these effects, $f_\text{peak}$ becomes lower as the symmetric mass ratio decreases.

In a short summary, the $m_0f_\text{peak}$--$\tilde{\Lambda}^{1/5}$ relation depends strongly on the symmetric mass ratio and the universal relations reported in Refs.~\cite{Read:2013zra} and \cite{Rezzolla:2016nxn} suffer from this systematics (see also Ref.~\cite{Kiuchi:2017pte}).
 This finding is consistent with a discussion in Ref.~\cite{Rezzolla:2016nxn}. They mentioned that the mass asymmetry could break the universality in the $m_0f_\text{peak}$--$\tilde{\Lambda}^{1/5}$ relation for a {\it possibly unrealistic} mass ratio.
 We find that the {\it realistic} value of the mass ratio breaks the universality as the symmetric mass ratio adopted in this paper is consistent with that in GW170817~\cite{TheLIGOScientific:2017qsa}.
 The scatter from the proposed universal relation in Ref.~\cite{Rezzolla:2016nxn} is as large as $\approx$ 18--19$\%$ at the maximum for  $0.244\le\eta\le 0.250$. 

\begin{figure}
  	 \includegraphics[width=.9\linewidth]{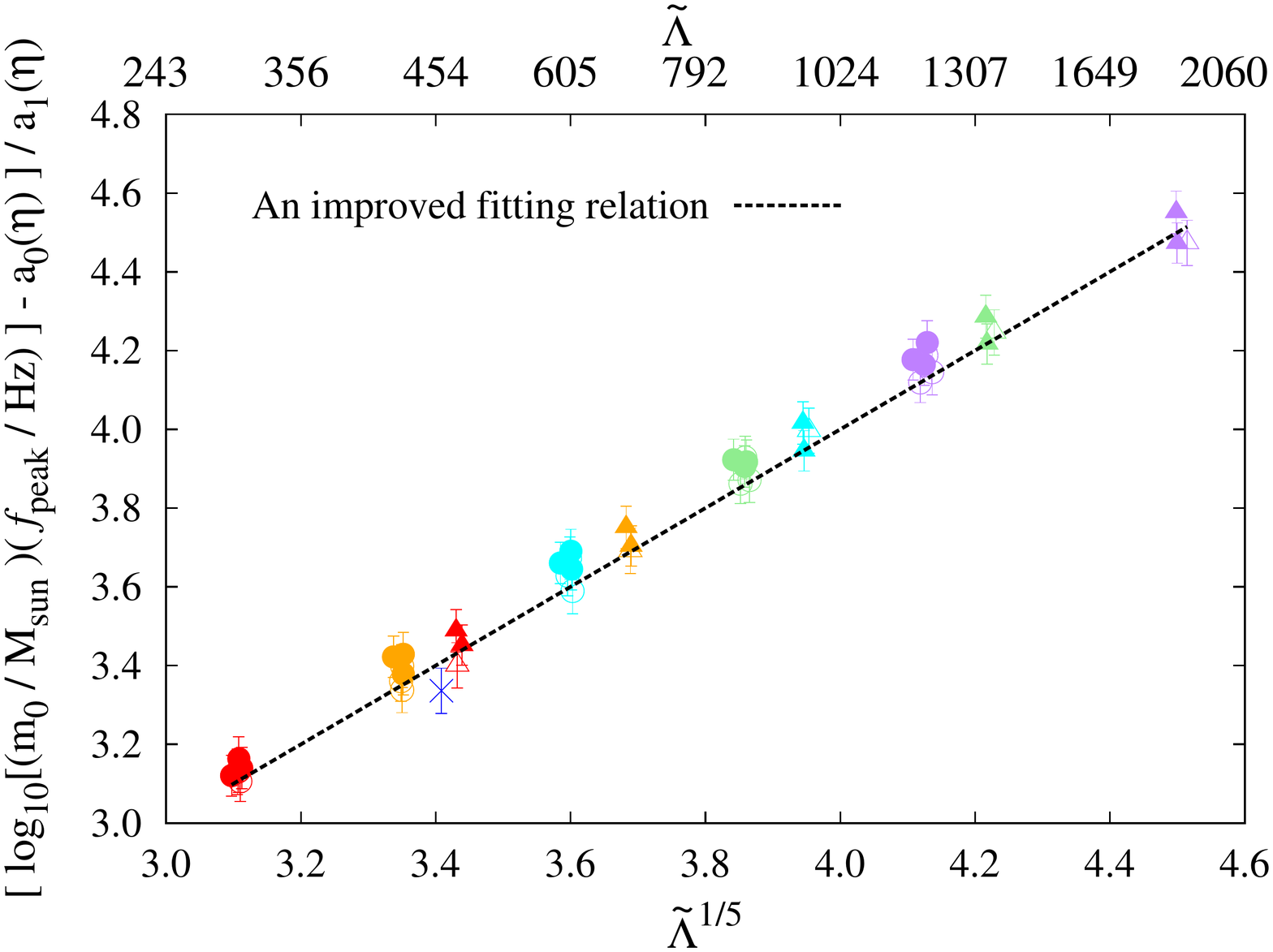}
 	 \caption{An improved $m_0f_\text{peak}$--$\tilde{\Lambda}^{1/5}$ relation with $a_0(\eta)$ and $a_1(\eta)$ in Eq.~(\ref{eq:fpeak}).
         }\label{fig:fpeak2}
\end{figure}

 We propose an improved fitting formula:
\begin{align}
  &\log_{10}\left[\left(\frac{f_\text{peak}}{\rm Hz}\right)\left(\frac{m_0}{M_\odot}\right)\right] = a_0(\eta) + a_1(\eta) \tilde{\Lambda}^{1/5},\nonumber\\
  &a_0(\eta) = 4.536 - 1.230 \eta,\nonumber\\
  &a_1(\eta) = - 0.929 + 3.120 \eta. \label{eq:fpeak}
\end{align}
With $\eta=0.2500$, $a_0(\eta)$ and $a_1(\eta)$ approximately reduce to be $a_0$ and $a_1$~\cite{footnote1} reported in Ref.~\cite{Rezzolla:2016nxn}. 
Figure~\ref{fig:fpeak2} plots the improved relation with the simulation data and we confirm that the relative error between the data and the fitting formula~(\ref{eq:fpeak}) is smaller than $3\%$.

We should keep in mind that this relation could still suffer from systematics associated with physical effects that are not taken into the simulation. 
Because of the spin-orbit coupling, high NS spin could change $f_\text{peak}$ compared to the non-spinning case. 
NS magnetic fields also could produce systematics in Eq.~(\ref{eq:fpeak}) because at the contact of the two NSs, which occurs before the peak time, the magnetic field could be exponentially amplified by the Kelvein-Helmholtz instability within a very short timescale $\ll 1$\,ms~\cite{Kiuchi:2014hja,Kiuchi:2015sga}
and the magnetic pressure could reach near the equipartition of the pressure locally, affecting the value of $f_\text{peak}$. These points should be explored in future work.

\subsubsection{Peak amplitude and binary tidal deformability relation}

\begin{figure*}
  	 \includegraphics[width=.41\linewidth]{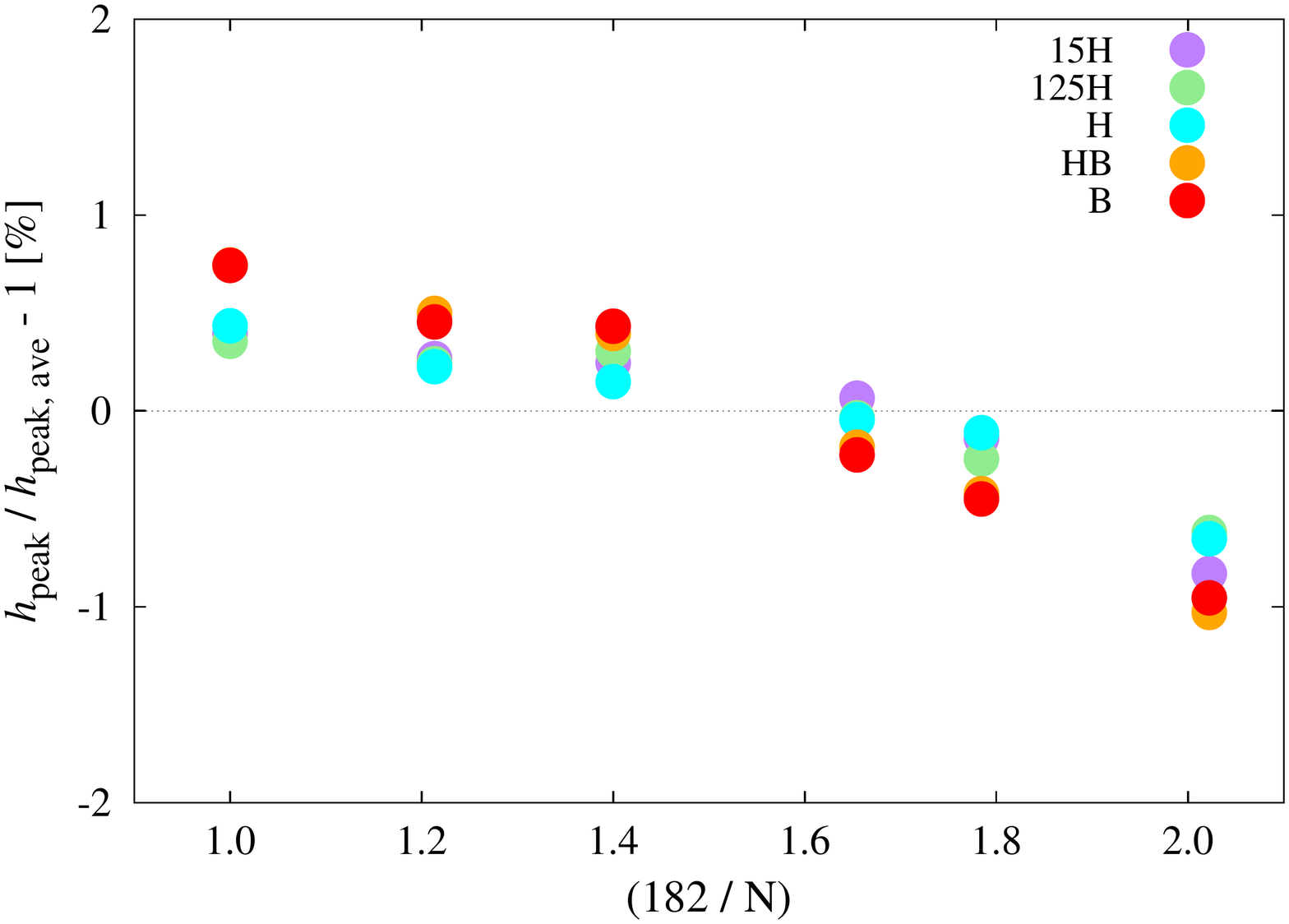} 
 	 \includegraphics[width=.45\linewidth]{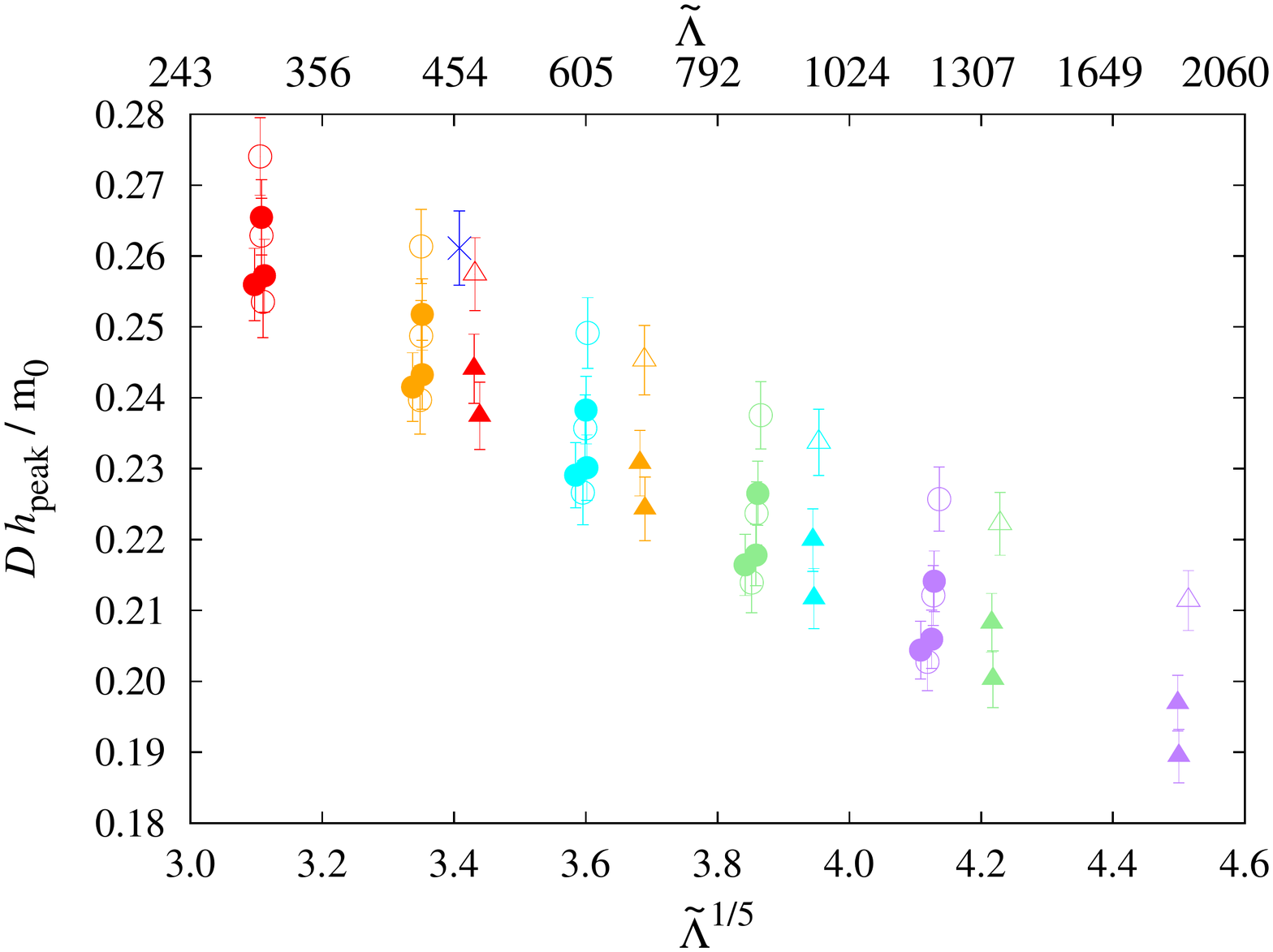}
 	 \caption{(Left) A deviation of the gravitational-wave amplitude at the peak time, $h_\text{peak}$, relative to $h_\text{peak,ave}$ as a function of $1/N$ for the binary systems with $m_1= 1.07M_\odot$ and $m_2= 1.46 M_\odot$. $h_\text{peak,ave}$ is an average of $h_\text{peak}$ over the results with different grid resolutions.  
           (Right) $D h_\text{peak}/m_0$ as a function of $\tilde{\Lambda}^{1/5}$. Meaning of the color and symbols is the same as Fig.~\ref{fig:model}. The error bar of $\pm 2\%$ comes from the uncertainty associated with the finite grid resolution in $h_\text{peak}$. 
         }\label{fig:hpeak}
\end{figure*}

References~\cite{Read:2013zra,Kiuchi:2017pte} reported that the gravitational-wave amplitude at the peak time, $h_\text{peak}$, correlates with $f_\text{peak}$, i.e., with $\tilde{\Lambda}^{1/5}$. Because we do not find perfectly convergent result for $h_\text{peak}$ with respect to the grid resolution, first, we assess deviation of $h_\text{peak}$ relative to the averaged value of $h_\text{peak}$ (average of the results with different grid resolutions) in the left panel of Fig.~\ref{fig:hpeak} for the binary systems with $m_1= 1.07 M_\odot$ and $m_2= 1.46 M_\odot$.
It is found that fluctuation around the averaged value is $\approx 1$--$2\%$. This is also the case for all the binary systems. Thus, we adopt $2\%$ as the systematics associated with the finite grid resolution in $h_\text{peak}$ and
summarize the values of $h_\text{peak}$ in Table~\ref{tb:fpeak}.

The right panel of Fig.~\ref{fig:hpeak} plots $D h_\text{peak}/m_0$ as a function of $\tilde{\Lambda}^{1/5}$. The error bar shows the systematics associated with the finite grid resolution in $h_\text{peak}$. This figure shows that the relation depends strongly on the symmetric mass ratio.
That is, the relation proposed in Refs.~\cite{Read:2013zra,Kiuchi:2017pte} is not in general satisfied.

\begin{figure}
  	 \includegraphics[width=.9\linewidth]{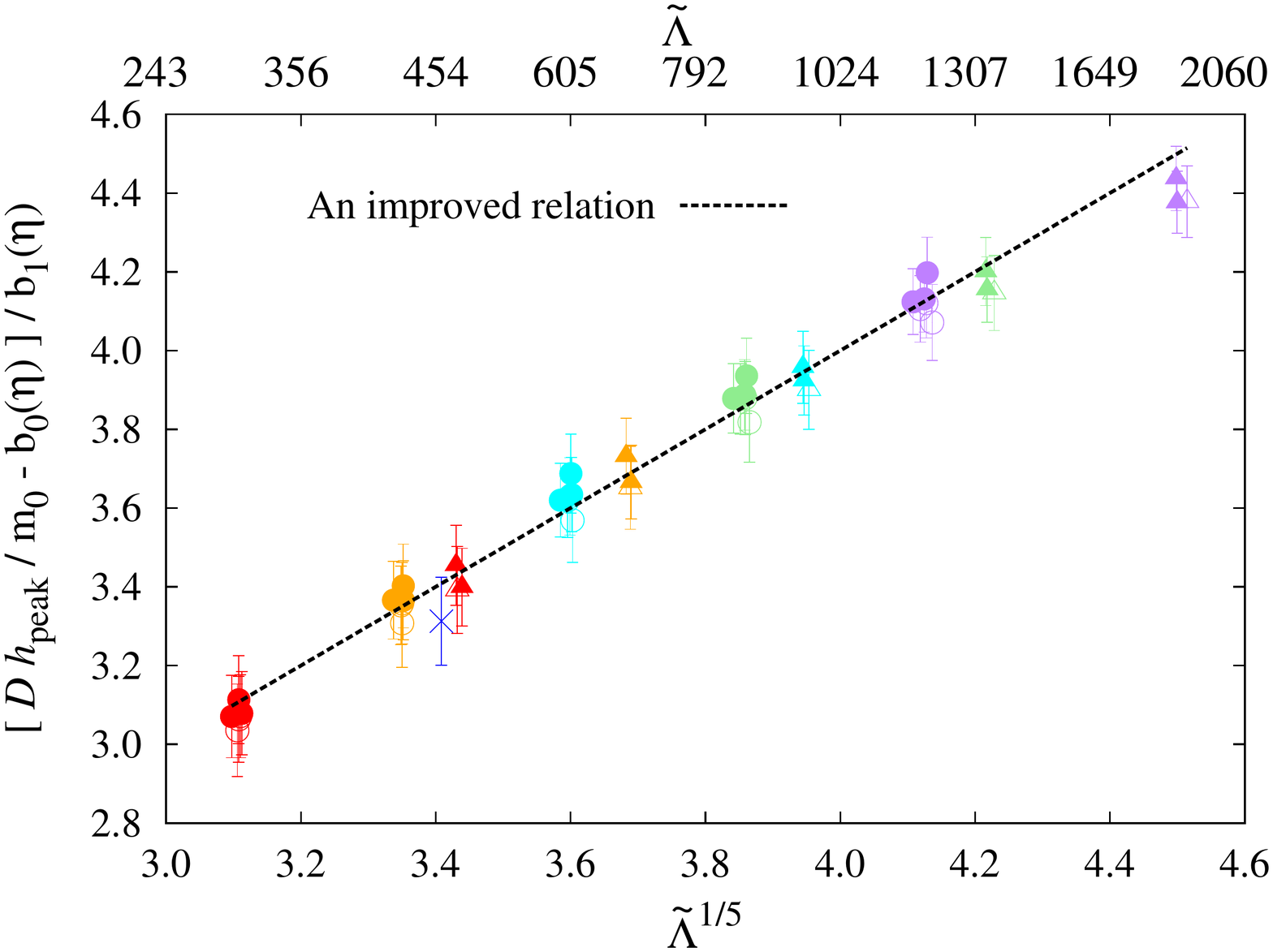}
 	 \caption{An improved $D h_\text{peak}/m_0$--$\tilde{\Lambda}^{1/5}$ relation with $b_0(\eta)$ and $b_1(\eta)$ in Eq.~(\ref{eq:hpeak}).
         }\label{fig:hpeak2}
\end{figure}

We propose a fitting formula for $D h_\text{peak}/m_0$:
\begin{align}
  &\frac{D h_\text{peak}}{m_0} = b_0(\eta) + b_1(\eta) \tilde{\Lambda}^{1/5},\nonumber\\
  &b_0(\eta) = -0.0583 + 1.896 \eta,\nonumber\\
  &b_1(\eta) = -0.1602 + 0.454 \eta. \label{eq:hpeak}
\end{align}
Figure~\ref{fig:hpeak2} plots the improved relation with the simulation data. We find that the relative error between the data and the fitting formula~(\ref{eq:hpeak}) is within $4\%$. Again note that this relation is calibrated in a limited class of the binary systems, i.e., non-magnetized non-spinning binary systems.
We should keep in mind this point in using this relation to infer the tidal deformability from observational data.

\begin{table*}[t]
\centering
\caption{Binary tidal deformability $\tilde{\Lambda}$, $f_\text{peak}$, $h_\text{peak}$, $f_2$, $E^{2,2}_{\rm GW,i}$, $E^{2,2}_{\rm GW,p}$, $J^{2,2}_{\rm GW,p}$, $J_\text{rem}$, and $m_0-M_\text{ADM,0}$.
  $M_\text{ADM,0}$ is the Arnowitt-Deser-Misner mass of the initial condition of the simulations. 
  We adopt 2$\%$ relative error for $f_{\rm peak}$ and $h_\text{peak}$ and $5\%$ relative error for $f_2$, respectively, as a typical value. For $f_2$, we exclude binary systems which collapse to a black hole within a few ms after the merger.
  For $E^{2,2}_{\rm GW,i}$ and $J_\text{rem}$, we adopt $2\%$ and $1\%$ relative error, respectively. 
  $E_\text{GW}$ and $m_0-M_\text{ADM,0}$ are given in units of $M_\odot$. $J_\text{GW}$ and $J_\text{rem}$ are in units of
  $M_\odot^2$. 
  }
\begin{tabular}{c|cccccccccc}
\hline\hline
System        & $\tilde{\Lambda}^{1/5}$ & $f_{\rm peak}$ [Hz] & $D h_\text{peak} / m_0$ & $f_2 [{\rm Hz}]$ & $E^{2,2}_{\text{GW},{\rm i}}$ & $E^{2,2}_{\text{GW},{\rm p}} $ & $J^{2,2}_{\text{GW},{\rm p}}$ & $J_\text{rem}$ & $m_0-M_\text{ADM,0}$ \\
\hline\
15H135-135   & 4.14 & 1503$\pm$30 & 0.226$\pm$0.005 & 2321$\pm$116 & (7.90$\pm 0.16)\times 10^{-3}$ & 1.35$\times 10^{-2}$ & 0.40  & 6.64$\pm 0.07$  & $1.65\times 10^{-2}$\\
125H135-135  & 3.87 & 1652$\pm$33 & 0.236$\pm$0.005 & 2517$\pm$126 & (9.04$\pm 0.18)\times 10^{-3}$ & 1.76$\times 10^{-2}$ & 0.48  & 6.54$\pm 0.07$  & $1.64\times 10^{-2}$\\
H135-135     & 3.60 & 1820$\pm$36 & 0.249$\pm$0.005 & 2790$\pm$139 & (1.03$\pm 0.02)\times 10^{-2}$ & 2.32$\times 10^{-2}$ & 0.56  & 6.46$\pm 0.06$  & $1.63\times 10^{-2}$\\
HB135-135    & 3.35 & 1986$\pm$40 & 0.261$\pm$0.005 & 3243$\pm$162 & (1.17$\pm 0.02)\times 10^{-2}$ & 2.89$\times 10^{-2}$ & 0.59  & 6.39$\pm 0.06$  & $1.64\times 10^{-2}$\\
B135-135     & 3.11 & 2133$\pm$43 & 0.274$\pm$0.005 & --           & (1.30$\pm 0.03)\times 10^{-2}$ & 7.39$\times 10^{-3}$ & 0.13  & 6.33$\pm 0.06$  & $1.65\times 10^{-2}$\\
15H121-151   & 4.13 & 1356$\pm$27 & 0.212$\pm$0.004 & 2261$\pm$163 & (7.47$\pm 0.15)\times 10^{-3}$ & 5.47$\times 10^{-3}$ & 0.17  & 6.66$\pm 0.07$  & $1.66\times 10^{-2}$\\
125H121-151  & 3.86 & 1490$\pm$30 & 0.224$\pm$0.004 & 2379$\pm$119 & (8.53$\pm 0.17)\times 10^{-3}$ & 8.24$\times 10^{-3}$ & 0.23  & 6.57$\pm 0.07$  & $1.66\times 10^{-2}$\\
H121-151     & 3.60 & 1637$\pm$33 & 0.236$\pm$0.005 & 2749$\pm$137 & (9.70$\pm 0.19)\times 10^{-3}$ & 1.05$\times 10^{-2}$ & 0.26  & 6.49$\pm 0.06$  & $1.66\times 10^{-2}$\\
HB121-151    & 3.35 & 1809$\pm$36 & 0.249$\pm$0.005 & 3268$\pm$161 & (1.10$\pm 0.02)\times 10^{-2}$ & 2.26$\times 10^{-2}$ & 0.48  & 6.41$\pm 0.06$  & $1.66\times 10^{-2}$\\
B121-151     & 3.11 & 1994$\pm$40 & 0.263$\pm$0.005 & --           & (1.23$\pm 0.02)\times 10^{-2}$ & 6.85$\times 10^{-3}$ & 0.13  & 6.35$\pm 0.06$  & $1.66\times 10^{-2}$\\
15H125-125   & 4.51 & 1450$\pm$29 & 0.211$\pm$0.004 & 2159$\pm$108 & (6.26$\pm 0.13)\times 10^{-3}$ & 7.98$\times 10^{-3}$ & 0.25  & 5.95$\pm 0.06$  & $1.53\times 10^{-2}$\\
125H125-125  & 4.23 & 1568$\pm$31 & 0.222$\pm$0.004 & 2350$\pm$118 & (7.19$\pm 0.14)\times 10^{-3}$ & 9.29$\times 10^{-3}$ & 0.27  & 5.87$\pm 0.06$  & $1.53\times 10^{-2}$\\
H125-125     & 3.95 & 1710$\pm$34 & 0.234$\pm$0.005 & 2749$\pm$137 & (8.15$\pm 0.16)\times 10^{-3}$ & 1.67$\times 10^{-2}$ & 0.42  & 5.80$\pm 0.06$  & $1.52\times 10^{-2}$\\
HB125-125    & 3.69 & 1900$\pm$38 & 0.245$\pm$0.005 & 2873$\pm$144 & (9.35$\pm 0.19)\times 10^{-3}$ & 1.66$\times 10^{-2}$ & 0.39  & 5.74$\pm 0.06$  & $1.53\times 10^{-2}$\\
B125-125     & 3.43 & 2099$\pm$42 & 0.257$\pm$0.005 & 3353$\pm$168 & (1.06$\pm 0.02))\times10^{-2}$ & 2.19$\times 10^{-2}$ & 0.44  & 5.69$\pm 0.06$  & $1.53\times 10^{-2}$\\
15H116-158   & 4.12 & 1273$\pm$26 & 0.205$\pm$0.004 & 2148$\pm$107 & (7.19$\pm 0.14)\times 10^{-3}$ & 4.63$\times 10^{-3}$ & 0.15  & 6.84$\pm 0.07$  & $1.65\times 10^{-2}$\\
125H116-158  & 3.85 & 1406$\pm$28 & 0.214$\pm$0.004 & 2276$\pm$124 & (8.20$\pm 0.16)\times 10^{-3}$ & 1.01$\times 10^{-2}$ & 0.28  & 6.76$\pm 0.07$  & $1.65\times 10^{-2}$\\
H116-158     & 3.60 & 1540$\pm$31 & 0.227$\pm$0.005 & 2767$\pm$138 & (9.30$\pm 0.19)\times 10^{-3}$ & 1.23$\times 10^{-2}$ & 0.31  & 6.69$\pm 0.07$  & $1.66\times 10^{-2}$\\
HB116-158    & 3.35 & 1709$\pm$34 & 0.240$\pm$0.005 & 3242$\pm$162 & (1.05$\pm 0.02)\times 10^{-2}$ & 1.40$\times 10^{-2}$ & 0.30  & 6.63$\pm 0.06$  & $1.65\times 10^{-2}$\\
B116-158     & 3.11 & 1885$\pm$37 & 0.254$\pm$0.005 & --           & (1.18$\pm 0.02)\times 10^{-2}$ & 4.64$\times 10^{-3}$ & 0.10  & 6.58$\pm 0.07$  & $1.65\times 10^{-2}$\\
15H125-146   & 4.13 & 1401$\pm$28 & 0.214$\pm$0.004 & 2336$\pm$117 & (7.62$\pm 0.02)\times 10^{-3}$ & 1.01$\times 10^{-2}$ & 0.30  & 6.81$\pm 0.07$  & $1.66\times 10^{-2}$\\
125H125-146  & 3.86 & 1560$\pm$31 & 0.226$\pm$0.005 & 2576$\pm$129 & (8.77$\pm 0.18)\times 10^{-3}$ & 1.26$\times 10^{-2}$ & 0.34  & 6.73$\pm 0.07$  & $1.66\times 10^{-2}$\\
H125-146     & 3.60 & 1691$\pm$34 & 0.238$\pm$0.003 & 2827$\pm$141 & (9.91$\pm 0.20)\times 10^{-3}$ & 1.89$\times 10^{-2}$ & 0.45  & 6.66$\pm 0.07$  & $1.66\times 10^{-2}$\\
HB125-146    & 3.35 & 1856$\pm$37 & 0.252$\pm$0.005 & 3251$\pm$163 & (1.12$\pm 0.20)\times 10^{-2}$ & 2.50$\times 10^{-2}$ & 0.52  & 6.60$\pm 0.07$  & $1.66\times 10^{-2}$\\
B125-146     & 3.11 & 2039$\pm$41 & 0.265$\pm$0.005 & --           & (1.26$\pm 0.25)\times 10^{-2}$ & 7.99$\times 10^{-3}$ & 0.14  & 6.56$\pm 0.06$  & $1.66\times 10^{-2}$\\
15H118-155   & 4.12 & 1308$\pm$26 & 0.206$\pm$0.004 & 2161$\pm$108 & (7.31$\pm 0.15)\times 10^{-3}$ & 5.72$\times 10^{-3}$ & 0.18  & 6.83$\pm 0.07$  & $1.66\times 10^{-2}$\\
125H118-155  & 3.86 & 1441$\pm$29 & 0.218$\pm$0.004 & 2358$\pm$118 & (8.35$\pm 0.17)\times 10^{-3}$ & 7.12$\times 10^{-3}$ & 0.21  & 6.75$\pm 0.07$  & $1.67\times 10^{-2}$\\
H118-155     & 3.60 & 1590$\pm$32 & 0.230$\pm$0.005 & 2782$\pm$139 & (9.49$\pm 0.19)\times 10^{-3}$ & 1.59$\times 10^{-2}$ & 0.39  & 6.68$\pm 0.07$  & $1.66\times 10^{-2}$\\
HB118-155    & 3.35 & 1759$\pm$35 & 0.243$\pm$0.005 & 3259$\pm$163 & (1.08$\pm 0.02)\times 10^{-2}$ & 2.03$\times 10^{-2}$ & 0.43  & 6.62$\pm 0.07$  & $1.66\times 10^{-2}$\\
B118-155     & 3.11 & 1942$\pm$39 & 0.257$\pm$0.005 & --           & (1.20$\pm 0.02)\times 10^{-2}$ & 5.54$\times 10^{-3}$ & 0.11  & 6.66$\pm 0.07$  & $1.66\times 10^{-2}$\\
15H117-156   & 4.11 & 1293$\pm$26 & 0.204$\pm$0.004 & 2161$\pm$108 & (7.26$\pm 0.15)\times 10^{-3}$ & 5.09$\times 10^{-3}$ & 0.17  & 6.83$\pm 0.07$  & $1.66\times 10^{-2}$\\
125H117-156  & 3.84 & 1425$\pm$29 & 0.216$\pm$0.004 & 2416$\pm$121 & (8.30$\pm 0.17)\times 10^{-3}$ & 8.09$\times 10^{-3}$ & 0.23  & 6.76$\pm 0.07$  & $1.66\times 10^{-2}$\\
H117-156     & 3.58 & 1574$\pm$32 & 0.229$\pm$0.005 & 2775$\pm$139 & (9.43$\pm 0.19)\times 10^{-3}$ & 1.39$\times 10^{-2}$ & 0.34  & 6.69$\pm 0.07$  & $1.66\times 10^{-2}$\\
HB117-156    & 3.34 & 1724$\pm$35 & 0.242$\pm$0.005 & 3201$\pm$160 & (1.06$\pm 0.02)\times 10^{-2}$ & 1.61$\times 10^{-2}$ & 0.35  & 6.62$\pm 0.07$  & $1.66\times 10^{-2}$\\
B117-156     & 3.10 & 1933$\pm$38 & 0.256$\pm$0.005 & --           & (1.20$\pm 0.02)\times 10^{-2}$ & 5.26$\times 10^{-3}$ & 0.11  & 6.58$\pm 0.06$  & $1.64\times 10^{-2}$\\
15H112-140   & 4.50 & 1281$\pm$26 & 0.197$\pm$0.004 & 2188$\pm$109 & (5.91$\pm 0.12)\times 10^{-3}$ & 5.37$\times 10^{-3}$ & 0.17  & 5.97$\pm 0.06$  & $1.49\times 10^{-2}$\\
125H112-140  & 4.21 & 1412$\pm$28 & 0.208$\pm$0.004 & 2269$\pm$113 & (6.80$\pm 0.14)\times 10^{-3}$ & 4.80$\times 10^{-3}$ & 0.15  & 5.89$\pm 0.06$  & $1.49\times 10^{-2}$\\
H112-140     & 3.94 & 1558$\pm$31 & 0.220$\pm$0.004 & 2470$\pm$123 & (7.78$\pm 0.16)\times 10^{-3}$ & 6.18$\times 10^{-3}$ & 0.17  & 5.82$\pm 0.06$  & $1.50\times 10^{-2}$\\
HB112-140    & 3.68 & 1717$\pm$34 & 0.231$\pm$0.005 & 2791$\pm$140 & (8.84$\pm 0.18)\times 10^{-3}$ & 9.52$\times 10^{-3}$ & 0.23  & 5.76$\pm 0.06$  & $1.50\times 10^{-2}$\\
B112-140     & 3.43 & 1890$\pm$38 & 0.244$\pm$0.005 & 3271$\pm$164 & (9.98$\pm 0.20)\times 10^{-3}$ & 1.59$\times 10^{-2}$ & 0.33  & 5.71$\pm 0.06$  & $1.52\times 10^{-2}$\\
15H107-146   & 4.50 & 1203$\pm$24 & 0.189$\pm$0.004 & 2054$\pm$103 & (5.70$\pm 0.11)\times 10^{-3}$ & 3.63$\times 10^{-3}$ & 0.13  & 5.99$\pm 0.06$  & $1.51\times 10^{-2}$\\
125H107-146  & 4.22 & 1328$\pm$27 & 0.200$\pm$0.004 & 2291$\pm$115 & (6.57$\pm 0.13)\times 10^{-3}$ & 4.56$\times 10^{-3}$ & 0.14  & 5.91$\pm 0.06$  & $1.50\times 10^{-2}$\\
H107-146     & 3.94 & 1475$\pm$30 & 0.212$\pm$0.004 & 2546$\pm$127 & (7.49$\pm 0.15)\times 10^{-3}$ & 7.82$\times 10^{-3}$ & 0.21  & 5.84$\pm 0.06$  & $1.49\times 10^{-2}$\\
HB107-146    & 3.69 & 1620$\pm$32 & 0.224$\pm$0.004 & 2870$\pm$143 & (8.51$\pm 0.17)\times 10^{-3}$ & 1.02$\times 10^{-2}$ & 0.25  & 5.78$\pm 0.06$  & $1.50\times 10^{-2}$\\
B107-146     & 3.44 & 1786$\pm$36 & 0.237$\pm$0.005 & 3298$\pm$165 & (9.60$\pm 0.19)\times 10^{-3}$ & 1.29$\times 10^{-2}$ & 0.27  & 5.73$\pm 0.06$  & $1.51\times 10^{-2}$\\
SFHo135-135  & 3.41 & 1987$\pm$40 & 0.261$\pm$0.005 & 3250$\pm$163 & (1.17$\pm 0.02)\times 10^{-2}$ & 2.91$\times 10^{-3}$ & 0.61  & 6.60$\pm 0.07$  & $1.68\times 10^{-2}$\\
\hline\hline
\end{tabular}\label{tb:fpeak}
\end{table*}

\subsubsection{$f_1,f_2$ and binary tidal deformability relation}

\begin{figure*}[t]
\hspace{-18.0mm}
\begin{minipage}{0.27\hsize}
\begin{center}
\includegraphics[width=4.5cm,angle=0]{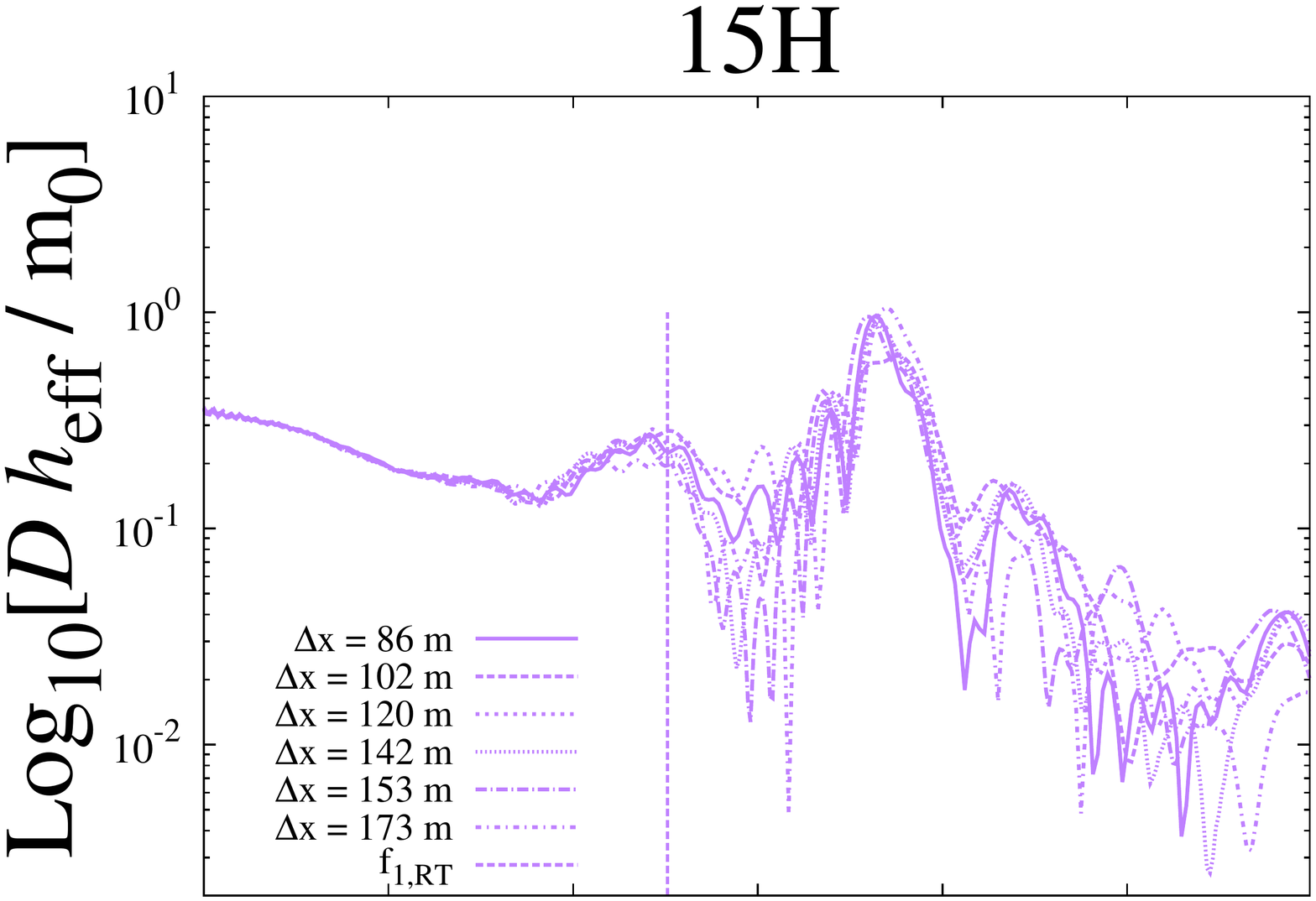}
\end{center}
\end{minipage}
\hspace{-13.35mm}
\begin{minipage}{0.27\hsize}
\begin{center}
\includegraphics[width=4.5cm,angle=0]{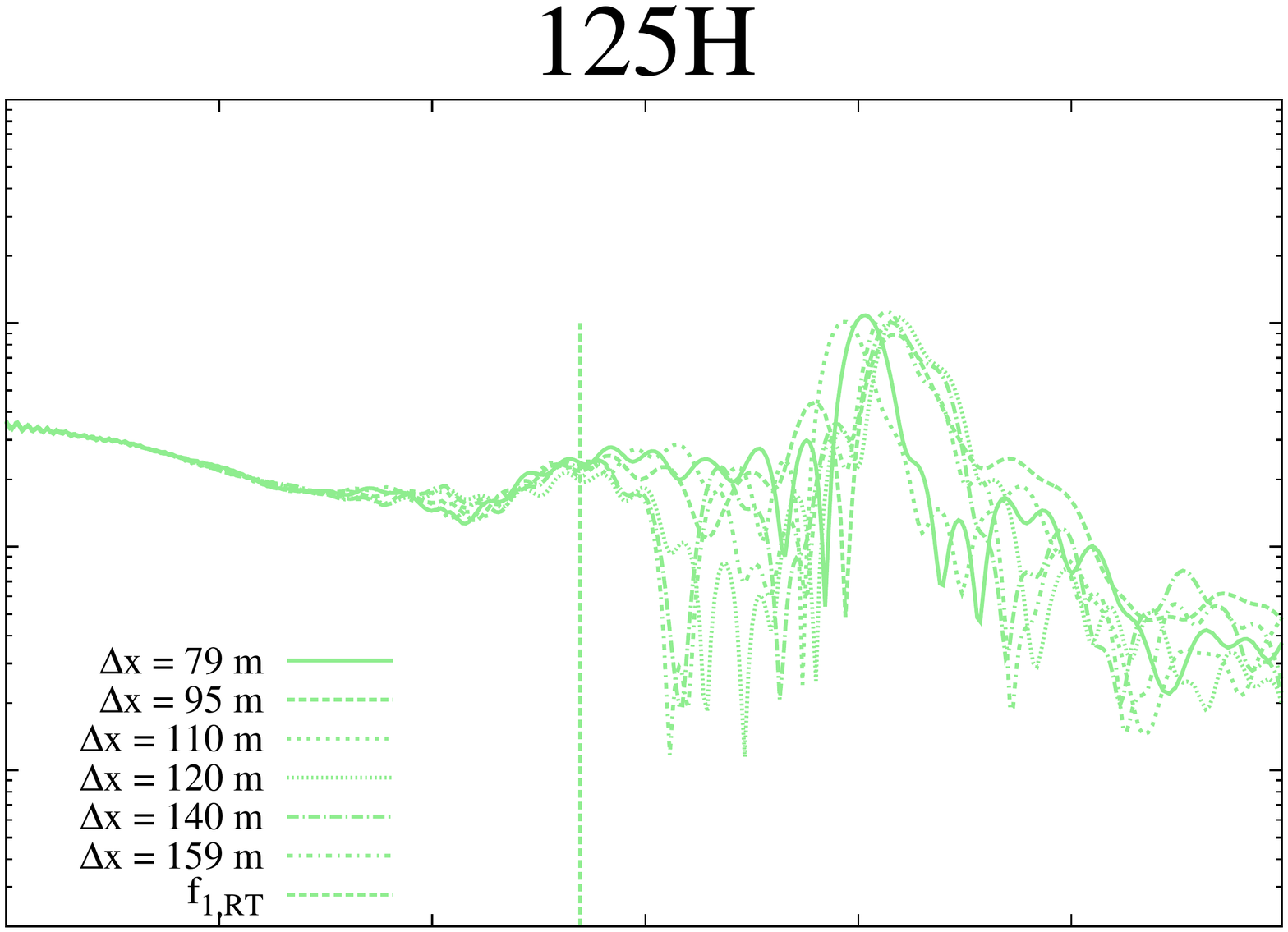}
\end{center}
\end{minipage}
\hspace{-13.35mm}
\begin{minipage}{0.27\hsize}
\begin{center}
\includegraphics[width=4.5cm,angle=0]{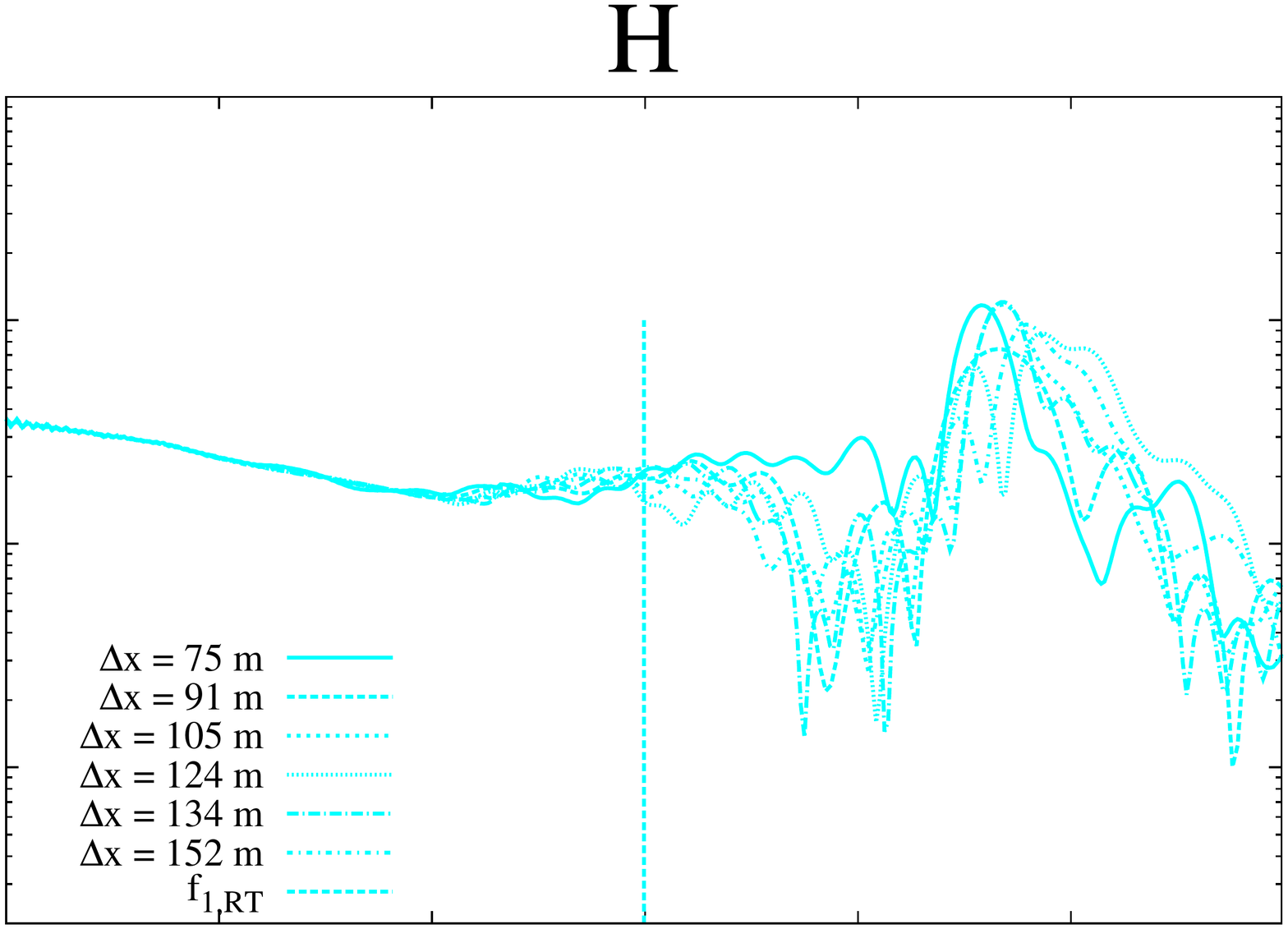}
\end{center}
\end{minipage}
\hspace{-13.35mm}
\begin{minipage}{0.27\hsize}
\begin{center}
\includegraphics[width=4.5cm,angle=0]{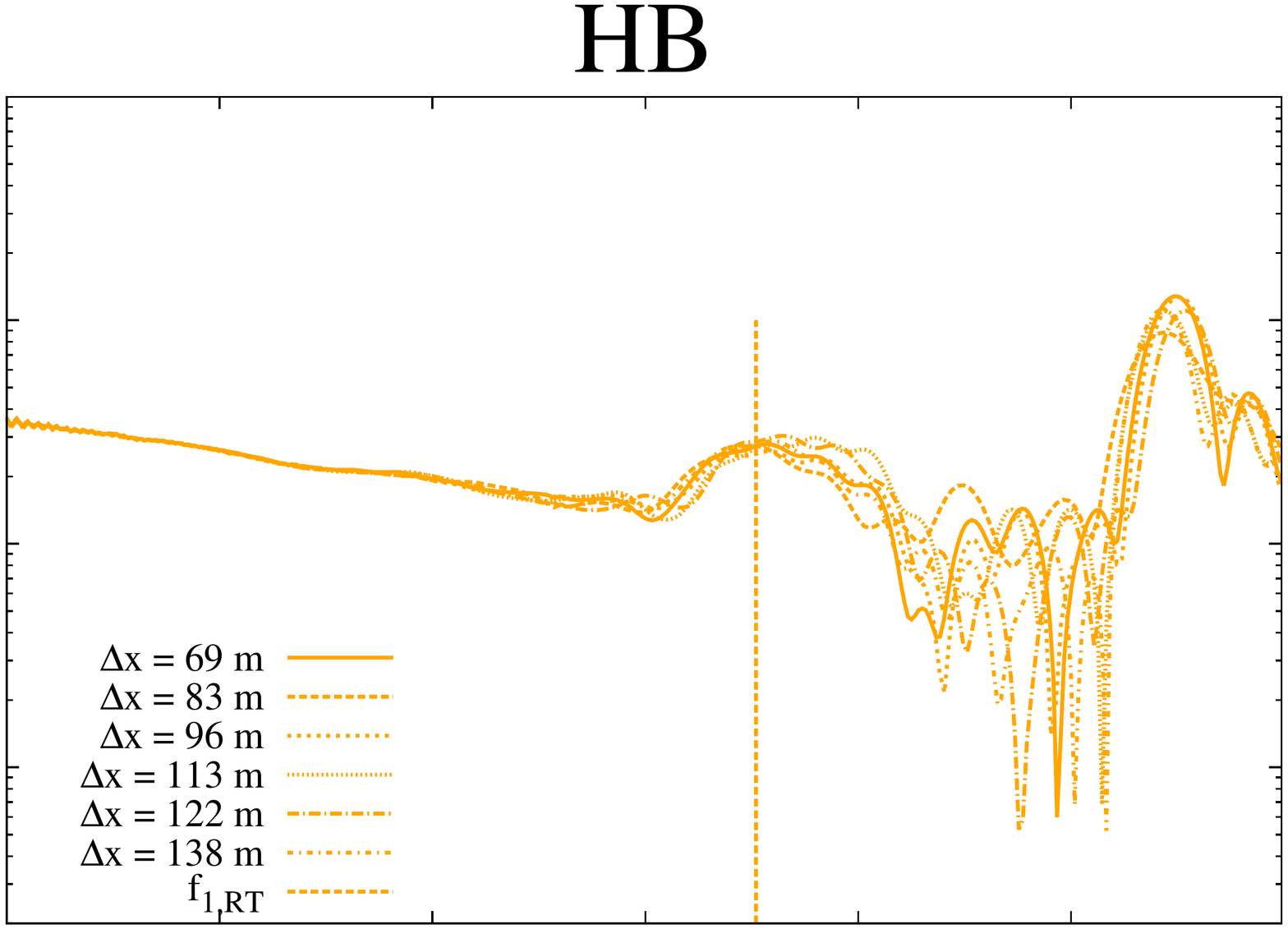}
\end{center}
\end{minipage}
\hspace{-13.35mm}
\begin{minipage}{0.27\hsize}
\begin{center}
\includegraphics[width=4.5cm,angle=0]{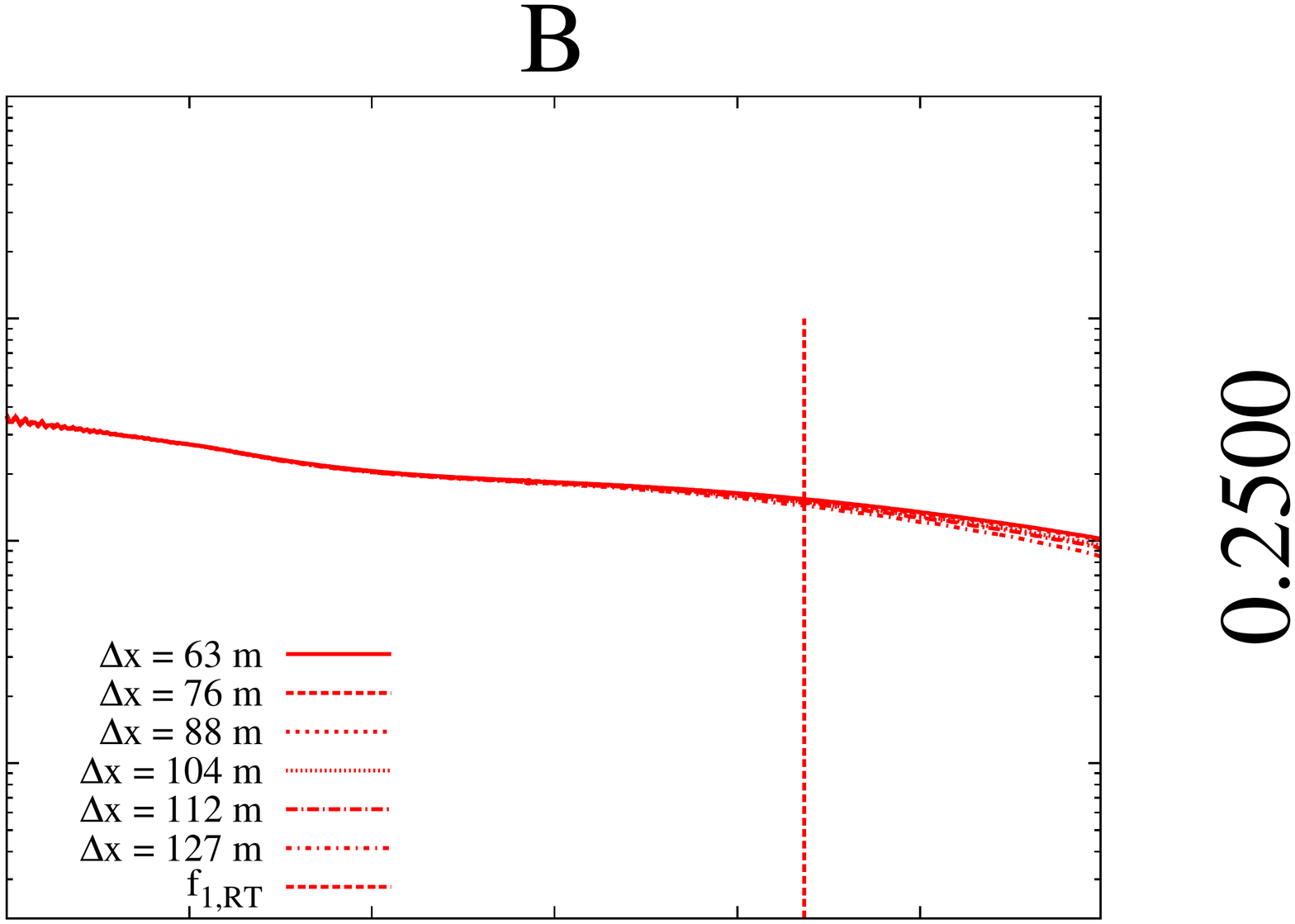}
\end{center}
\end{minipage}\\
\vspace{-9mm}
\hspace{-18.0mm}
\begin{minipage}{0.27\hsize}
\begin{center}
\includegraphics[width=4.5cm,angle=0]{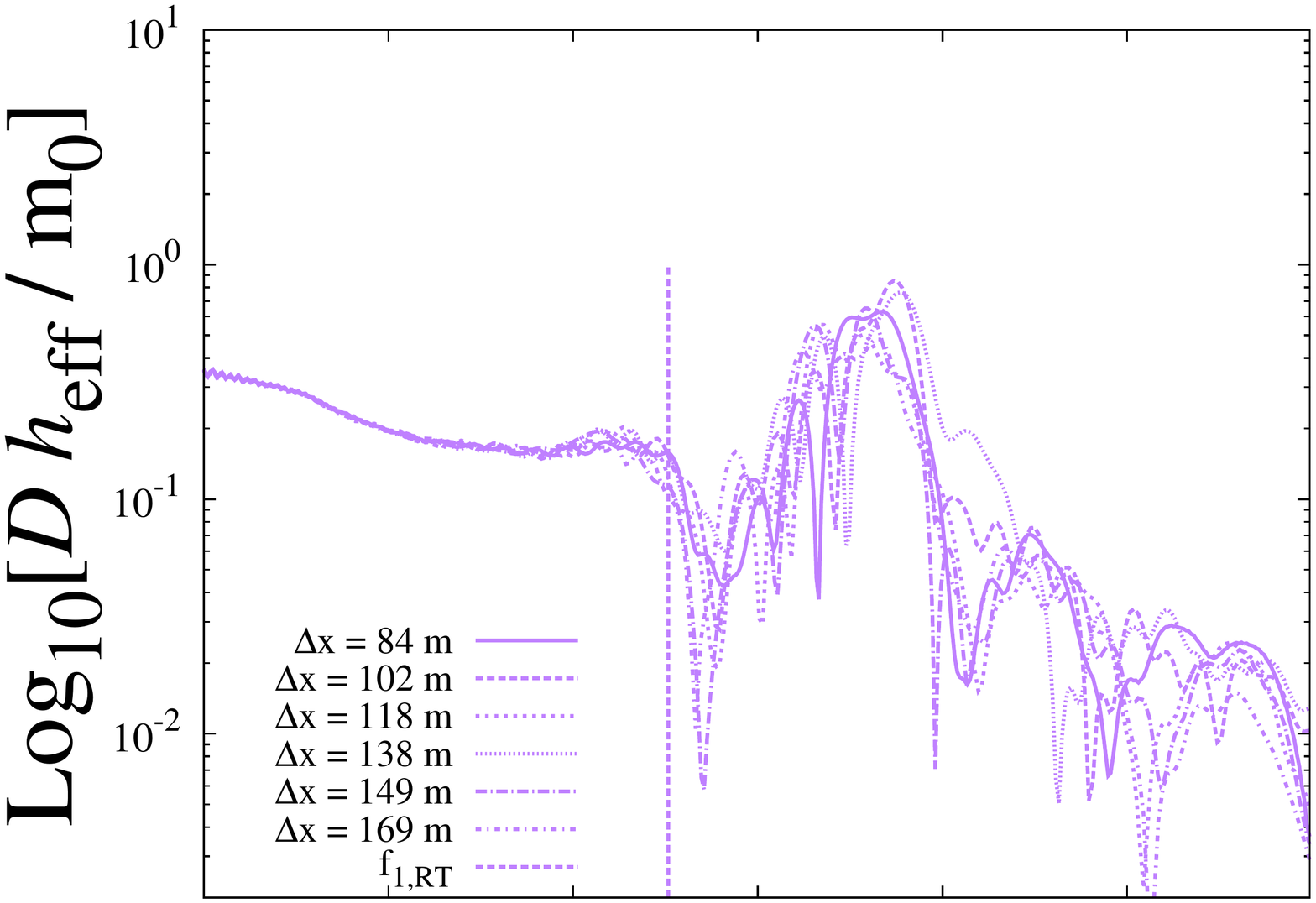}
\end{center}
\end{minipage}
\hspace{-13.35mm}
\begin{minipage}{0.27\hsize}
\begin{center}
\includegraphics[width=4.5cm,angle=0]{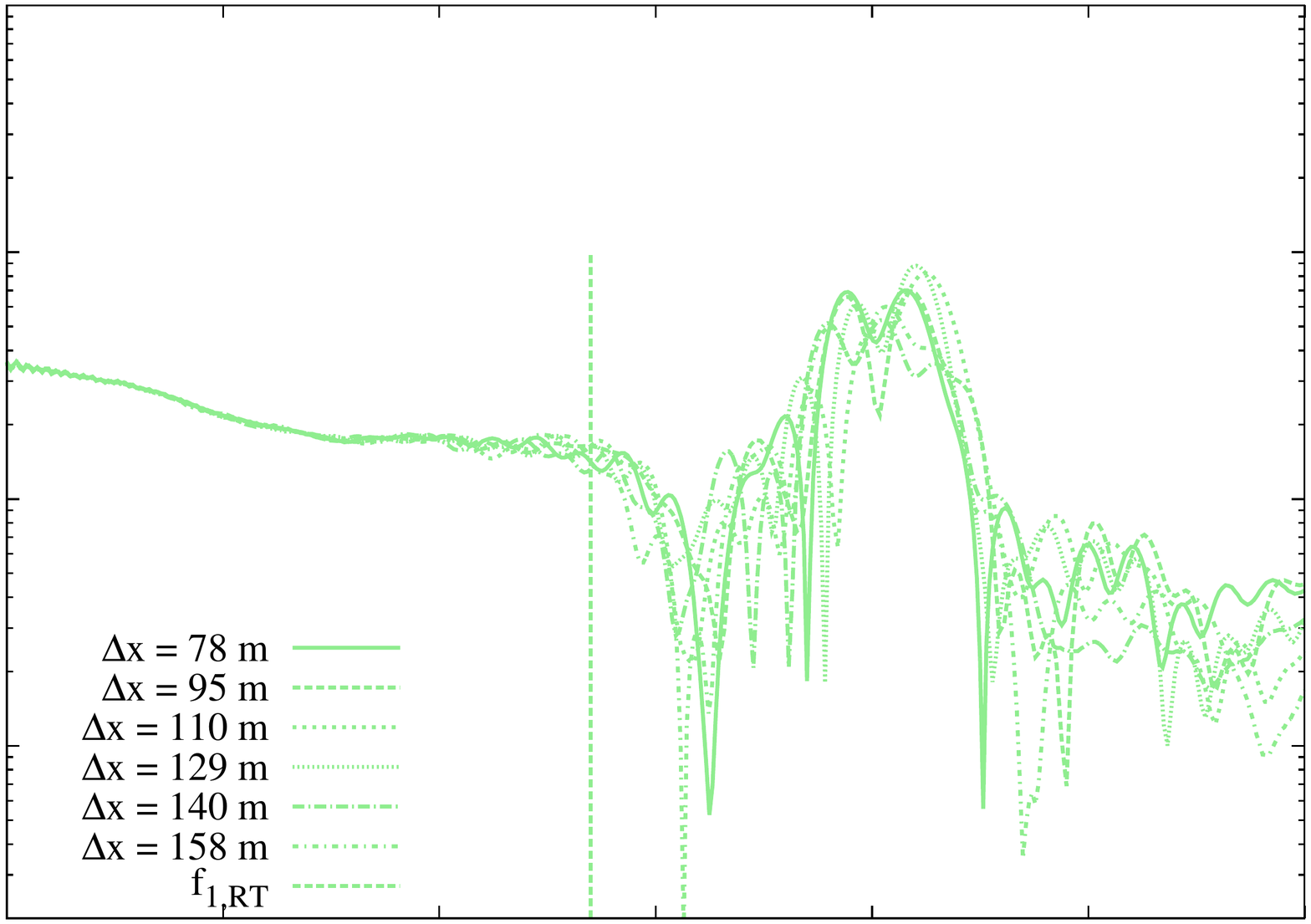}
\end{center}
\end{minipage}
\hspace{-13.35mm}
\begin{minipage}{0.27\hsize}
\begin{center}
\includegraphics[width=4.5cm,angle=0]{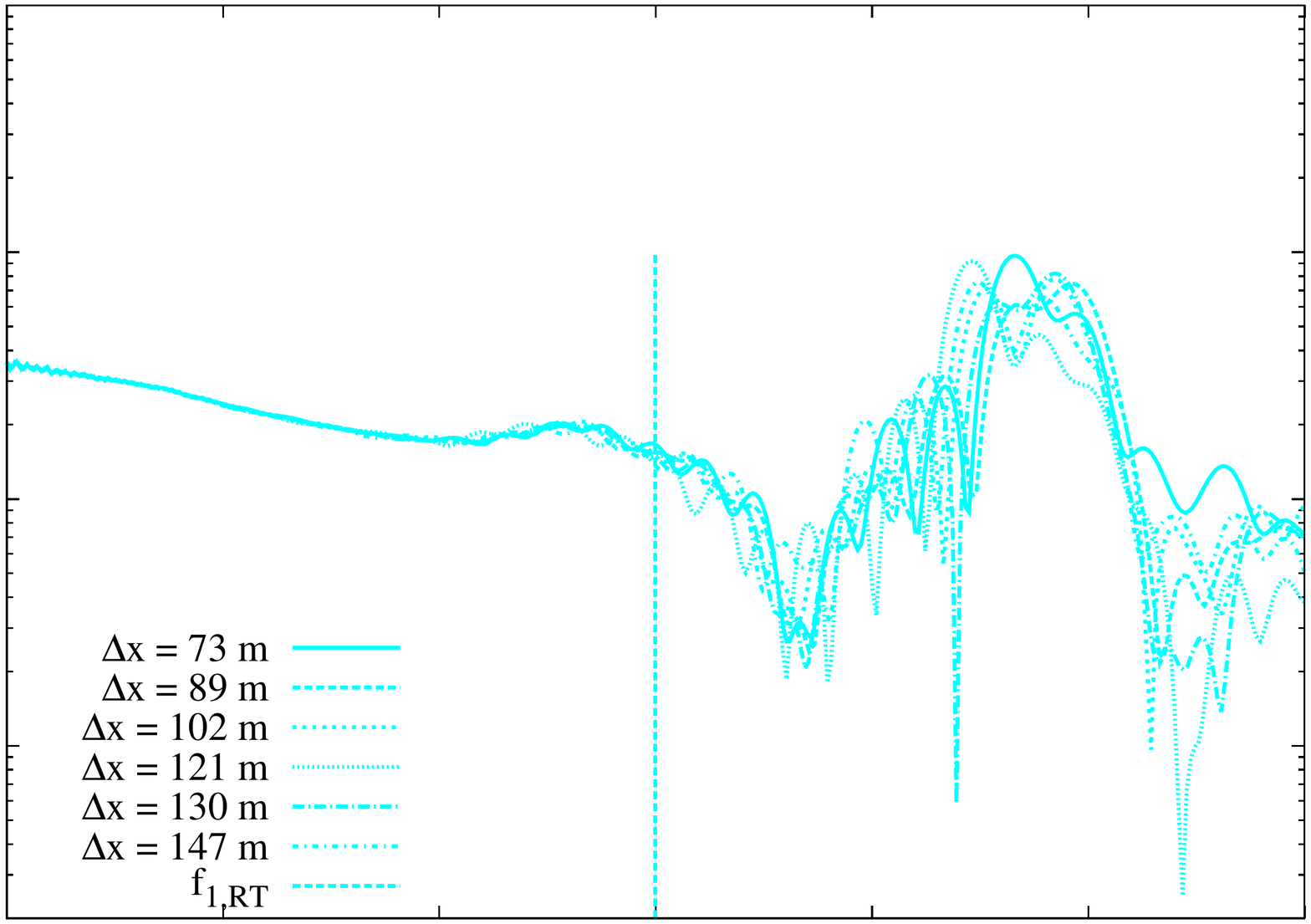}
\end{center}
\end{minipage}
\hspace{-13.35mm}
\begin{minipage}{0.27\hsize}
\begin{center}
\includegraphics[width=4.5cm,angle=0]{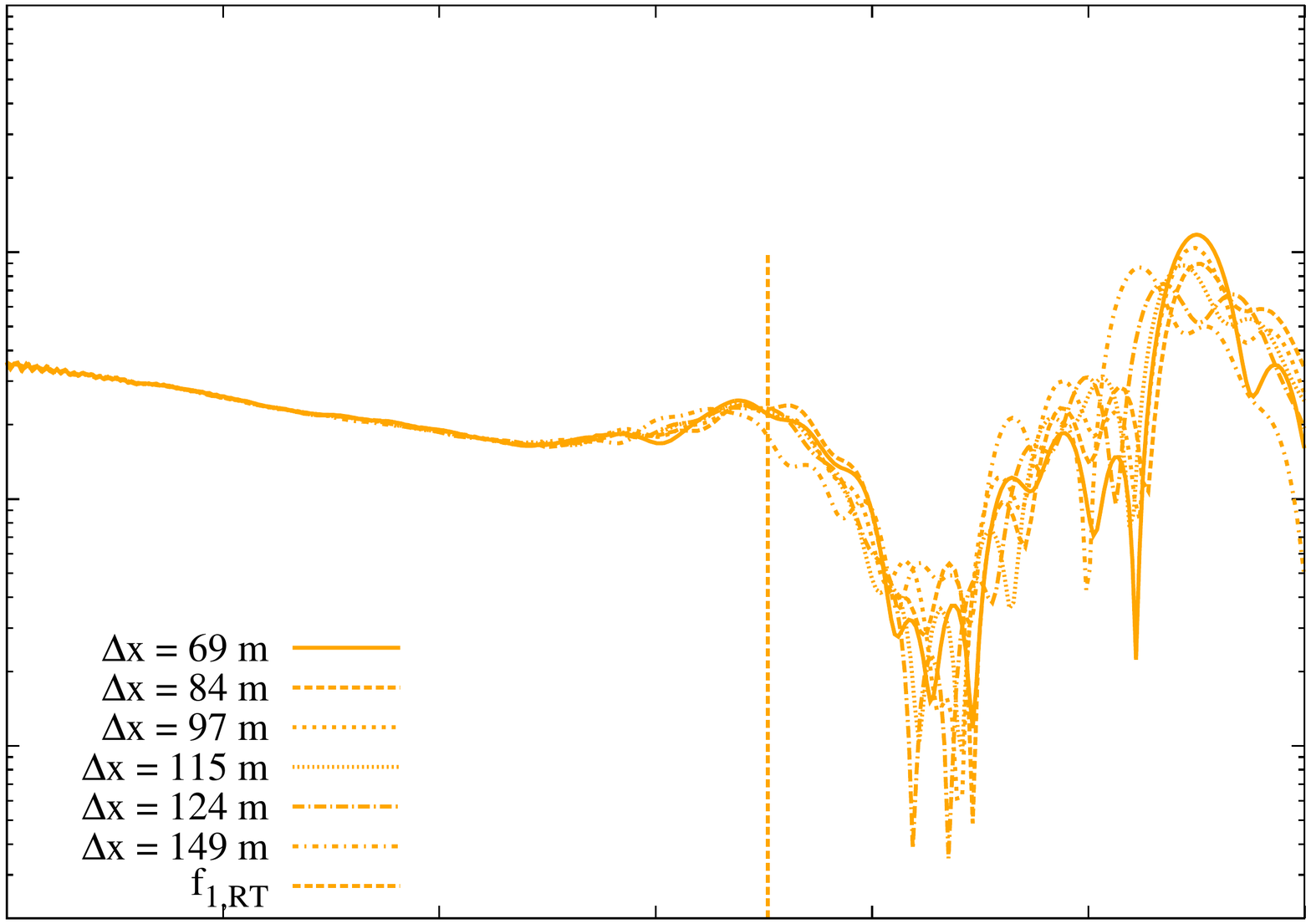}
\end{center}
\end{minipage}
\hspace{-13.35mm}
\begin{minipage}{0.27\hsize}
\begin{center}
\includegraphics[width=4.5cm,angle=0]{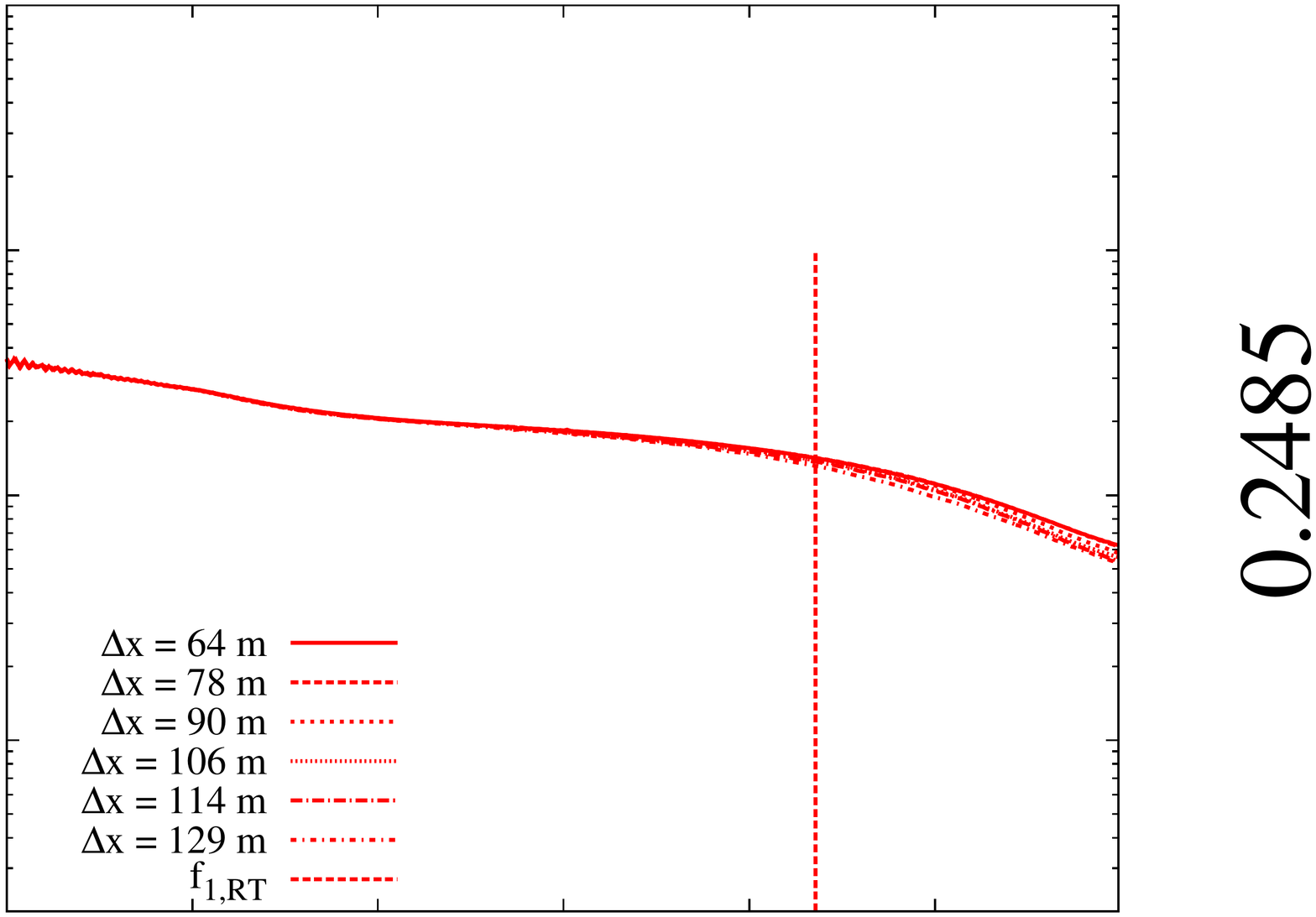}
\end{center}
\end{minipage}\\
\vspace{-9mm}
\hspace{-18.0mm}
\begin{minipage}{0.27\hsize}
\begin{center}
\includegraphics[width=4.5cm,angle=0]{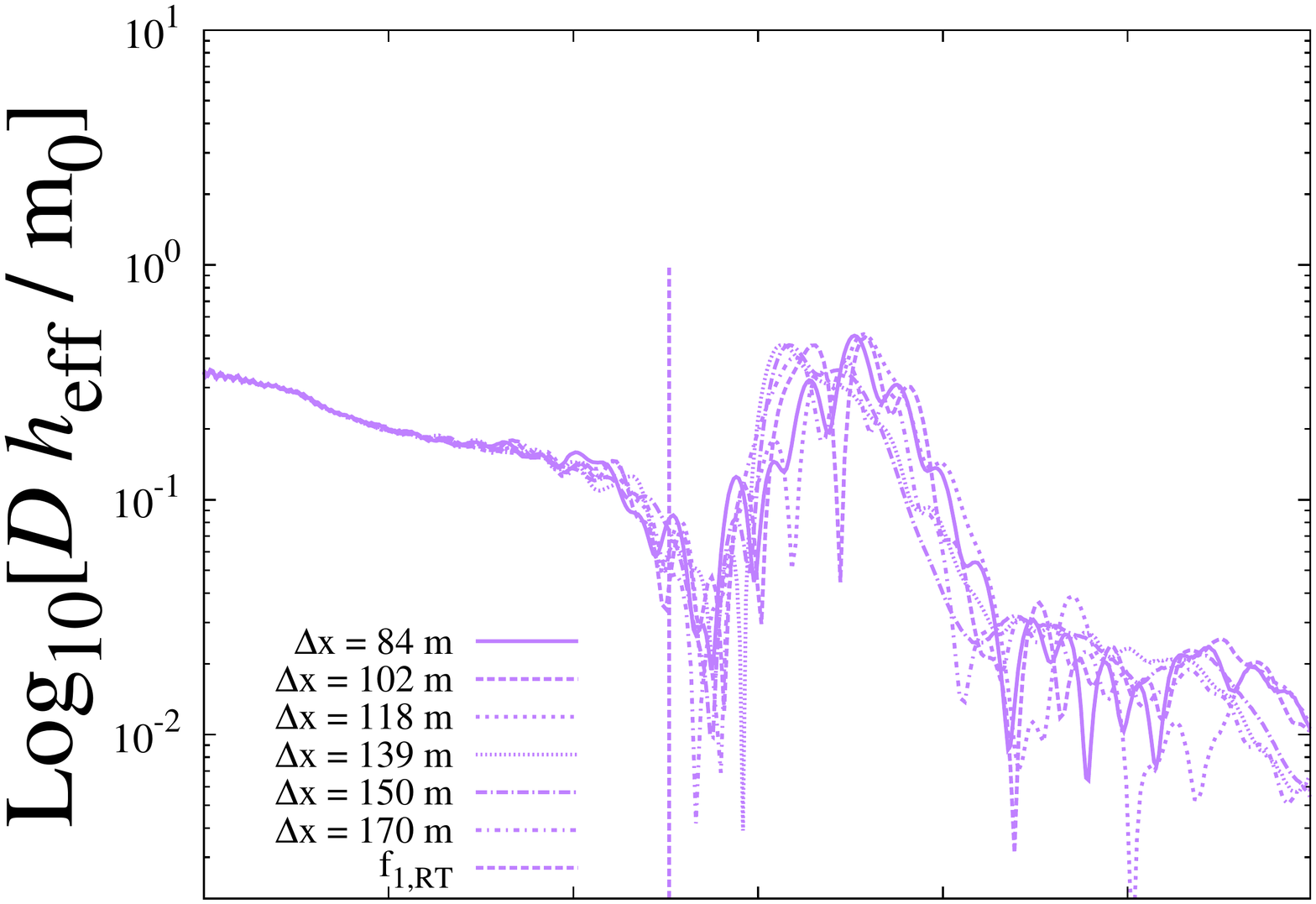}
\end{center}
\end{minipage}
\hspace{-13.35mm}
\begin{minipage}{0.27\hsize}
\begin{center}
\includegraphics[width=4.5cm,angle=0]{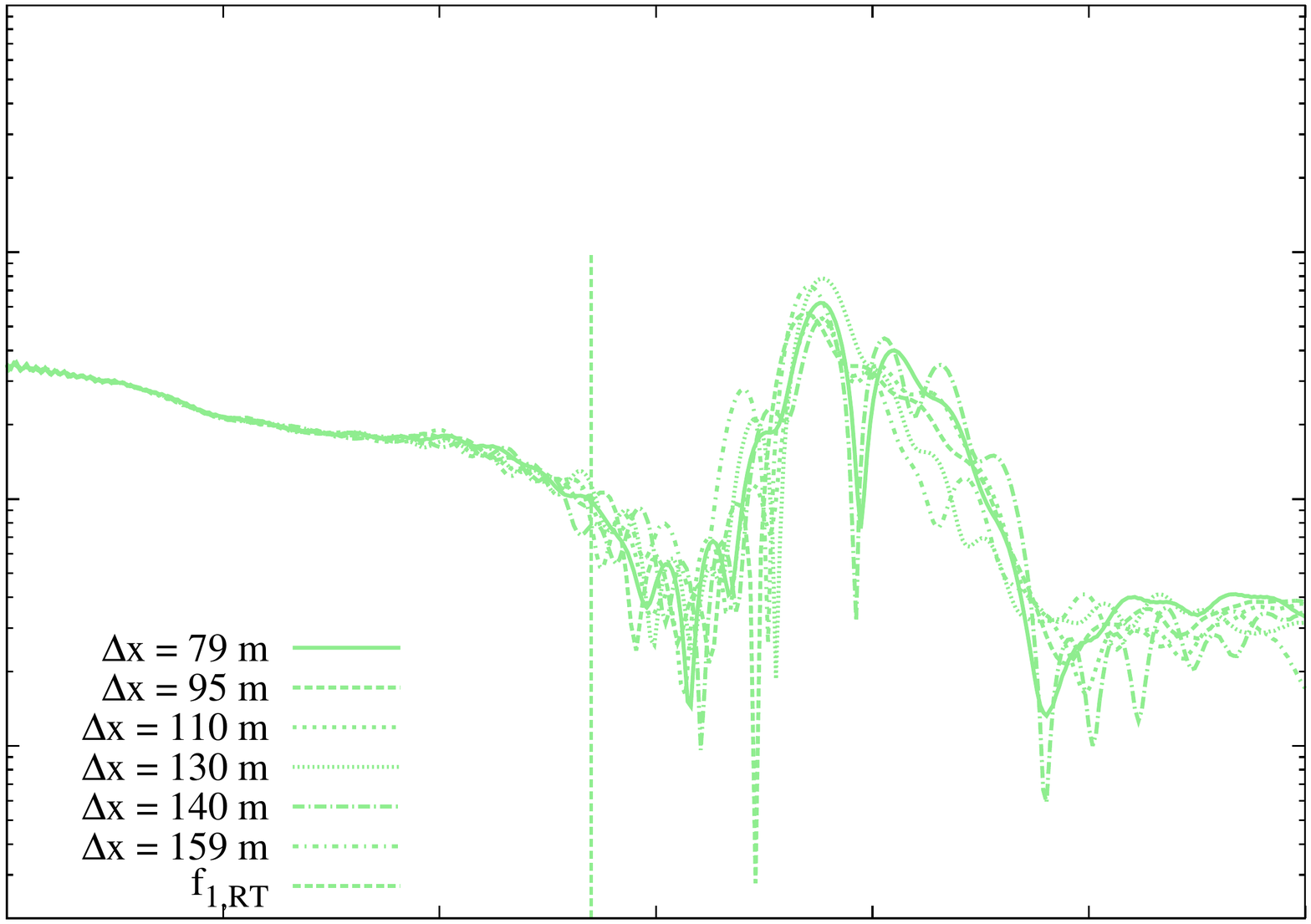}
\end{center}
\end{minipage}
\hspace{-13.35mm}
\begin{minipage}{0.27\hsize}
\begin{center}
\includegraphics[width=4.5cm,angle=0]{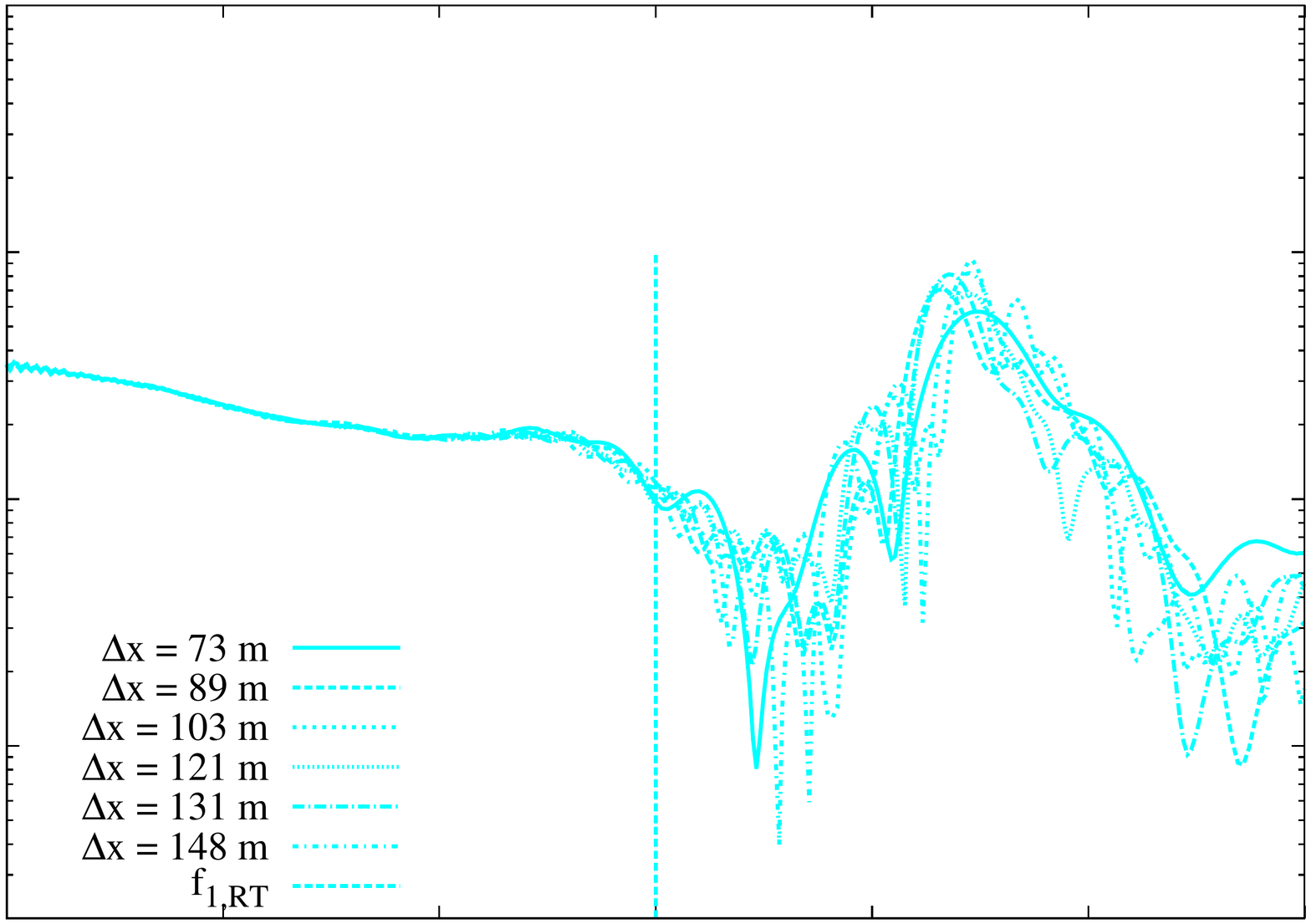}
\end{center}
\end{minipage}
\hspace{-13.35mm}
\begin{minipage}{0.27\hsize}
\begin{center}
\includegraphics[width=4.5cm,angle=0]{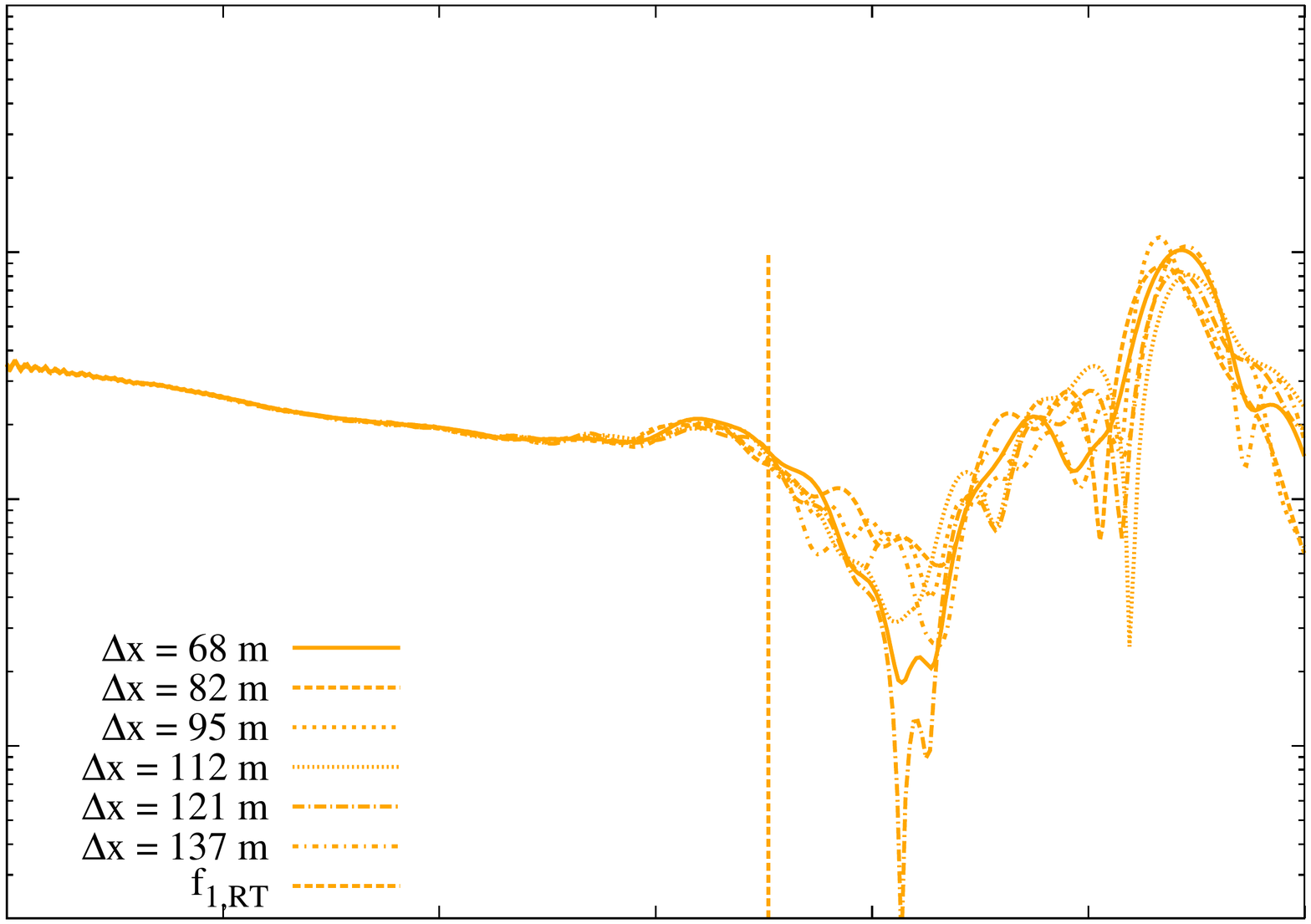}
\end{center}
\end{minipage}
\hspace{-13.35mm}
\begin{minipage}{0.27\hsize}
\begin{center}
\includegraphics[width=4.5cm,angle=0]{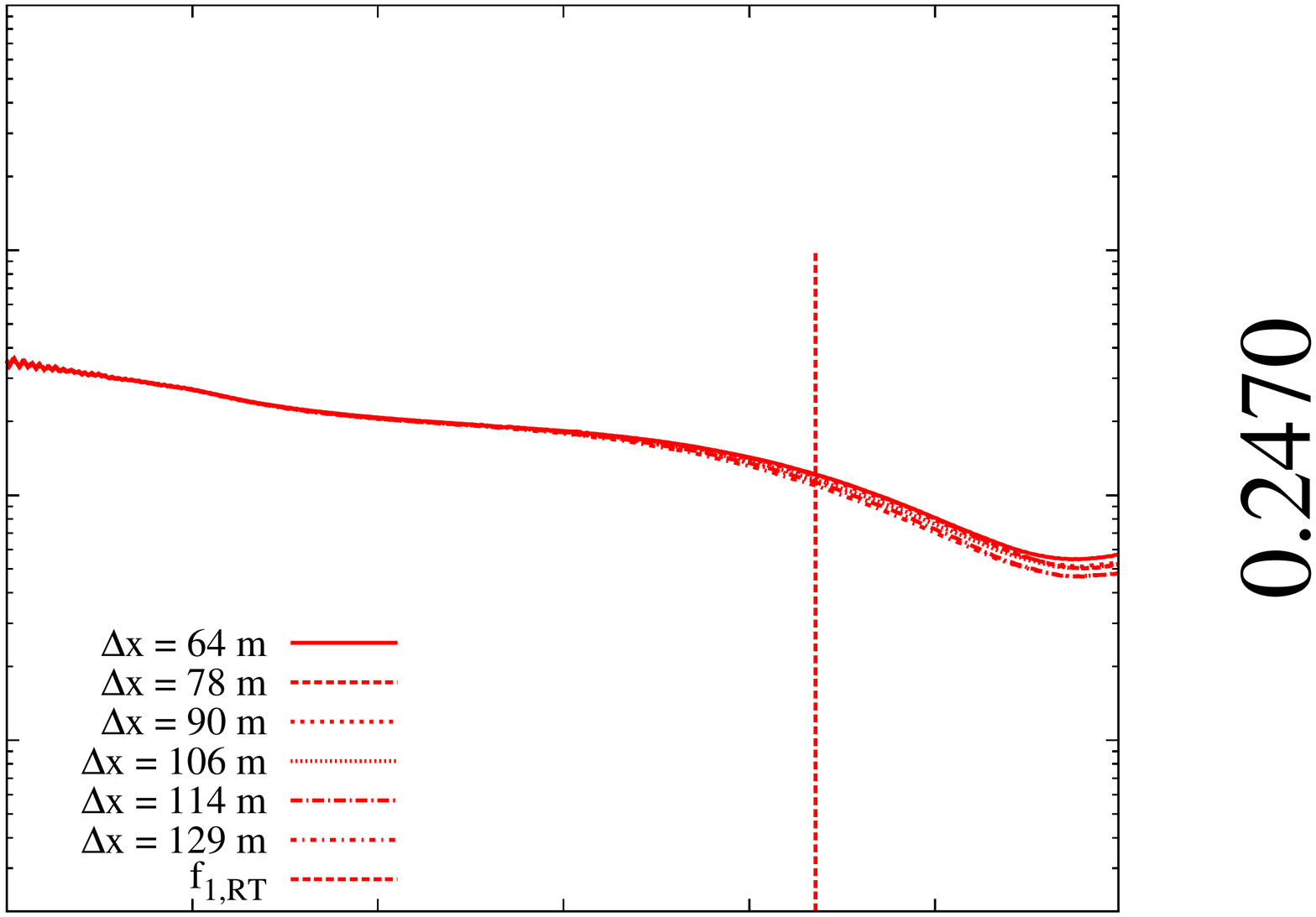}
\end{center}
\end{minipage}\\
\vspace{-9mm}
\hspace{-18.0mm}
\begin{minipage}{0.27\hsize}
\begin{center}
\includegraphics[width=4.5cm,angle=0]{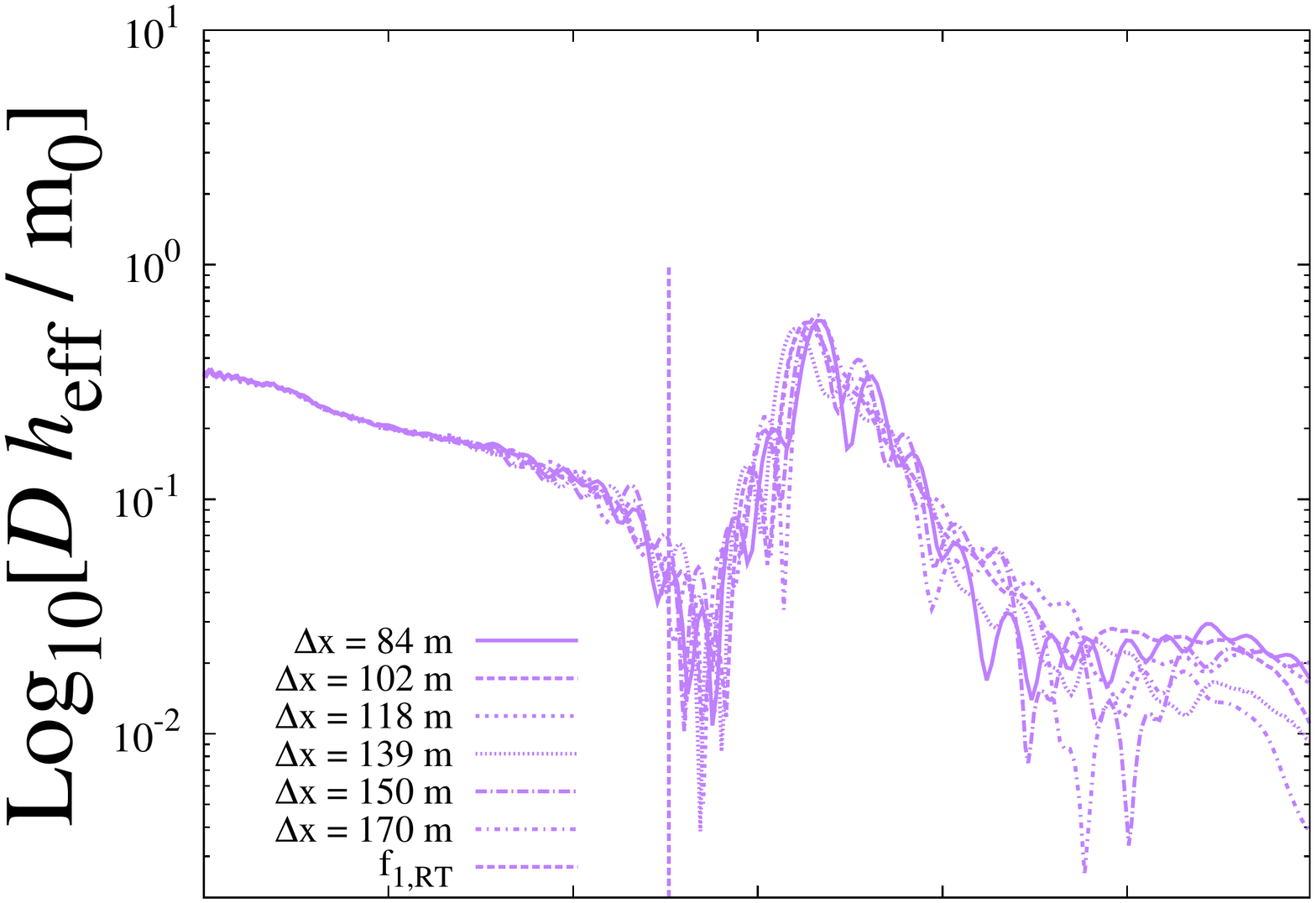}
\end{center}
\end{minipage}
\hspace{-13.35mm}
\begin{minipage}{0.27\hsize}
\begin{center}
\includegraphics[width=4.5cm,angle=0]{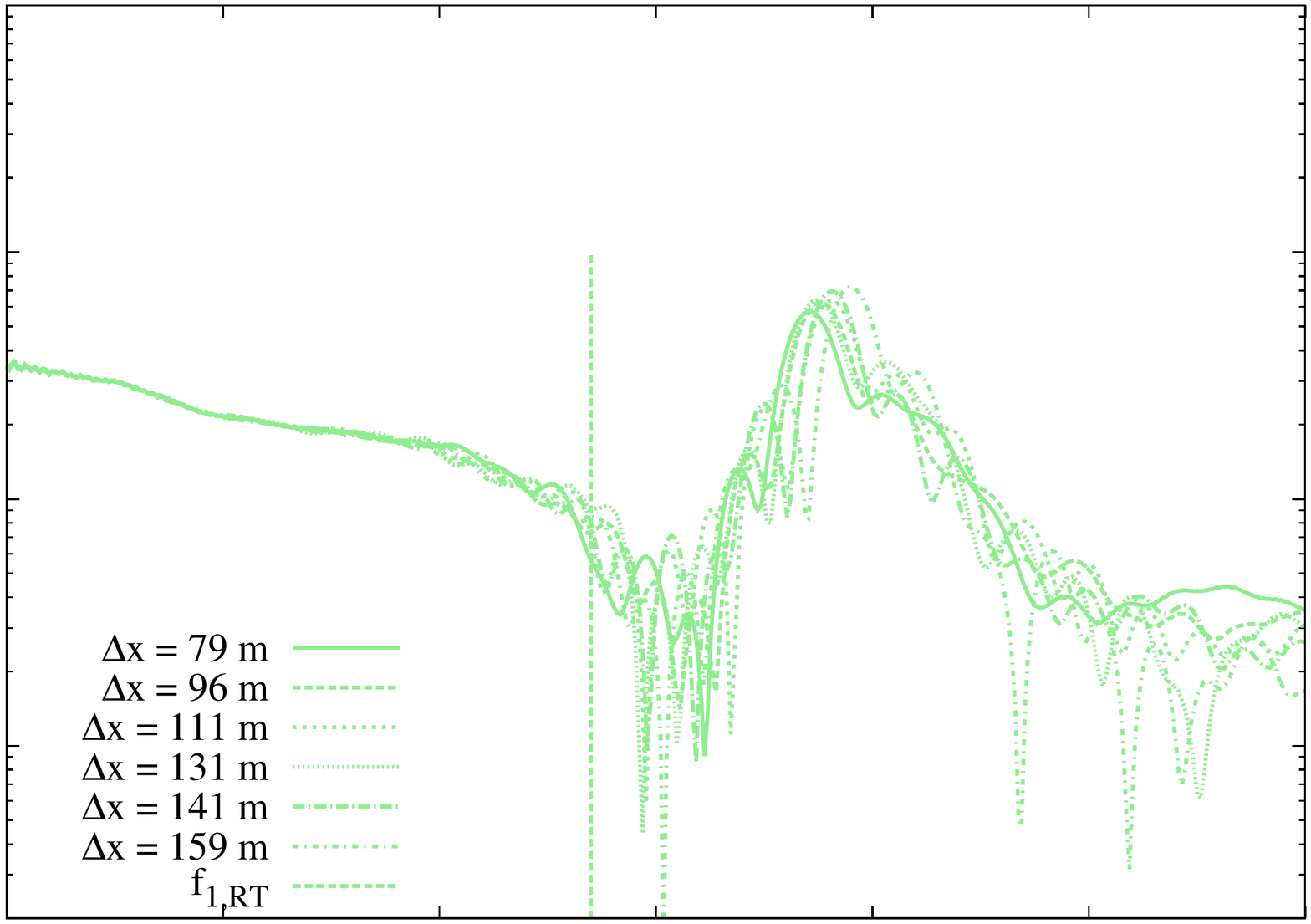}
\end{center}
\end{minipage}
\hspace{-13.35mm}
\begin{minipage}{0.27\hsize}
\begin{center}
\includegraphics[width=4.5cm,angle=0]{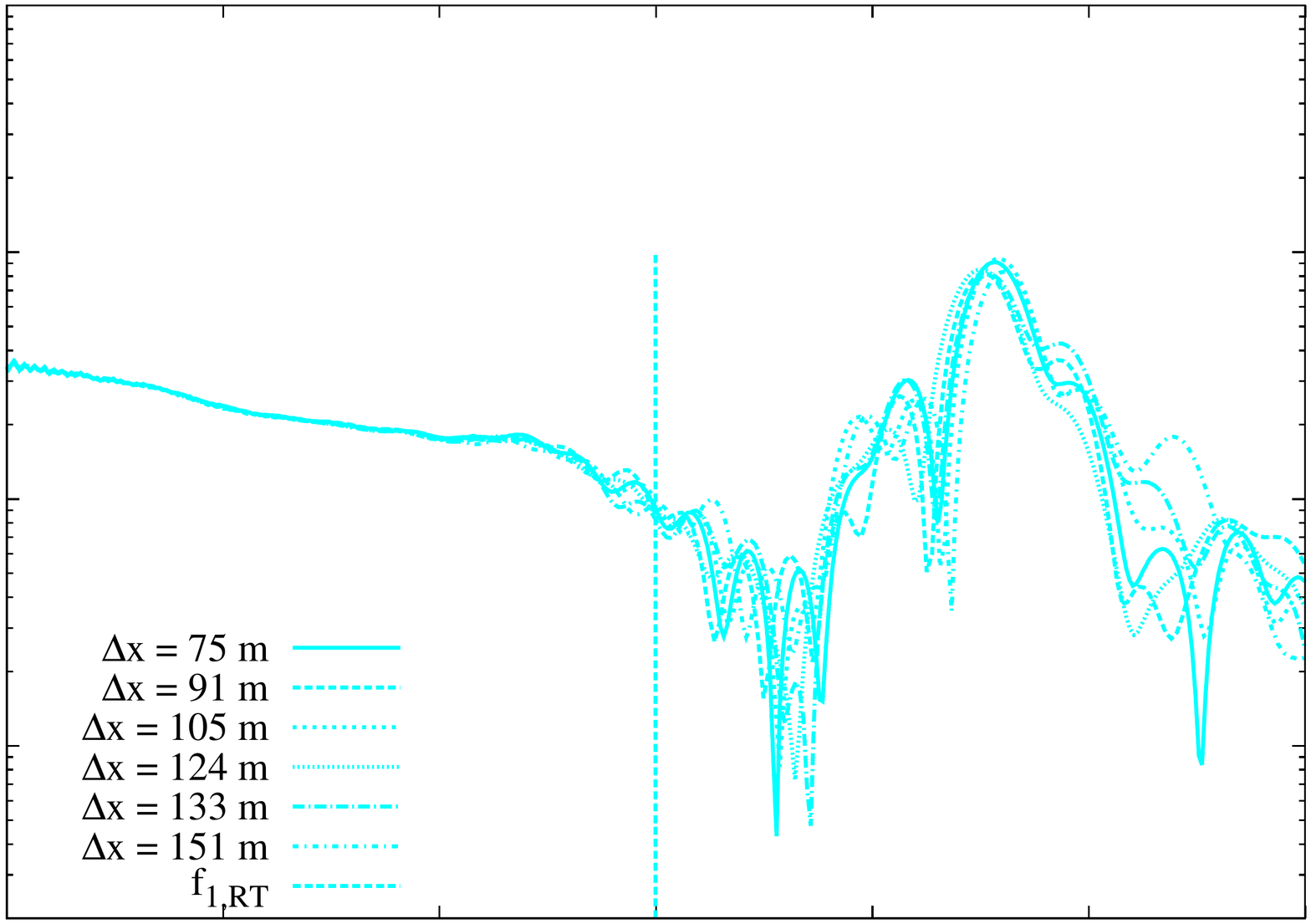}
\end{center}
\end{minipage}
\hspace{-13.35mm}
\begin{minipage}{0.27\hsize}
\begin{center}
\includegraphics[width=4.5cm,angle=0]{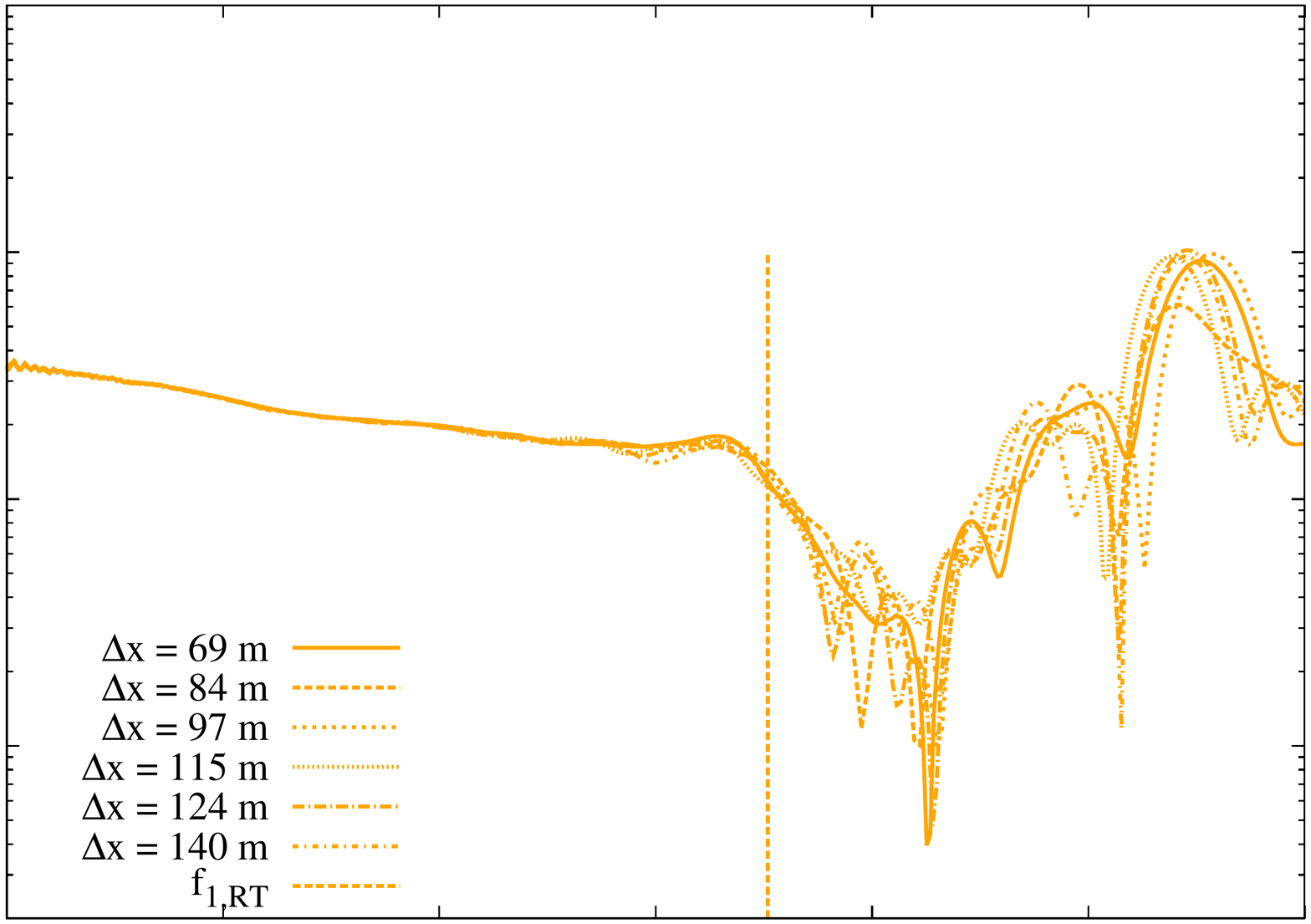}
\end{center}
\end{minipage}
\hspace{-13.35mm}
\begin{minipage}{0.27\hsize}
\begin{center}
\includegraphics[width=4.5cm,angle=0]{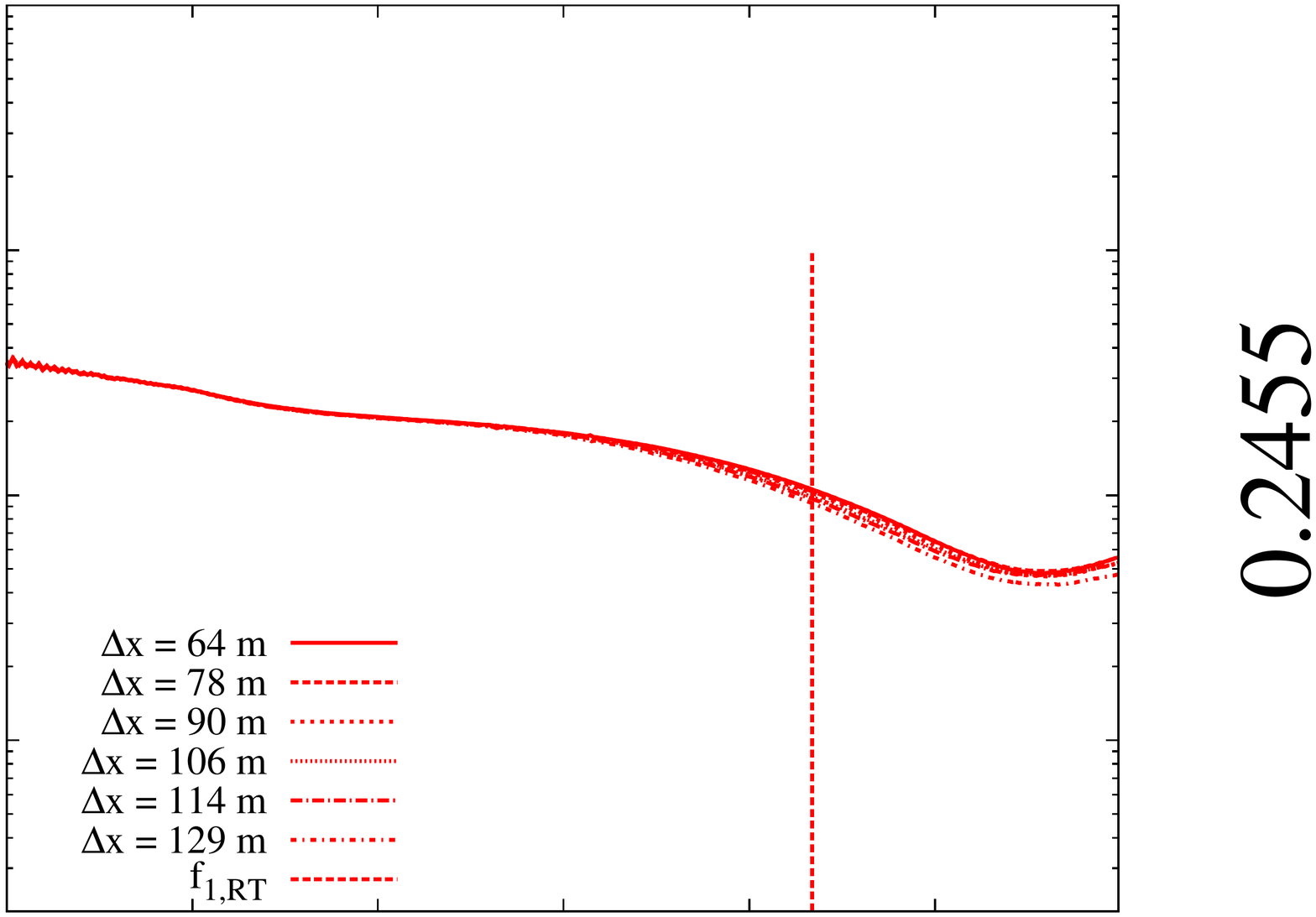}
\end{center}
\end{minipage}\\
\vspace{-9mm}
\hspace{-18.0mm}
\begin{minipage}{0.27\hsize}
\begin{center}
\includegraphics[width=4.5cm,angle=0]{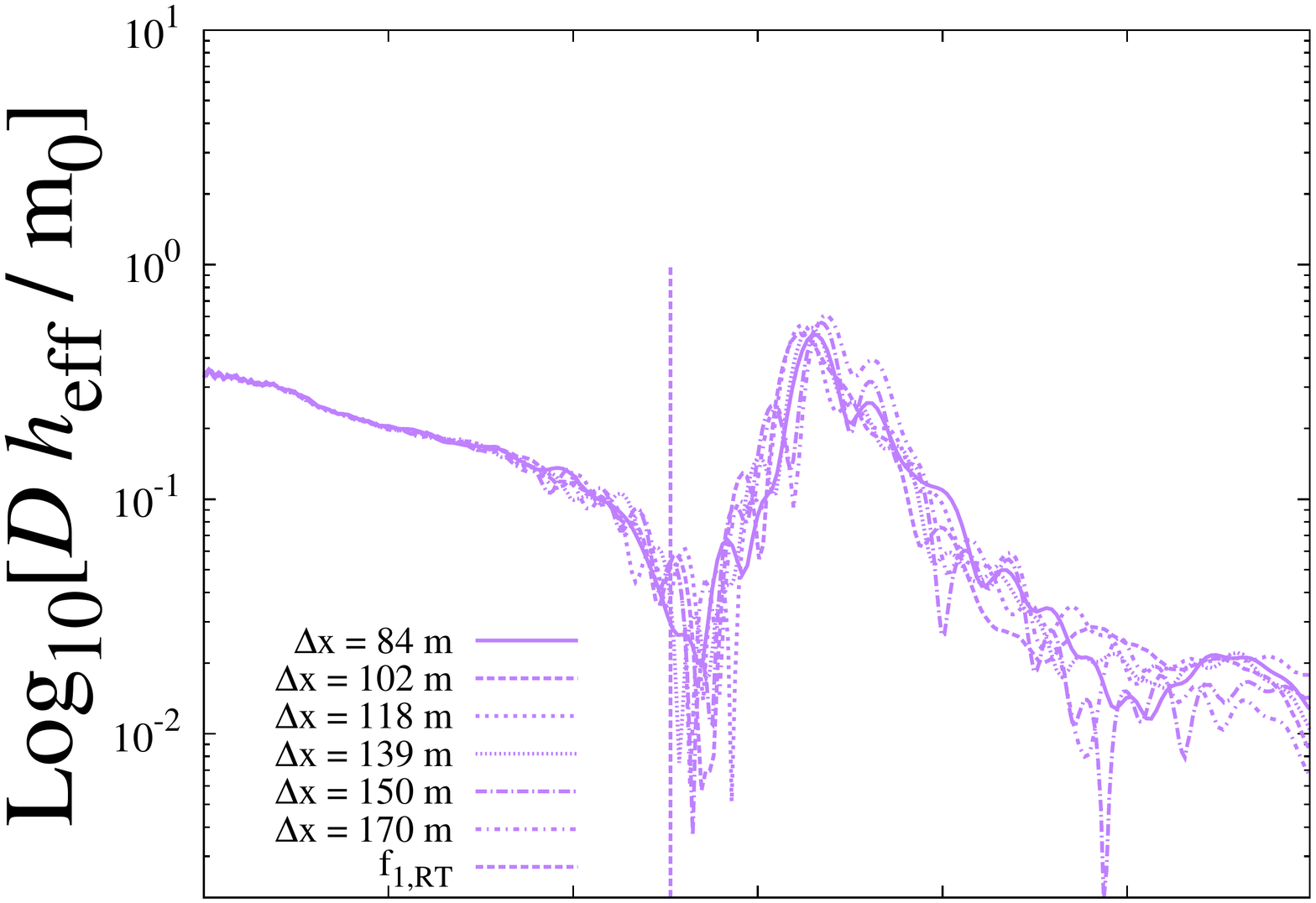}
\end{center}
\end{minipage}
\hspace{-13.35mm}
\begin{minipage}{0.27\hsize}
\begin{center}
\includegraphics[width=4.5cm,angle=0]{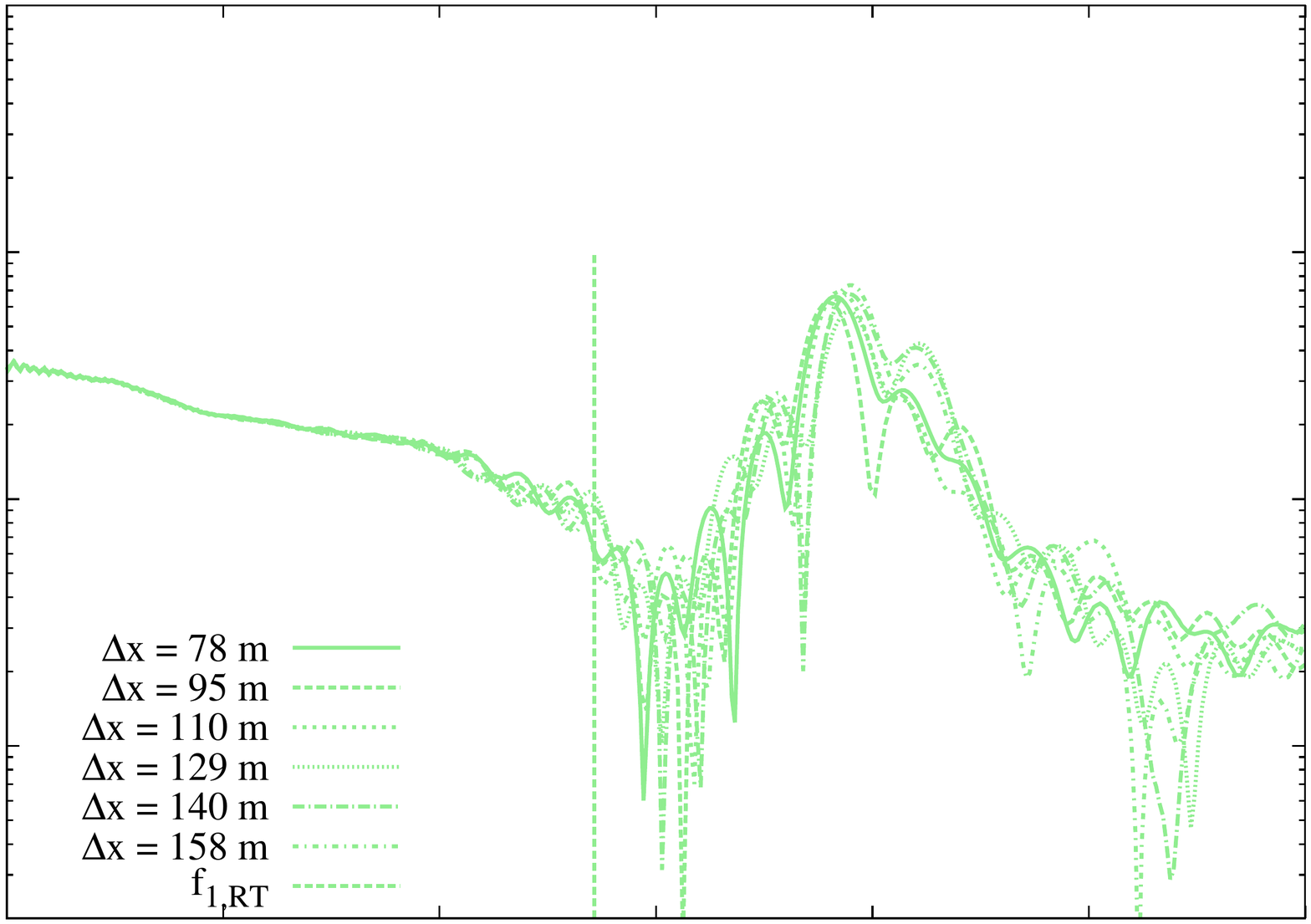}
\end{center}
\end{minipage}
\hspace{-13.35mm}
\begin{minipage}{0.27\hsize}
\begin{center}
\includegraphics[width=4.5cm,angle=0]{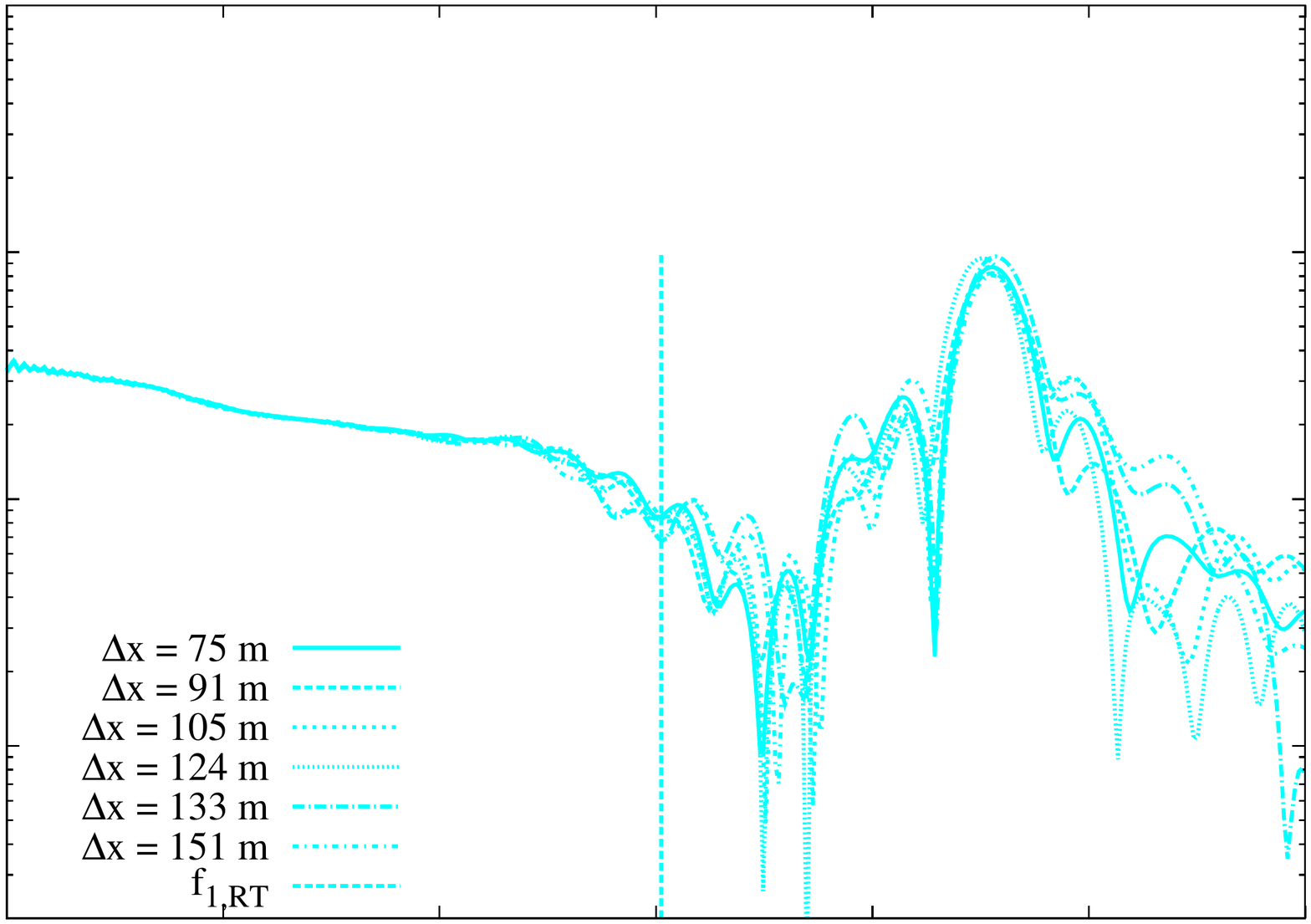}
\end{center}
\end{minipage}
\hspace{-13.35mm}
\begin{minipage}{0.27\hsize}
\begin{center}
\includegraphics[width=4.5cm,angle=0]{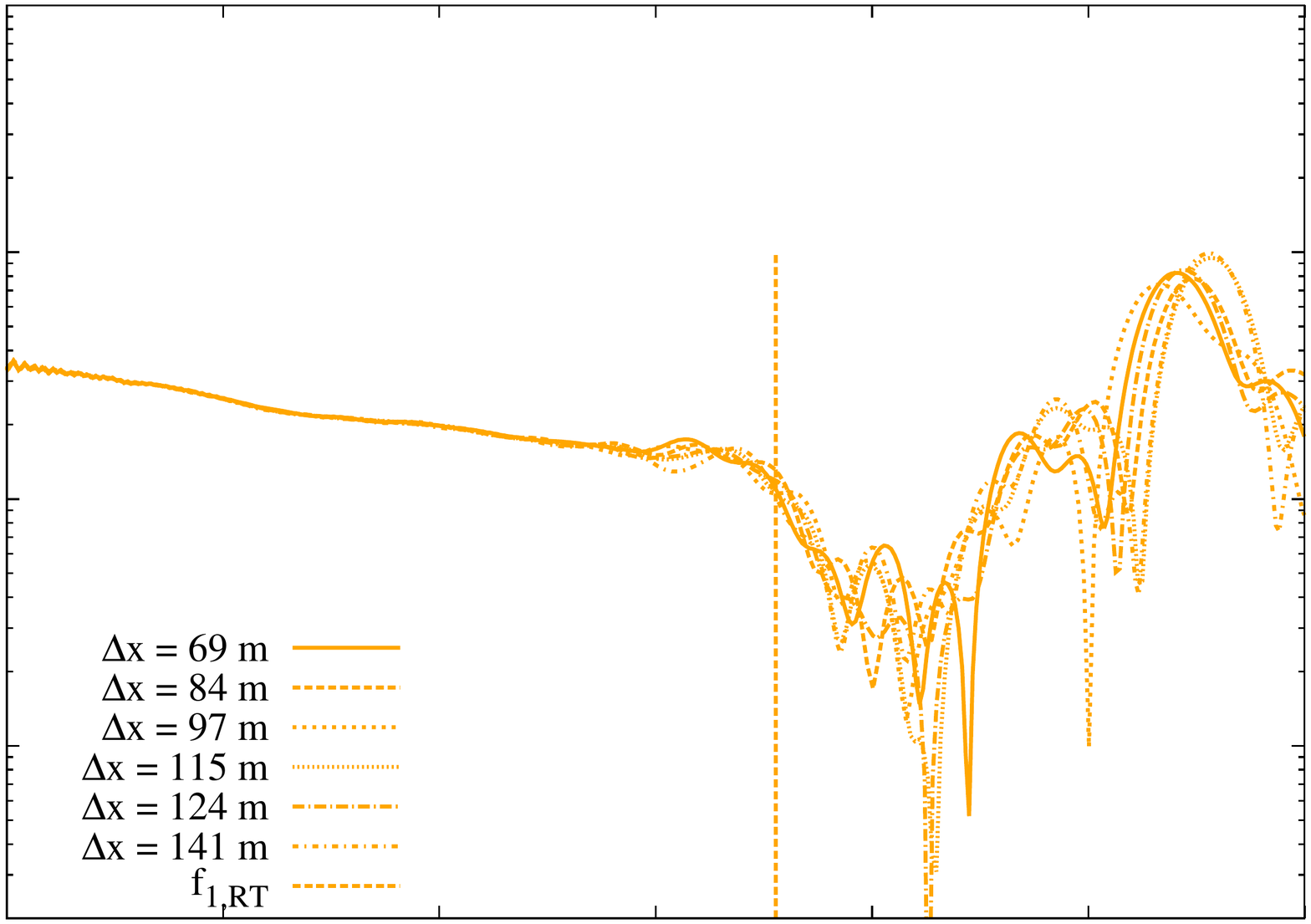}
\end{center}
\end{minipage}
\hspace{-13.35mm}
\begin{minipage}{0.27\hsize}
\begin{center}
\includegraphics[width=4.5cm,angle=0]{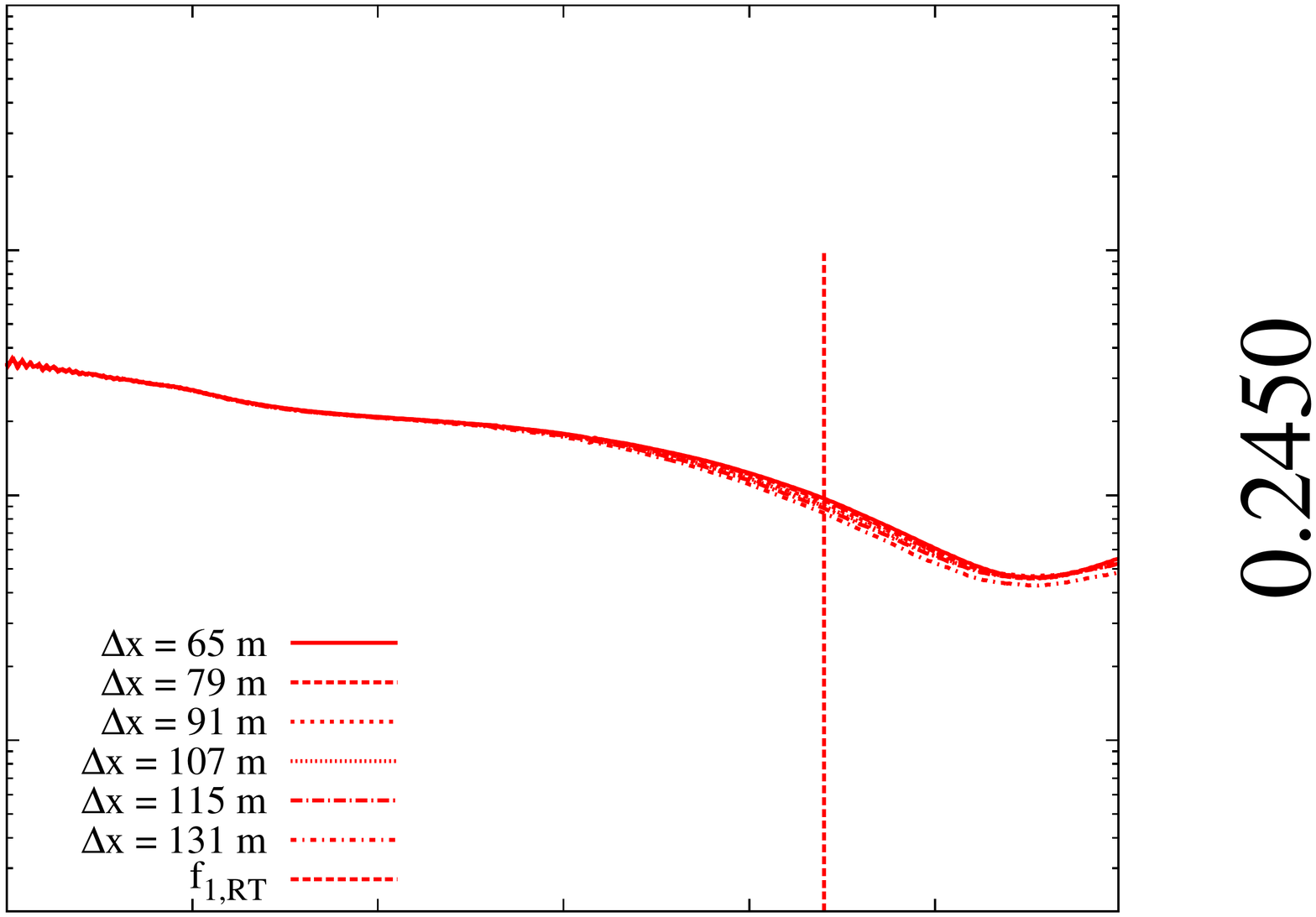}
\end{center}
\end{minipage}\\
\vspace{-9mm}
\hspace{-18.0mm}
\begin{minipage}{0.27\hsize}
\begin{center}
\includegraphics[width=4.5cm,angle=0]{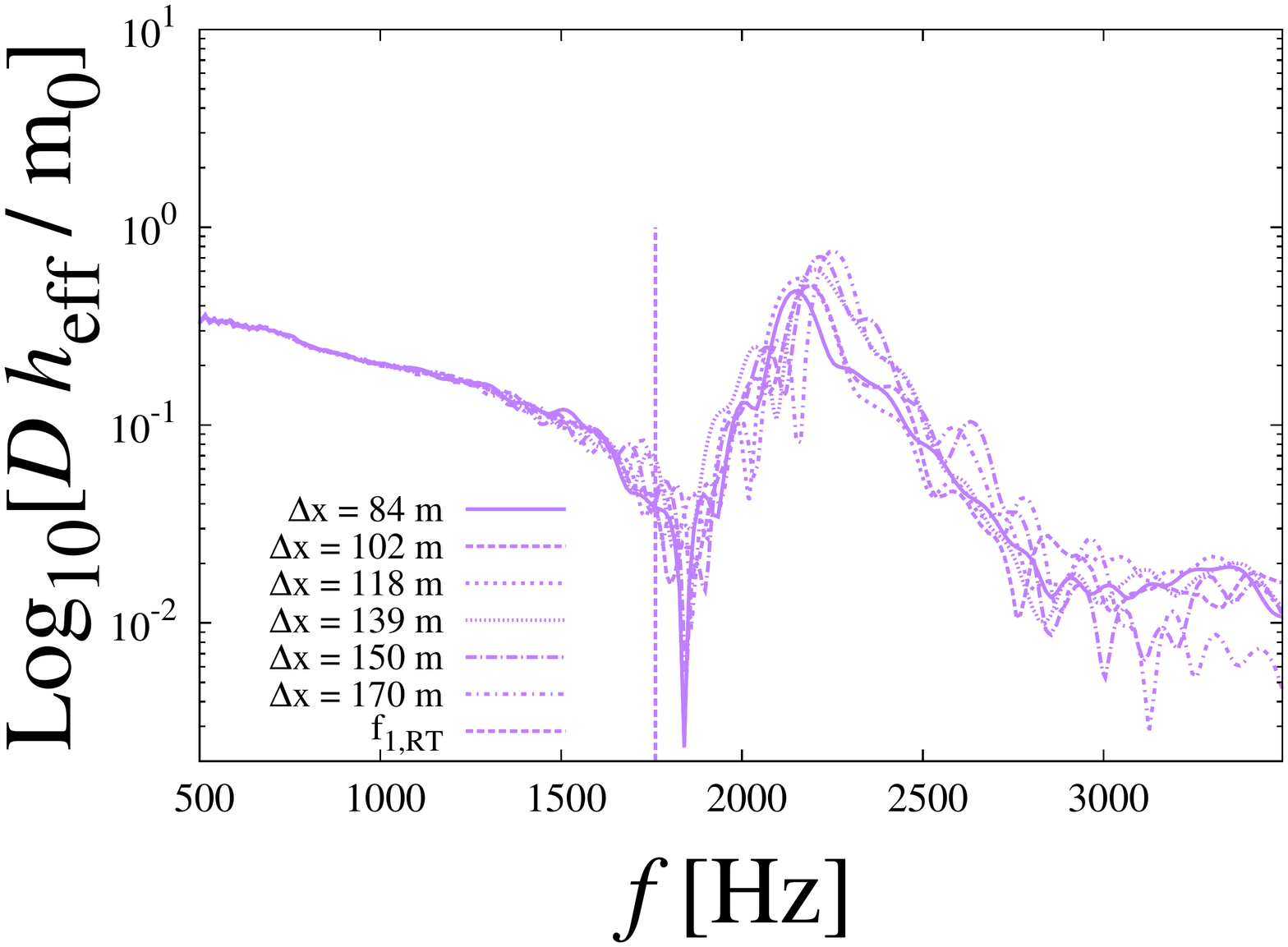}
\end{center}
\end{minipage}
\hspace{-13.35mm}
\begin{minipage}{0.27\hsize}
\begin{center}
\includegraphics[width=4.5cm,angle=0]{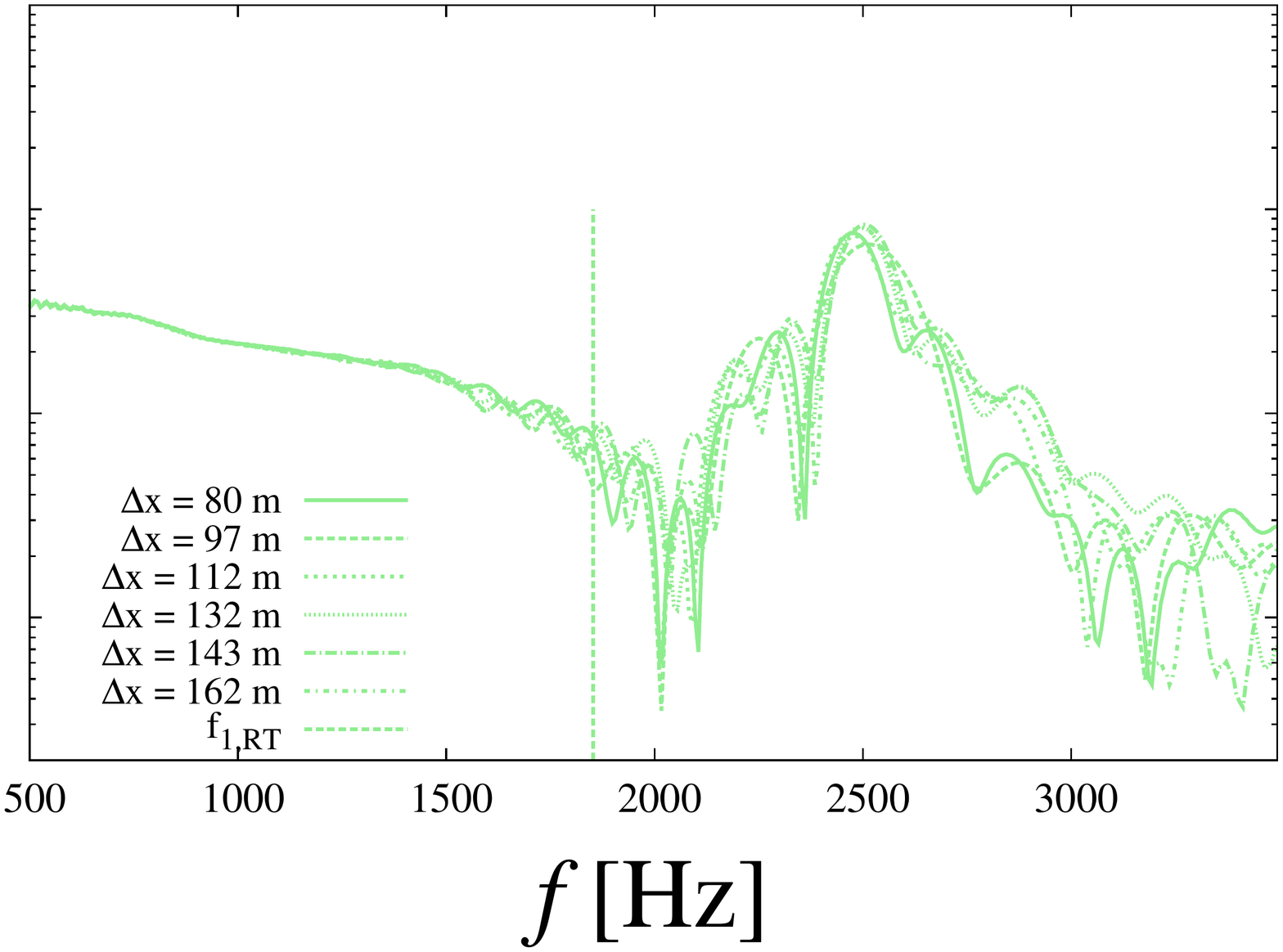}
\end{center}
\end{minipage}
\hspace{-13.35mm}
\begin{minipage}{0.27\hsize}
\begin{center}
\includegraphics[width=4.5cm,angle=0]{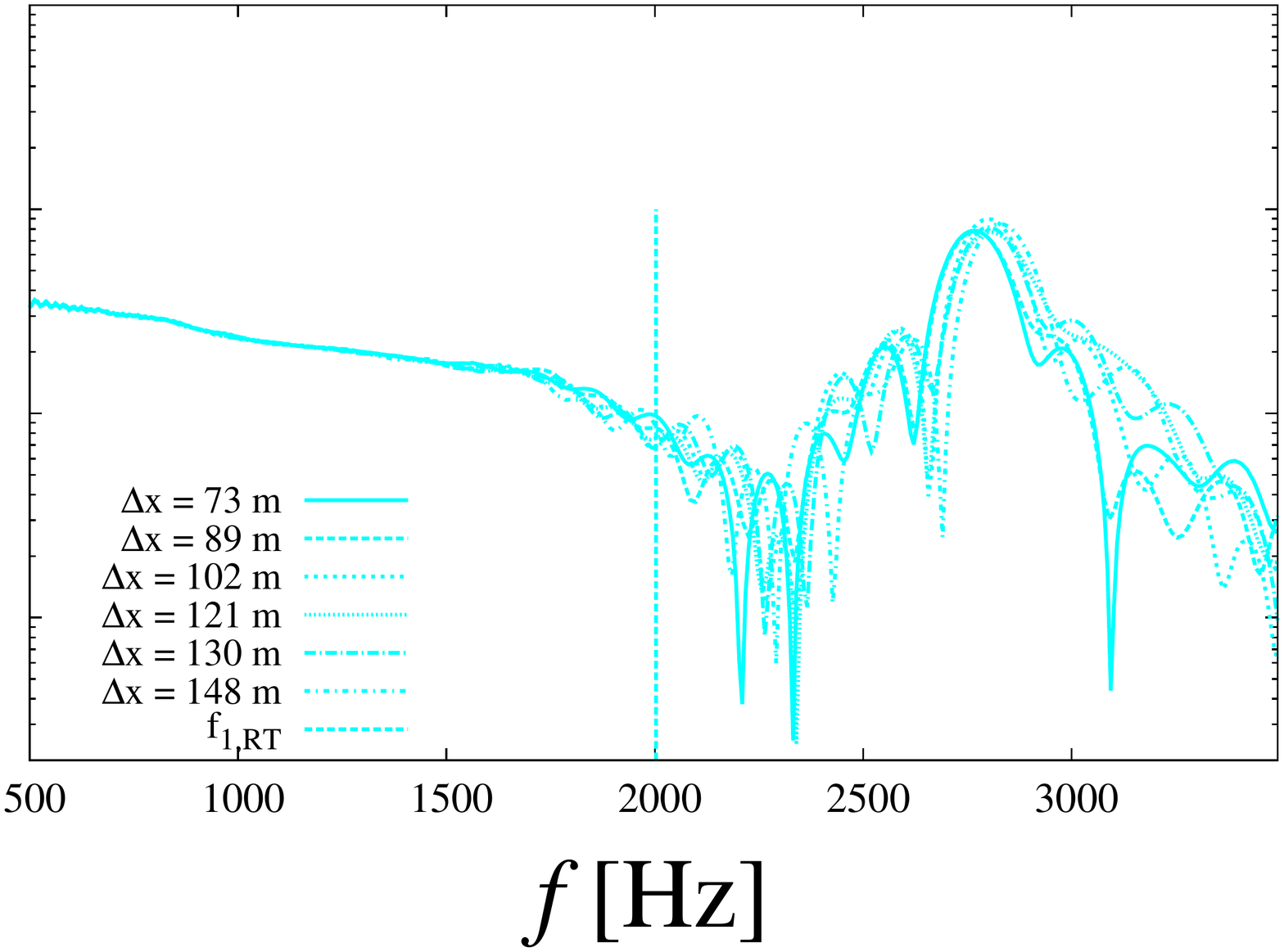}
\end{center}
\end{minipage}
\hspace{-13.35mm}
\begin{minipage}{0.27\hsize}
\begin{center}
\includegraphics[width=4.5cm,angle=0]{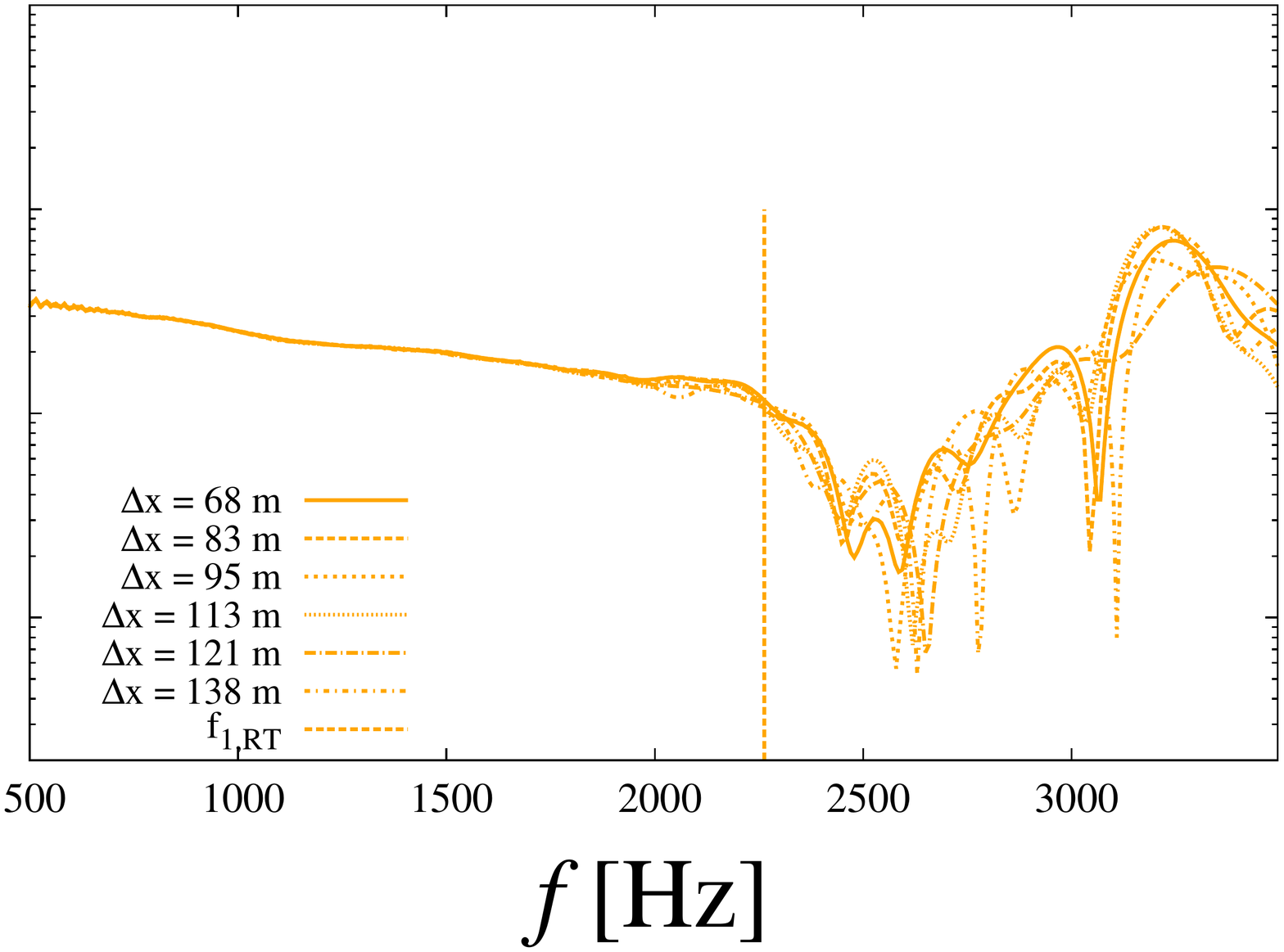}
\end{center}
\end{minipage}
\hspace{-13.35mm}
\begin{minipage}{0.27\hsize}
\begin{center}
\includegraphics[width=4.5cm,angle=0]{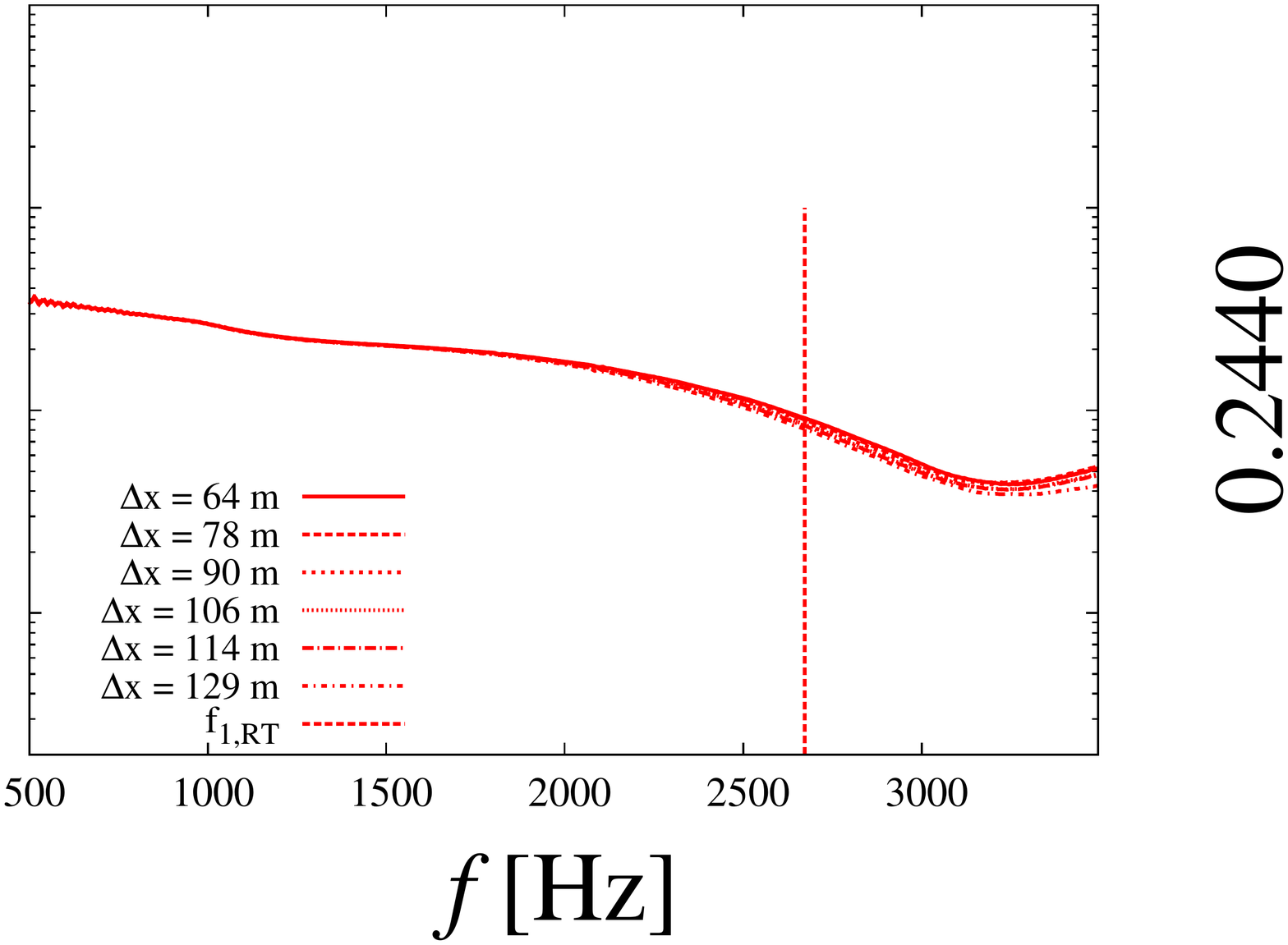}
\end{center}
\end{minipage}\\
\caption{\label{fig:PSD}Spectrum amplitudes of gravitational waves for the binary systems with ${\cal M}_c= 1.1752M_\odot$. The number attached in the right-hand side vertical axis is the symmetric mass ratio $\eta$. 
  We also show $f_1$ frequency proposed in Ref.~\cite{Rezzolla:2016nxn} with vertical dashed lines. 
  For completeness, we also show the systems reported in Refs.~\cite{Kiuchi:2017pte,Kawaguchi:2018gvj}.
}
\end{figure*}

\begin{figure*}[t]
\hspace{-18.0mm}
\begin{minipage}{0.27\hsize}
\begin{center}
\includegraphics[width=4.5cm,angle=0]{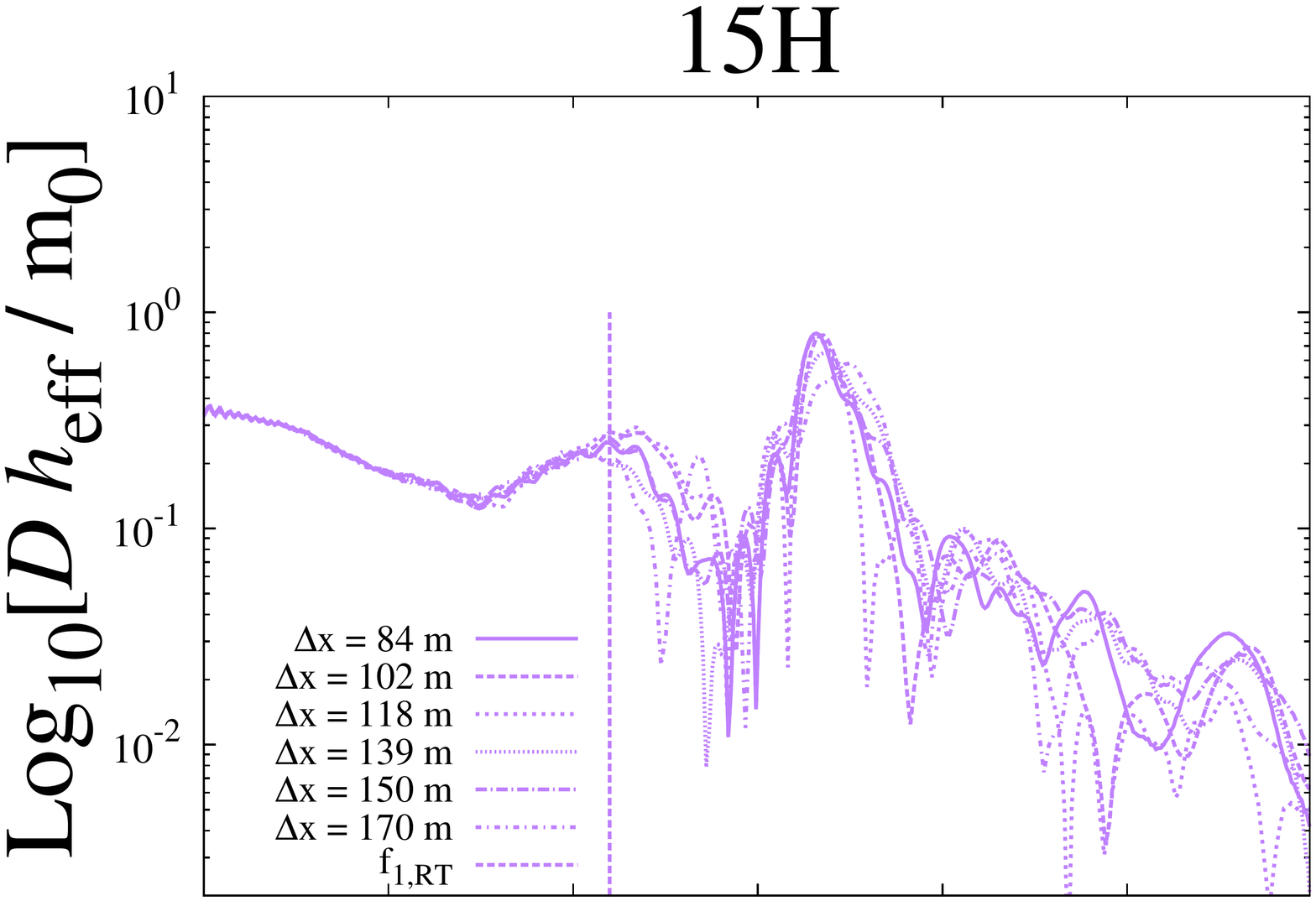}
\end{center}
\end{minipage}
\hspace{-13.35mm}
\begin{minipage}{0.27\hsize}
\begin{center}
\includegraphics[width=4.5cm,angle=0]{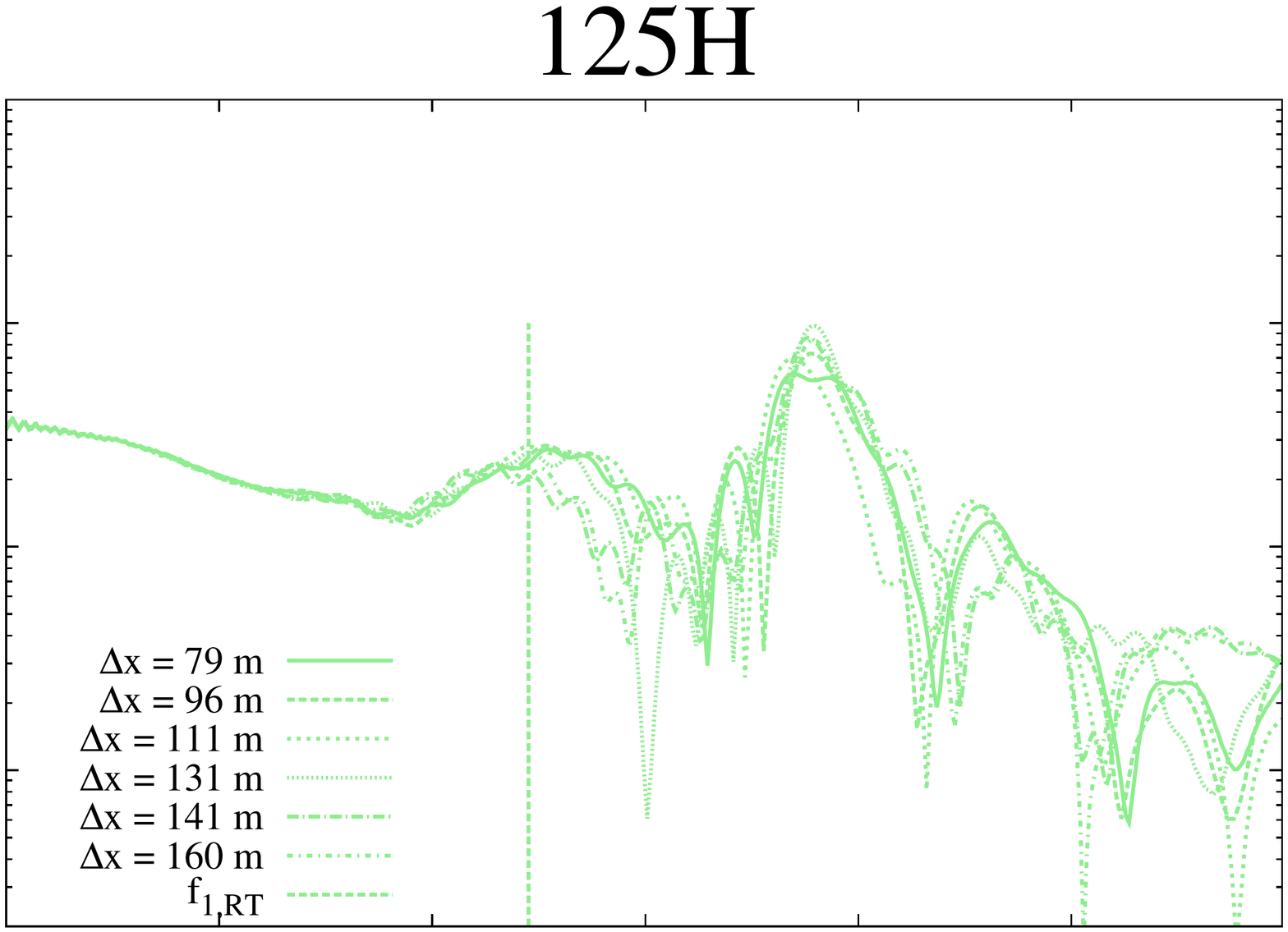}
\end{center}
\end{minipage}
\hspace{-13.35mm}
\begin{minipage}{0.27\hsize}
\begin{center}
\includegraphics[width=4.5cm,angle=0]{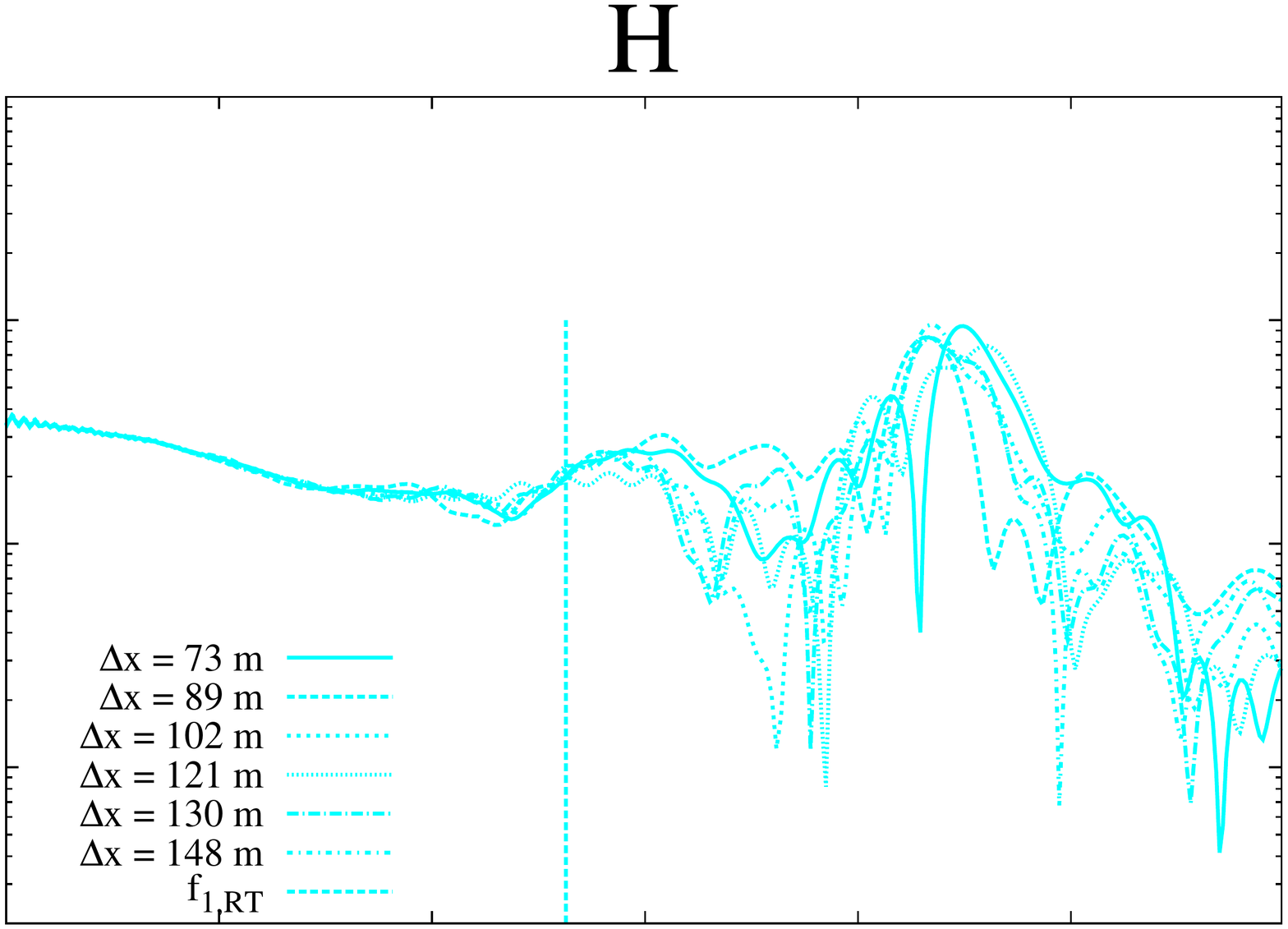}
\end{center}
\end{minipage}
\hspace{-13.35mm}
\begin{minipage}{0.27\hsize}
\begin{center}
\includegraphics[width=4.5cm,angle=0]{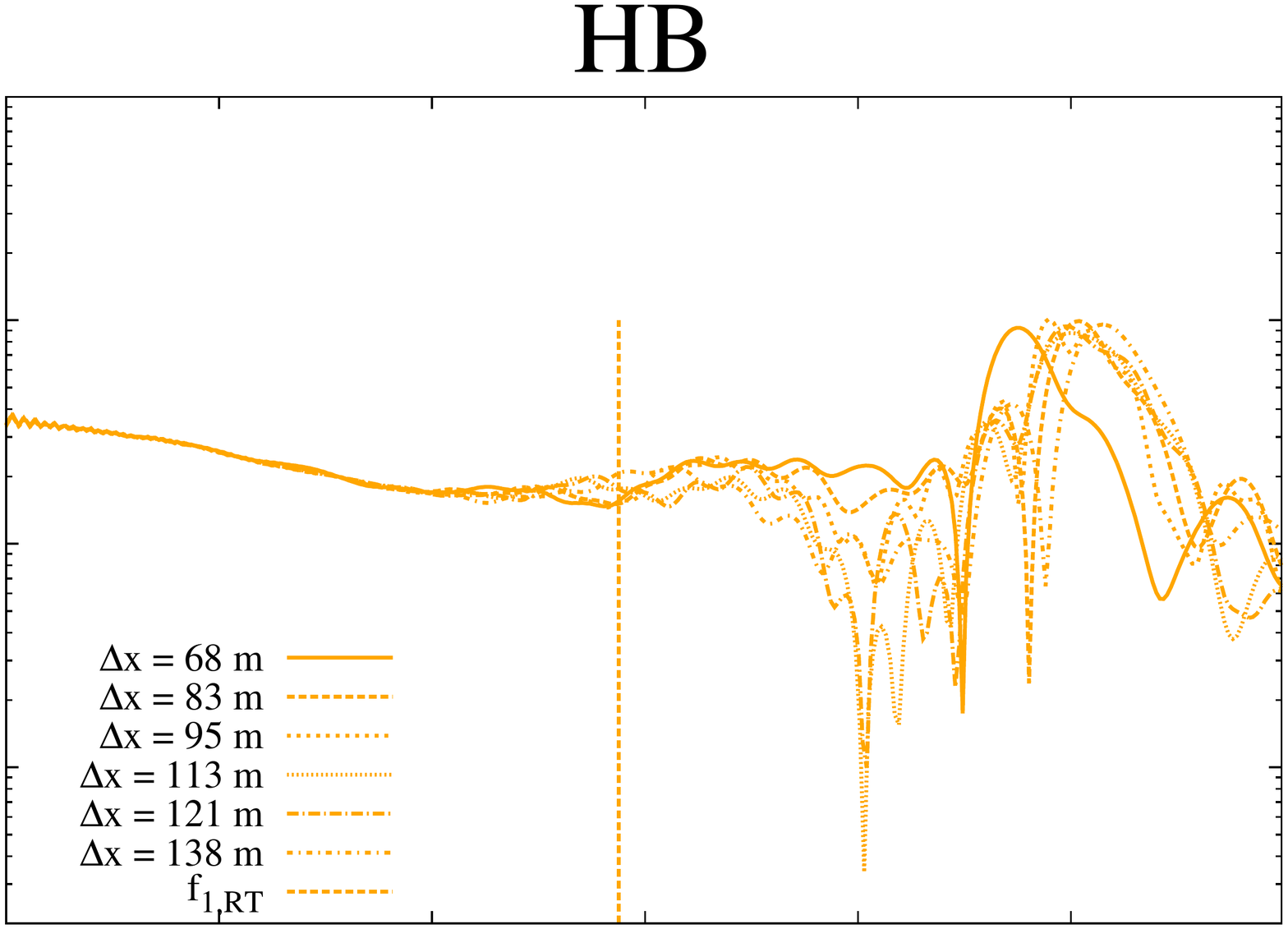}
\end{center}
\end{minipage}
\hspace{-13.35mm}
\begin{minipage}{0.27\hsize}
\begin{center}
\includegraphics[width=4.5cm,angle=0]{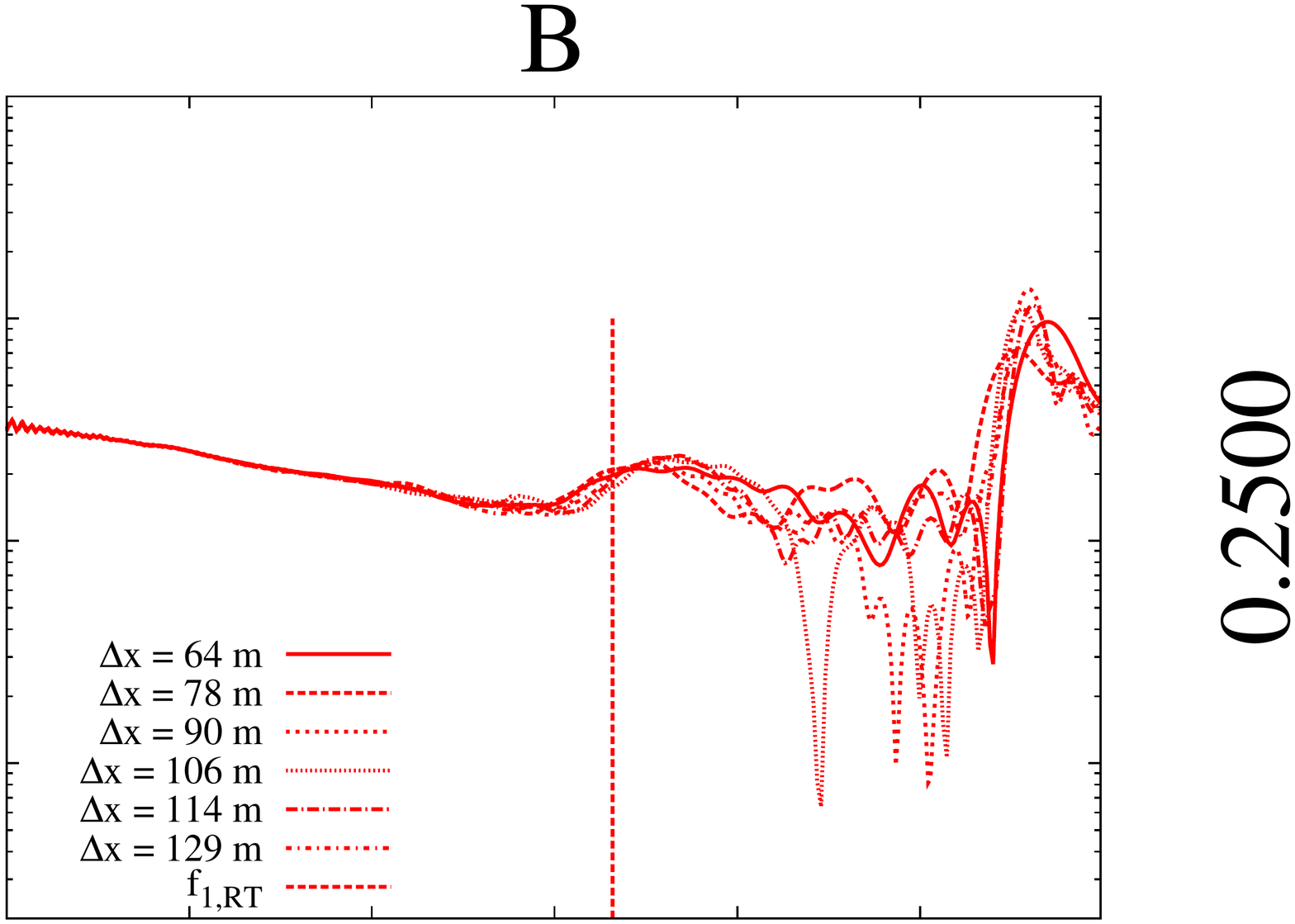}
\end{center}
\end{minipage}\\
\vspace{-9mm}
\hspace{-18.0mm}
\begin{minipage}{0.27\hsize}
\begin{center}
\includegraphics[width=4.5cm,angle=0]{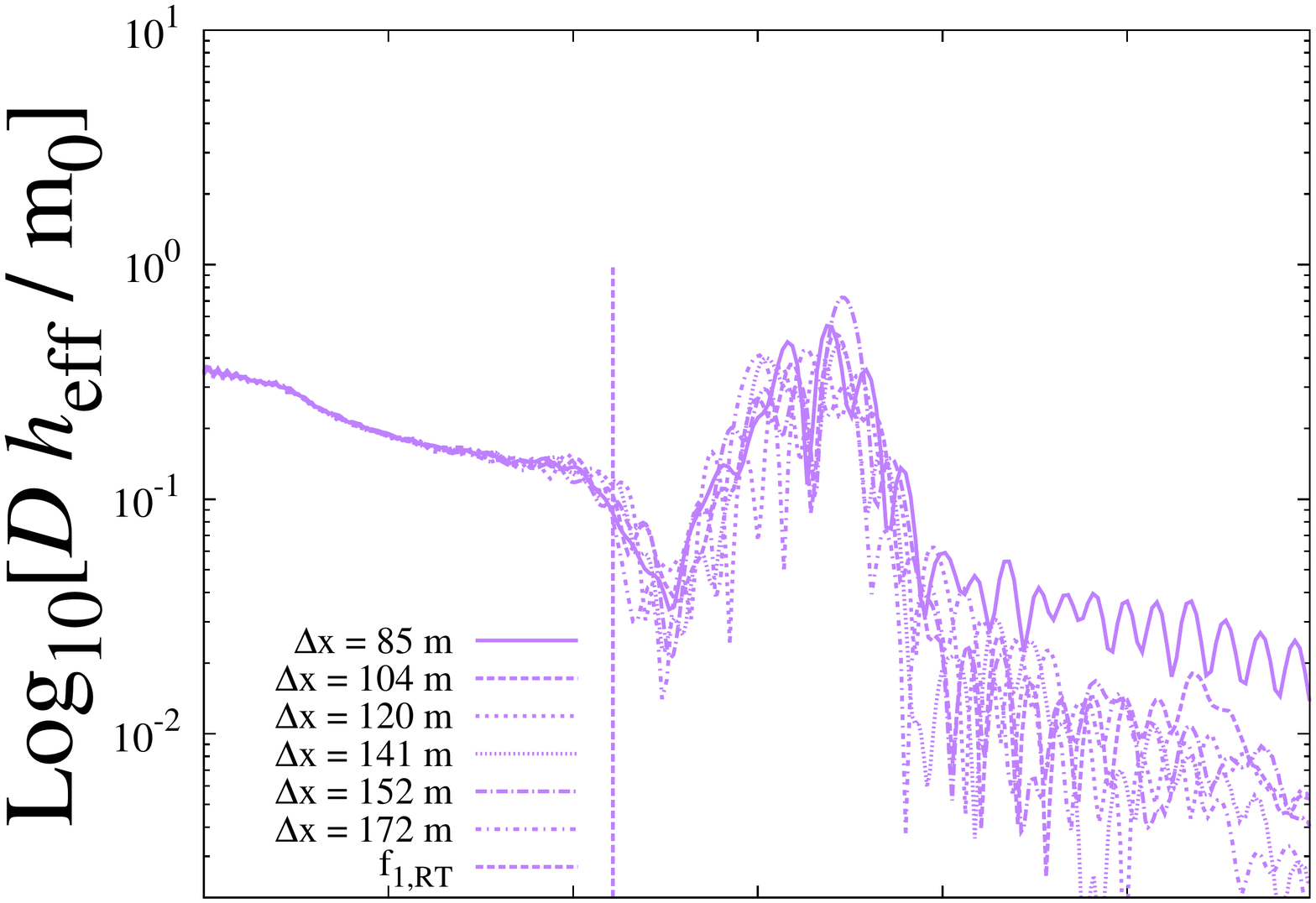}
\end{center}
\end{minipage}
\hspace{-13.35mm}
\begin{minipage}{0.27\hsize}
\begin{center}
\includegraphics[width=4.5cm,angle=0]{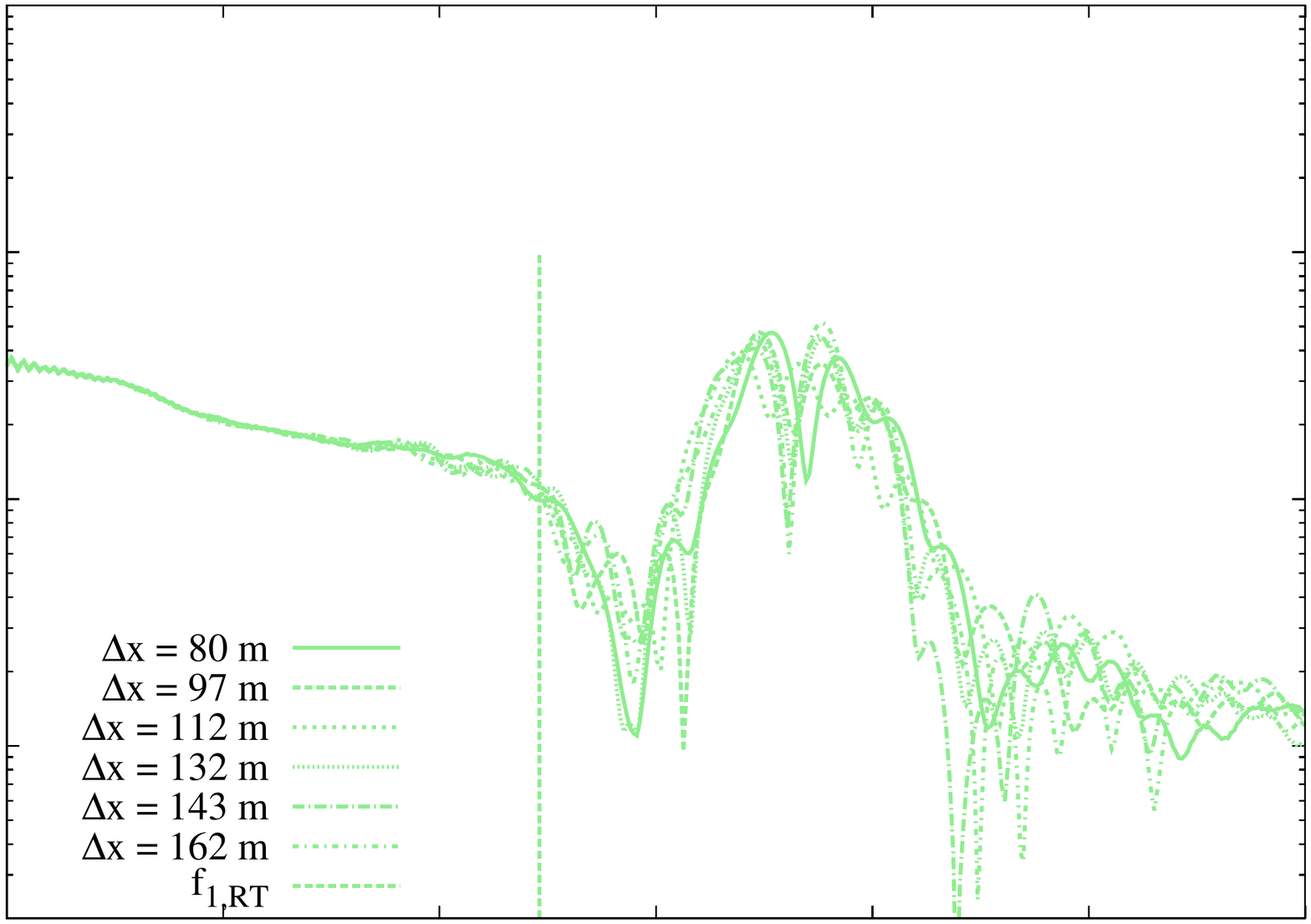}
\end{center}
\end{minipage}
\hspace{-13.35mm}
\begin{minipage}{0.27\hsize}
\begin{center}
\includegraphics[width=4.5cm,angle=0]{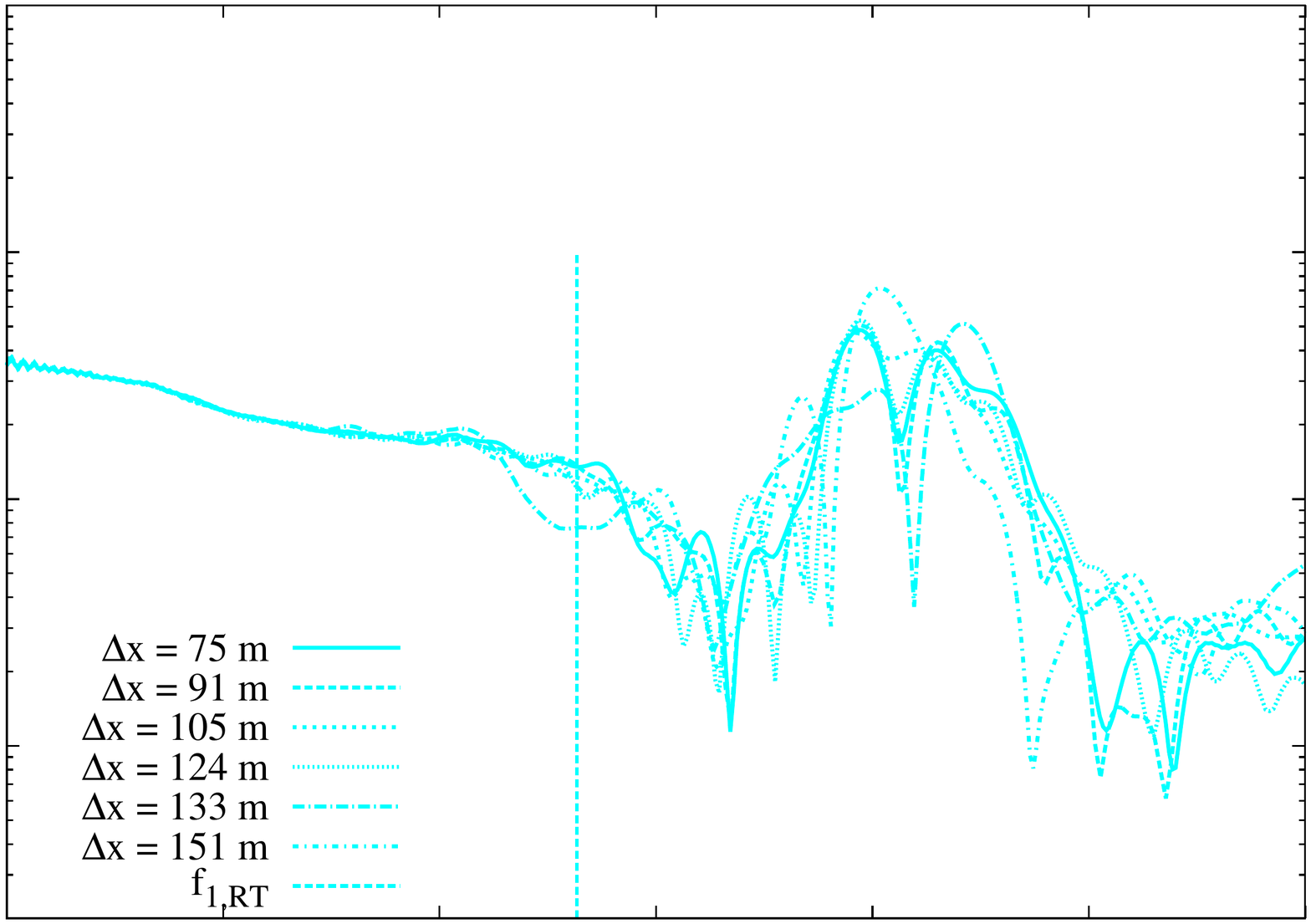}
\end{center}
\end{minipage}
\hspace{-13.35mm}
\begin{minipage}{0.27\hsize}
\begin{center}
\includegraphics[width=4.5cm,angle=0]{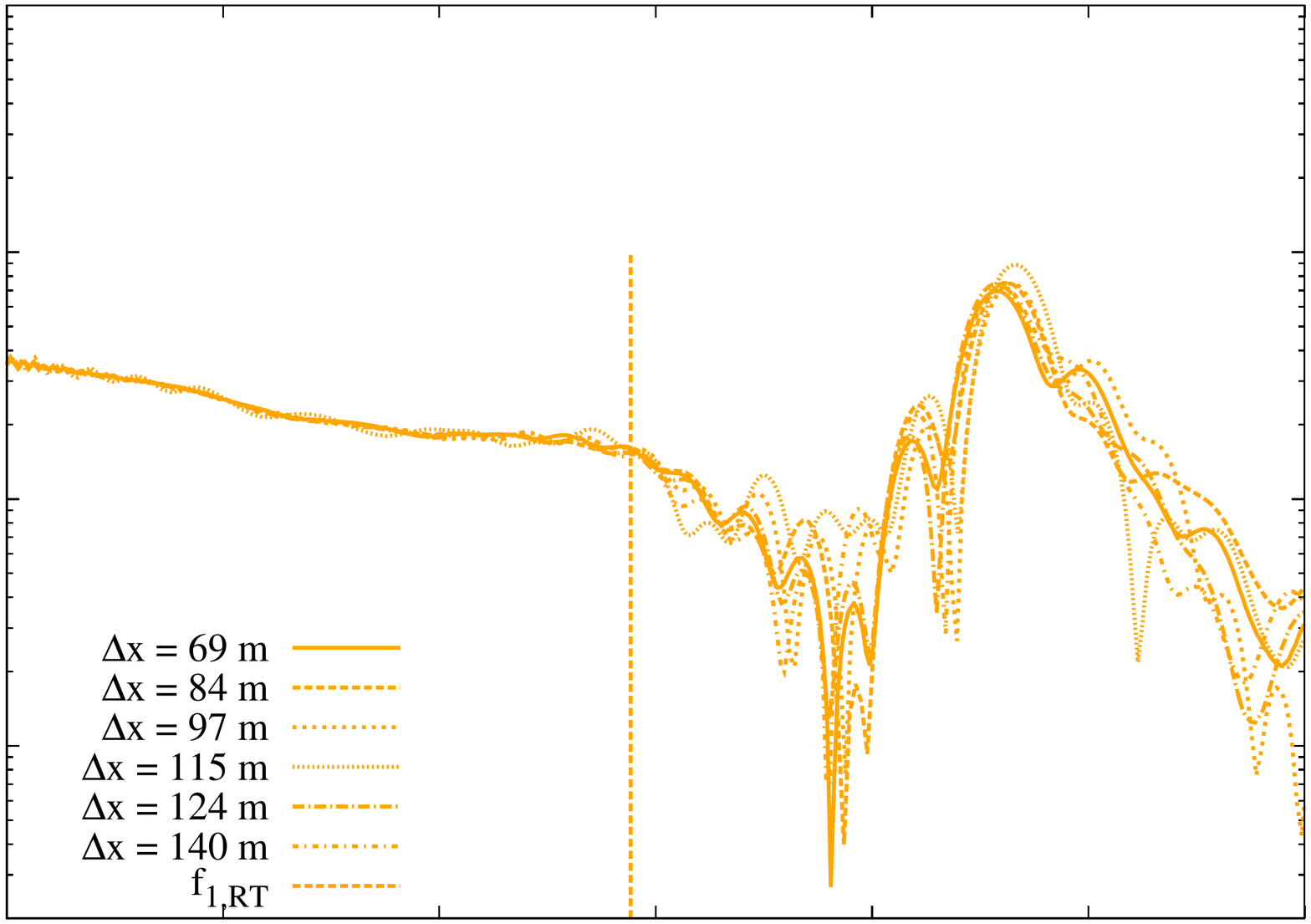}
\end{center}
\end{minipage}
\hspace{-13.35mm}
\begin{minipage}{0.27\hsize}
\begin{center}
\includegraphics[width=4.5cm,angle=0]{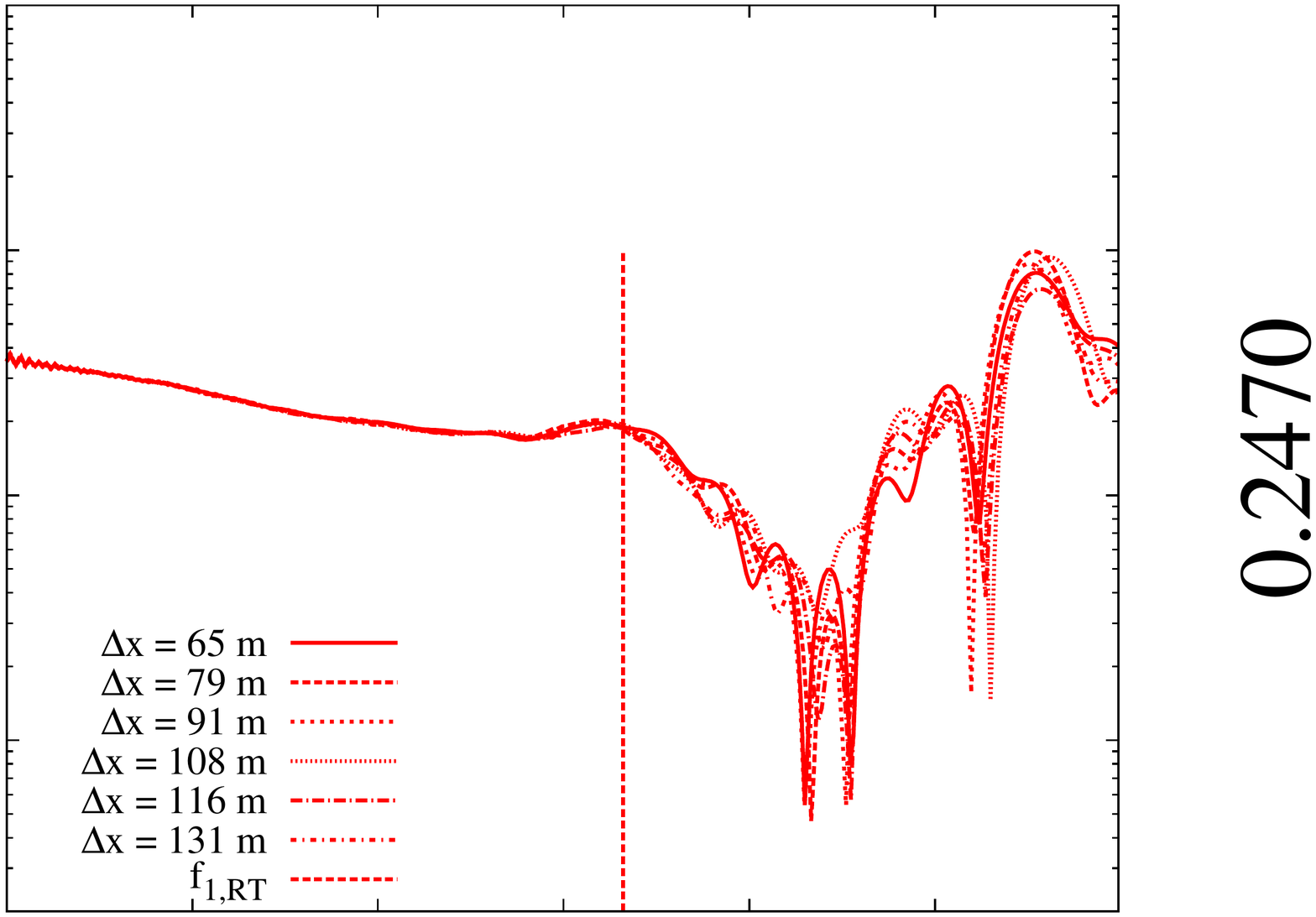}
\end{center}
\end{minipage}\\
\vspace{-9mm}
\hspace{-18.0mm}
\begin{minipage}{0.27\hsize}
\begin{center}
\includegraphics[width=4.5cm,angle=0]{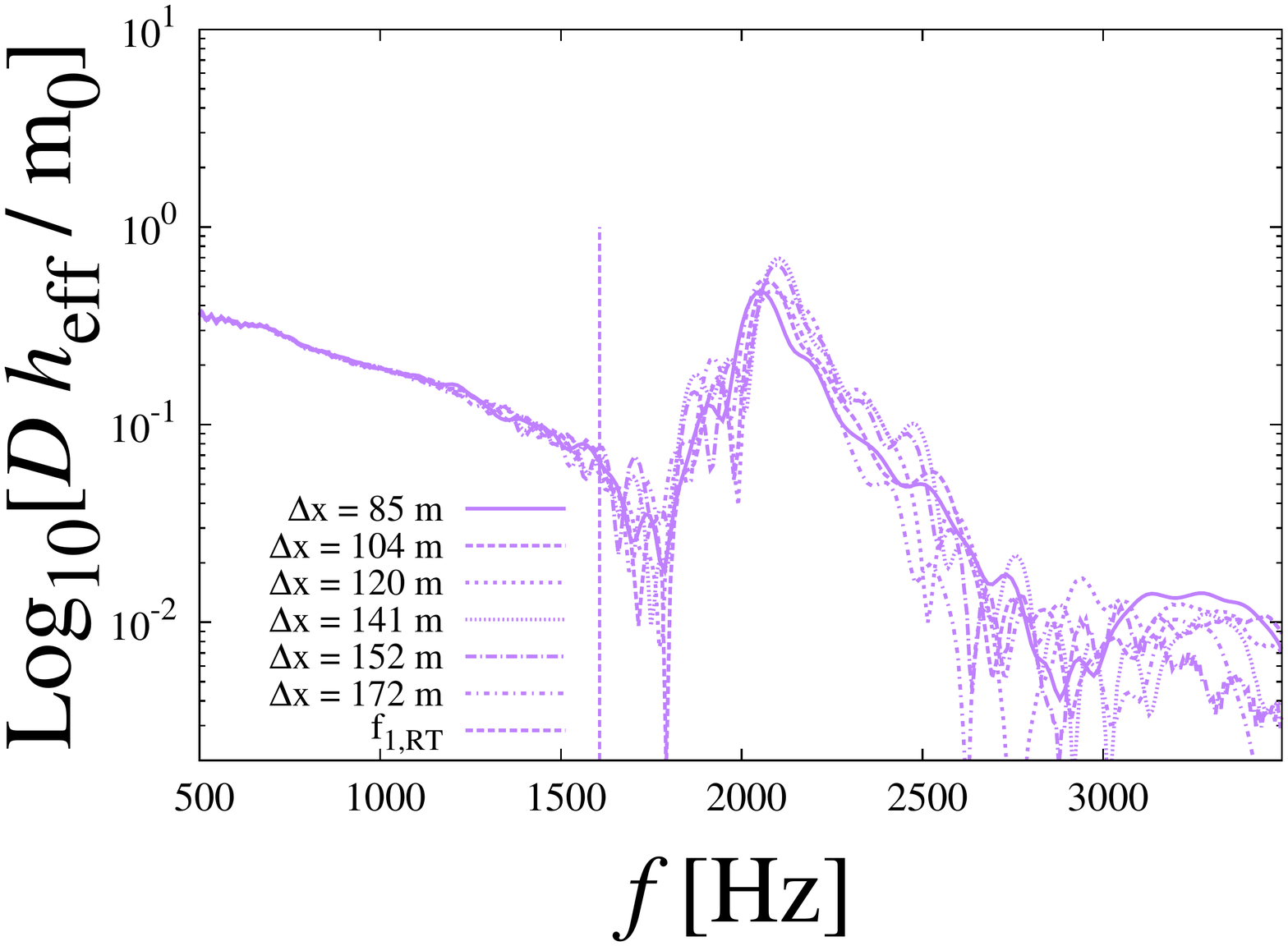}
\end{center}
\end{minipage}
\hspace{-13.35mm}
\begin{minipage}{0.27\hsize}
\begin{center}
\includegraphics[width=4.5cm,angle=0]{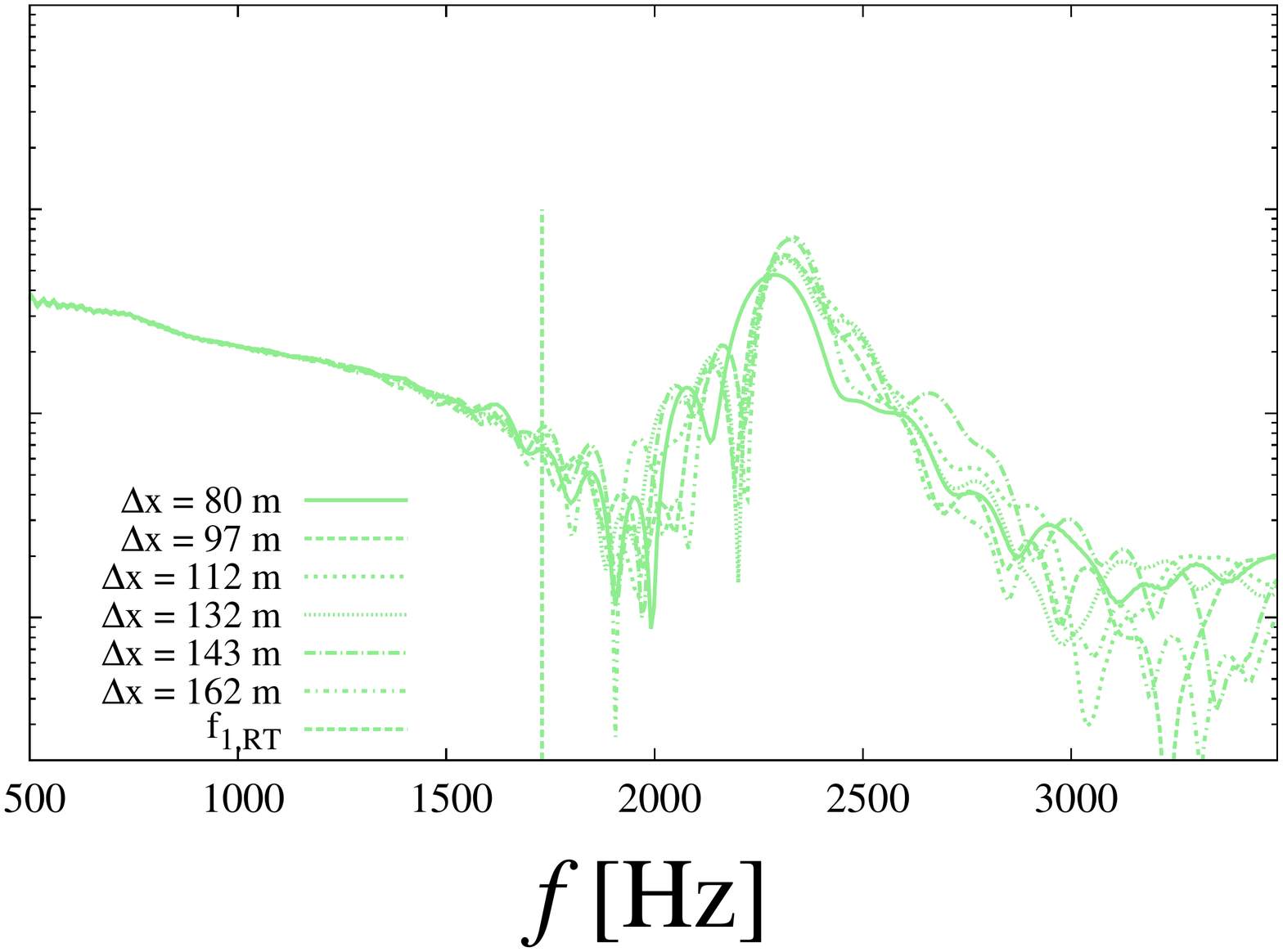}
\end{center}
\end{minipage}
\hspace{-13.35mm}
\begin{minipage}{0.27\hsize}
\begin{center}
\includegraphics[width=4.5cm,angle=0]{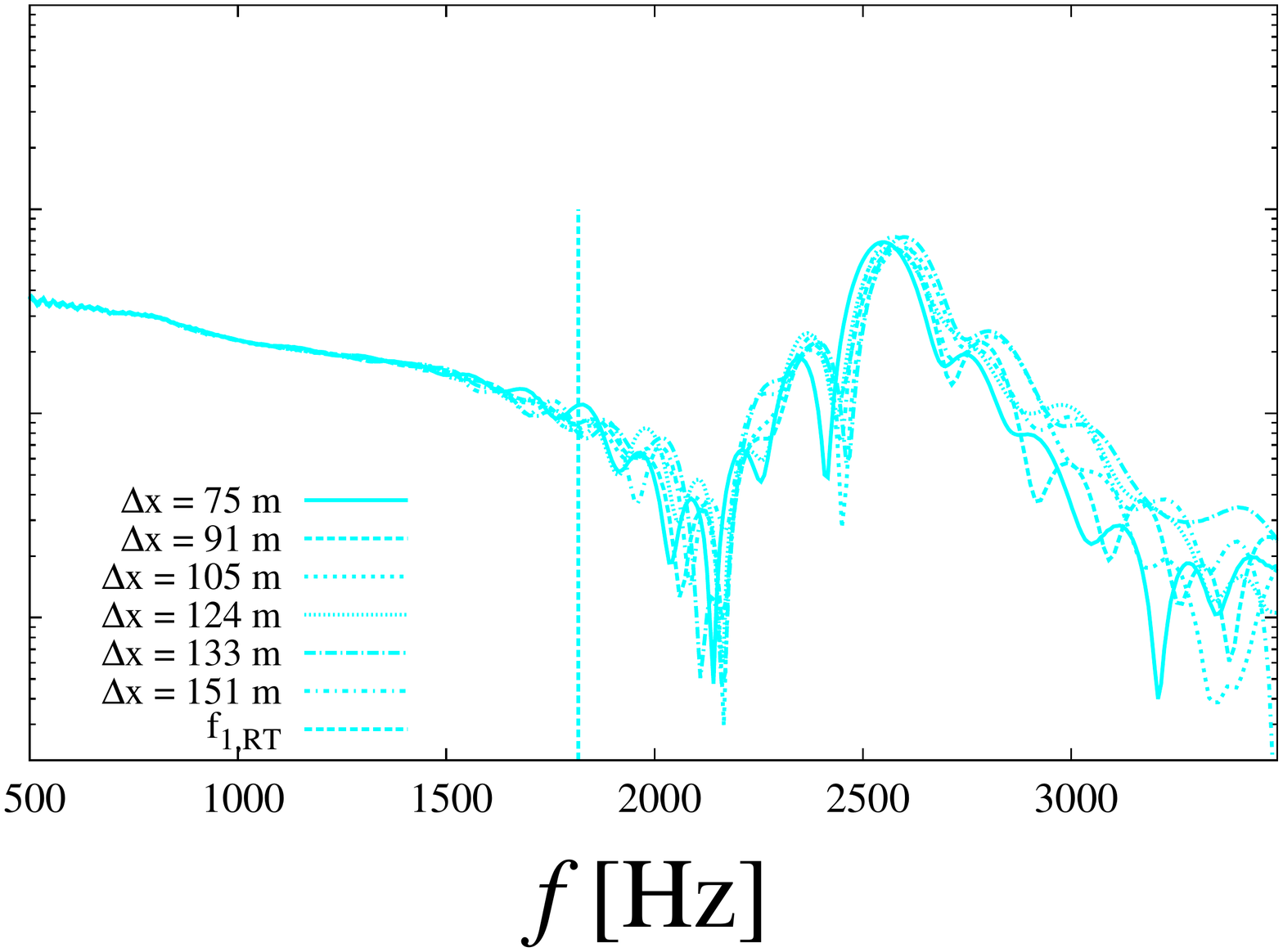}
\end{center}
\end{minipage}
\hspace{-13.35mm}
\begin{minipage}{0.27\hsize}
\begin{center}
\includegraphics[width=4.5cm,angle=0]{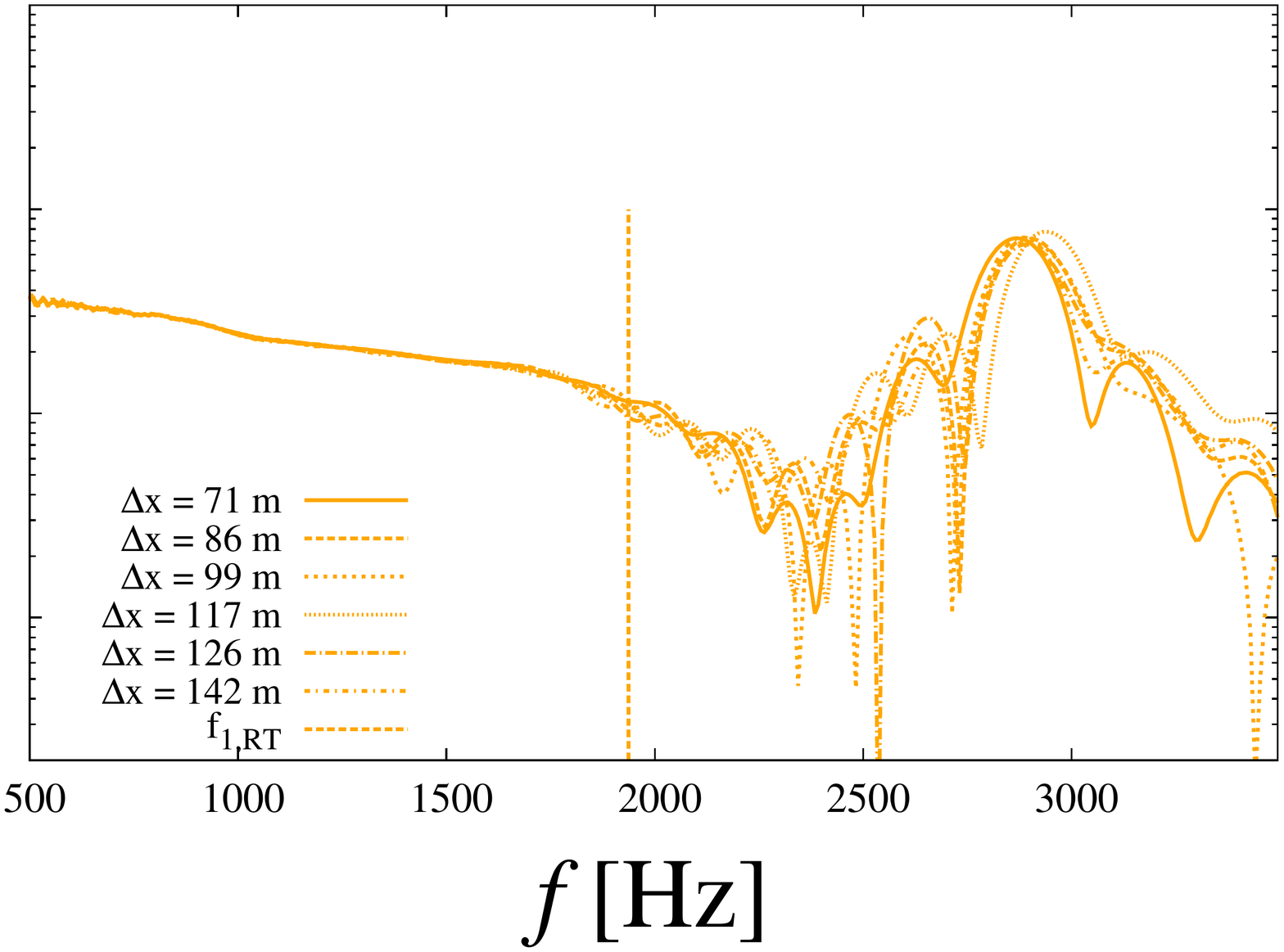}
\end{center}
\end{minipage}
\hspace{-13.35mm}
\begin{minipage}{0.27\hsize}
\begin{center}
\includegraphics[width=4.5cm,angle=0]{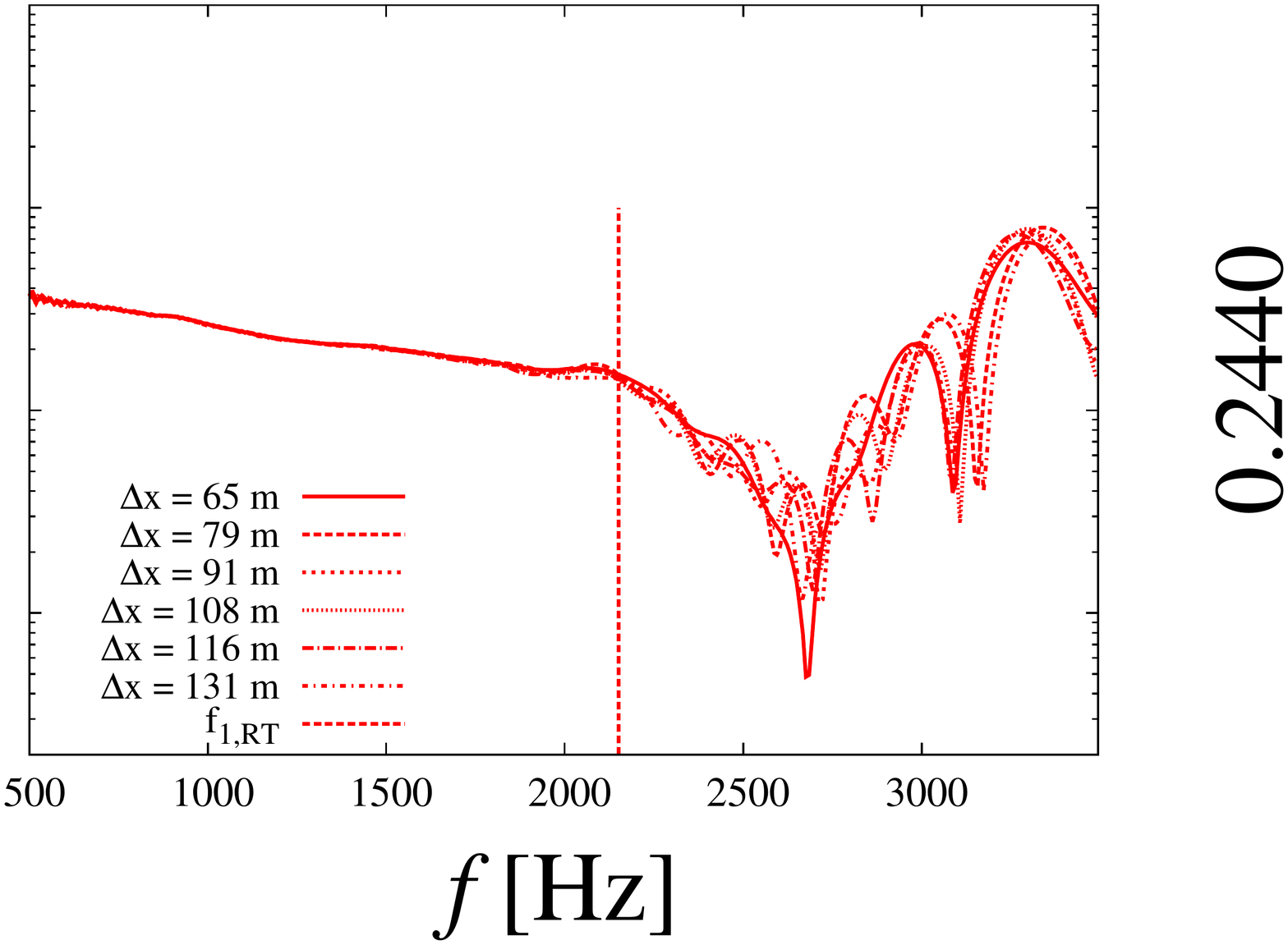}
\end{center}
\end{minipage}\\
\caption{\label{fig:PSDb}The same as Fig.~\ref{fig:PSD}, but for the binary systems with ${\cal M}_c= 1.0882M_\odot$. 
}
\end{figure*}

\begin{figure}[t]
\includegraphics[width=.9\linewidth]{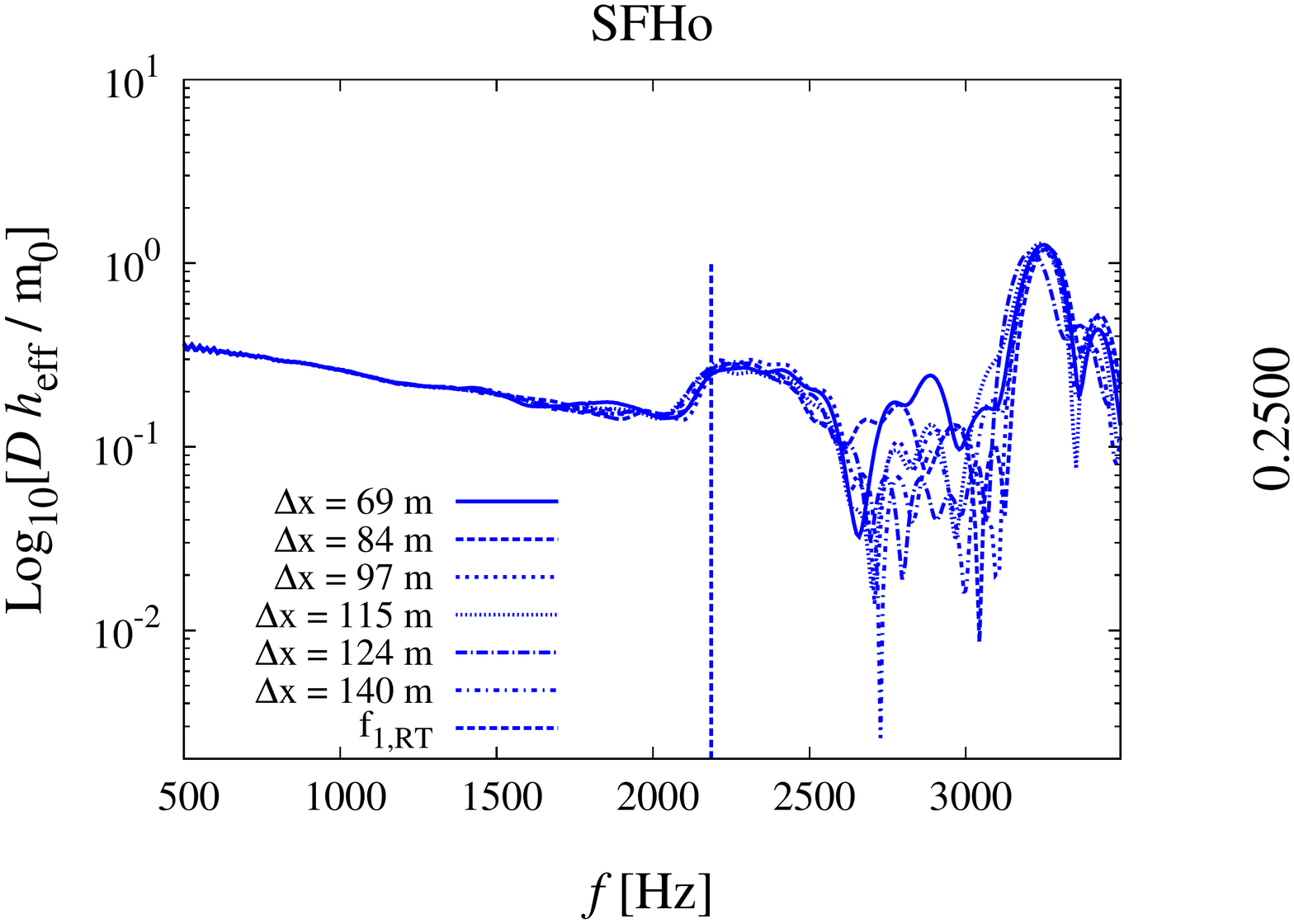}
\caption{\label{fig:PSD2}The same as Fig.~\ref{fig:PSD}, but for the SFHo (tabulated) EOS case. 
}
\end{figure}

Reference~\cite{Rezzolla:2016nxn} reported that several gravitational-wave frequencies associated with the main peaks in the spectrum amplitude for post-merger gravitational waves correlate with the tidal coupling constant. 
Figures~\ref{fig:PSD}--\ref{fig:PSD2} show the spectrum amplitudes for the quadrupole mode of gravitational waves for all the systems defined by
\begin{align}
  h_\text{eff}(f)=f\sqrt{\frac{|\tilde{h}_+(f)|^2+|\tilde{h}_\times(f)|^2}{2}}, \label{eq:deffreqdom2}
\end{align}
with $\tilde{h}_+(f)$ and $\tilde{h}_\times(f)$ in Eq.~(\ref{eq:deffreqdom}). 
In Figs.~\ref{fig:PSD}--\ref{fig:PSD2}, the vertical dashed lines indicate the so-called $f_1$ frequency for the fitting formula in Ref.~\cite{Rezzolla:2016nxn}. This peak is a side-band peak of the main peak of $f=f_2$, and it is naturally understood as a result of the modulation of the main peak. 
    According to Ref.~\cite{Takami:2014tva}, the remnant might be represented by a mechanical toy model composed of a rotating disk with two spheres. In this model, the two spheres, which mimic the double dense cores appearing after merger, 
    are connected with a spring and oscillate freely (see their Fig.~17). 
    $f_1$ frequency corresponds to the spin frequency when the separation between the two spheres becomes largest if we assume the angular momentum conservation. They claimed this scenario for the interpretation of $f_1$ frequency. 

    In Ref.~\cite{Rezzolla:2016nxn}, $f_1$ frequency is determined by
    identifying one of the main peaks in the spectrum amplitude and the spectrogram of post-merger gravitational waves. 
    For the symmetric binary systems, $f_1$ peak could be identified in our numerical results using the same methods. However, the structure of the spectrum amplitude around $f=f_1$ depends highly on the grid resolution (see  125H135-135 and H135-135 systems for example).
    For a sequence with the fixed EOS and chirp mass, e.g., 15H135-135, 15H125-146, 15H121-151, 15H118-155, 15H117-156, and 15H116-158, 
 we find it more difficult to identify $f_1$ peak as the symmetric mass ratio decreases. This was also pointed out in Ref.~\cite{Dietrich:2015iva} although their grid resolution was much lower than those in our present study and the resolution study on the spectrum amplitude of gravitational waves is not performed (see their Fig.~13). 
    As demonstrated in Fig.~\ref{fig:PSD}, $f_1$ peak cannot be clearly identified for the asymmetric binary systems. Figure~\ref{fig:PSDb} shows that this is also the case for binary systems of relatively small mass $\sim 2.5M_\odot$ as discussed in Refs.~\cite{Foucart:2015gaa,Bauswein:2015yca,Rezzolla:2016nxn}.

We also analyze the spectrogram of post-merger gravitational waves and confirm that there is no prominent peak around $f_\text{GW}=f_1$ for the asymmetric binary systems. 
Therefore, we conclude that the universal relation for $f_1$ could be only applicable to nearly symmetric binary systems: essentially no universal relation is present. 
We speculate that for the asymmetric binary systems, the mechanical toy model proposed in Ref.~\cite{Takami:2014tva} could not describe the merger remnant because the less massive NS is tidally disrupted before merger and there is no prominent double dense cores. 
We also note that the method for constraining the EOS proposed in Ref.~\cite{Takami:2014zpa} could not be applied unless the symmetric mass ratio is measured precisely to be $0.25$ because this method relies on $f_1$ universal relation.

\begin{figure*}
  	 \includegraphics[width=.41\linewidth]{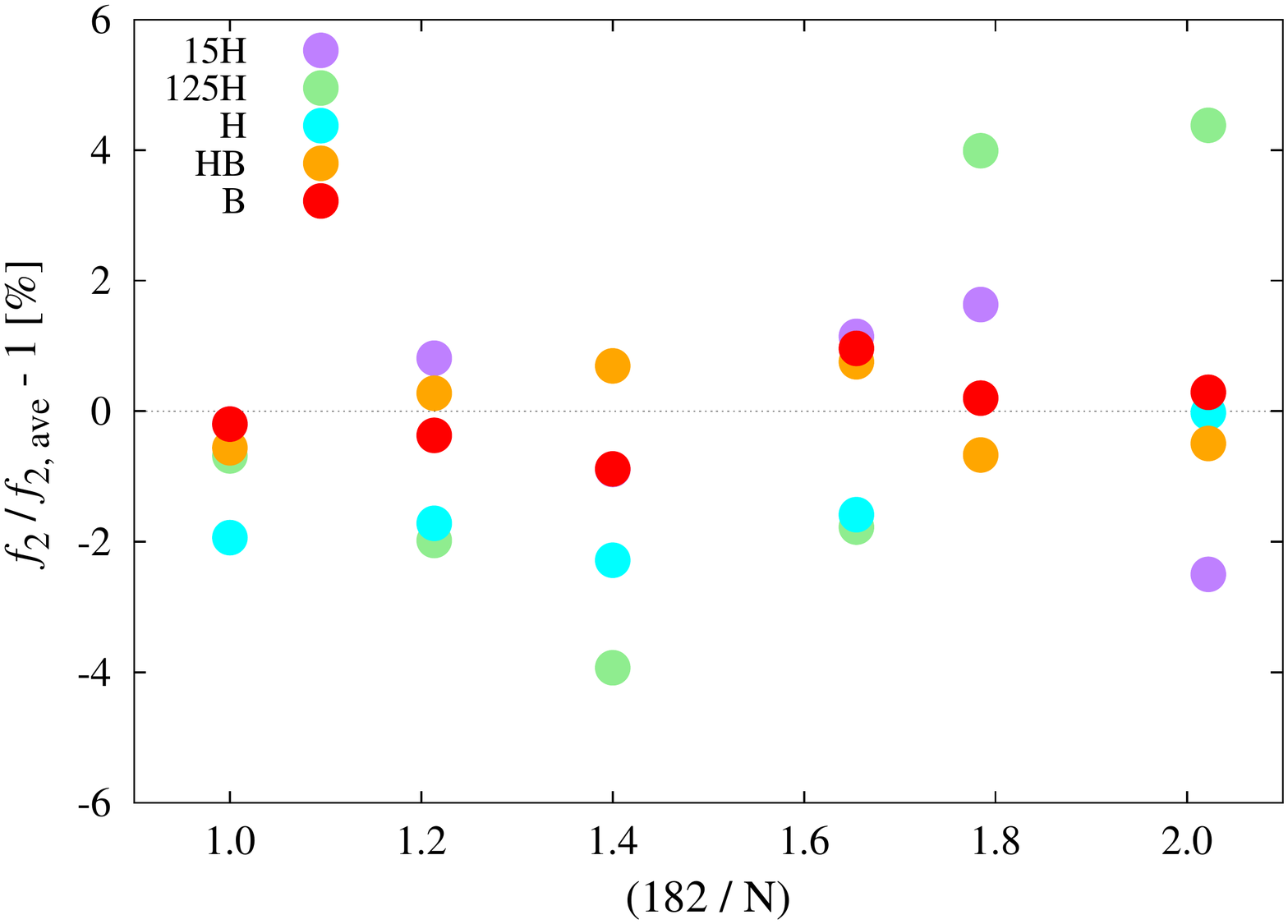} 
 	 \includegraphics[width=.45\linewidth]{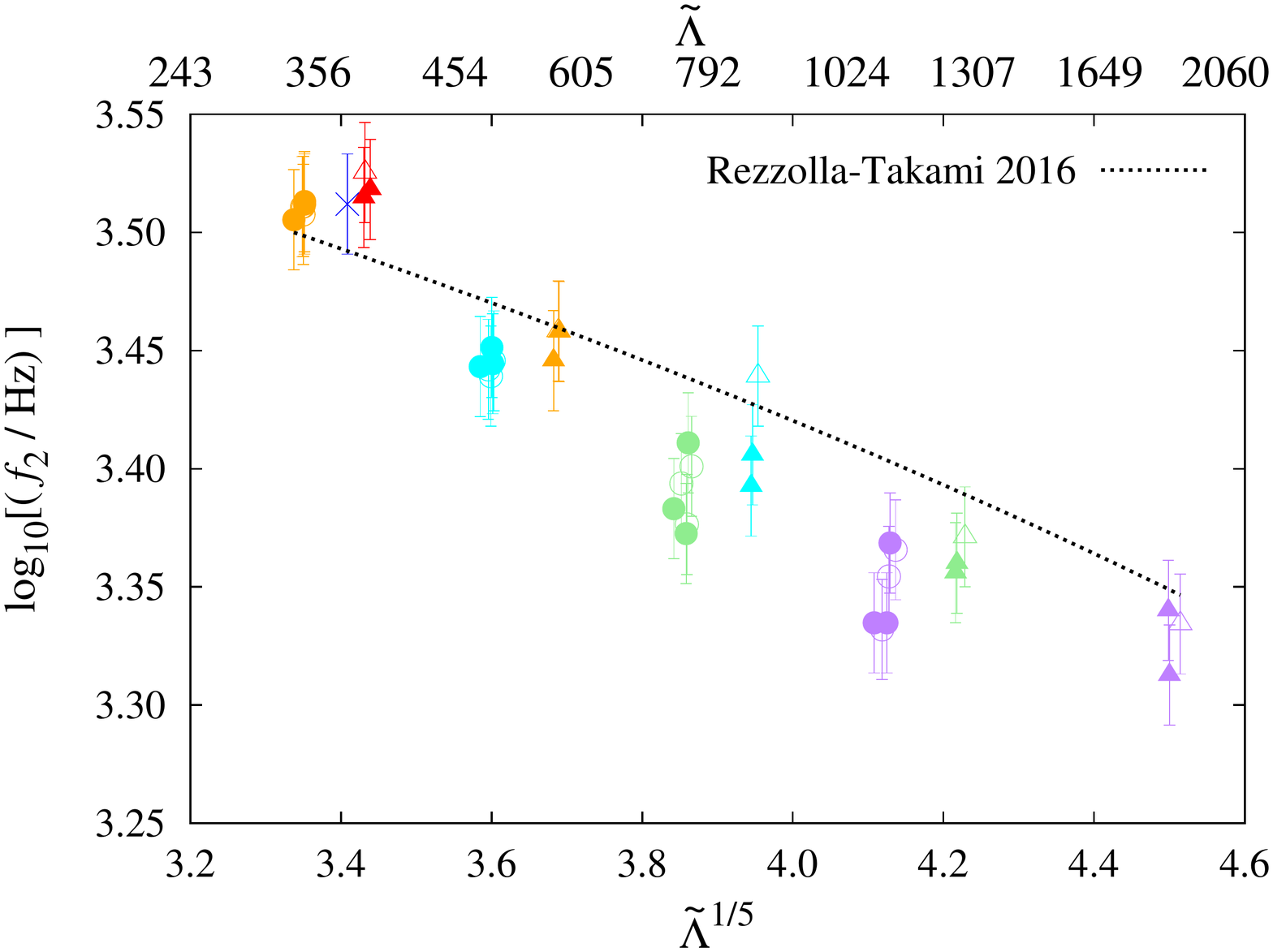}
 	 \caption{(Left) A deviation of $f_2$ frequency in the spectrum amplitude relative to $f_\text{2,ave}$ as a function of $1/N$ for the binary systems with $m_1= 1.12 M_\odot$ and $m_2= 1.40 M_\odot$. $f_\text{2,ave}$ is an average of $f_{2}$ over the results with different grid resolutions. 
           (Right)
         $f_2$--$\tilde{\Lambda}^{1/5}$ relation for the binary systems except for those which collapse to a black hole within a few ms after merger. The error bar of $\pm 5\%$ comes from the systematics associated with the finite grid resolution in $f_2$.
         }\label{fig:f2}
\end{figure*}

\begin{figure}
  	 \includegraphics[width=.9\linewidth]{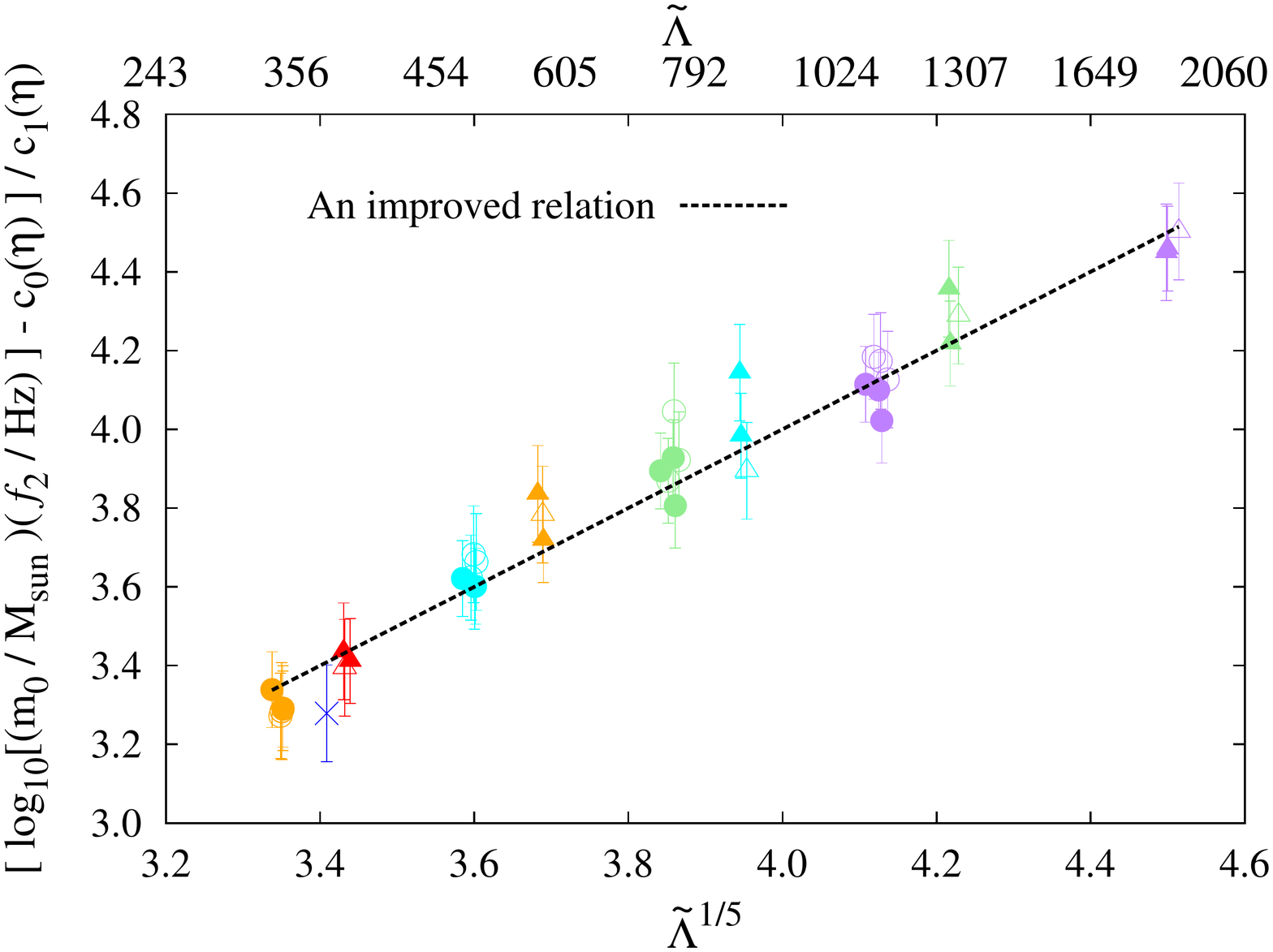}
 	 \caption{An improved $m_0 f_2$--$\tilde{\Lambda}^{1/5}$ relation with $c_0(\eta)$ and $c_1(\eta)$ in Eq.~(\ref{eq:f2}).
         }\label{fig:f2v2}
\end{figure}

In Ref.~\cite{Rezzolla:2016nxn}, the peak frequency, $f_2$, in the spectrum amplitude~\cite{footnote2} is reported to have a correlation with the tidal coupling constant. This peak frequency approximately corresponds to the f--mode oscillation of the remnant massive NS (see also Refs.~\cite{Shibata:2005xz,Hotokezaka:2013iia,Bauswein:2011tp,Shibata:2005ss}).
The left panel of Fig.~\ref{fig:f2} plots fluctuation around the averaged value of $f_2$ (average of the results with different grid resolutions) for the binary systems with $m_1= 1.12M_\odot$ and $m_2= 1.40 M_\odot$.
We measure $f_2$ in the spectrum amplitude as a prominent peak for $f \ge 2$ kHz. 
The fluctuation is within $\approx 4$--$5\%$ and we find that this is also the case for all the binary systems. 
Thus, we adopt $5\%$ as a relative error of $f_2$ (see also Table~\ref{tb:fpeak}). The right panel of Fig.~\ref{fig:f2} shows $f_2$ as a function of $\tilde{\Lambda}^{1/5}$. We exclude the systems which collapse to a black hole within a few ms after merger because the peak associated with $f_2$ is not prominent or absent in the spectrum amplitude.
We also overplot the fitting formula proposed in Ref.~\cite{Rezzolla:2016nxn}. It is found that with this fitting formula, the scatter is $\approx 14\%$ at the maximum. Thus, we propose an improved fitting formula for $m_0f_2$; 
\begin{align}
  &\log_{10}\left[\left(\frac{f_2}{\rm Hz}\right)\left(\frac{m_0}{M_\odot}\right)\right] = c_0(\eta) + c_1(\eta) \tilde{\Lambda}^{1/5},\nonumber\\
  &c_0(\eta) = 11.363 - 27.418 \eta,\nonumber\\
  &c_1(\eta) = -2.158 + 7.941 \eta. \label{eq:f2}
\end{align}
Even with this formula, the relative error is as large as $9\%$ (see also Fig.~\ref{fig:f2v2}).
This implies that even if the value of $f_2$ is determined precisely in the data analysis of gravitational waves, $\tilde{\Lambda}^{1/5}$ will be constrained with the error of $\approx \pm 0.1$. 

\subsubsection{$f_2$ and NS radius with $1.6M_\odot$ relation}

References~\cite{Bauswein:2011tp,Bauswein:2012ya} reported that $f_2$ frequency has a tight correlation with the NS radius of $1.6M_\odot$ (see Eq.~(3) in Ref.~\cite{Bauswein:2012ya}).
In Ref.~\cite{Hotokezaka:2013iia}, we assessed their relation by using our numerical-relativity results and found that the scatter in the relation is larger than that reported in Ref.~\cite{Bauswein:2012ya}. 
We revisit this assessment because the initial orbital eccentricity reduction was not implemented in Ref.~\cite{Hotokezaka:2013iia}.
In addition, the grid resolution in Ref.~\cite{Hotokezaka:2013iia} is much lower than that in this paper. These ingredients could modify the post-merger dynamics and the resulting gravitational waveforms.

Because the relation in Ref.~\cite{Bauswein:2012ya} holds only for symmetric binary systems of $m_0 = 2.7M_\odot$, we first assess this relation by employing binary systems of $(m_1,m_2)=(1.35M_\odot,1.35M_\odot)$ and found that the error is $\approx 6\%$~\cite{crust_comment}.
Second, we assess the relation by employing binary systems of $(m_1,m_2)=(1.25M_\odot,1.46M_\odot)$, $(1.21M_\odot,1.51M_\odot)$, $(1.18M_\odot,1.55M_\odot)$, $(1.17M_\odot,1.56M_\odot)$, and $(1.16M_\odot,1.58M_\odot)$. We found that the scatter from their fitting formula is $\approx 10\%$. 
Therefore, the scatter larger than that reported in Ref.~\cite{Bauswein:2012ya} stems from the mass asymmetry of the binary.
Our numerical results suggest that the fitting formula in Ref.~\cite{Bauswein:2012ya} could infer the radius of the $1.6M_\odot$ NS within the $1$ km accuracy only if the symmetric mass ratio is well constrained to be $0.25$.
Otherwise, we constrain the radius of the $1.6M_\odot$ NS with the accuracy of $\approx \pm 1$ km if the value of $f_2$ is determined precisely,

In Table~\ref{tb:universal}, we summarize to what extent the so-called universal relations hold. 

\subsection{Energy and angular momentum} \label{subsec:energy}

Using Eqs.~(\ref{eq:EGW})--(\ref{eq:JGW2}), we calculate the energy and angular momentum carried by gravitational waves. We define $E_{\rm GW,i}^\text{tot}$ and $E_{\rm GW,p}~(J_{\rm GW,p})$ as the energy (angular momentum) emitted in the inspiral stage and in the post-merger stage, respectively.
The subscripts $\rm i$ and $\rm p$ in these quantities denote the inspiral and the post-merger stage, respectively. 
The peak time introduced in Sec.~\ref{subsec:overview} defines the boundary between the inspiral and post-merger stages. In the following we summarize the energy and angular momentum emitted in each stage for all the systems. Their values are presented in Table~\ref{tb:fpeak}.

\subsubsection{inspiral stage}

\begin{figure*}
         \includegraphics[width=.45\linewidth]{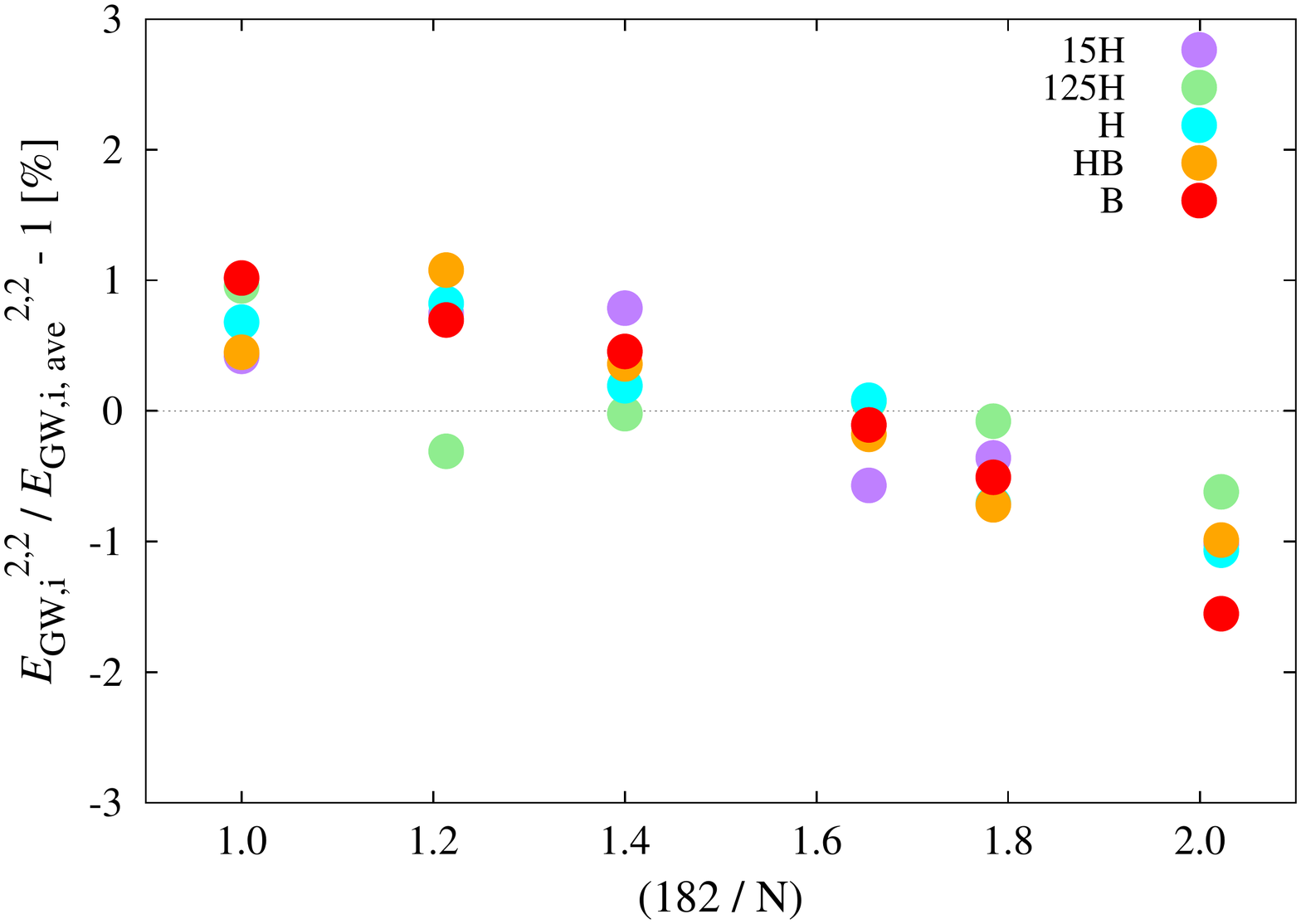}
    	 \includegraphics[width=.45\linewidth]{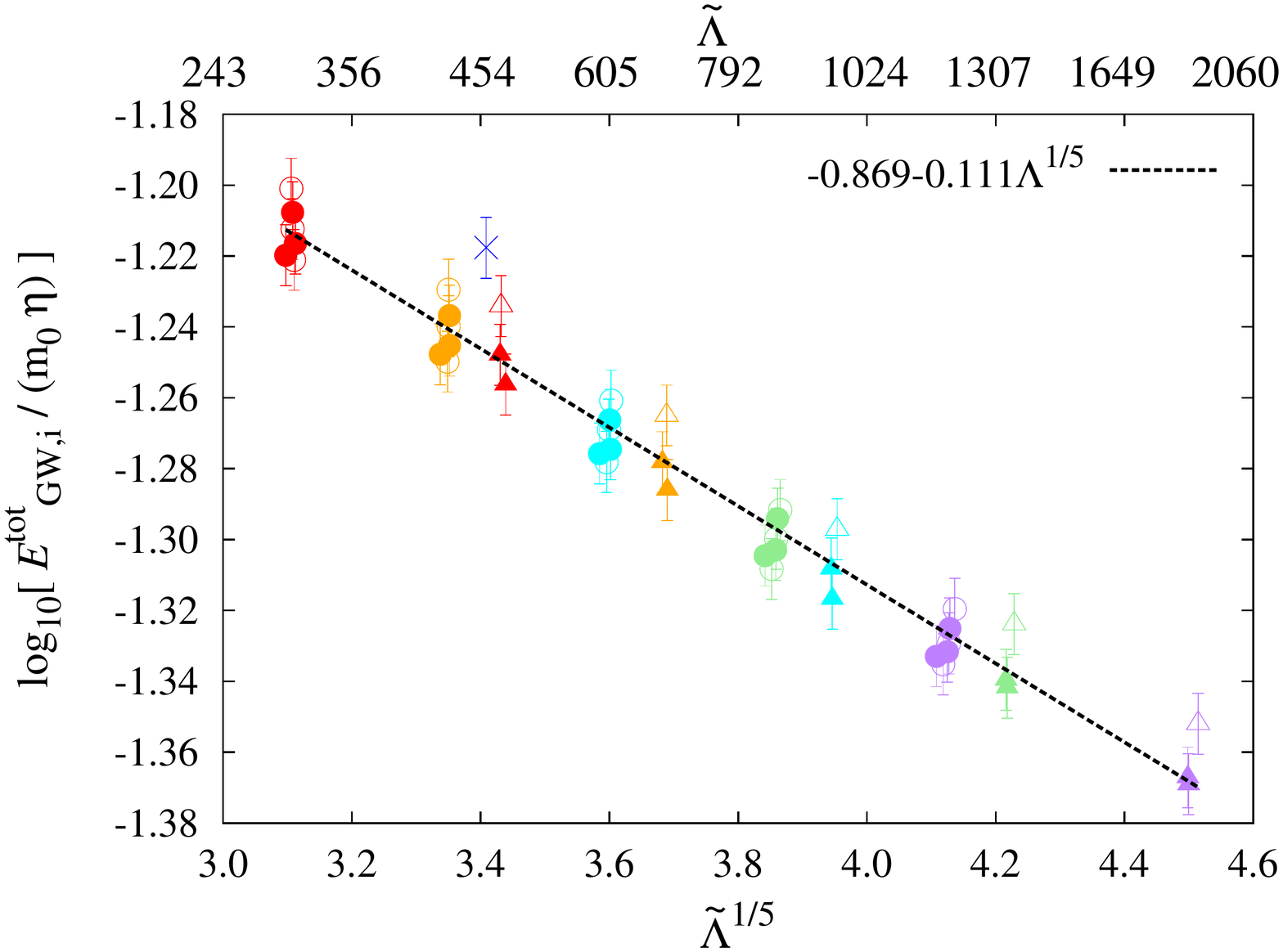} 
 	 \caption{(Left) A deviation of $E_{\rm GW,i}^{2,2}$ relative to $E_{\rm GW,i,ave}^{2,2}$ as a function of $1/N$ for binary systems with $m_1= 1.25M_\odot$ and $m_2= 1.46 M_\odot$.  $E^{2,2}_{\rm GW,i,ave}$ is an average of $E^{2,2}_{\rm GW,i}$ over the results with different grid resolutions. 
           (Right) $E^\text{tot}_{\rm GW,i}/(m_0\eta)$--$\tilde{\Lambda}^{1/5}$ relation with a fitting formula~(\ref{eq:EGWi}). In the right panel, the error bar of $\pm 2\%$ comes from the systematics associated with the finite grid resolution in $E_{\rm GW,i}^{2,2}$.
         }\label{fig:EGWi}
\end{figure*}

Table~\ref{tb:fpeak} and Fig.~\ref{fig:EGWi} show the energy, $E_{\rm GW,i}^{2,2}$, carried by gravitational waves with $(l,m)=(2,2)$ mode during the inspiral stage. 
We measure the relative error with respect to the averaged value in the left panel of Fig.~\ref{fig:EGWi} and find that the error relative to its averaged value of $E^{2,2}_{\rm GW,i}$ (average of the results with different grid resolutions)
never exceeds $2\%$ for a wide range of the grid resolution. This is also the case for all the binary systems. 
Thus, we adopt this fluctuation as an error in $E_{\rm GW,i}^{2,2}$. Note that the other modes such as $(l,m)=(2,1)$ and $(3,3)$ are $\lesssim 0.1\%$ and $\lesssim 0.5\%$, respectively,  of $E^{2.2}_{\rm GW,i}$. 

The right panel of Fig.~\ref{fig:EGWi} plots $E^\text{tot}_{\rm GW,i}/(m_0\eta)$ as a function of $\tilde{\Lambda}^{1/5}$. We include the contribution due to the gravitational-wave emission during evolution from
infinite separation to the initial orbital separation of the simulation, $m_0-M_\text{ADM,0}$ in Table~\ref{tb:fpeak},
by $E^\text{tot}_{\rm GW,i} \approx 2E_{\rm GW,i}^{2,2}+m_0-M_\text{ADM,0}$. $M_\text{ADM,0}$ is the Arnowitt-Deser-Misner mass of the initial condition of the simulations. 
As proposed in Ref.~\cite{Zappa:2017xba}, this quantity correlates with the tidal coupling constant. We explicitly derive a fitting formula with the binary tidal deformability as
\begin{align}
  &\log_{10} \left[\frac{E^\text{tot}_{\rm GW,i}}{m_0\eta}\right] = - 0.869 - 0.111 \tilde{\Lambda}^{1/5}. \label{eq:EGWi}
\end{align}
    It is reasonable that $E^\text{tot}_{\rm GW,i}$ decreases as $\tilde{\Lambda}$ increases because the binary systems with larger values of $\tilde{\Lambda}$ merge earlier than those with smaller values of $\tilde{\Lambda}$. 
    This fitting formula reproduces the simulation data of $E^\text{tot}_{\rm GW,i}$ within an error of $\approx 4\%$. 
    In the limit to a binary black hole merger $(\tilde{\Lambda}\to 0)$, the fitting formula predicts $E^\text{tot}_\text{GW,i} \approx 0.034m_0$ for $\eta=0.250$ and $E^\text{tot}_\text{GW,i} \approx 0.033m_0$ for $\eta=0.244$, respectively.
      On the other hand, high-precision binary black hole merger simulations for non-spinning system suggests $E^\text{tot}_\text{GW,i} \approx 0.03 m_0$ for $0.247 \le \eta \le 0.250$~\cite{Blackman:2017dfb,Boyle:2019kee}.
     We conclude that the fitting formula Eq.~(\ref{eq:EGWi}) reproduces the BBH result with $\approx 10\%$ error. 

\subsubsection{Post-merger stage}

\begin{figure*}
         \includegraphics[width=.45\linewidth]{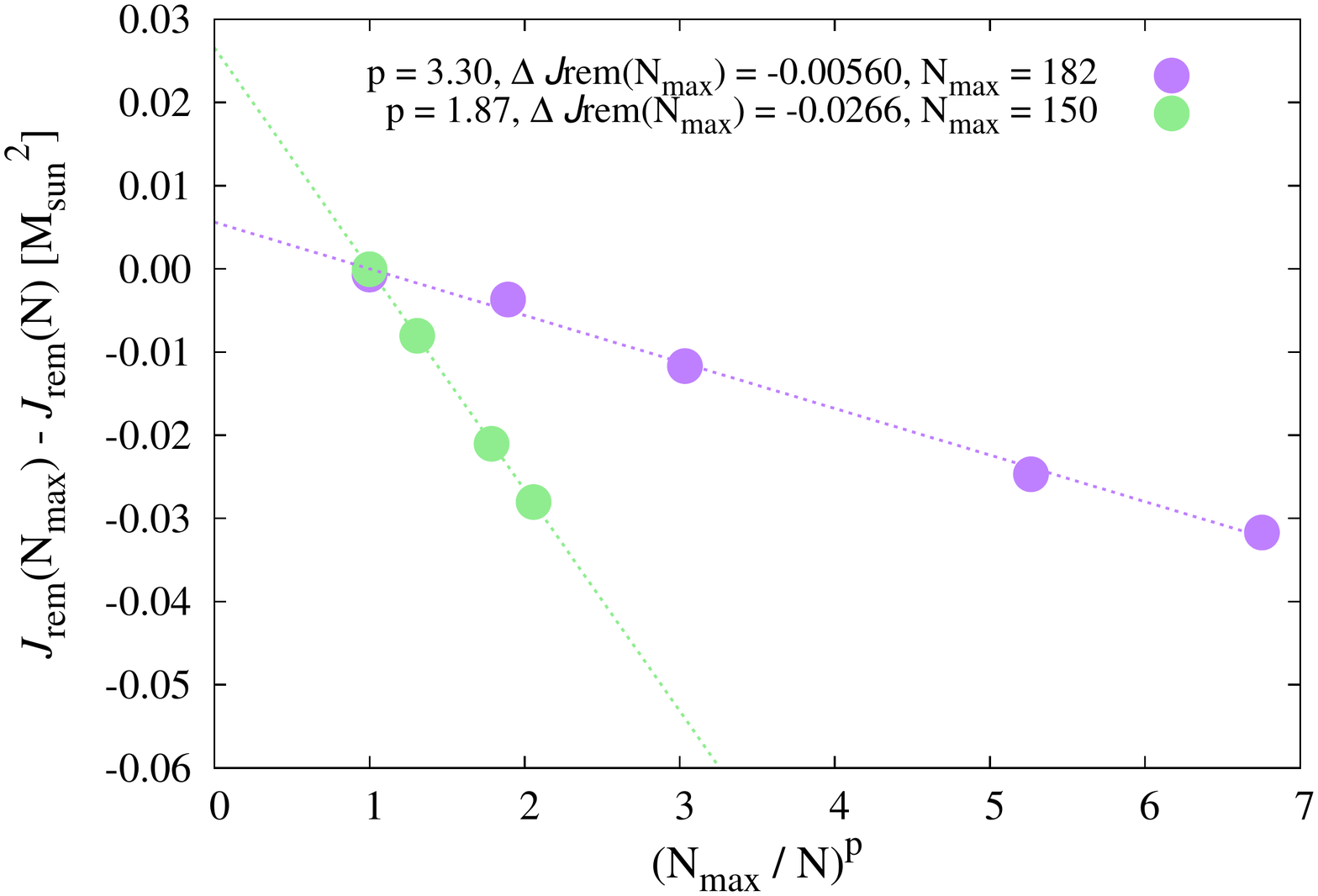}
  	 \includegraphics[width=.45\linewidth]{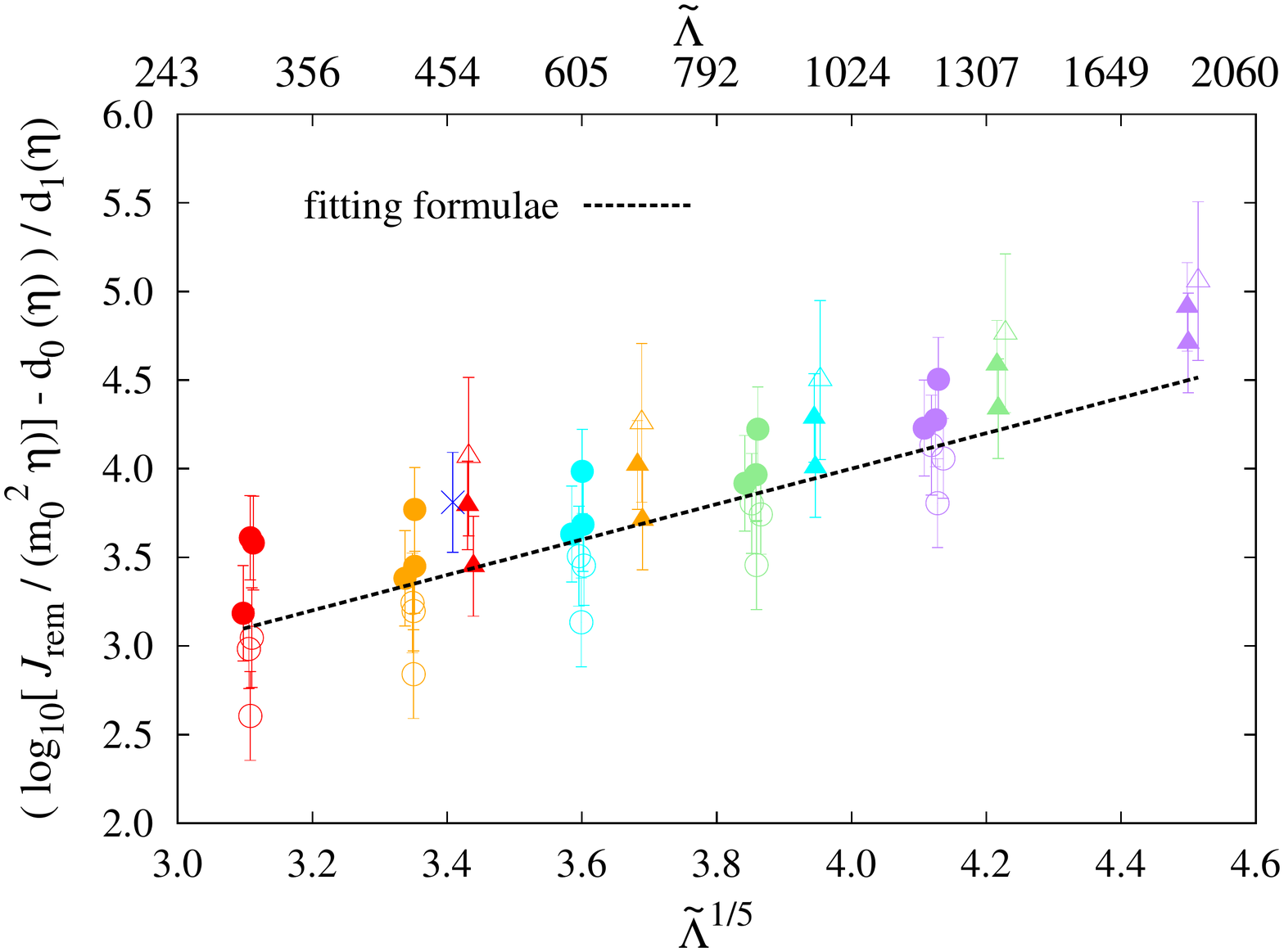}
 	 \caption{(Left) Convergence of $J_\text{rem}$ with respect to the grid resolution for HB$118$--$155$. (Right) $J_\text{rem}/(m_0^2\eta)$--$\tilde{\Lambda}^{1/5}$ relation with $d_0(\eta)$ and $d_1(\eta)$ in Eq.~(\ref{eq:Jrem}). The error bar of $\pm1\%$ comes from the systematics associated with the finite grid resolution in $J_\text{rem}$.
         }\label{fig:jrem}
\end{figure*}

We estimate angular momentum of the remnant, $J_\text{rem}$ at the peak time of the gravitational-wave amplitude in the retarded time~(\ref{eq:tret}) by performing a surface integral on the sphere of $r=r_0$; 
\begin{align}
  J_\text{rem} = \frac{1}{8\pi}\epsilon^{zjk}\oint_{r=r_0} x_j ({K^l}_k-K{\delta^l}_k)dS_l. \label{eq:Jrem2}
\end{align}
    $K_{ij}$, $K$, ${\delta^i}_j$, and $dS_l$ are the extrinsic curvature, its trace part, the Kronecker delta, and an element of the surface integral, respectively.
We typically integrate it on the sphere of $r_0 = 200 m_0$ and $214m_0$ for the binary systems with ${\cal M}_c=1.1752M_\odot$ and $1.0882M_\odot$, respectively. Table~\ref{tb:fpeak} and Fig.~\ref{fig:jrem} show the result. 
In the left panel of Fig.~\ref{fig:jrem}, we estimate the residual error in $J_\text{rem}$ for HB$118$--$155$. We again assume that the numerical result obeys the following form;
\begin{align}
  J_\text{rem}(N) = J_\text{rem}^\infty(N_\text{max}) - \Delta J_\text{rem}(N_\text{max}) \left(\frac{N_\text{max}}{N}\right)^p, \label{eq:jrem_num}
\end{align}
where $J_\text{rem}^\infty(N_\text{max})$ is the angular momentum of the remnant in the continuum limit of the finite difference.
We estimate three unknowns, $J_\text{rem}^\infty(N_\text{max})$, $\Delta J_\text{rem}(N_\text{max})$, and $p$ by fitting the numerical data with $N=90,102,\cdots,$ and $N_\text{max}$ with Eq.~(\ref{eq:jrem_num}).
By comparing $N_\text{max}=150$ and $182$ cases, we confirm that adding a result of the higher resolution simulation reduces the residual error (see the legend of Fig.~\ref{fig:jrem} for $p$ and $\Delta J_\text{rem}(N_\text{max})$).
We find that $\Delta J_\text{rem}(N_\text{max})$ is $\lesssim 1\%$ of the continuum limit, $J_\text{rem}^{\infty}(N_\text{max})$, for $N_\text{max}=182$. This is also the case for all the binary systems. Thus, we adopt $1\%$ as a systematics associated with the finite grid resolution in $J_\text{rem}$. 

Because $J_\text{rem}$ could correlate with $\tilde{\Lambda}^{1/5}$, we propose a fitting formula of $J_\text{rem}/(m_0^2\eta)$:
\begin{align}
  &\log_{10}\left[\frac{J_\text{rem}}{m_0^2\eta}\right] = d_0(\eta) + d_1(\eta) \tilde{\Lambda}^{1/5},\nonumber\\
  &d_0(\eta) = 1.552 - 4.275 \eta,\nonumber\\
  &d_1(\eta) = -0.141+0.642 \eta. \label{eq:Jrem}
\end{align}
The right panel of Fig.~\ref{fig:jrem} plots this relation and we confirm that it is accurate within $3\%$ error.

\begin{figure*}
         \includegraphics[width=.45\linewidth]{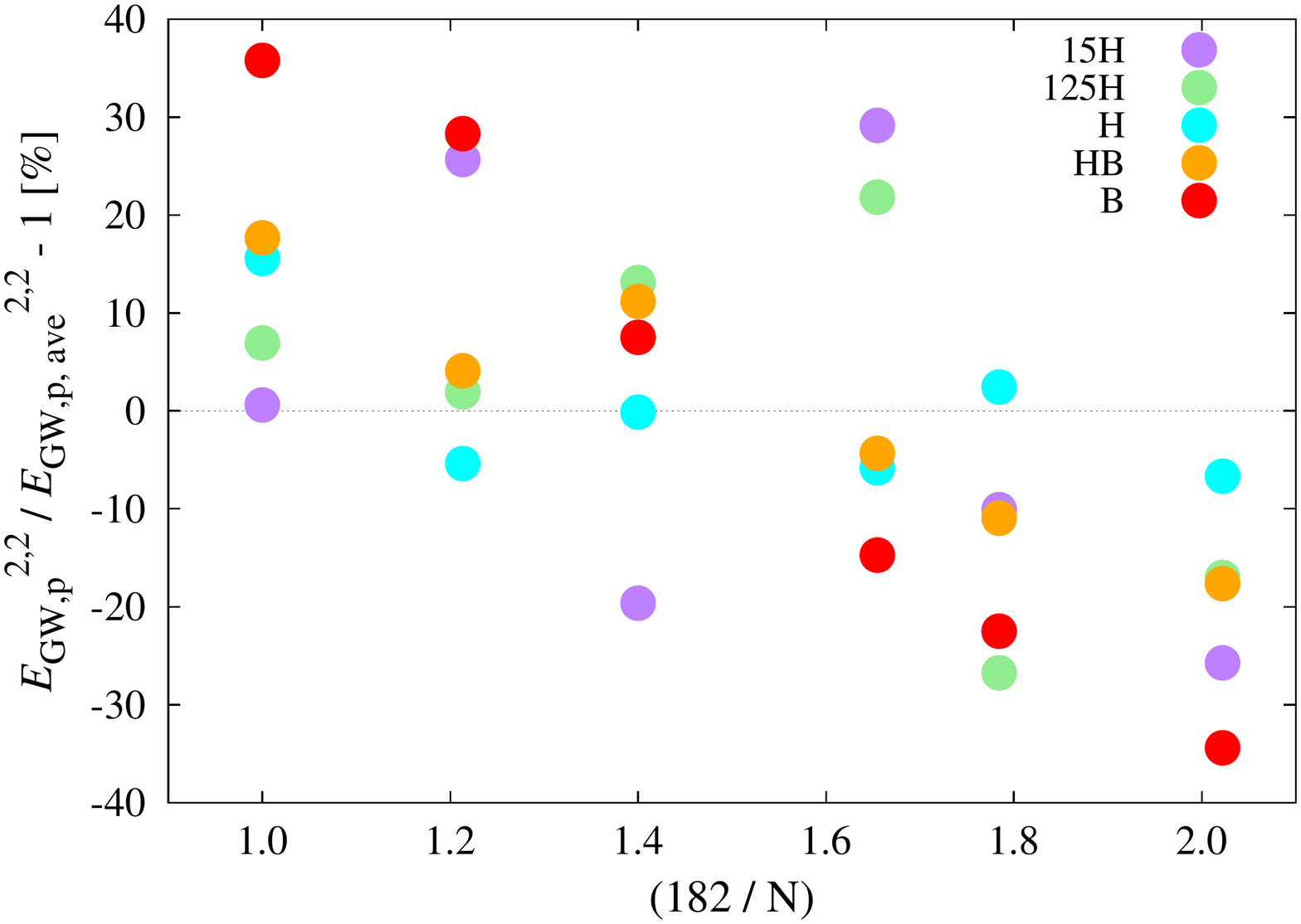}
    	 \includegraphics[width=.45\linewidth]{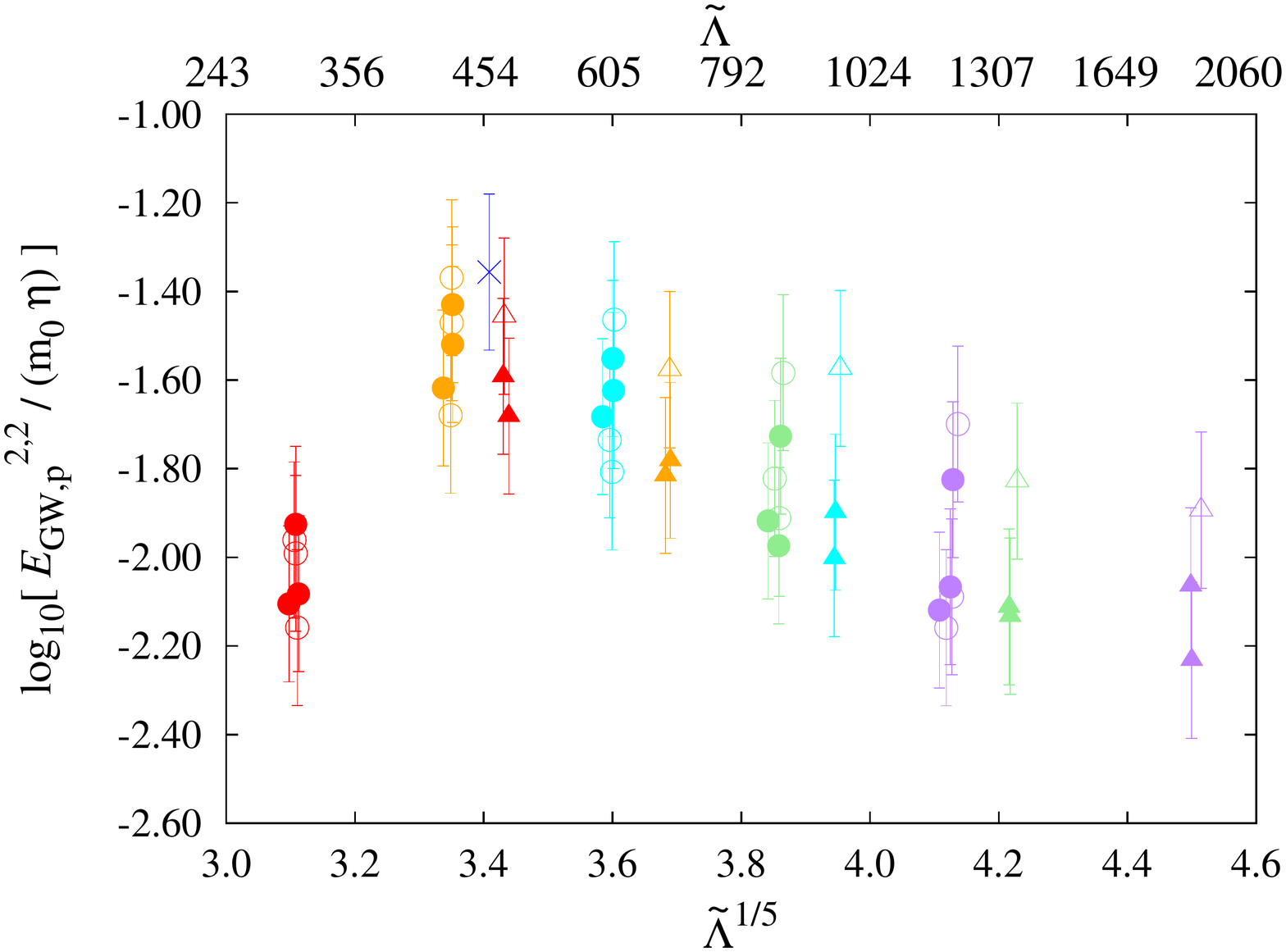} 
 	 \caption{(Left) A deviation of $E_{\rm GW,p}^{2,2}$ relative to $E_\text{GW,p,ave}^{2,2}$ as a function of $1/N$ for binary systems with $m_1= 1.25M_\odot$ and $m_2= 1.46 M_\odot$.  $E^{2,2}_\text{GW,p,ave}$ is an average of $E^{2,2}_{\rm GW,p}$ over the results with different grid resolutions. 
           (Right) $E_{\rm GW,p}^{2,2}/(m_0\eta)$--$\tilde{\Lambda}^{1/5}$ relation. In the right panel, the error bar of $\pm 50\%$ comes from the systematics associated with the finite grid resolution in $E_{\rm GW,p}^{2,2}$.
         }\label{fig:EGWp}
\end{figure*}

\begin{figure*}
  \includegraphics[width=.45\linewidth]{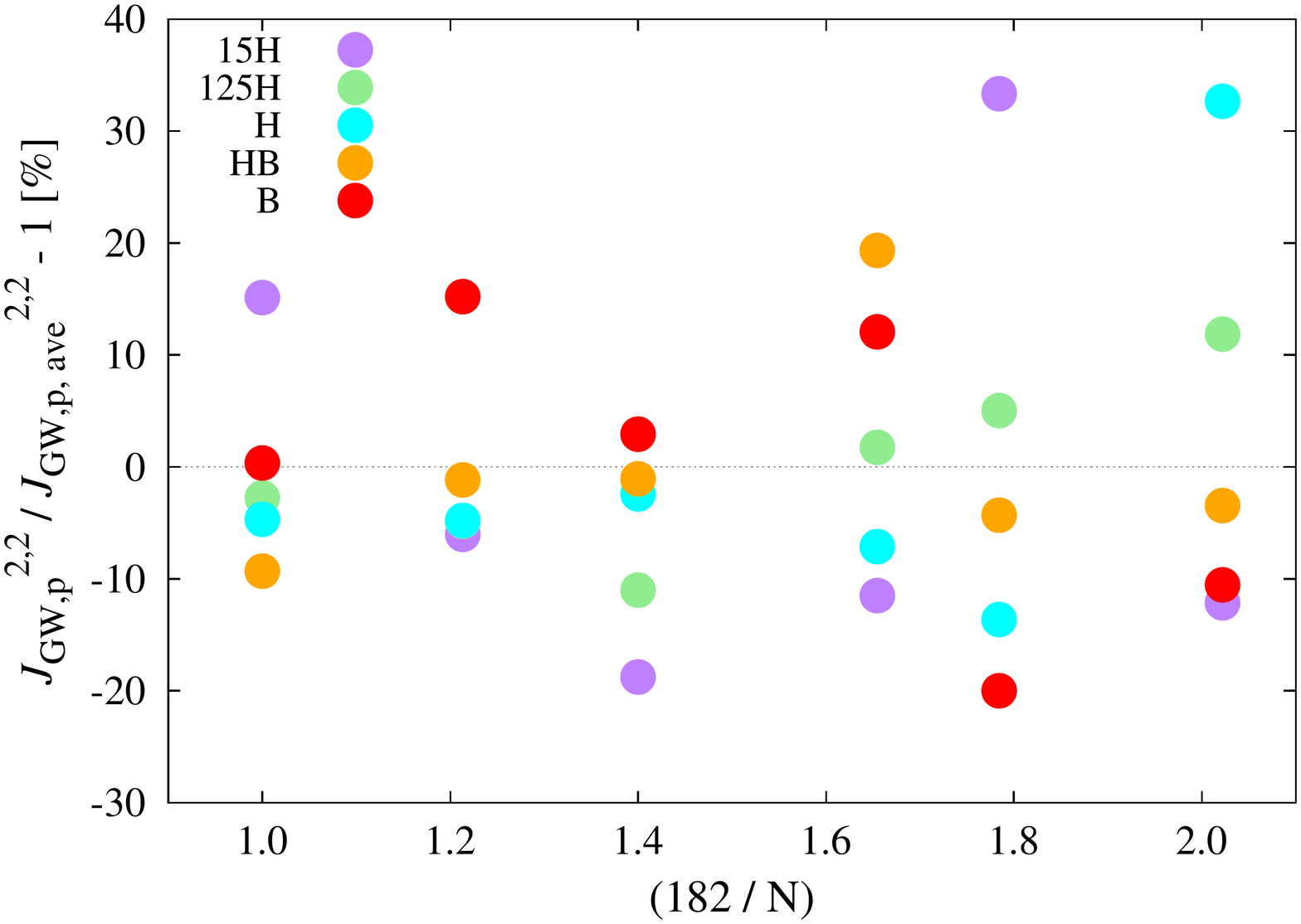}
    	 \includegraphics[width=.45\linewidth]{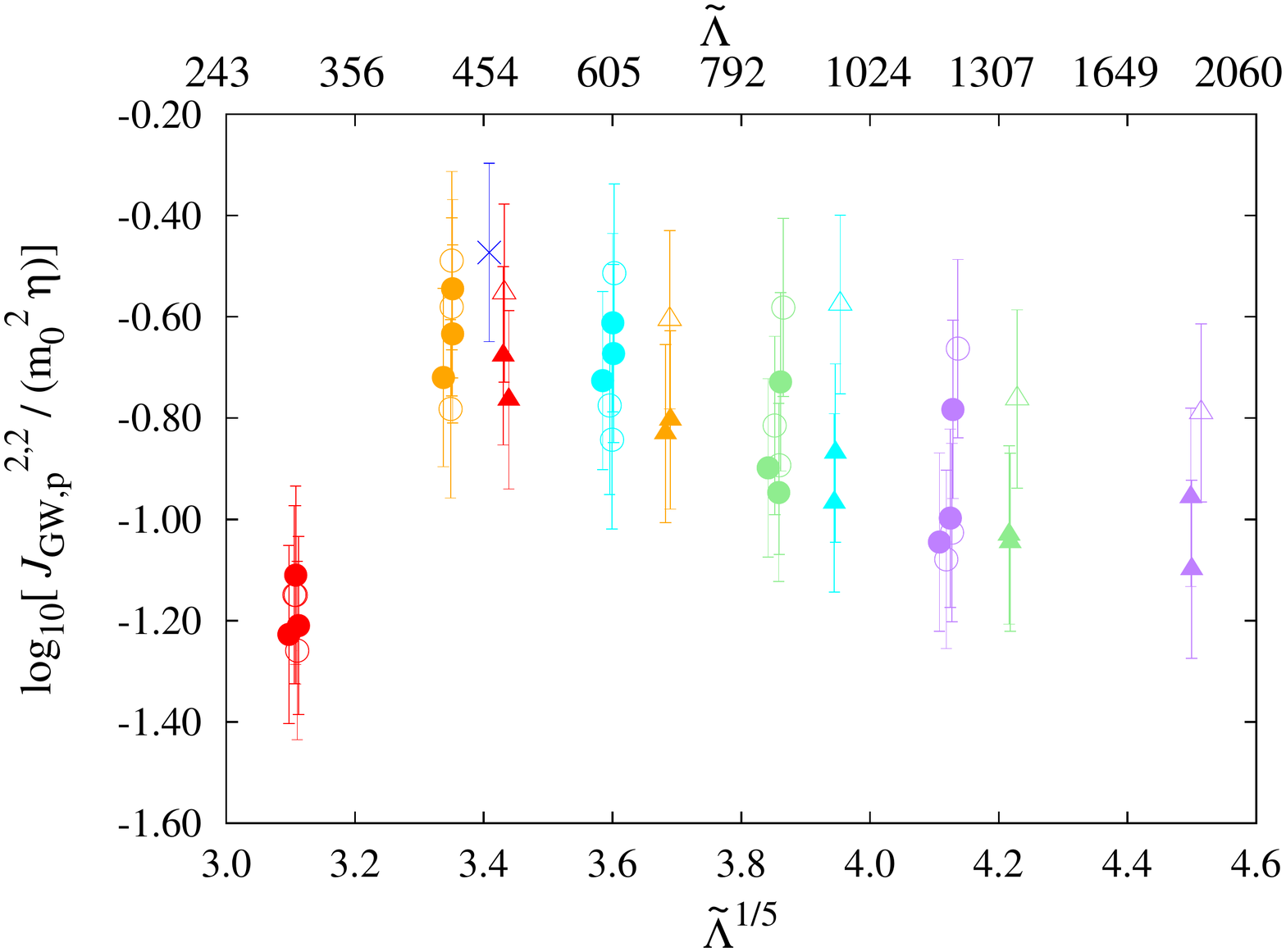} 
 	 \caption{The same as Fig.~\ref{fig:EGWp}, but for $J_{\rm GW,p}^{2,2}$. The left panel is for the binary systems with $m_1= 1.12 M_\odot$ and $m_2= 1.40 M_\odot$. 
         }\label{fig:JGWp}
\end{figure*}

Figures~\ref{fig:EGWp} and \ref{fig:JGWp} plot $E_{\rm GW,p}^{2,2}$ and $J_{\rm GW,p}^{2,2}$ emitted in the post-merger stage. It is worth noting that energy and angular momentum radiated by gravitational waves in $(l,m)=(2,1)$ and $(3,3)$ modes are $\lesssim 2.5\%$ of $E^{2,2}_{\rm GW,p}$ and $\lesssim 2.4\%$ of $J^{2,2}_{\rm GW,p}$, respectively, even for the highly asymmetric binary systems, e.g., 15H107-146 (see also the upper panel of Fig.~\ref{fig:dEdJ3}). 
The left panels in these figures show that it is hard to achieve a perfect convergence and the scatter is rather large compared to $E_{\rm GW,i}^{2,2}$, although the scatter never exceeds $50\%$ in $E_{\rm GW,p}^{2,2}$ and $J_{\rm GW,p}^{2,2}$. This is also the case for all the binary systems. 
The right panels in Figs.~\ref{fig:EGWp} and \ref{fig:JGWp} show $E_{\rm GW,p}^{2,2}/(m_0\eta)$ and $J_{\rm GW,p}^{2,2}/(m_0^2\eta)$ as a function of $\tilde{\Lambda}^{1/5}$.
As discussed in Ref.~\cite{Zappa:2017xba}, the energy and angular momentum radiated in the post-merger stage peak around $\tilde{\Lambda} \approx 400$ because the binary systems with $\tilde{\Lambda} \lesssim 350$ collapse to a black hole within a few ms after the peak time.
However, $\tilde{\Lambda}$ at the peak in $E_{\rm GW,p}^{2,2}$ and $J_{\rm GW,p}^{2,2}$ could decrease for general EOSs because as discussed in Ref.~\cite{Kiuchi:2019lls} the remnant would survive for more than $20$ ms after the peak time even for the binary systems with $\tilde{\Lambda} \lesssim 300$.
For $\tilde{\Lambda}\gtrsim 400$, correlation between $E_{\rm GW,p}^{2,2}$ and the binary tidal deformability is not as tight as that in $E^\text{tot}_{\rm GW,i}/(m_0\eta)$--$\tilde{\Lambda}^{1/5}$. 
For  $J_{\rm GW,p}^{2,2}$, the correlation with the binary tidal deformability is also not very tight. 

Note that $E_{\rm GW,p}^{2,2}$ and $J_{\rm GW,p}^{2,2}$ could increase from the values listed in Table~\ref{tb:fpeak} because we artificially terminated the simulations at $10$--$15$ ms after the peak time. At that moment, the gravitational-wave amplitude is still comparable to that in the late inspiral stage except for the systems which collapse to a black hole within a few ms after the peak time.

We also should keep in mind that we might miss relevant physics such as effective turbulent viscosity generated by the magneto-hydrodynamical instabilities during the merger~\cite{Kiuchi:2014hja,Kiuchi:2015sga,Kiuchi:2017pte} and/or the neutrino cooling~\cite{Sekiguchi:2011zd,Foucart:2015gaa} for modeling the post-merger signal.
Reference \cite{Shibata:2017xht} suggests that the post-merger signal could be significantly suppressed in the presence of efficient angular momentum transport by the viscous effect inside the remnant NS. 

\begin{figure*}[t]
\hspace{-18.0mm}
\begin{minipage}{0.27\hsize}
\begin{center}
\includegraphics[width=4.5cm,angle=0]{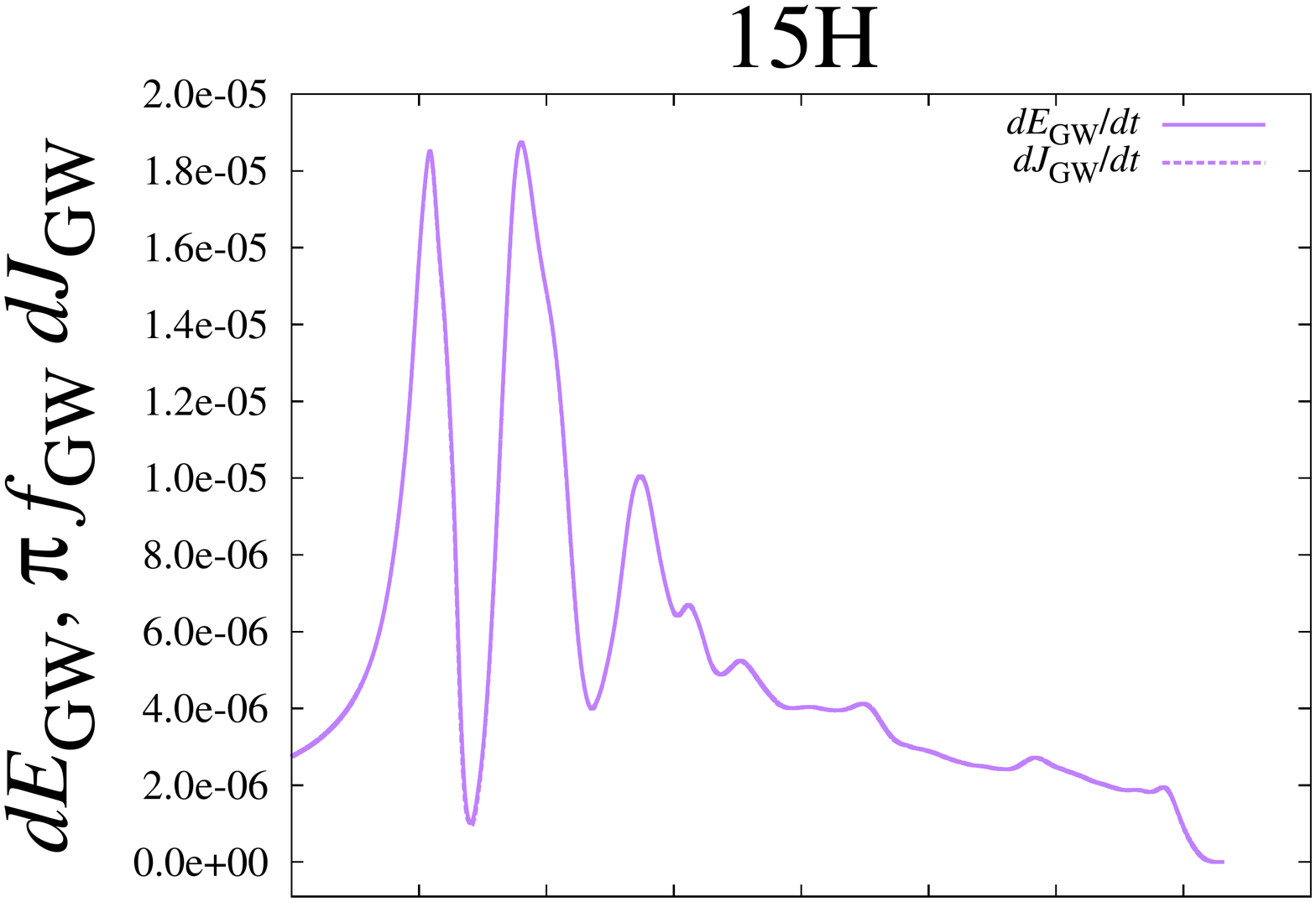}
\end{center}
\end{minipage}
\hspace{-13.35mm}
\begin{minipage}{0.27\hsize}
\begin{center}
\includegraphics[width=4.5cm,angle=0]{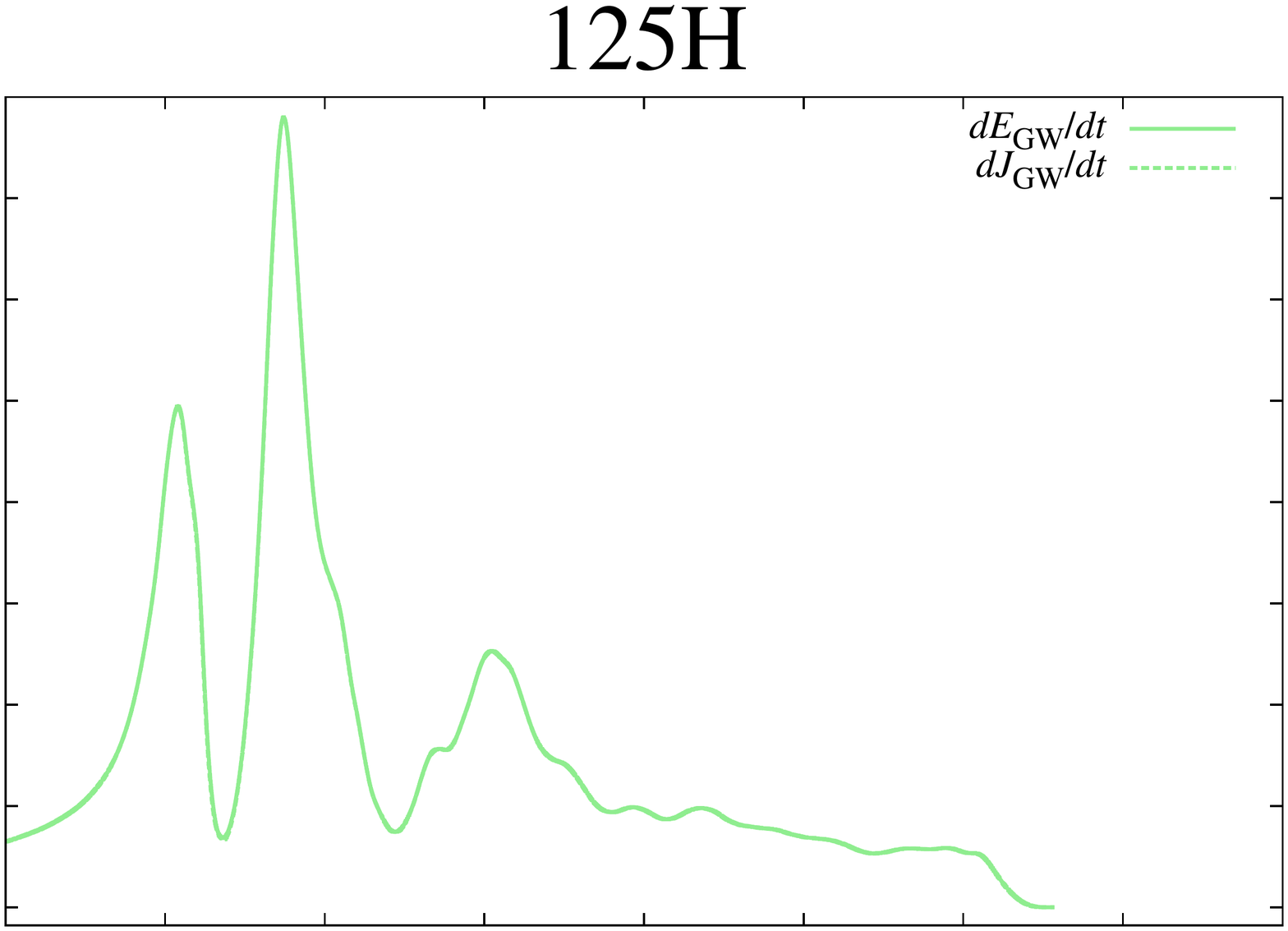}
\end{center}
\end{minipage}
\hspace{-13.35mm}
\begin{minipage}{0.27\hsize}
\begin{center}
\includegraphics[width=4.5cm,angle=0]{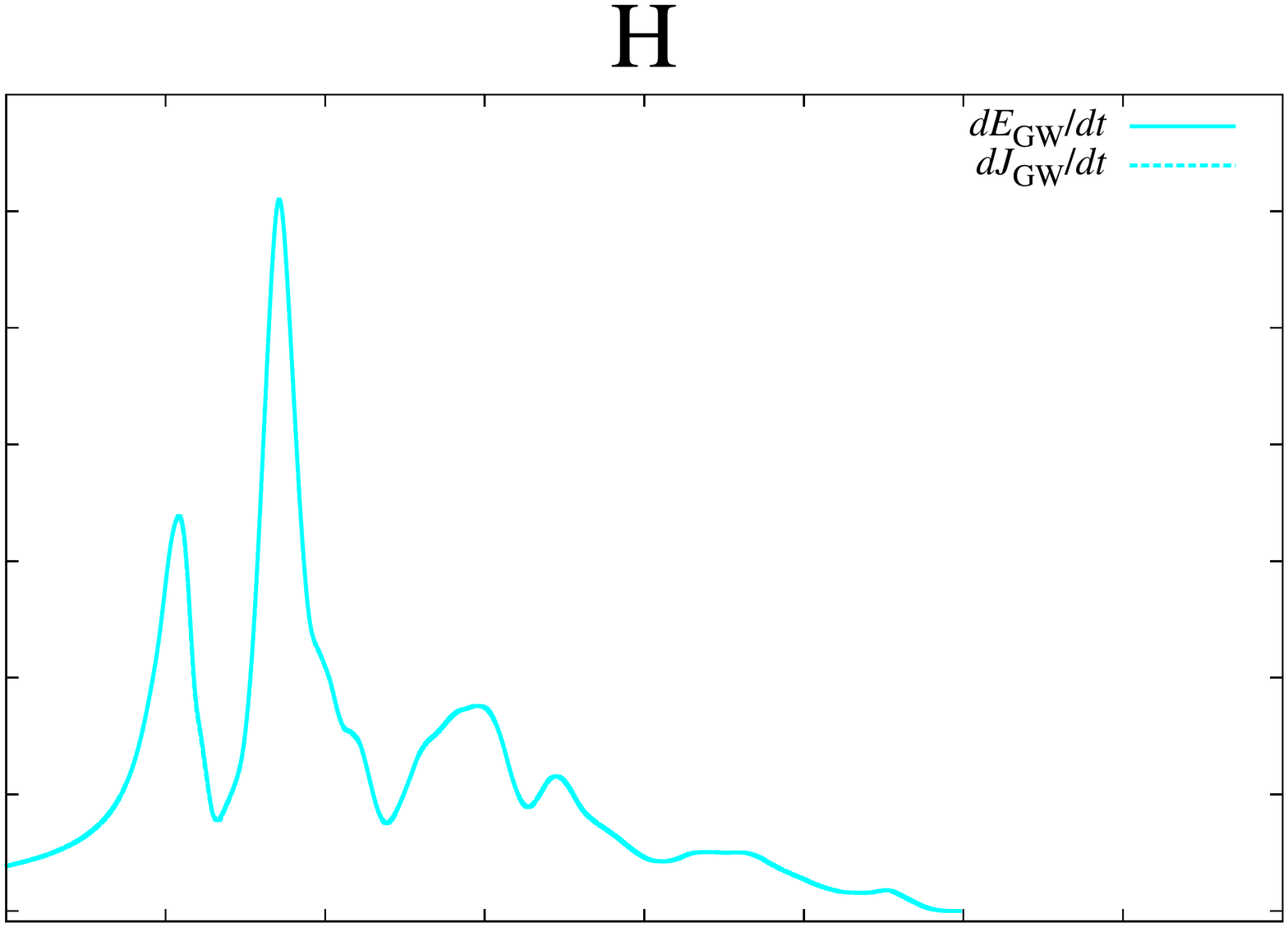}
\end{center}
\end{minipage}
\hspace{-13.35mm}
\begin{minipage}{0.27\hsize}
\begin{center}
\includegraphics[width=4.5cm,angle=0]{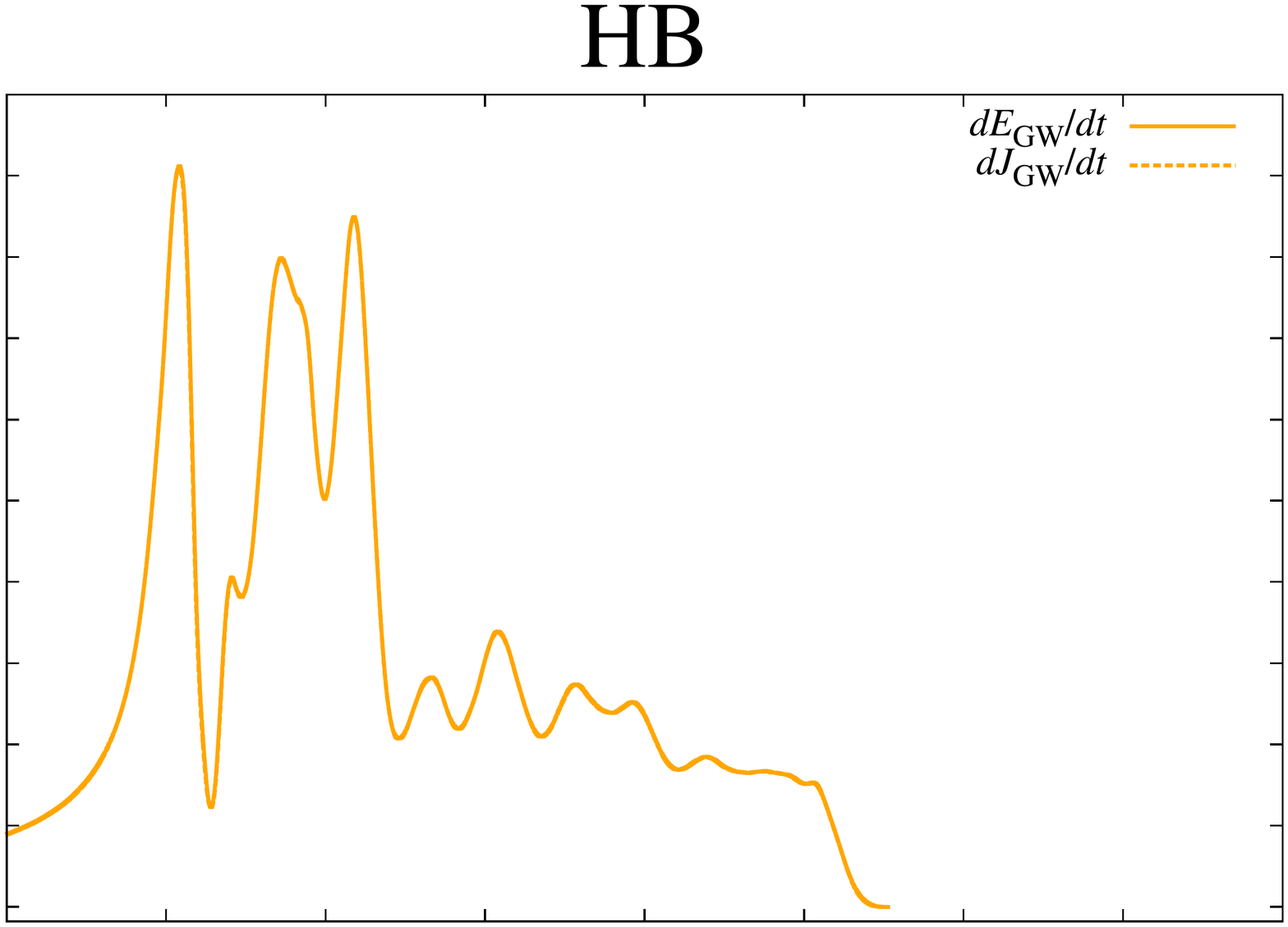}
\end{center}
\end{minipage}
\hspace{-13.35mm}
\begin{minipage}{0.27\hsize}
\begin{center}
\includegraphics[width=4.5cm,angle=0]{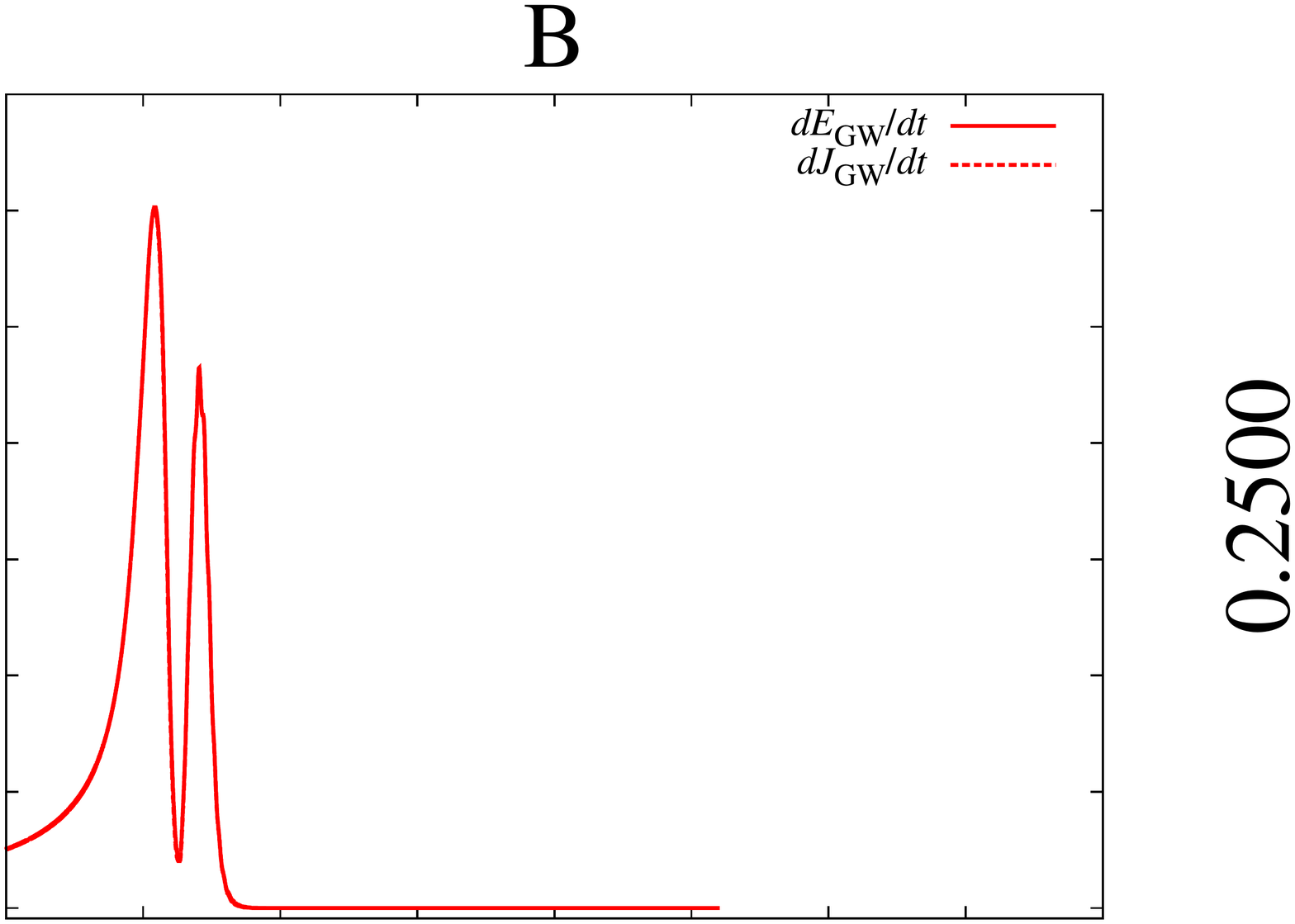}
\end{center}
\end{minipage}\\
\vspace{-9mm}
\hspace{-18.0mm}
\begin{minipage}{0.27\hsize}
\begin{center}
\includegraphics[width=4.5cm,angle=0]{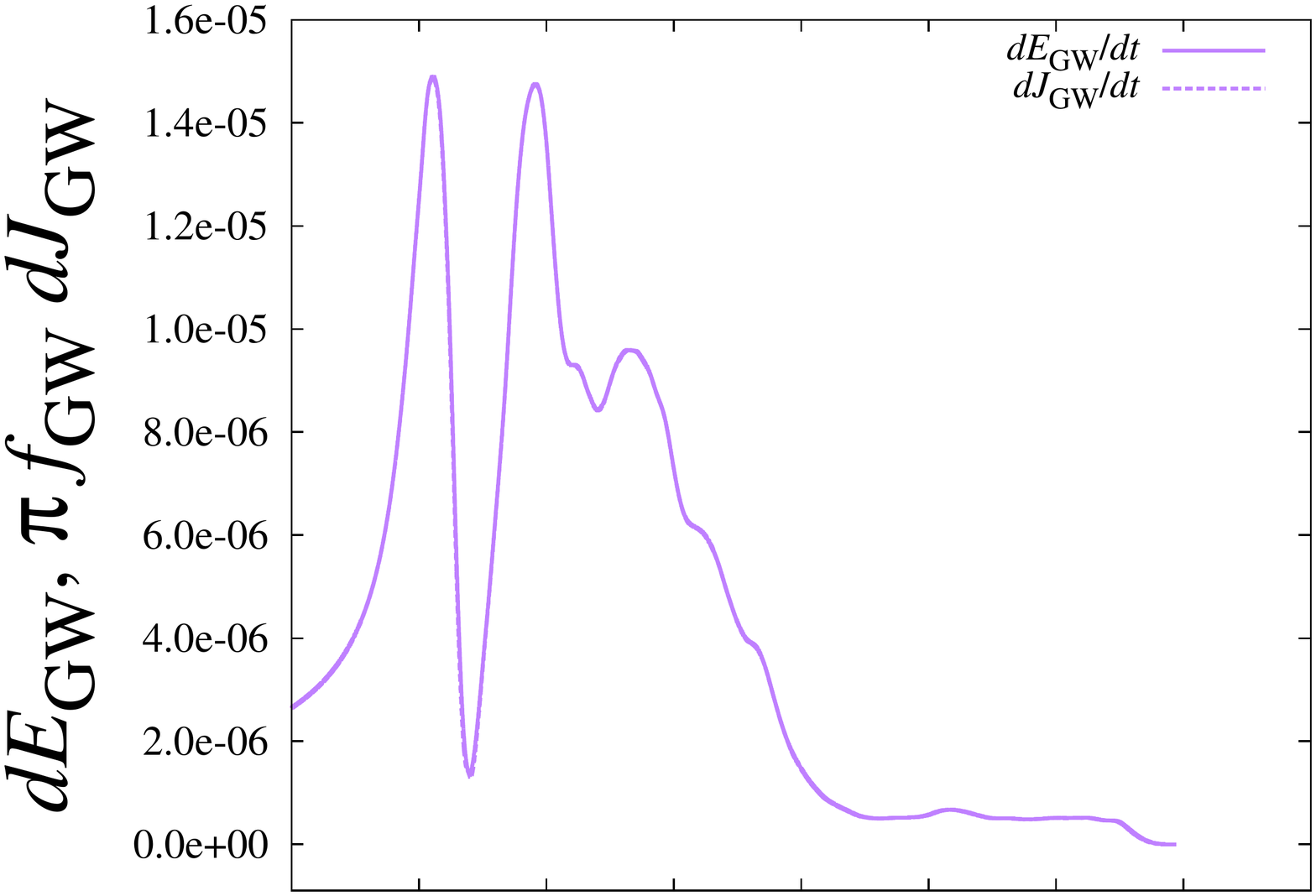}
\end{center}
\end{minipage}
\hspace{-13.35mm}
\begin{minipage}{0.27\hsize}
\begin{center}
\includegraphics[width=4.5cm,angle=0]{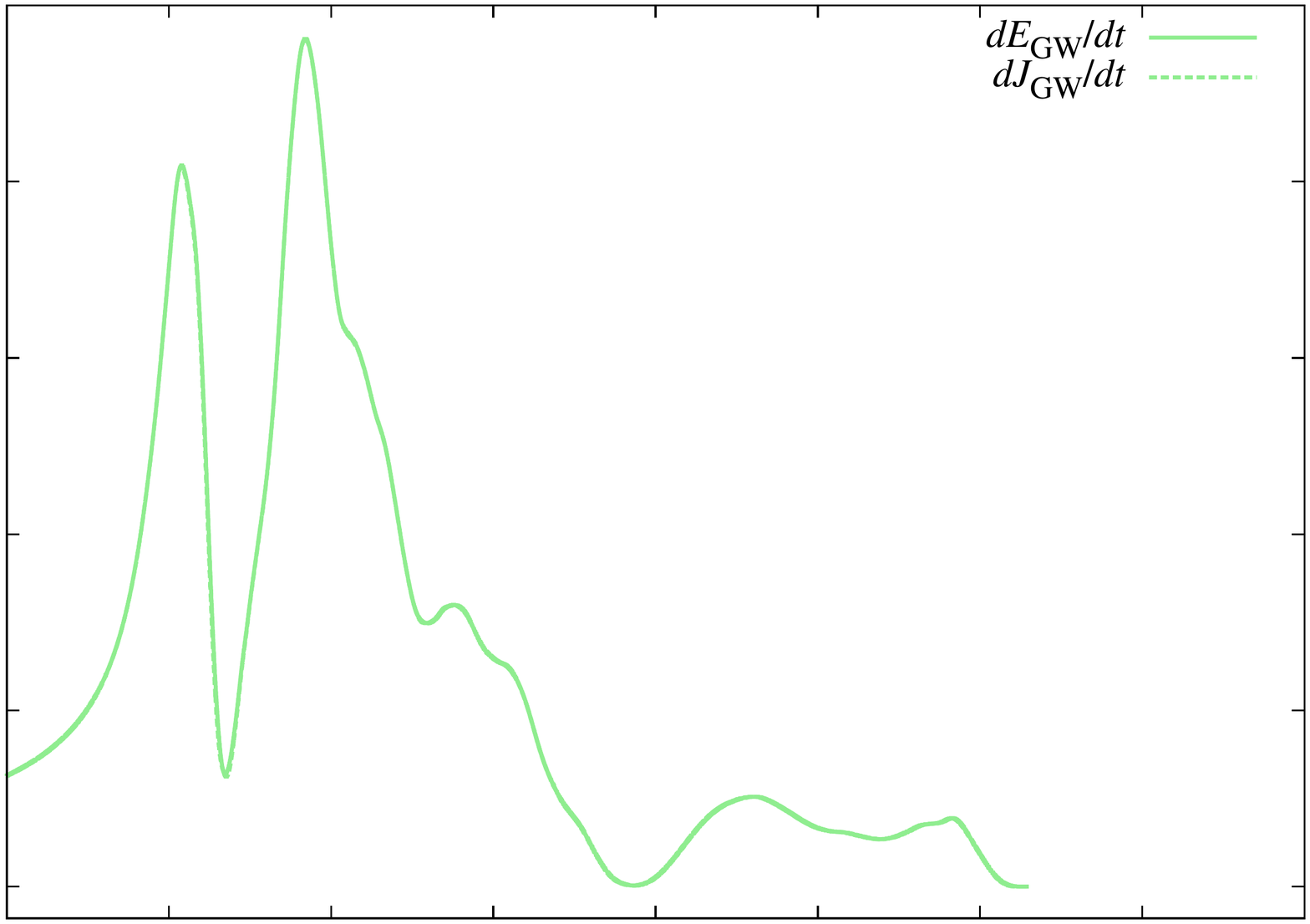}
\end{center}
\end{minipage}
\hspace{-13.35mm}
\begin{minipage}{0.27\hsize}
\begin{center}
\includegraphics[width=4.5cm,angle=0]{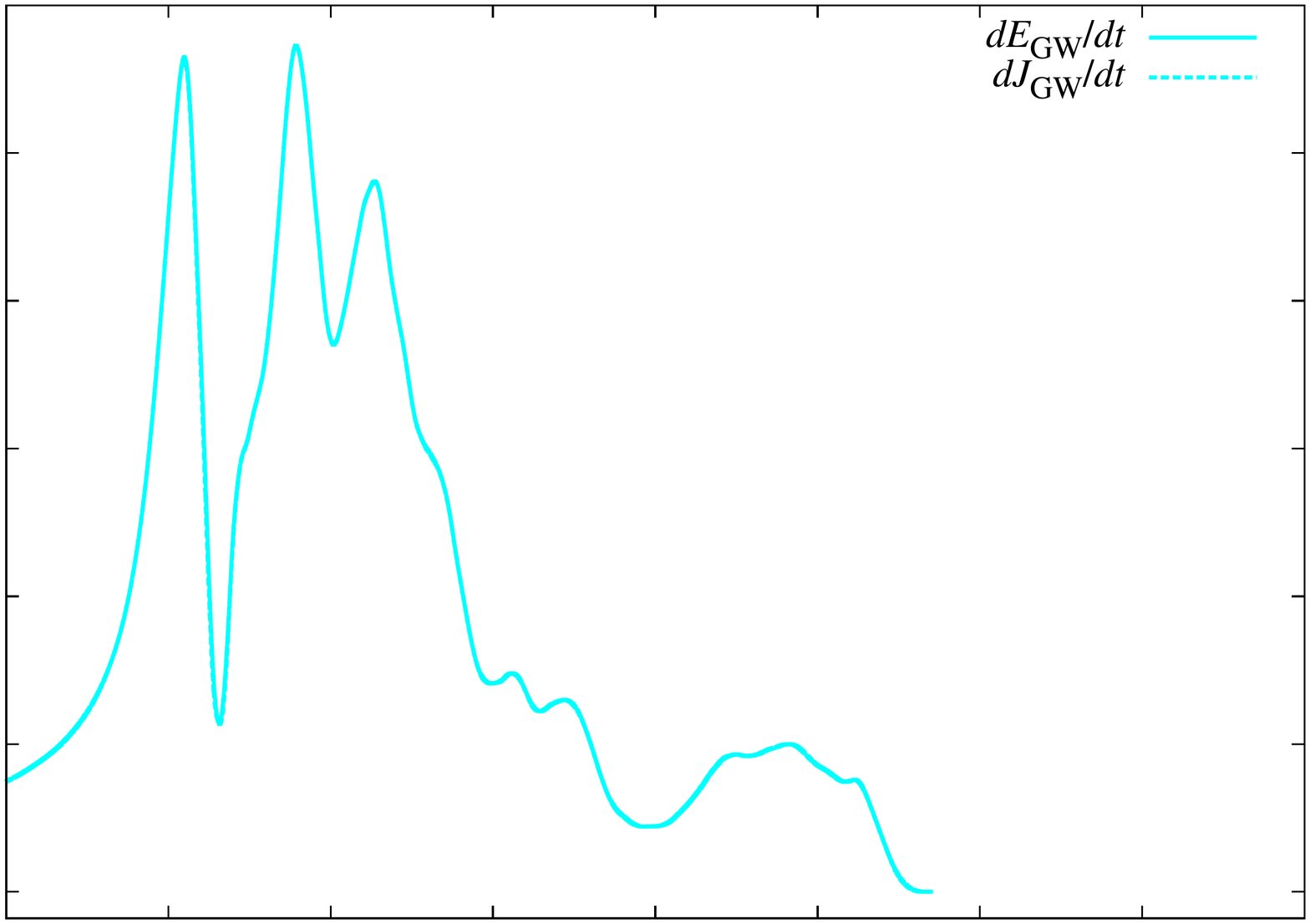}
\end{center}
\end{minipage}
\hspace{-13.35mm}
\begin{minipage}{0.27\hsize}
\begin{center}
\includegraphics[width=4.5cm,angle=0]{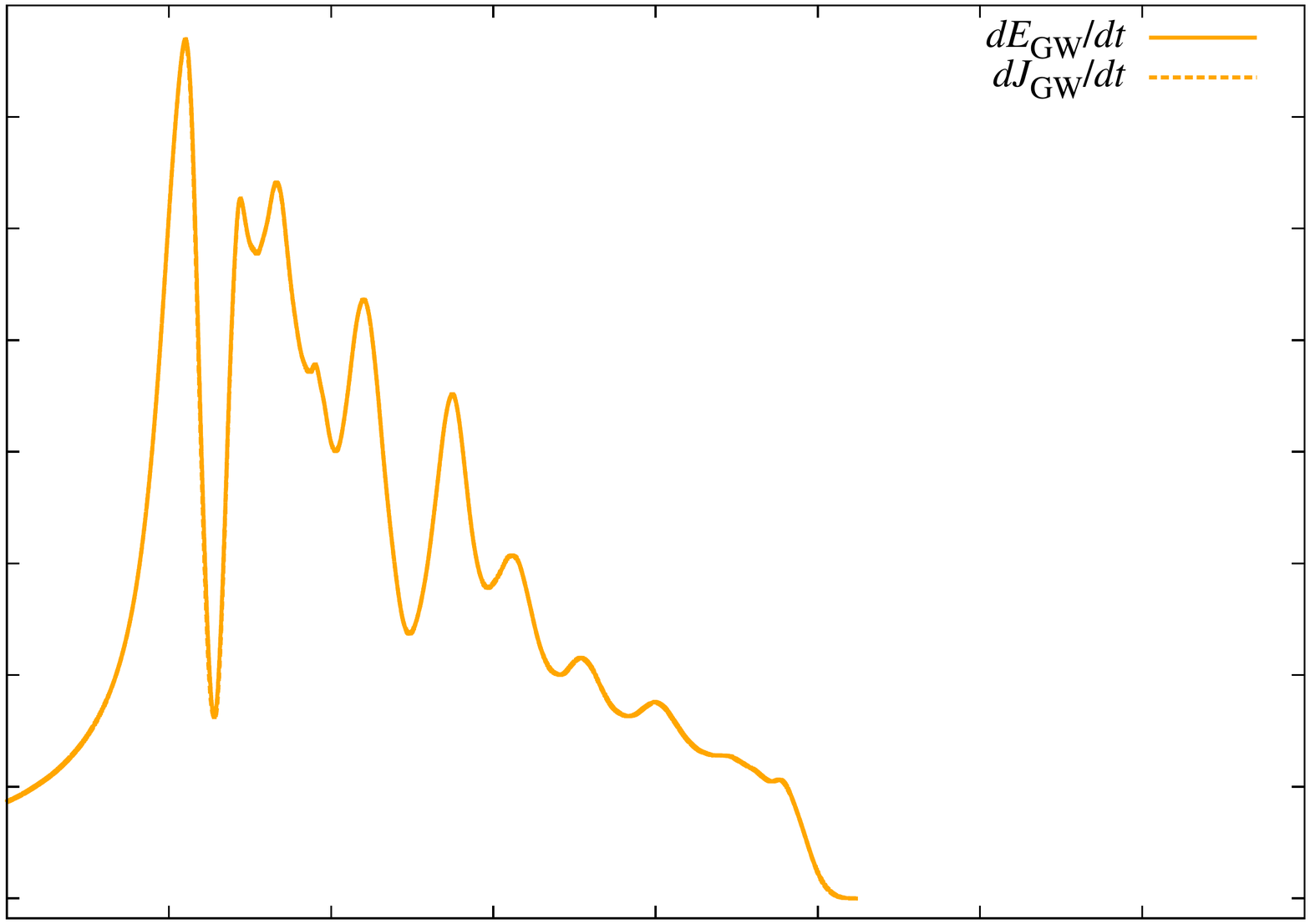}
\end{center}
\end{minipage}
\hspace{-13.35mm}
\begin{minipage}{0.27\hsize}
\begin{center}
\includegraphics[width=4.5cm,angle=0]{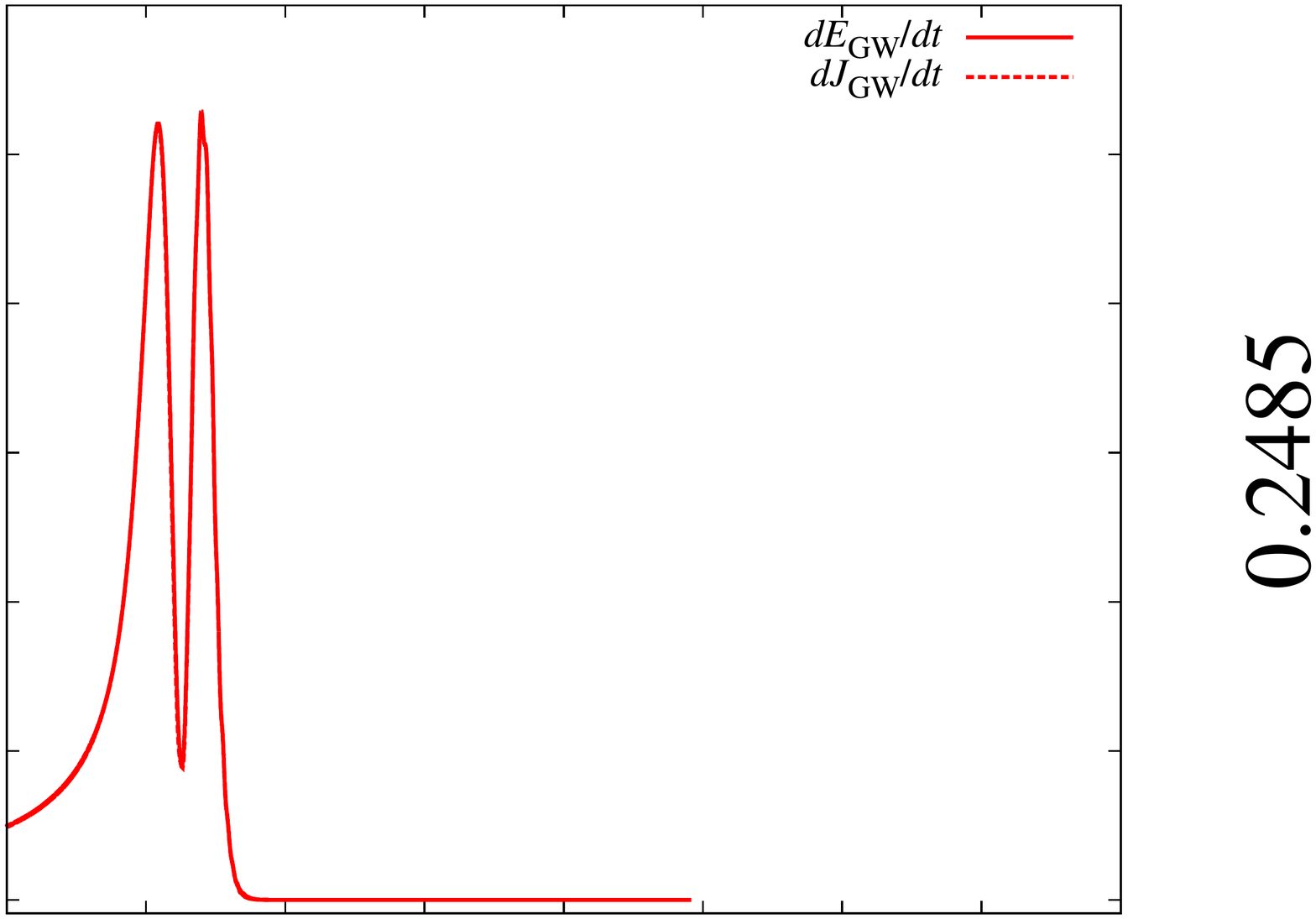}
\end{center}
\end{minipage}\\
\vspace{-9mm}
\hspace{-18.0mm}
\begin{minipage}{0.27\hsize}
\begin{center}
\includegraphics[width=4.5cm,angle=0]{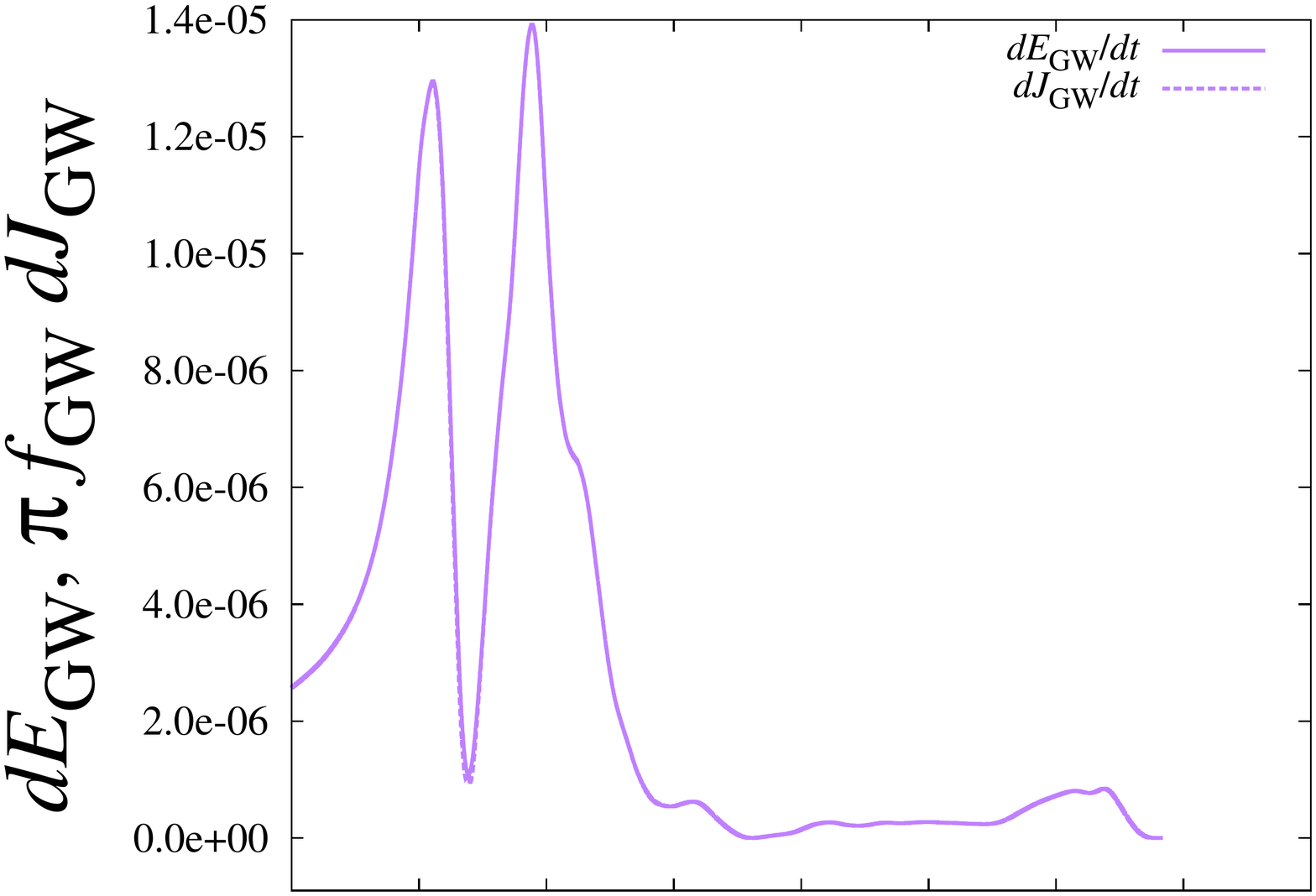}
\end{center}
\end{minipage}
\hspace{-13.35mm}
\begin{minipage}{0.27\hsize}
\begin{center}
\includegraphics[width=4.5cm,angle=0]{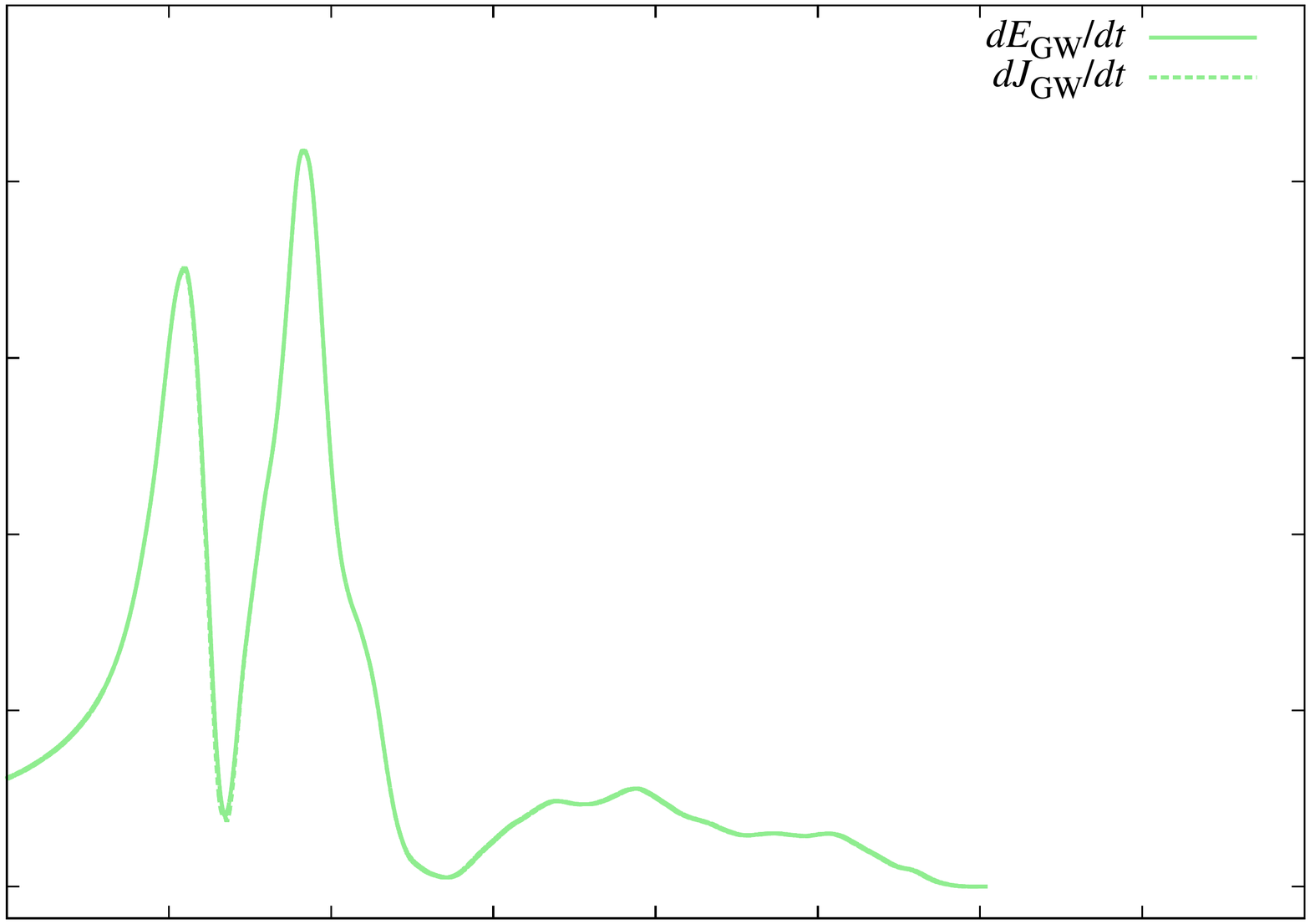}
\end{center}
\end{minipage}
\hspace{-13.35mm}
\begin{minipage}{0.27\hsize}
\begin{center}
\includegraphics[width=4.5cm,angle=0]{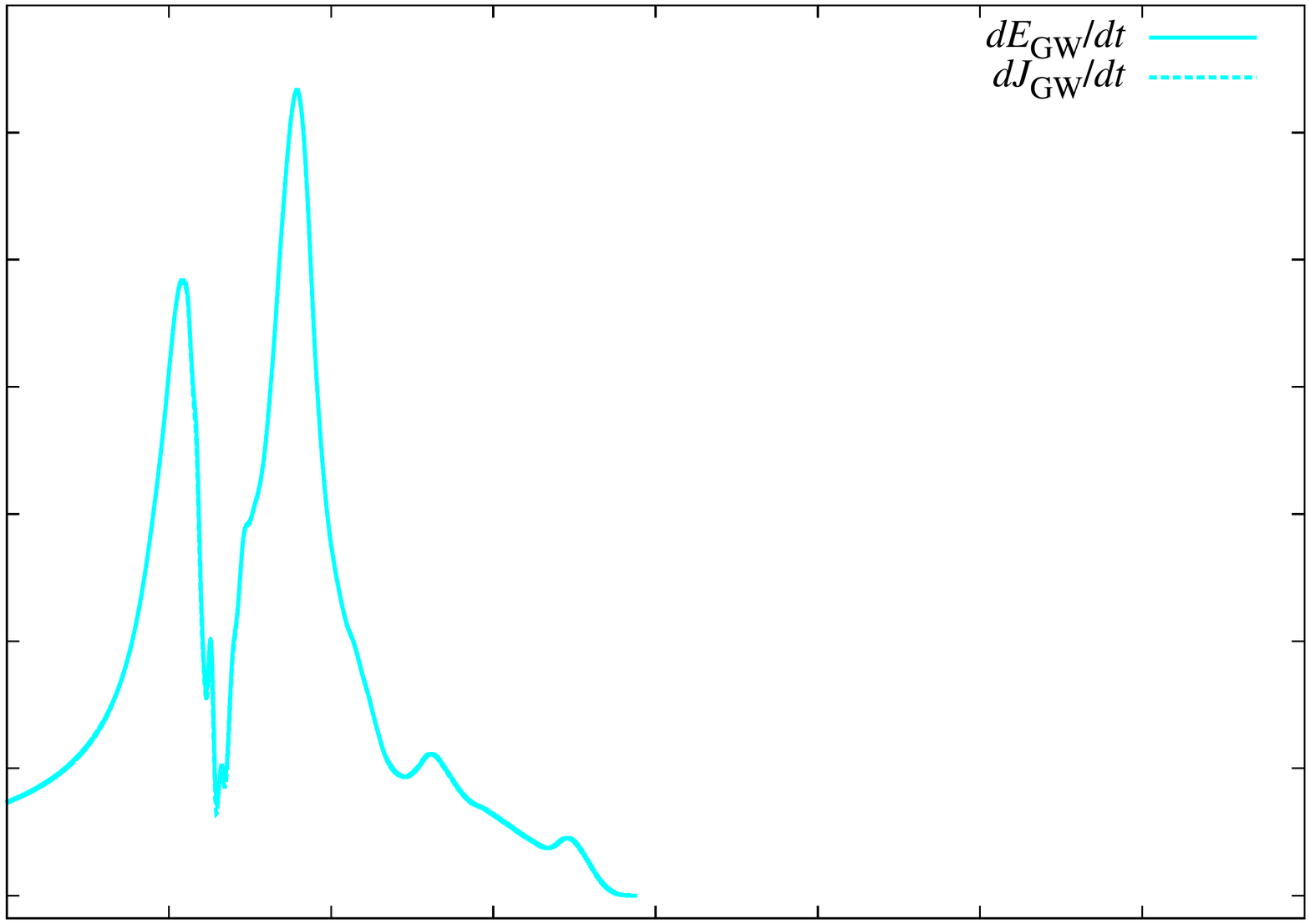}
\end{center}
\end{minipage}
\hspace{-13.35mm}
\begin{minipage}{0.27\hsize}
\begin{center}
\includegraphics[width=4.5cm,angle=0]{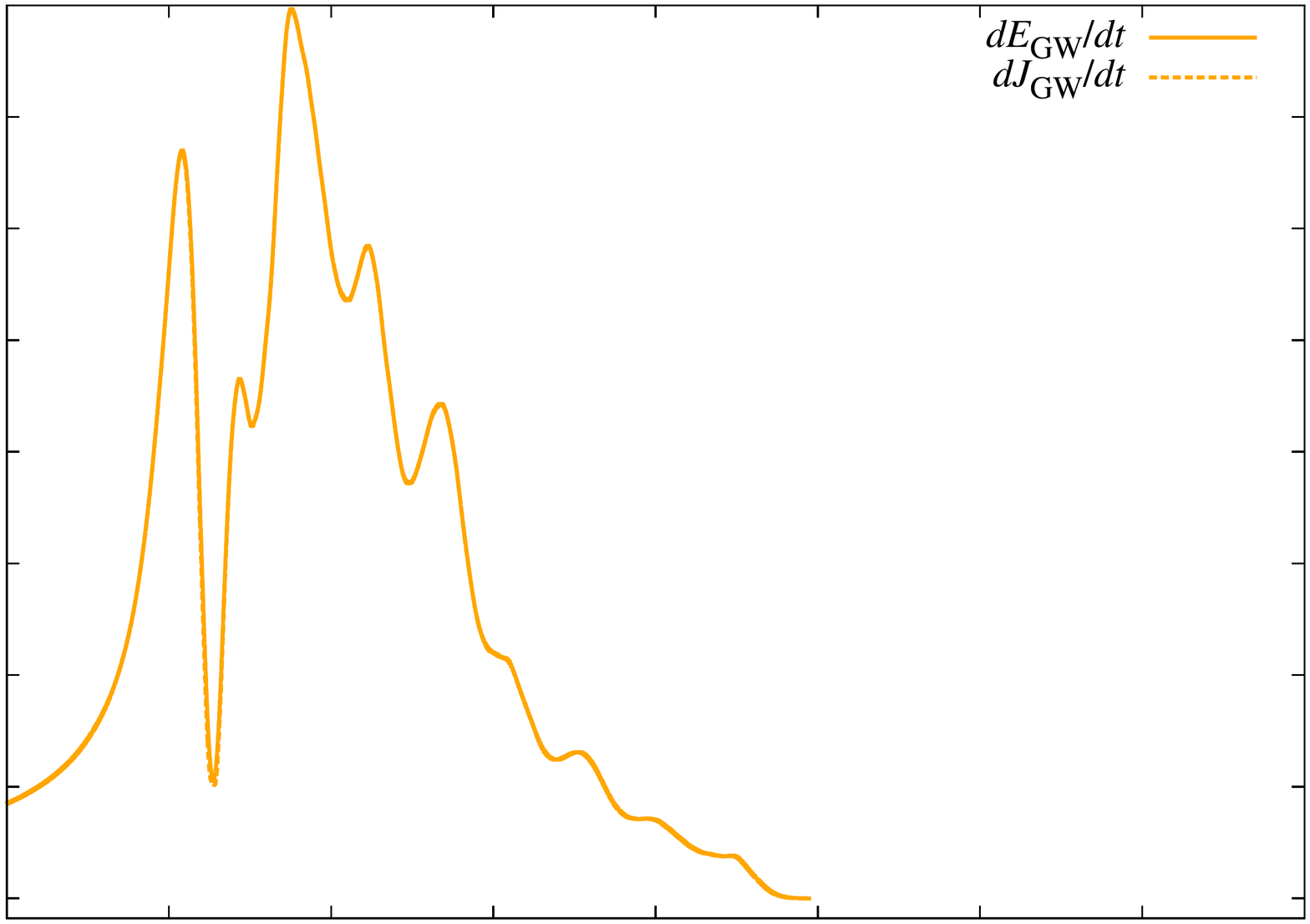}
\end{center}
\end{minipage}
\hspace{-13.35mm}
\begin{minipage}{0.27\hsize}
\begin{center}
\includegraphics[width=4.5cm,angle=0]{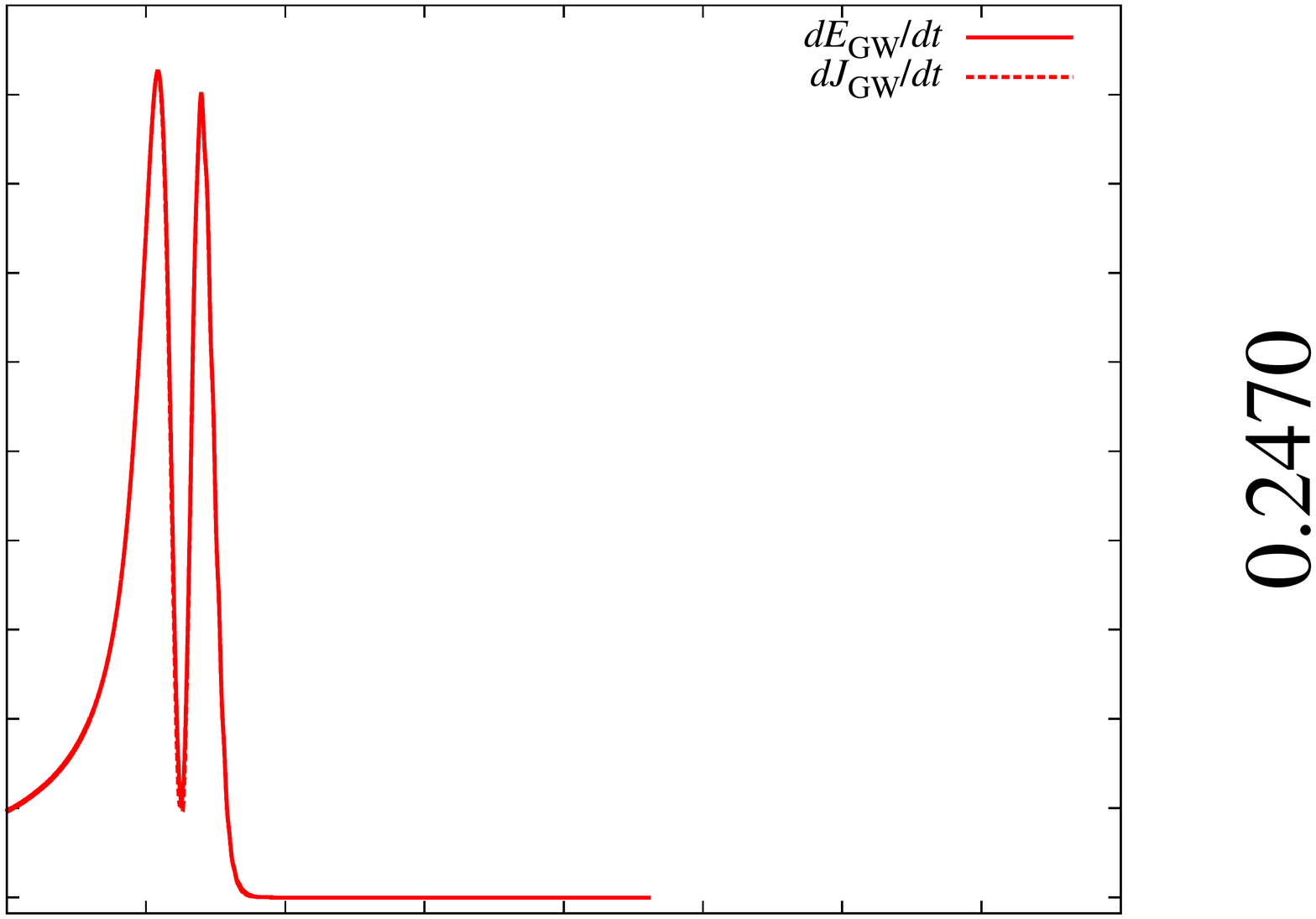}
\end{center}
\end{minipage}\\
\vspace{-9mm}
\hspace{-18.0mm}
\begin{minipage}{0.27\hsize}
\begin{center}
\includegraphics[width=4.5cm,angle=0]{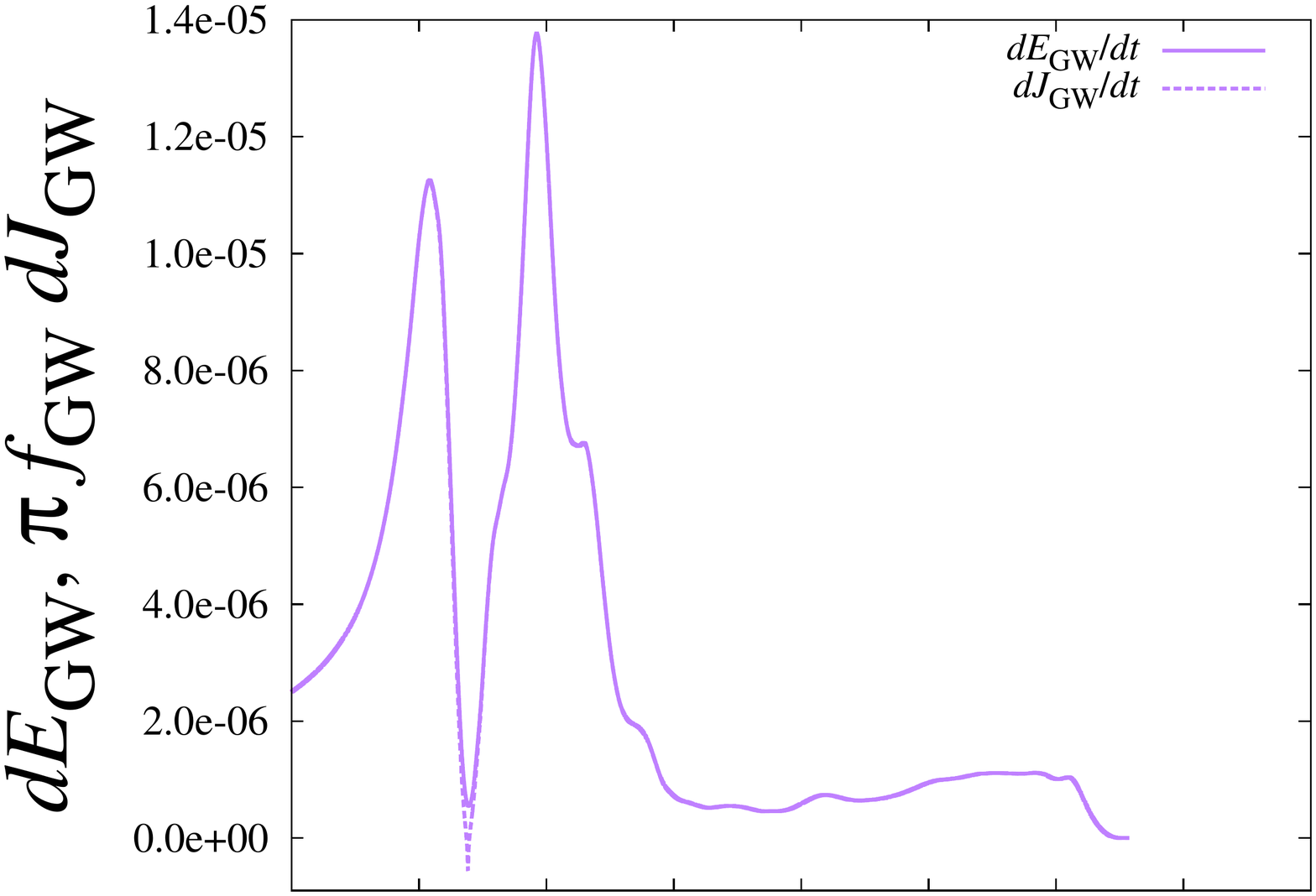}
\end{center}
\end{minipage}
\hspace{-13.35mm}
\begin{minipage}{0.27\hsize}
\begin{center}
\includegraphics[width=4.5cm,angle=0]{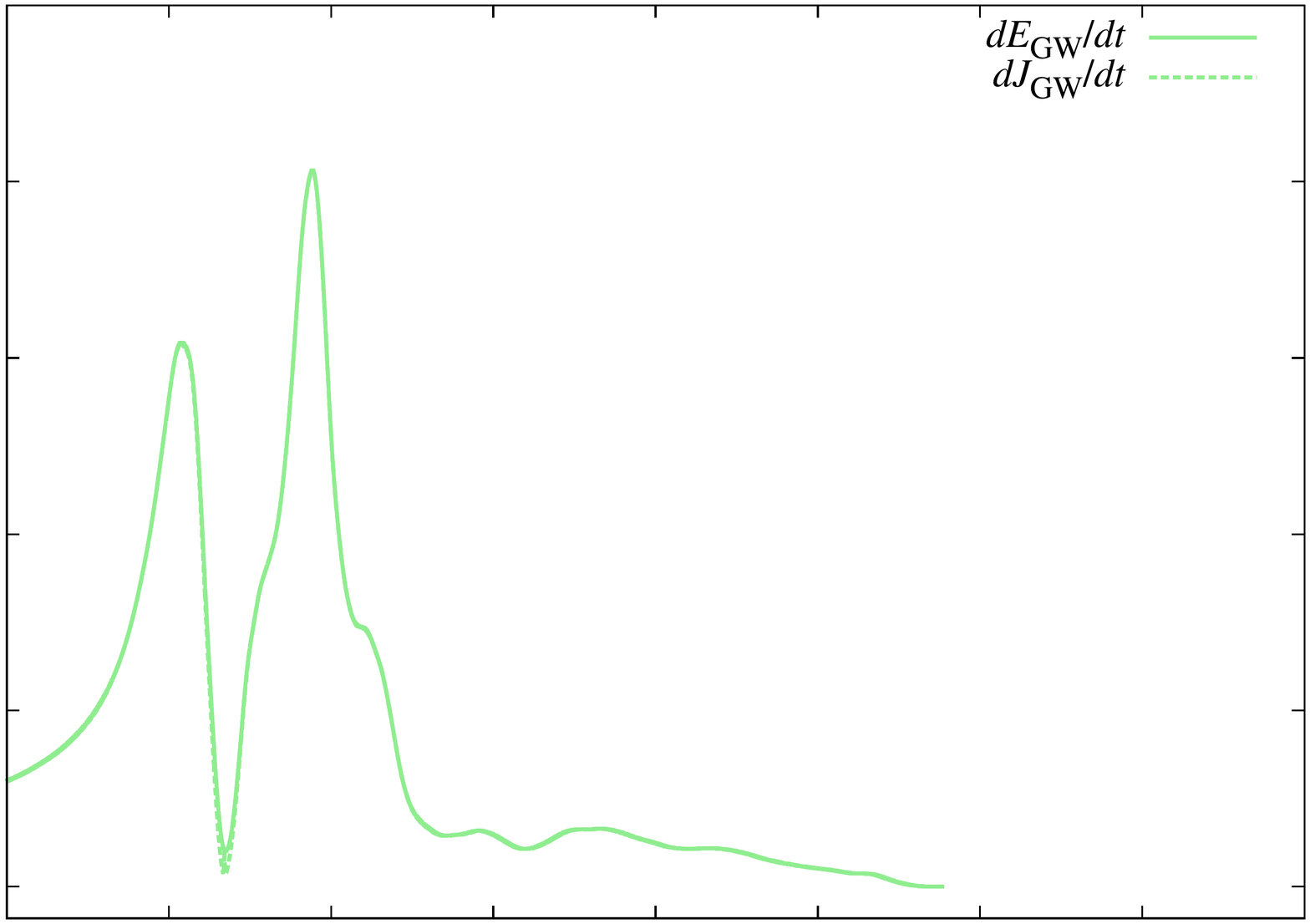}
\end{center}
\end{minipage}
\hspace{-13.35mm}
\begin{minipage}{0.27\hsize}
\begin{center}
\includegraphics[width=4.5cm,angle=0]{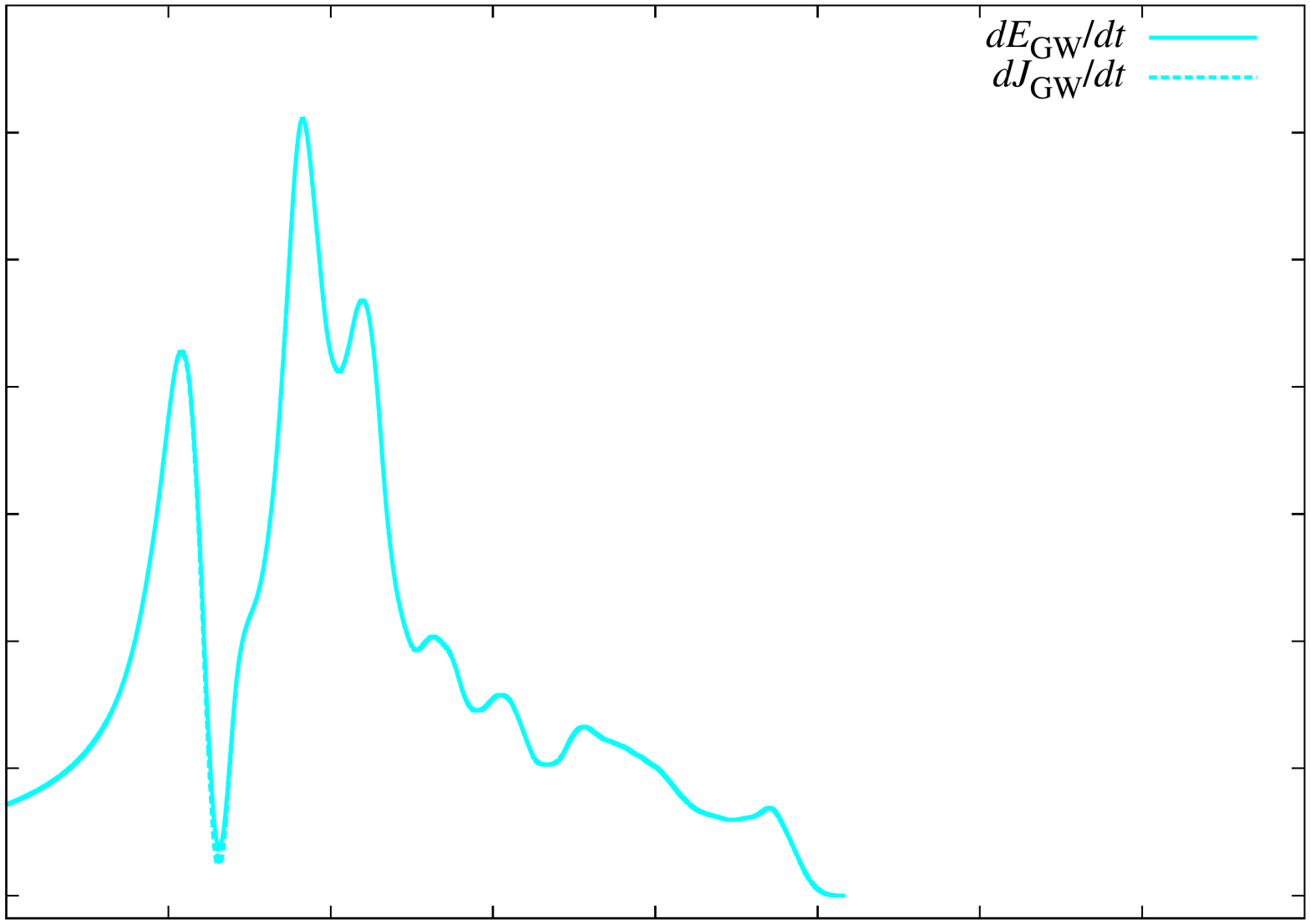}
\end{center}
\end{minipage}
\hspace{-13.35mm}
\begin{minipage}{0.27\hsize}
\begin{center}
\includegraphics[width=4.5cm,angle=0]{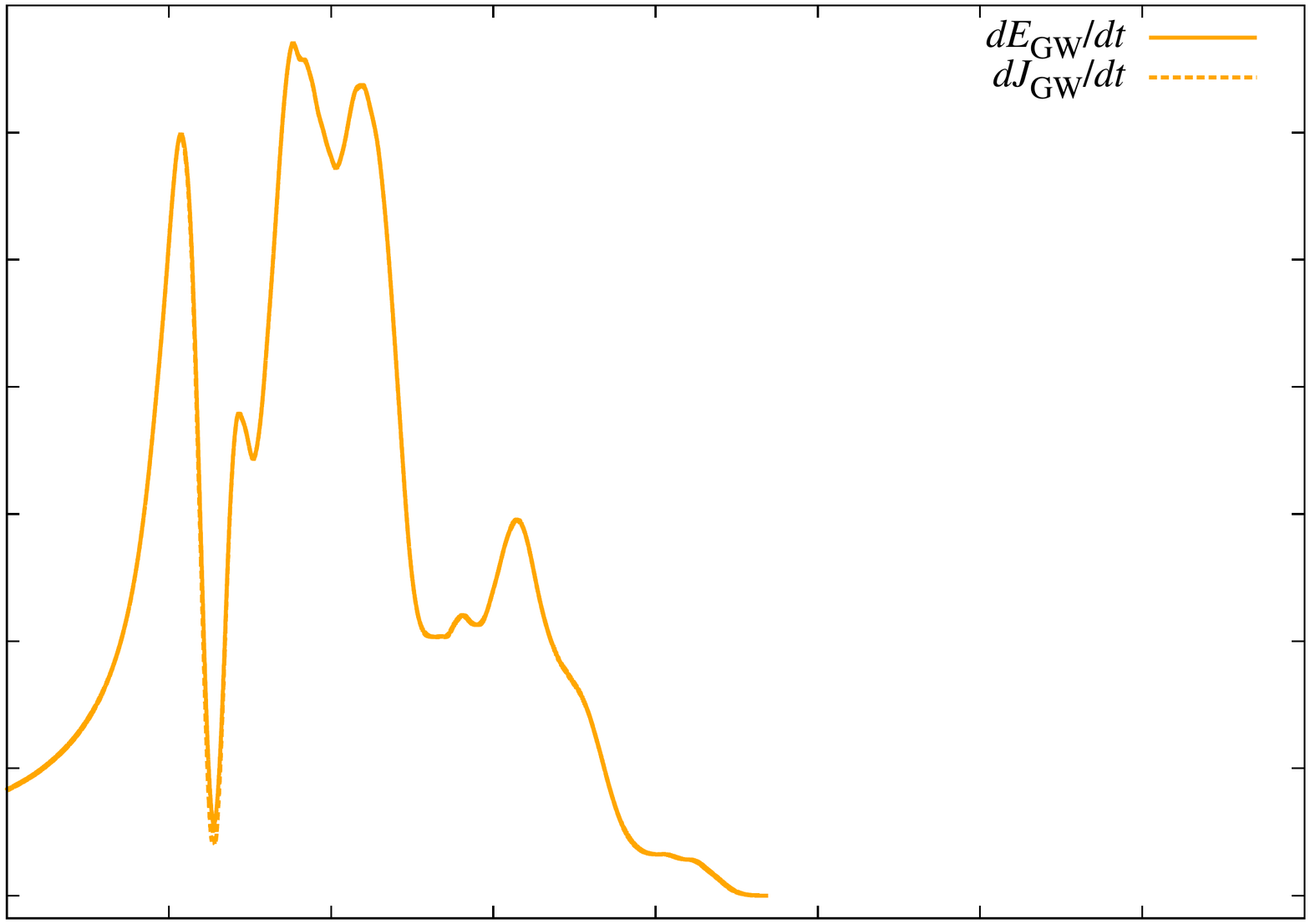}
\end{center}
\end{minipage}
\hspace{-13.35mm}
\begin{minipage}{0.27\hsize}
\begin{center}
\includegraphics[width=4.5cm,angle=0]{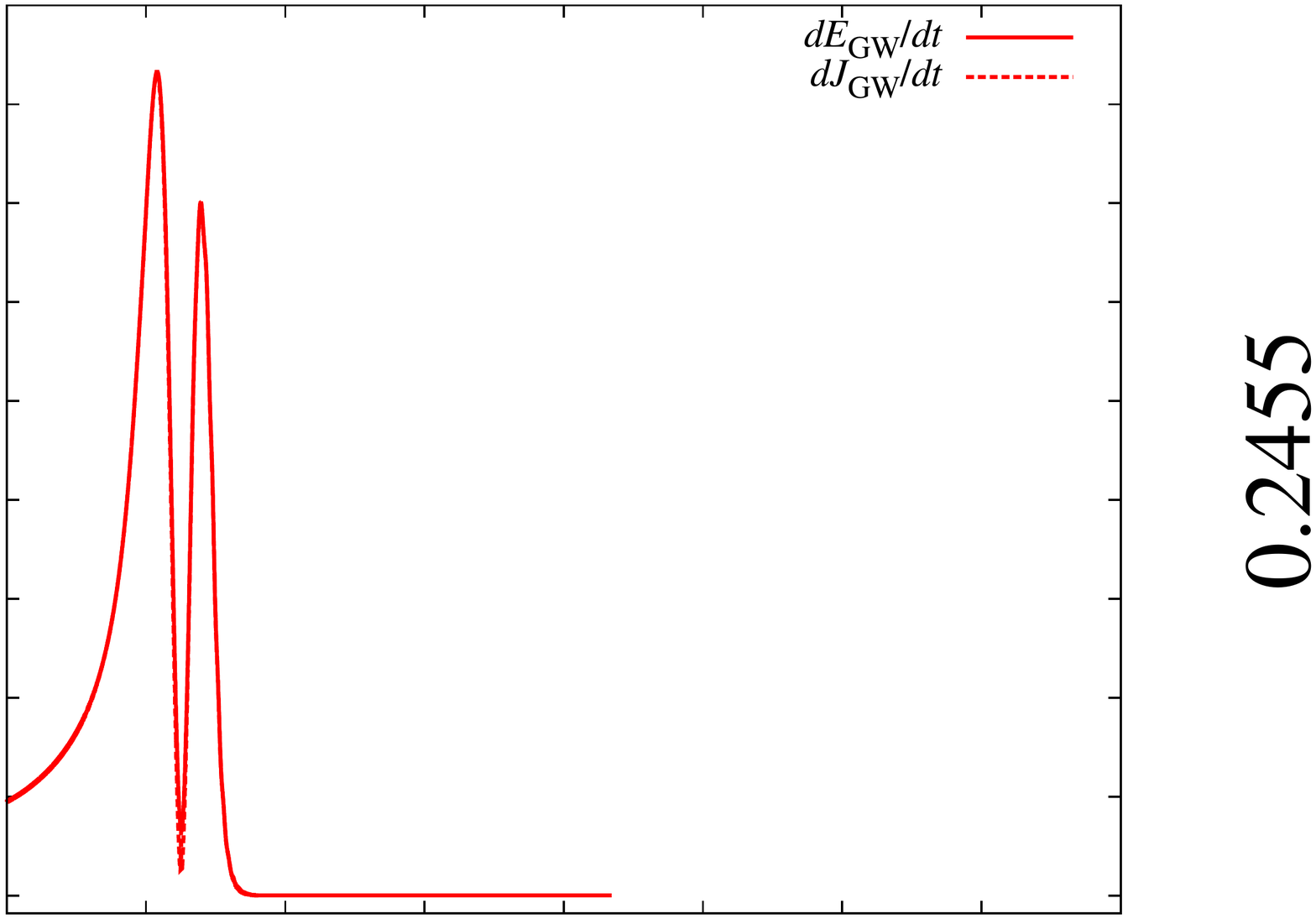}
\end{center}
\end{minipage}\\
\vspace{-9mm}
\hspace{-18.0mm}
\begin{minipage}{0.27\hsize}
\begin{center}
\includegraphics[width=4.5cm,angle=0]{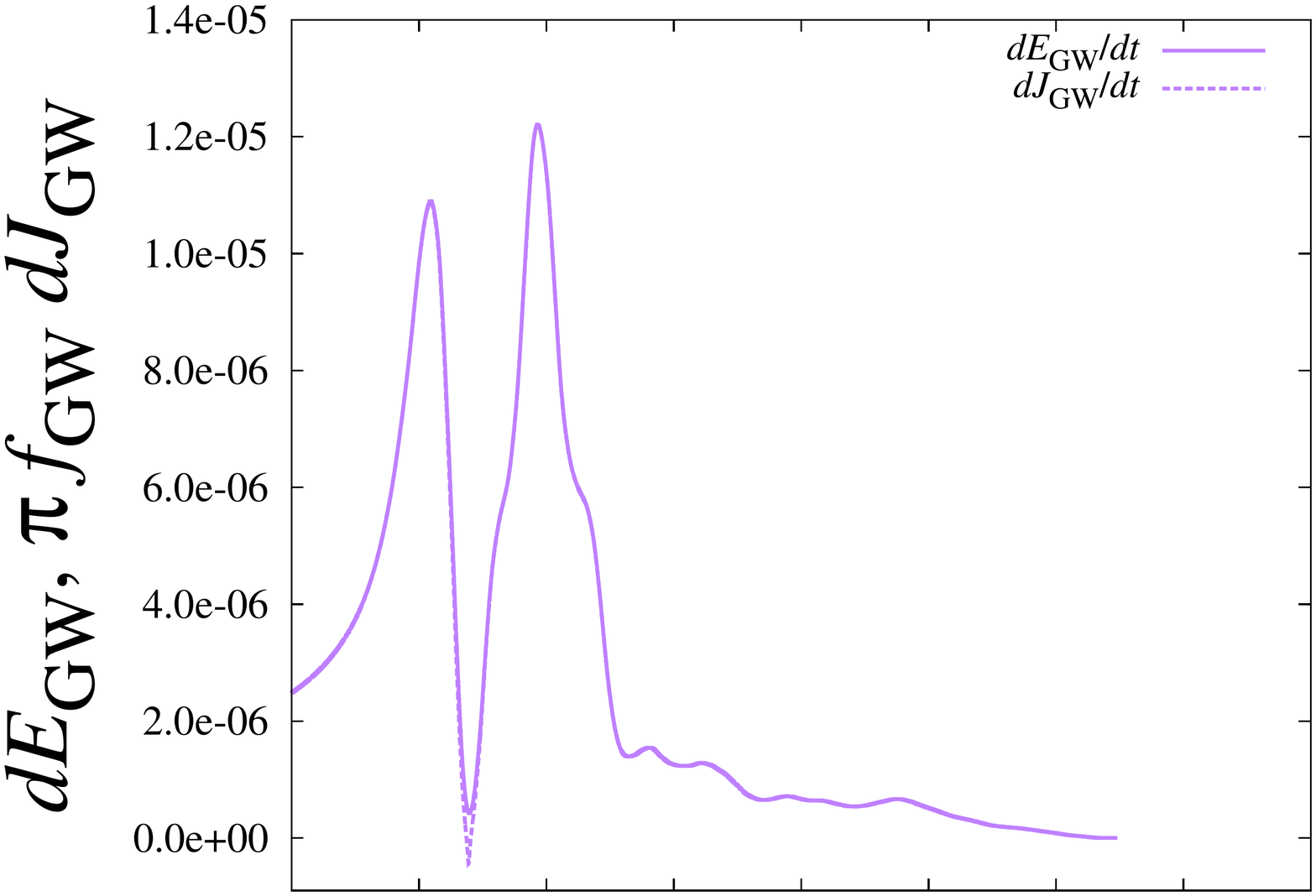}
\end{center}
\end{minipage}
\hspace{-13.35mm}
\begin{minipage}{0.27\hsize}
\begin{center}
\includegraphics[width=4.5cm,angle=0]{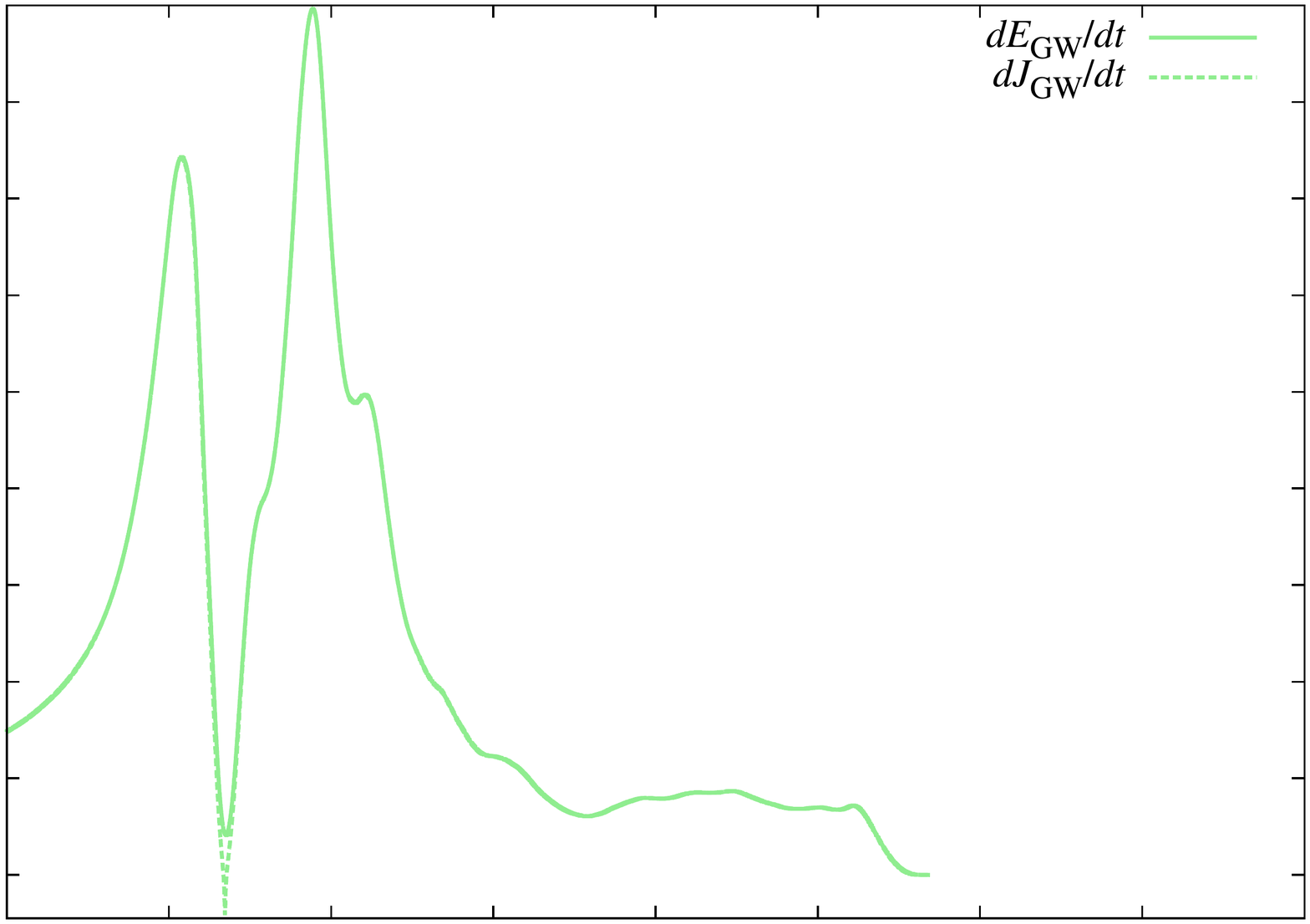}
\end{center}
\end{minipage}
\hspace{-13.35mm}
\begin{minipage}{0.27\hsize}
\begin{center}
\includegraphics[width=4.5cm,angle=0]{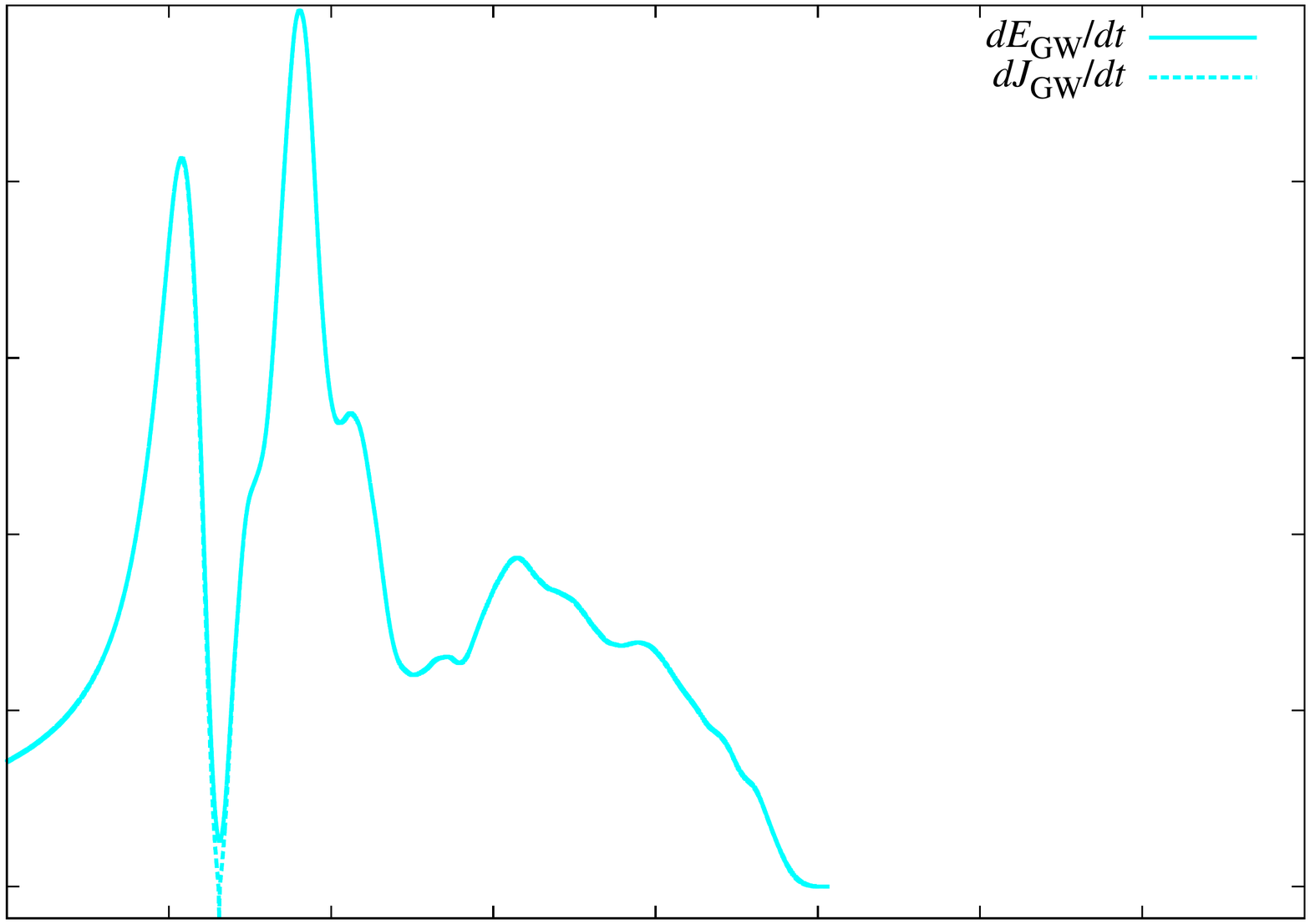}
\end{center}
\end{minipage}
\hspace{-13.35mm}
\begin{minipage}{0.27\hsize}
\begin{center}
\includegraphics[width=4.5cm,angle=0]{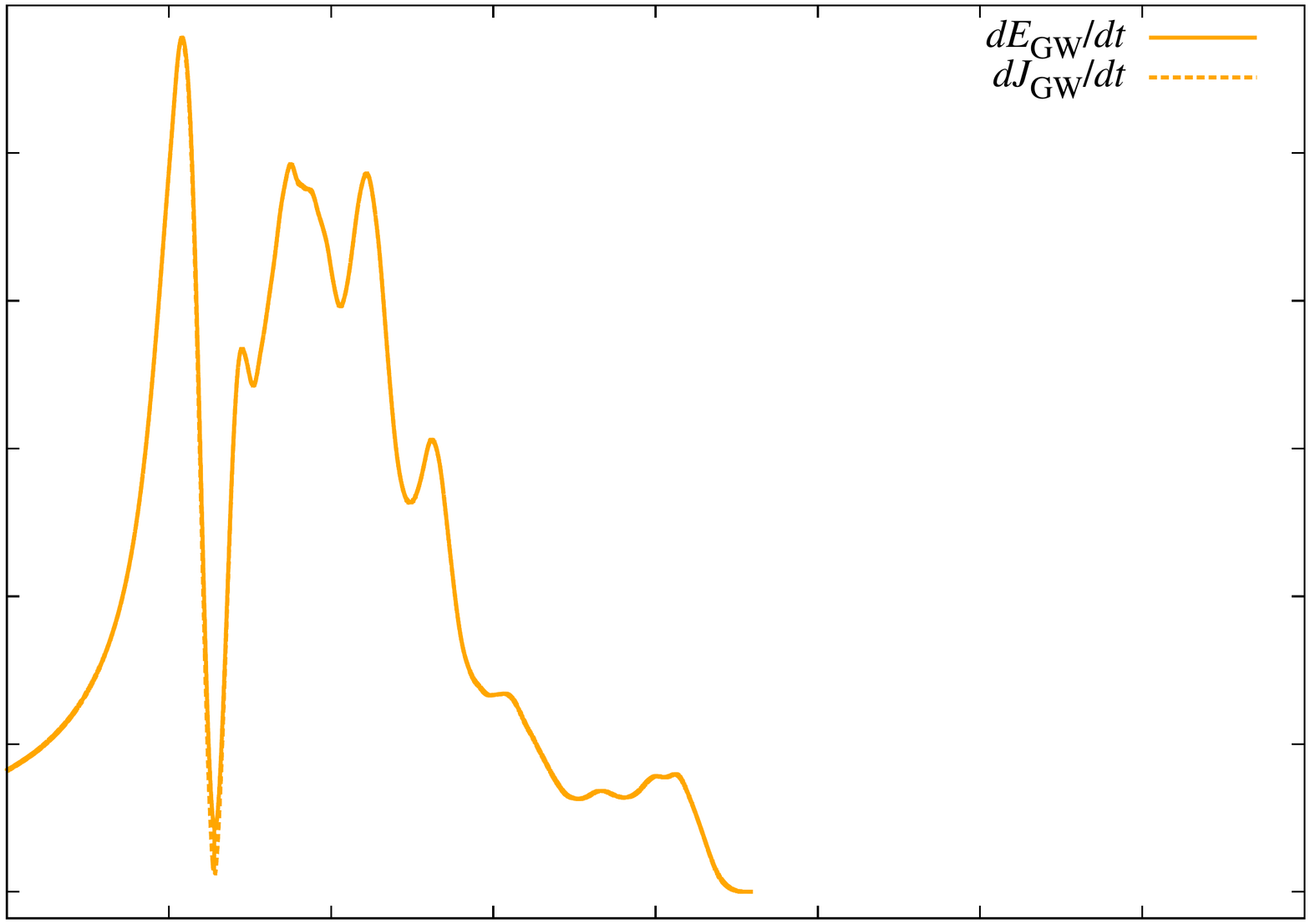}
\end{center}
\end{minipage}
\hspace{-13.35mm}
\begin{minipage}{0.27\hsize}
\begin{center}
\includegraphics[width=4.5cm,angle=0]{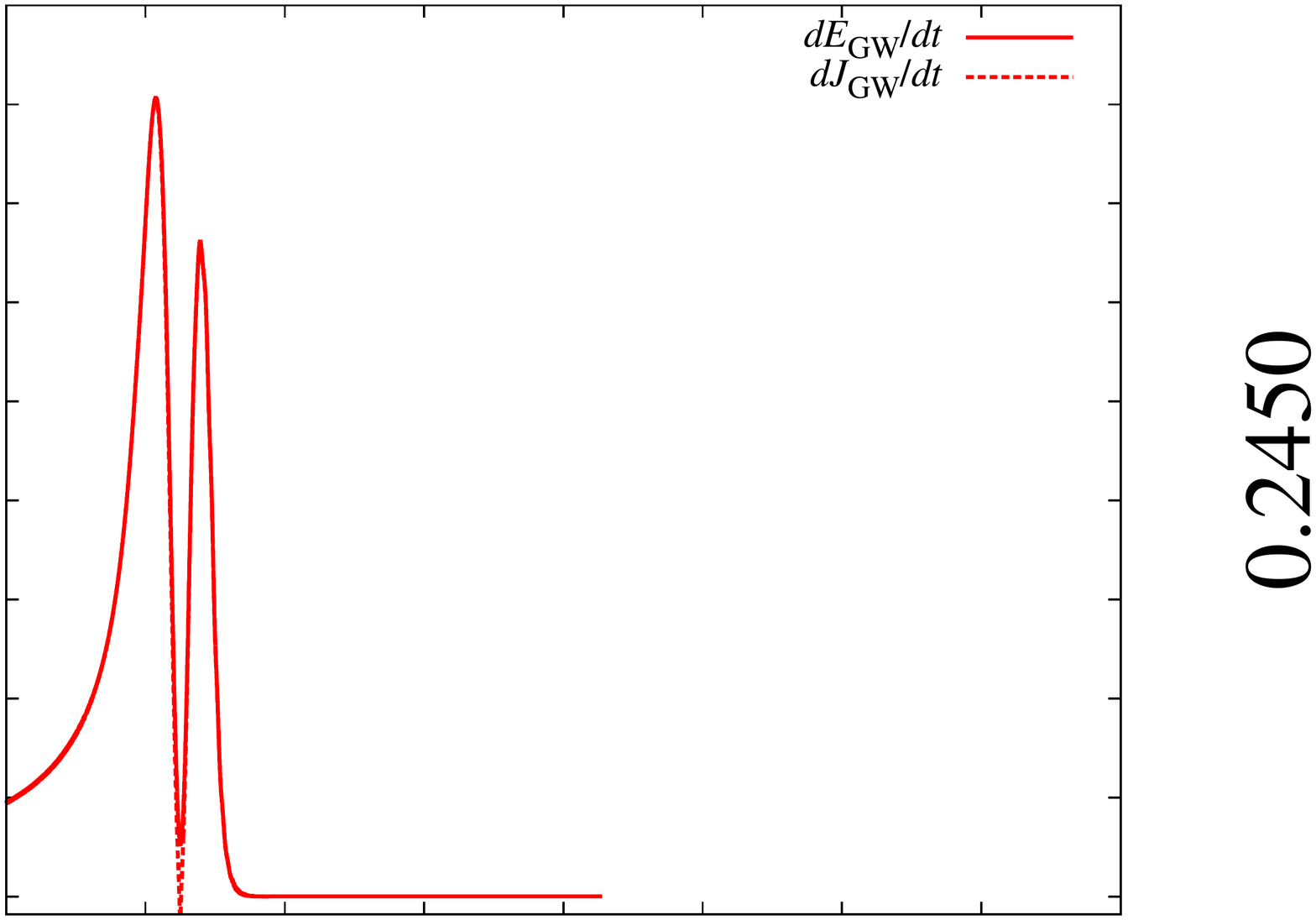}
\end{center}
\end{minipage}\\
\vspace{-9mm}
\hspace{-18.0mm}
\begin{minipage}{0.27\hsize}
\begin{center}
\includegraphics[width=4.5cm,angle=0]{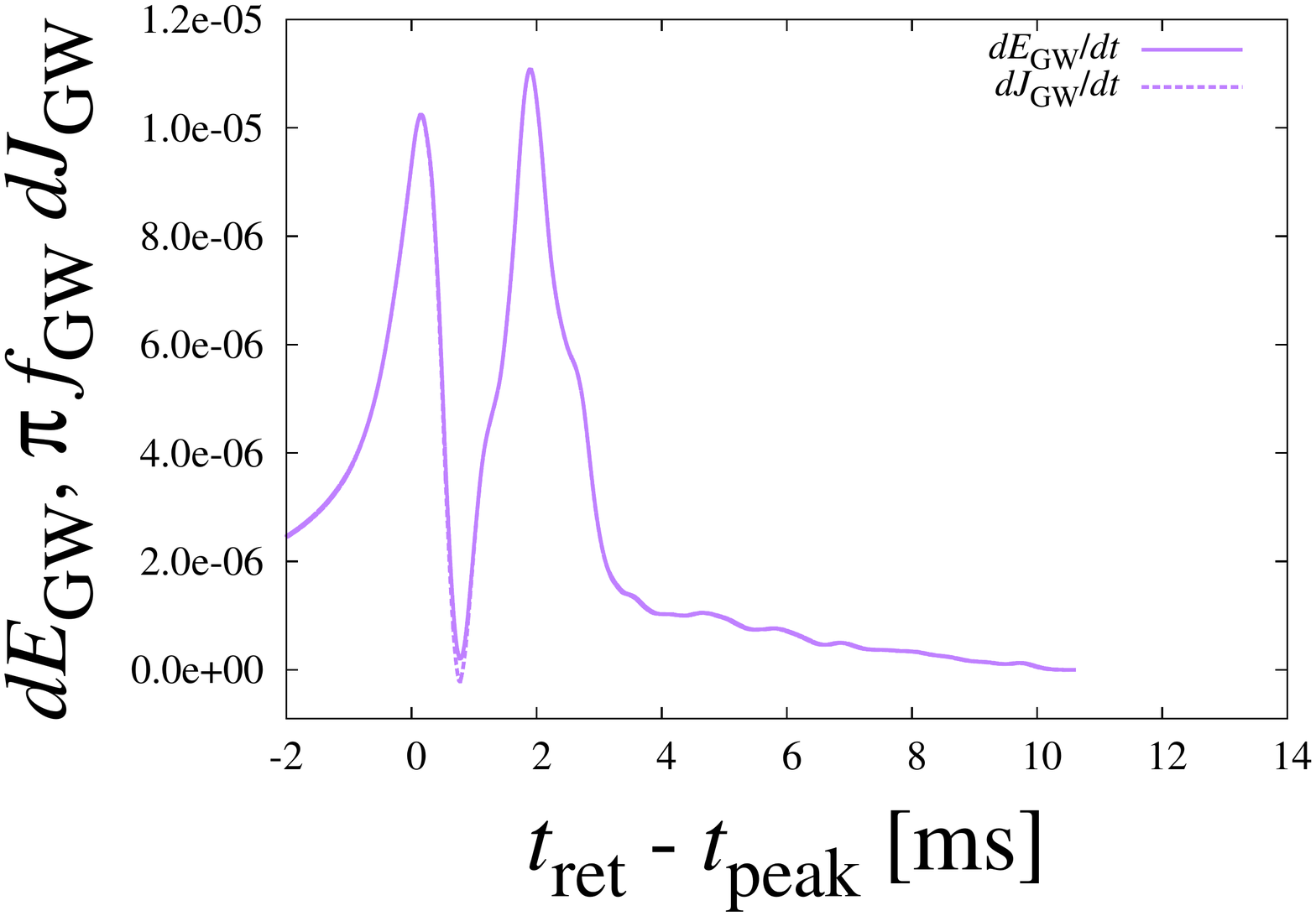}
\end{center}
\end{minipage}
\hspace{-13.35mm}
\begin{minipage}{0.27\hsize}
\begin{center}
\includegraphics[width=4.5cm,angle=0]{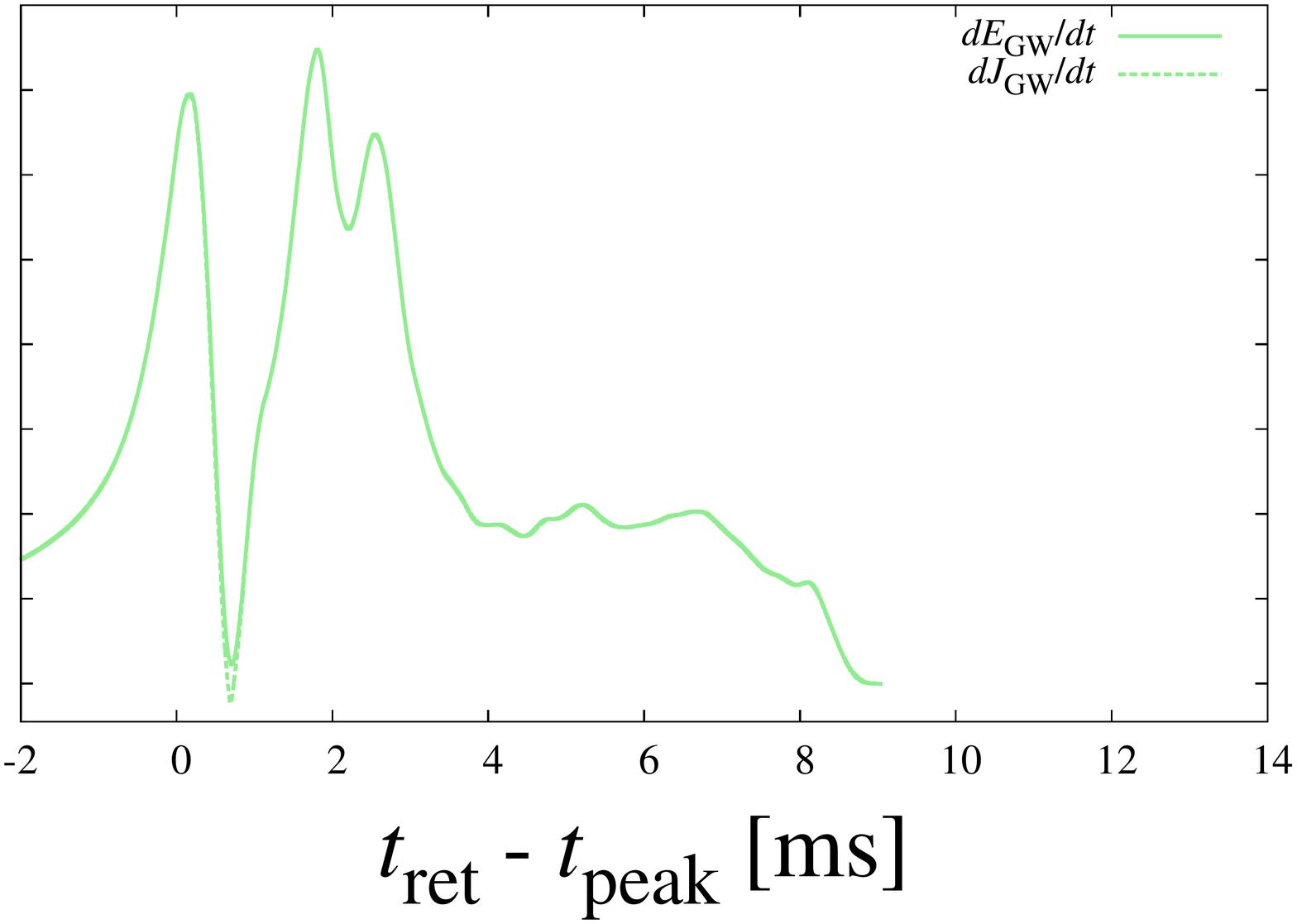}
\end{center}
\end{minipage}
\hspace{-13.35mm}
\begin{minipage}{0.27\hsize}
\begin{center}
\includegraphics[width=4.5cm,angle=0]{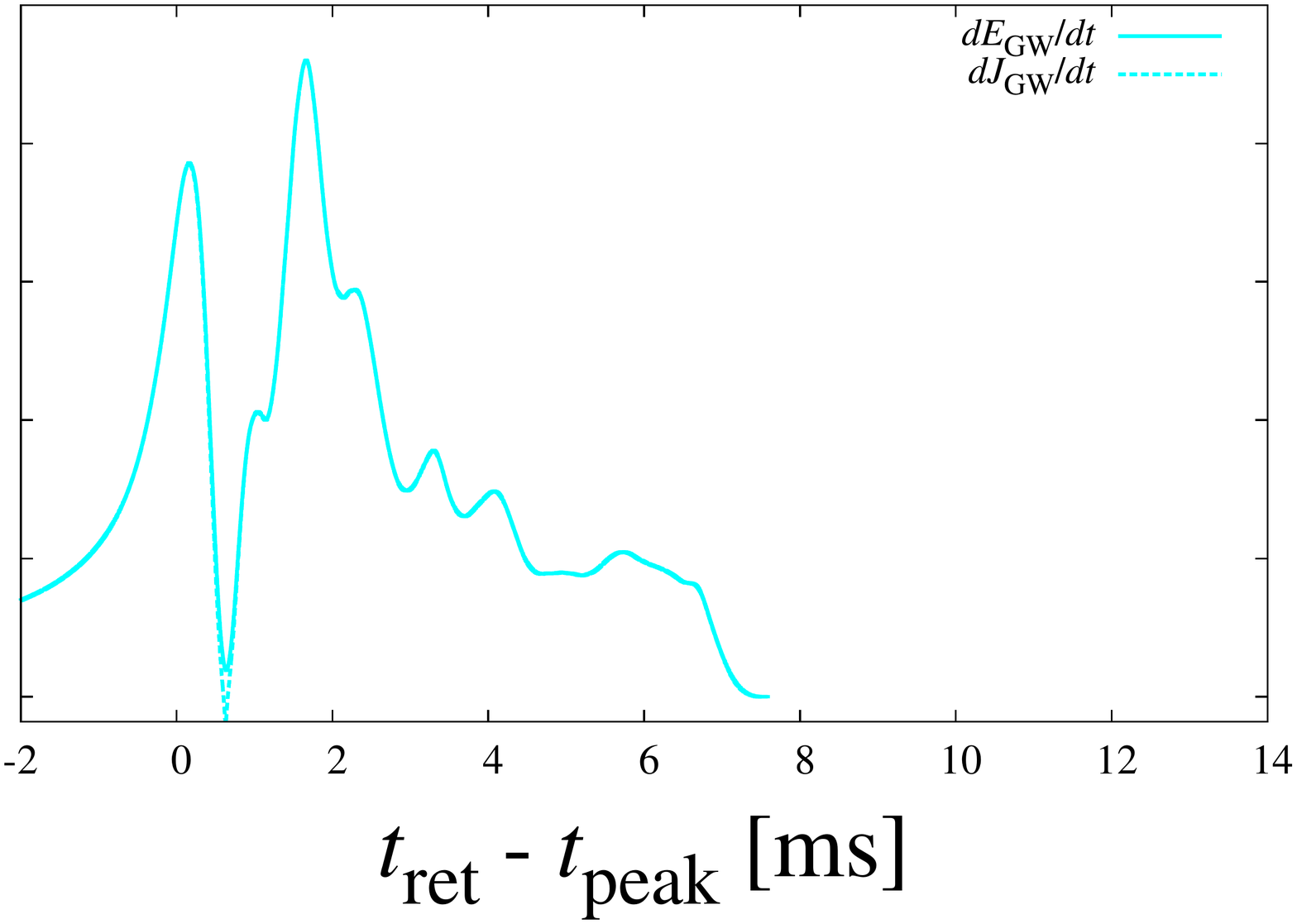}
\end{center}
\end{minipage}
\hspace{-13.35mm}
\begin{minipage}{0.27\hsize}
\begin{center}
\includegraphics[width=4.5cm,angle=0]{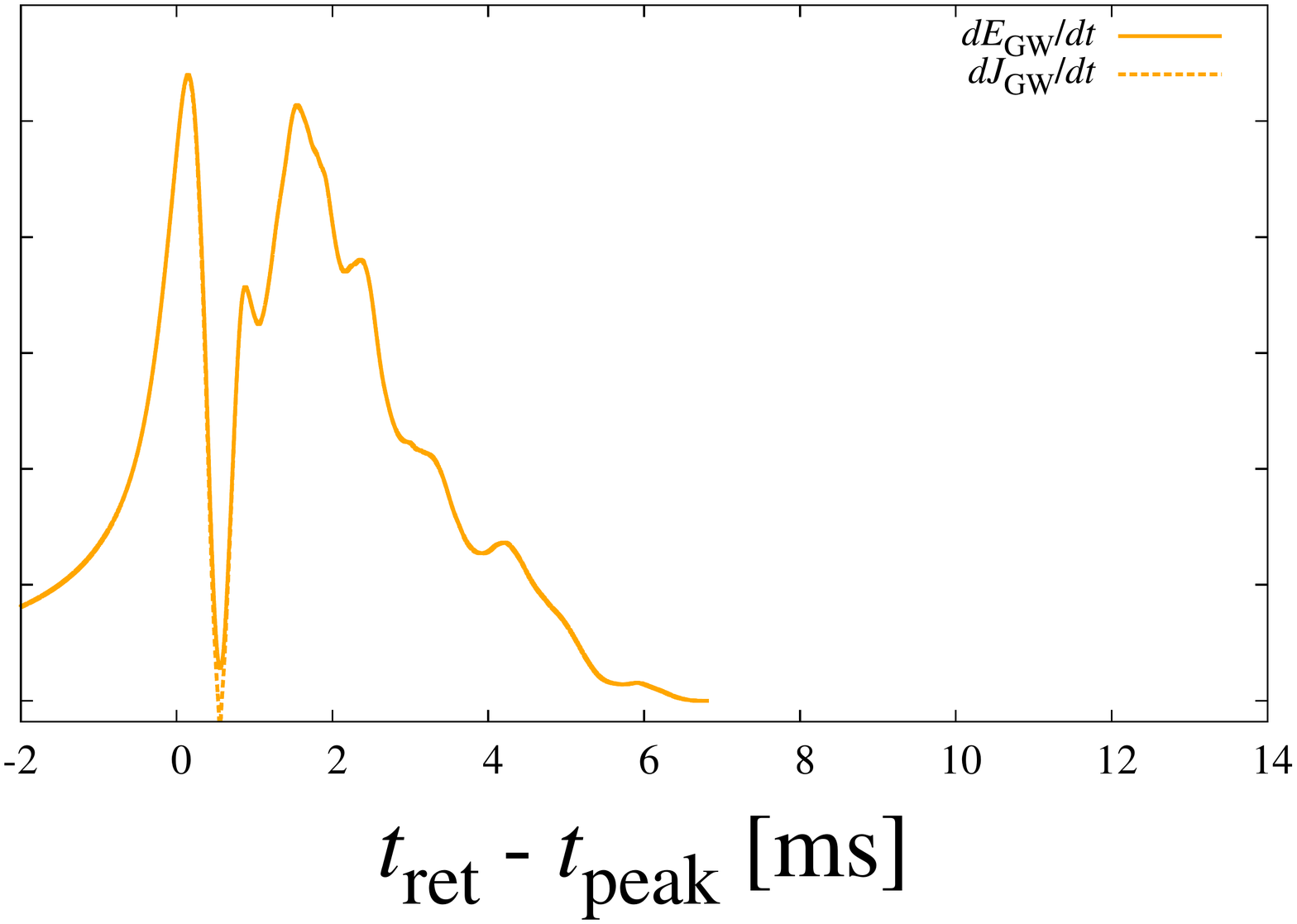}
\end{center}
\end{minipage}
\hspace{-13.35mm}
\begin{minipage}{0.27\hsize}
\begin{center}
\includegraphics[width=4.5cm,angle=0]{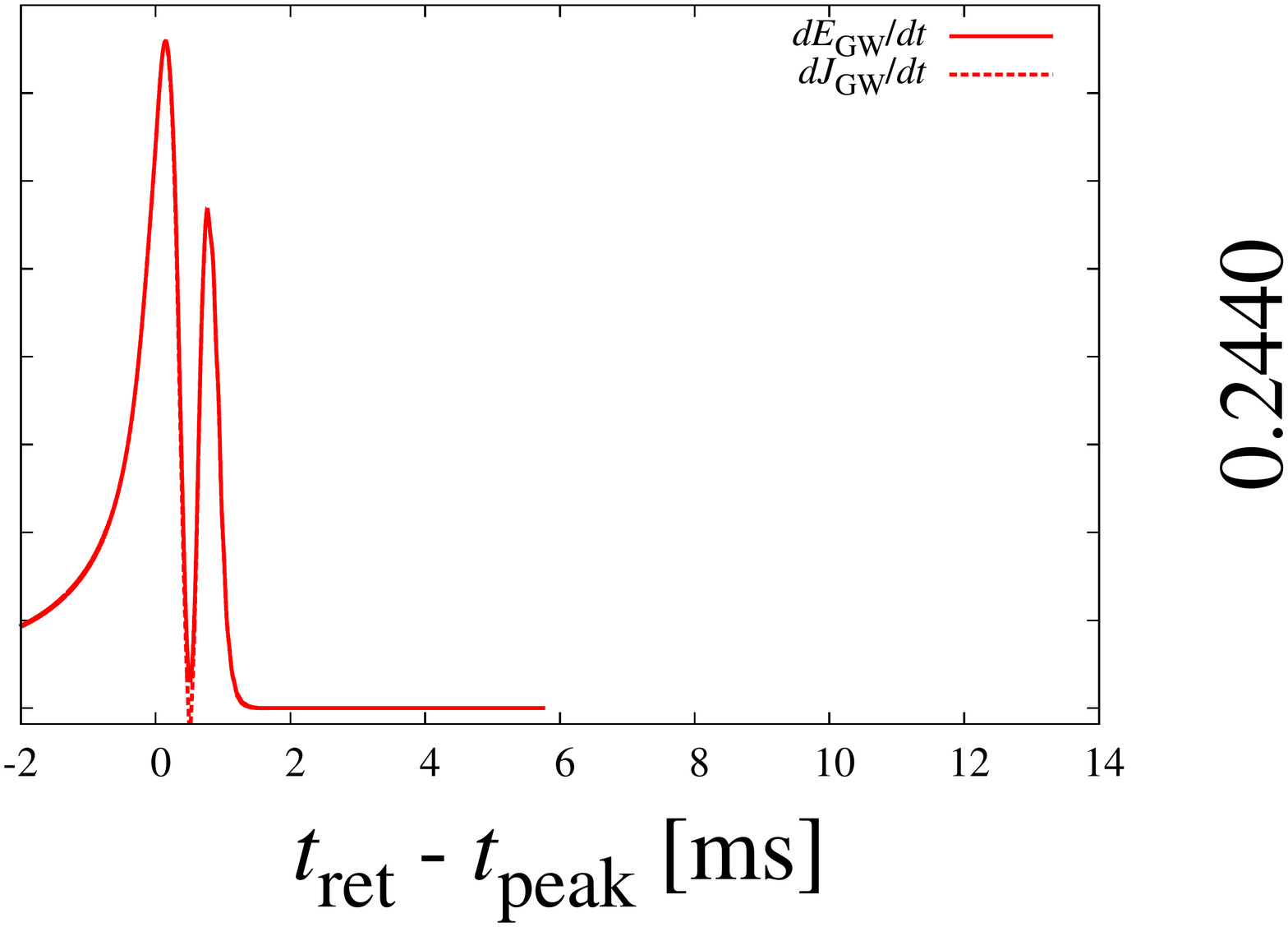}
\end{center}
\end{minipage}\\
\caption{\label{fig:dEdJ}
  Energy (solid) and angular momentum (dashed) emission rate by gravitational waves~(\ref{eq:dEdJ}) for the binary systems with ${\cal M}_c= 1.1752M_\odot$.  The time axis is set to be zero at the peak time of the gravitational-wave amplitude. 
  For completeness, we also show the systems reported in Refs.~\cite{Kiuchi:2017pte,Kawaguchi:2018gvj}.
}
\end{figure*}

\begin{figure*}[t]
\hspace{-18.0mm}
\begin{minipage}{0.27\hsize}
\begin{center}
\includegraphics[width=4.5cm,angle=0]{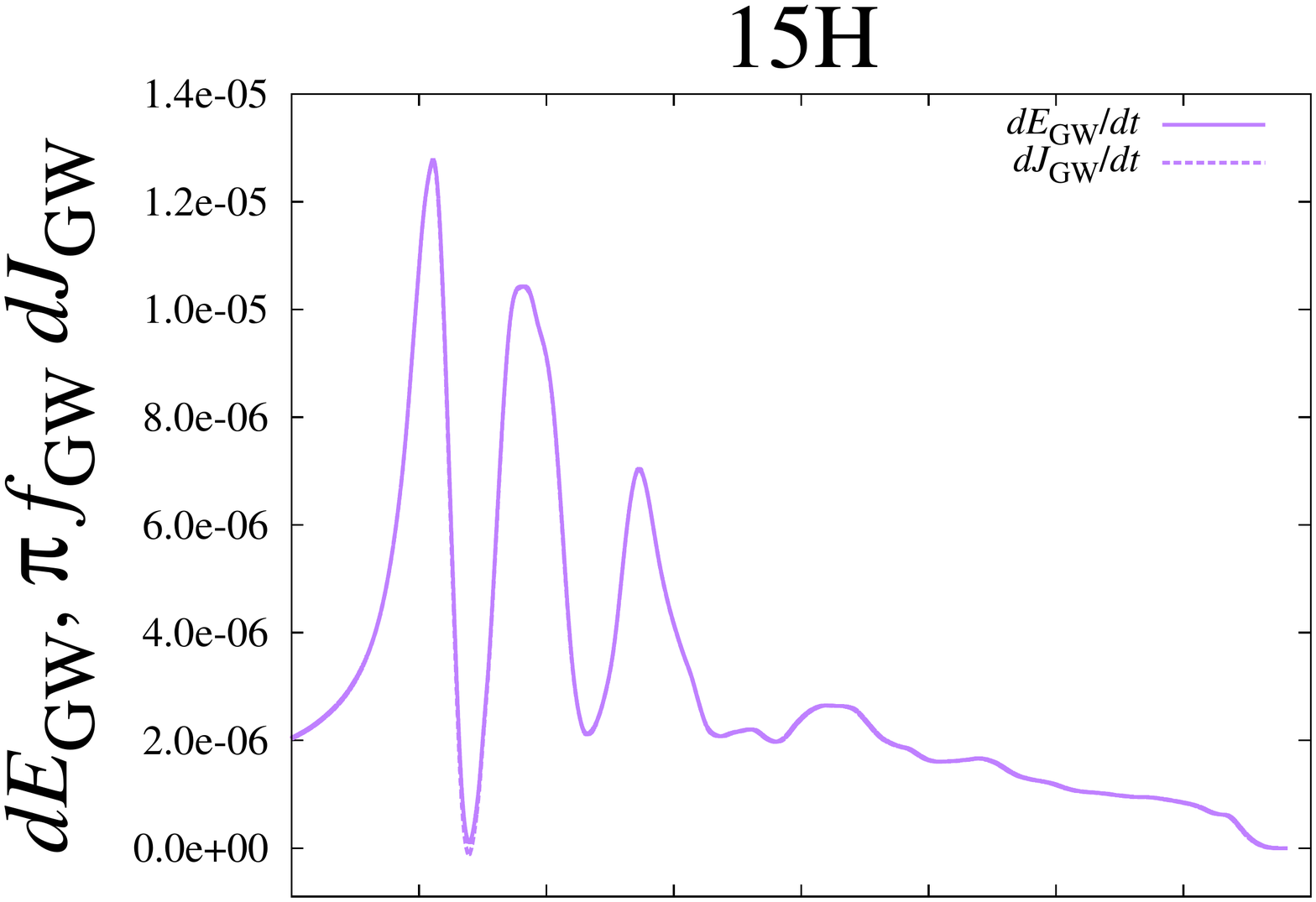}
\end{center}
\end{minipage}
\hspace{-13.35mm}
\begin{minipage}{0.27\hsize}
\begin{center}
\includegraphics[width=4.5cm,angle=0]{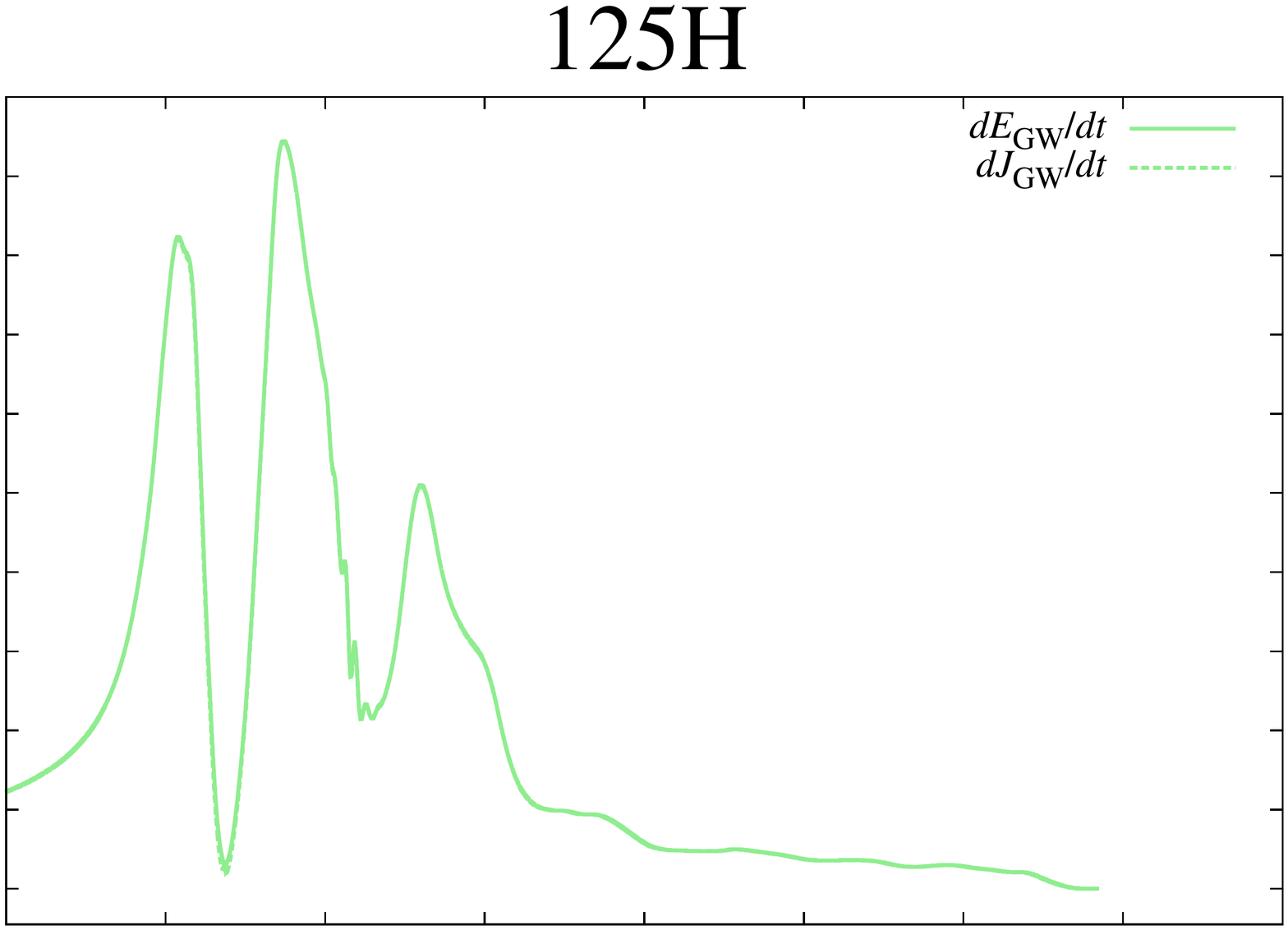}
\end{center}
\end{minipage}
\hspace{-13.35mm}
\begin{minipage}{0.27\hsize}
\begin{center}
\includegraphics[width=4.5cm,angle=0]{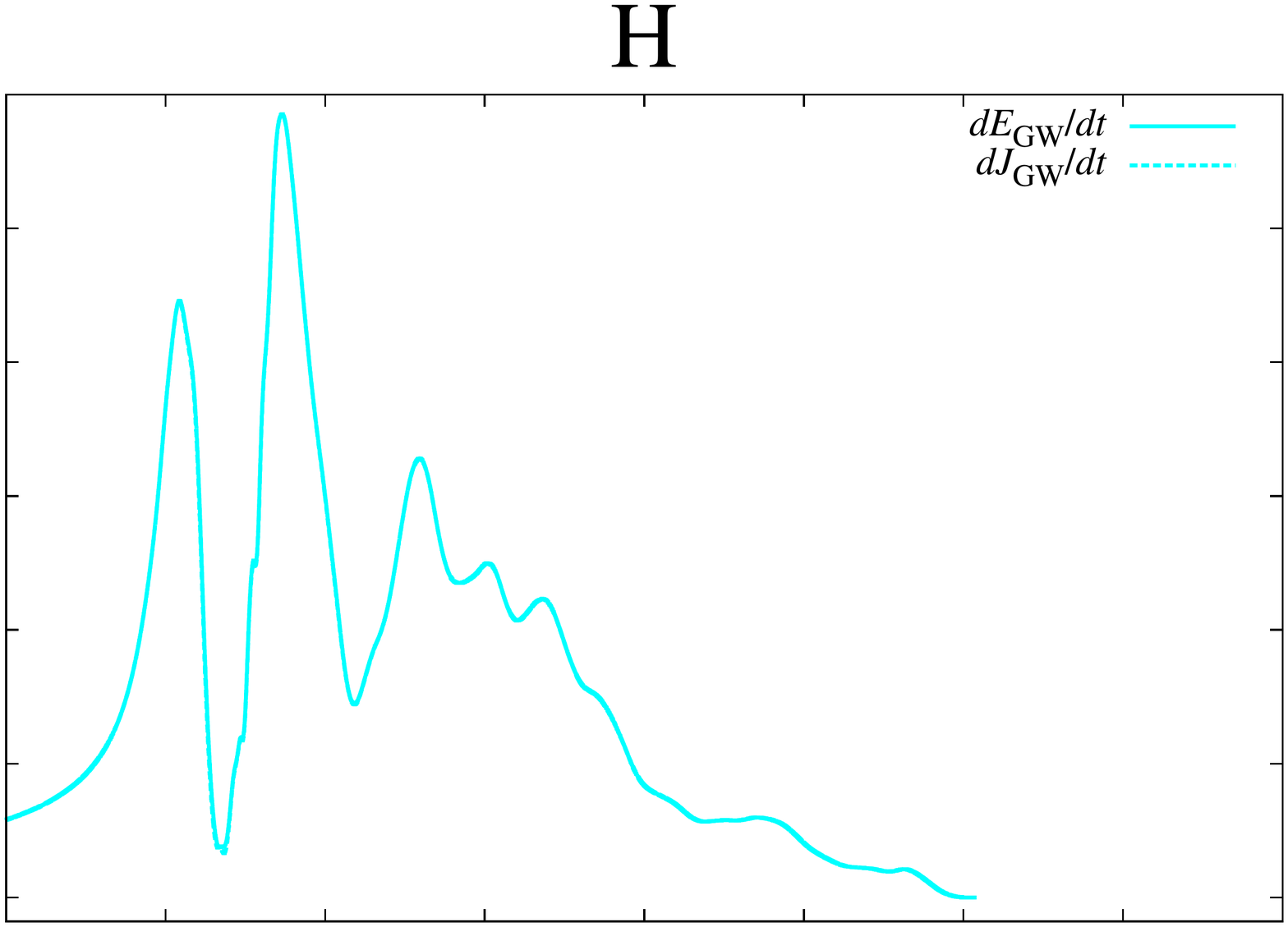}
\end{center}
\end{minipage}
\hspace{-13.35mm}
\begin{minipage}{0.27\hsize}
\begin{center}
\includegraphics[width=4.5cm,angle=0]{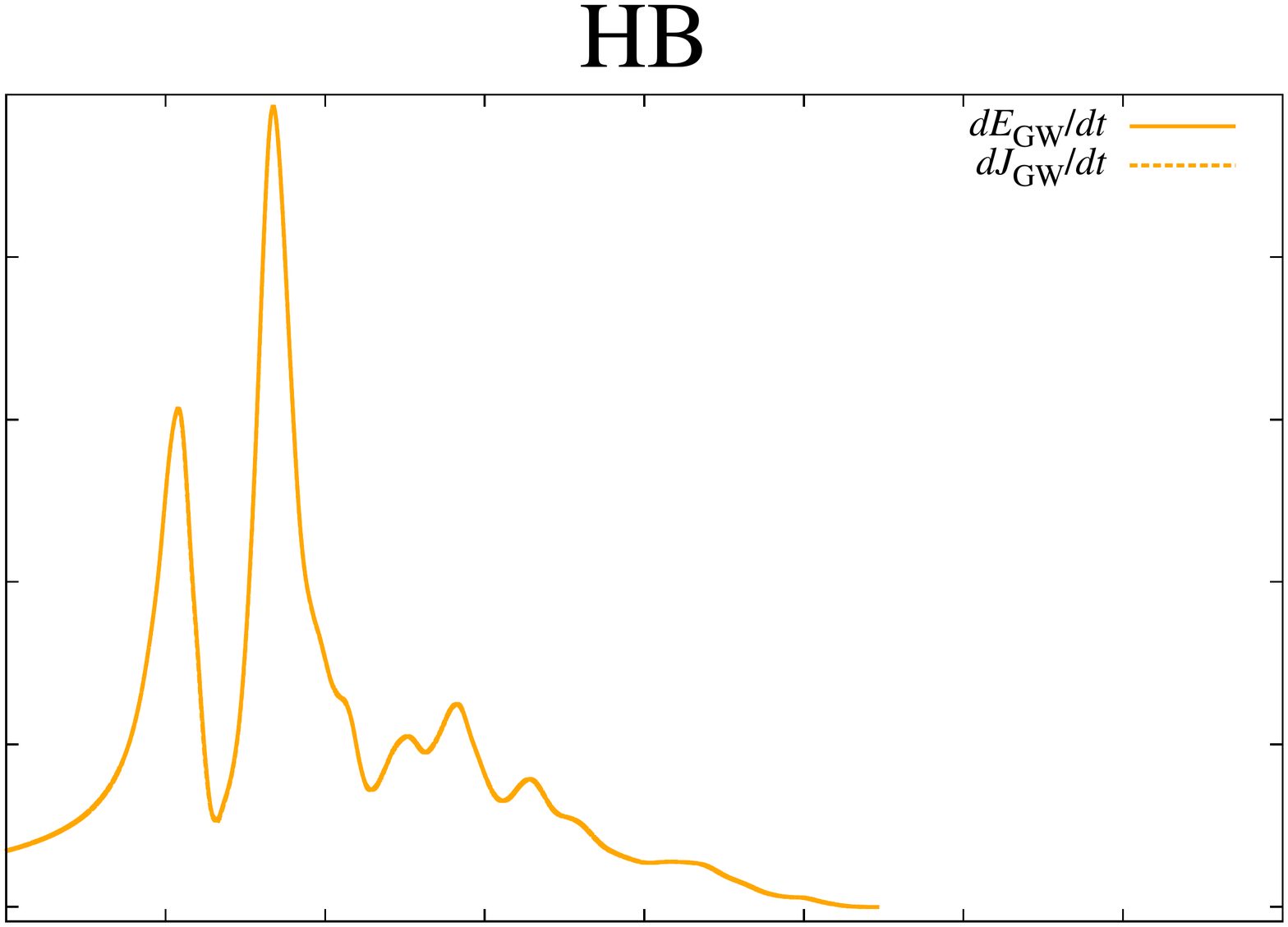}
\end{center}
\end{minipage}
\hspace{-13.35mm}
\begin{minipage}{0.27\hsize}
\begin{center}
\includegraphics[width=4.5cm,angle=0]{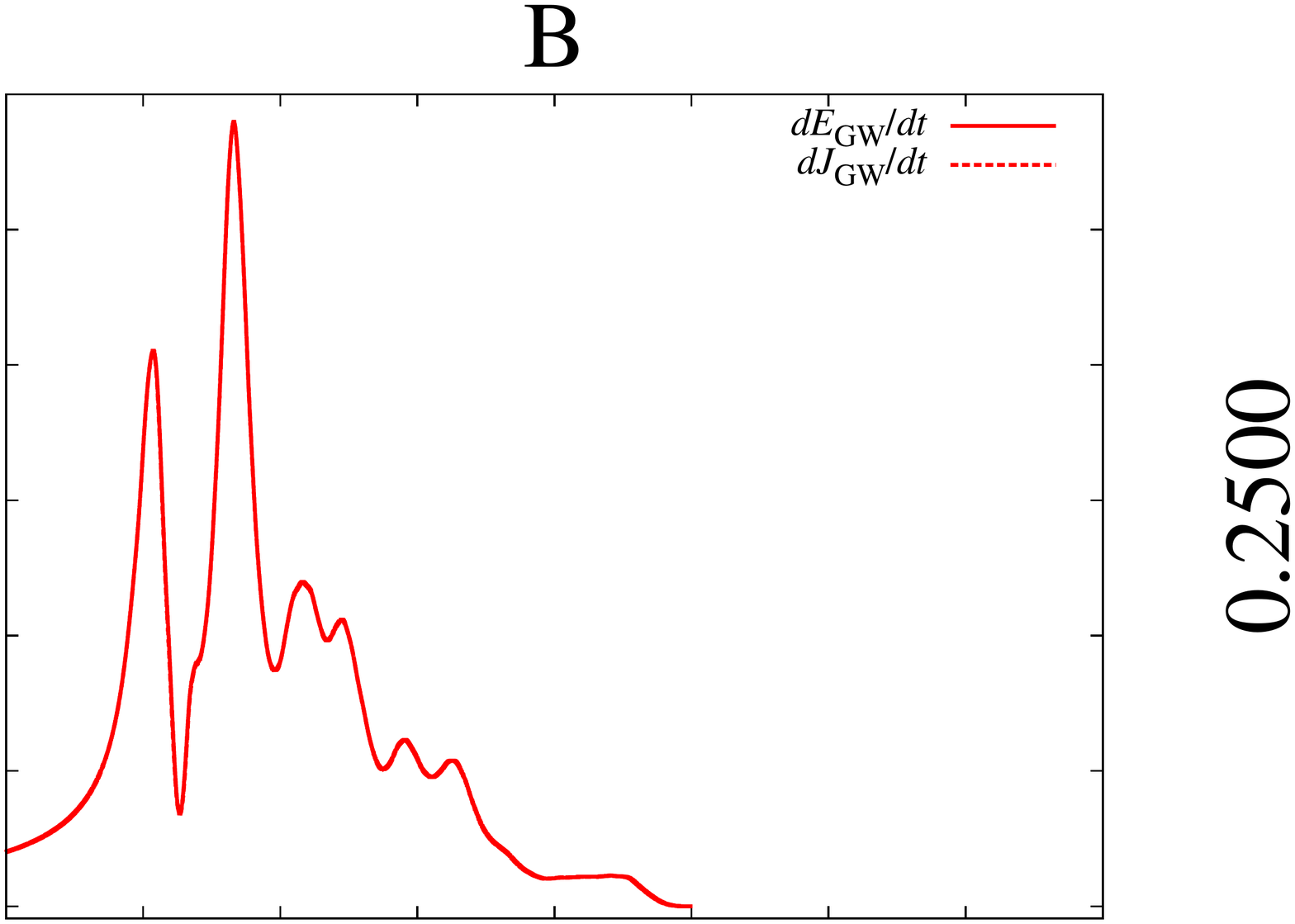}
\end{center}
\end{minipage}\\
\vspace{-9mm}
\hspace{-18.0mm}
\begin{minipage}{0.27\hsize}
\begin{center}
\includegraphics[width=4.5cm,angle=0]{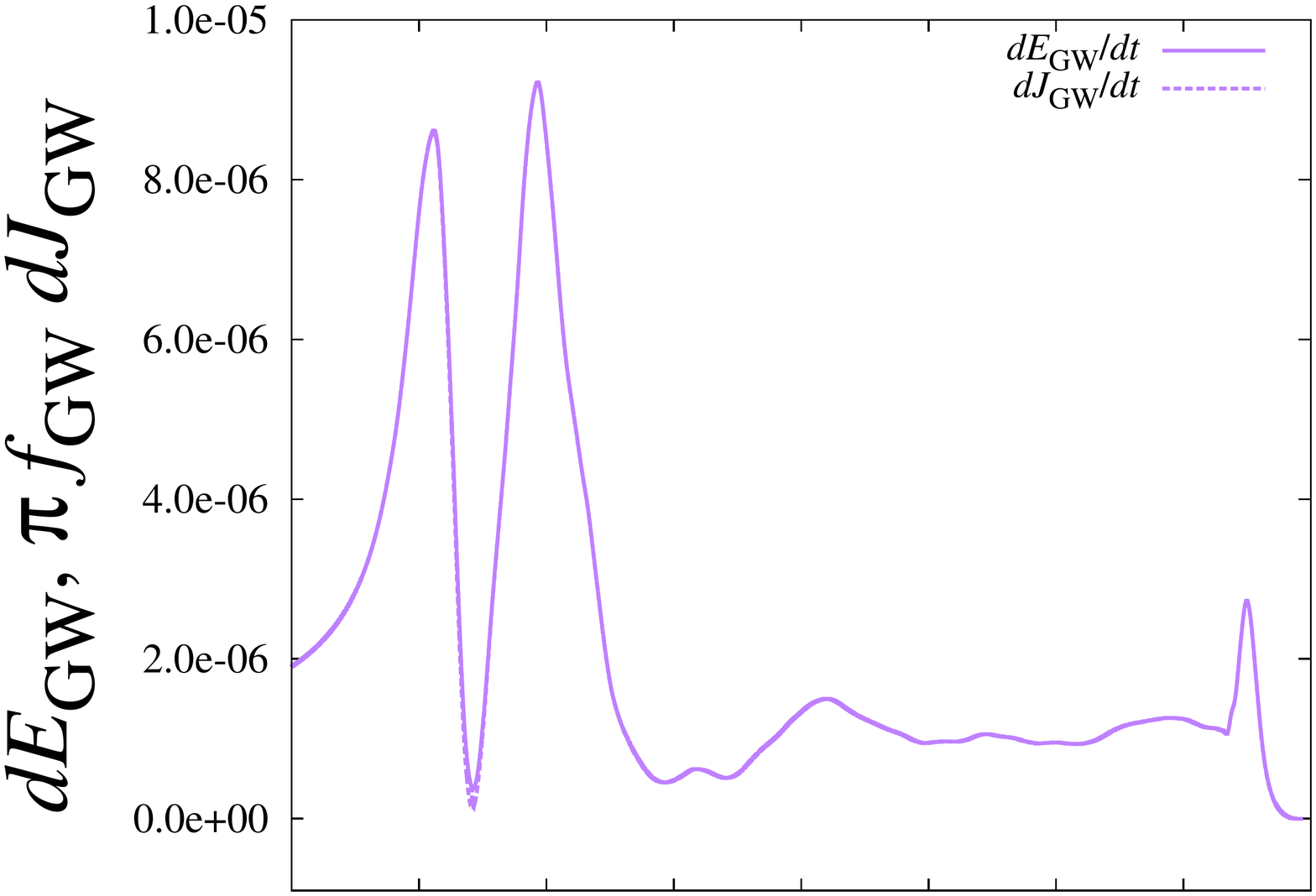}
\end{center}
\end{minipage}
\hspace{-13.35mm}
\begin{minipage}{0.27\hsize}
\begin{center}
\includegraphics[width=4.5cm,angle=0]{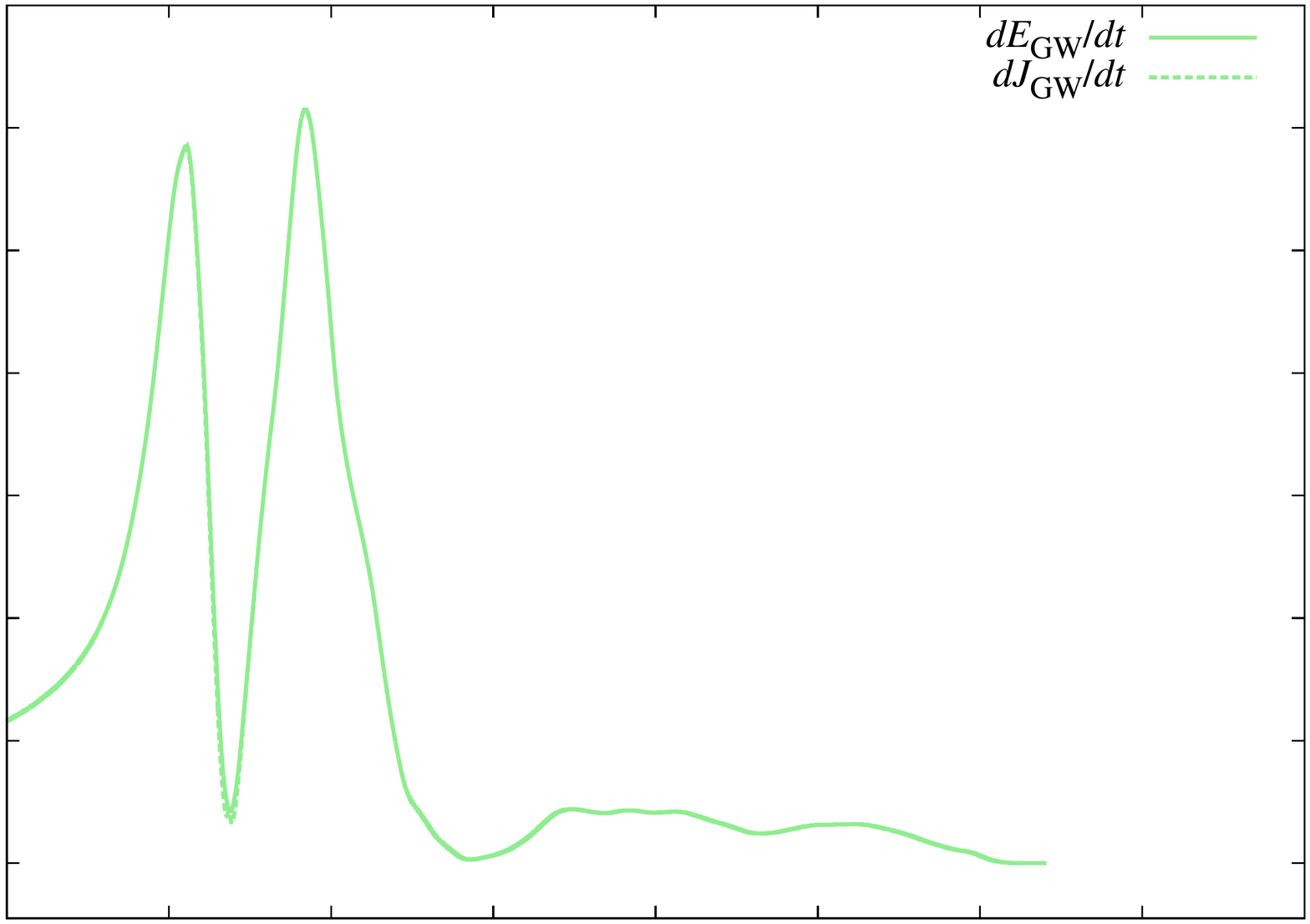}
\end{center}
\end{minipage}
\hspace{-13.35mm}
\begin{minipage}{0.27\hsize}
\begin{center}
\includegraphics[width=4.5cm,angle=0]{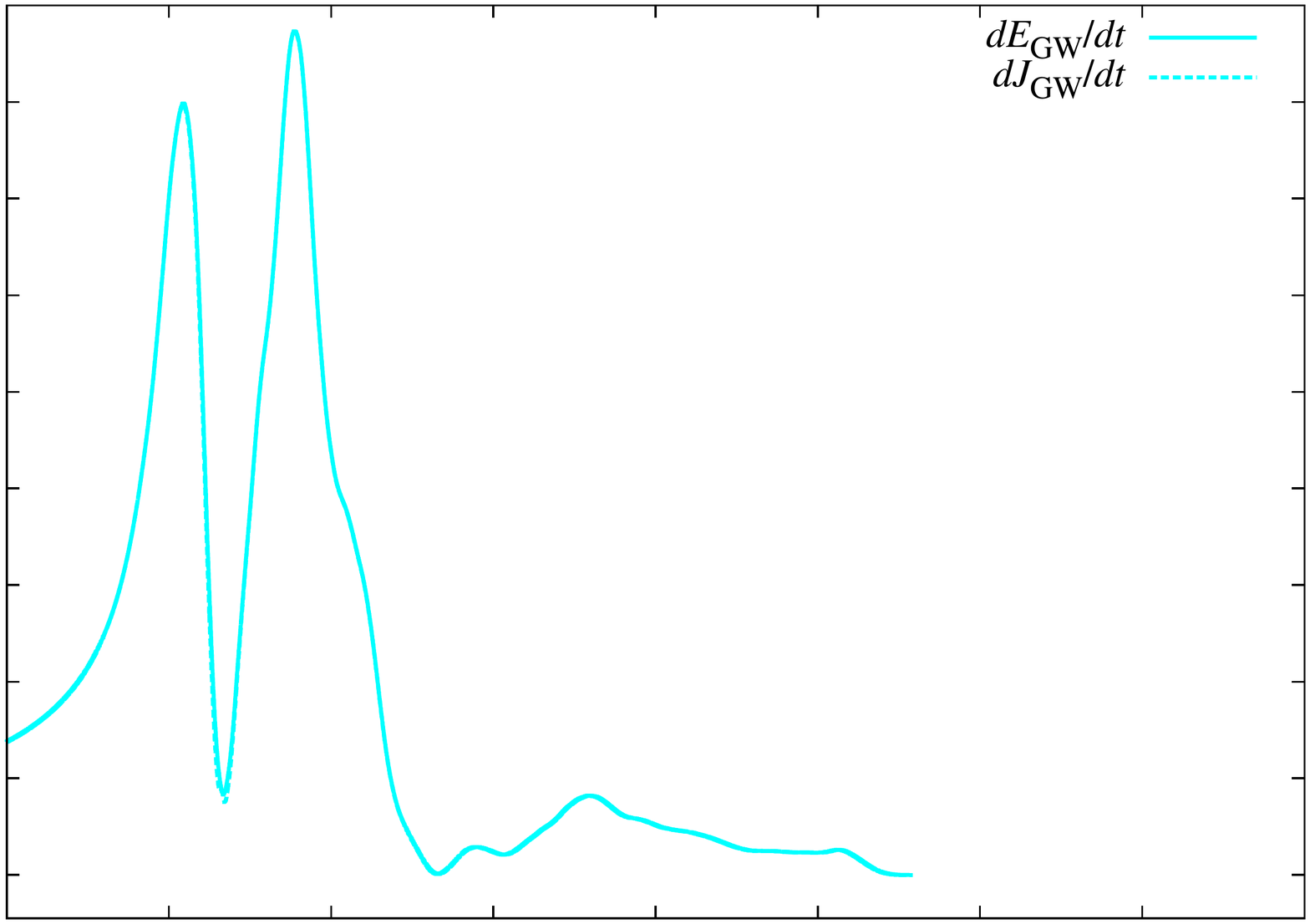}
\end{center}
\end{minipage}
\hspace{-13.35mm}
\begin{minipage}{0.27\hsize}
\begin{center}
\includegraphics[width=4.5cm,angle=0]{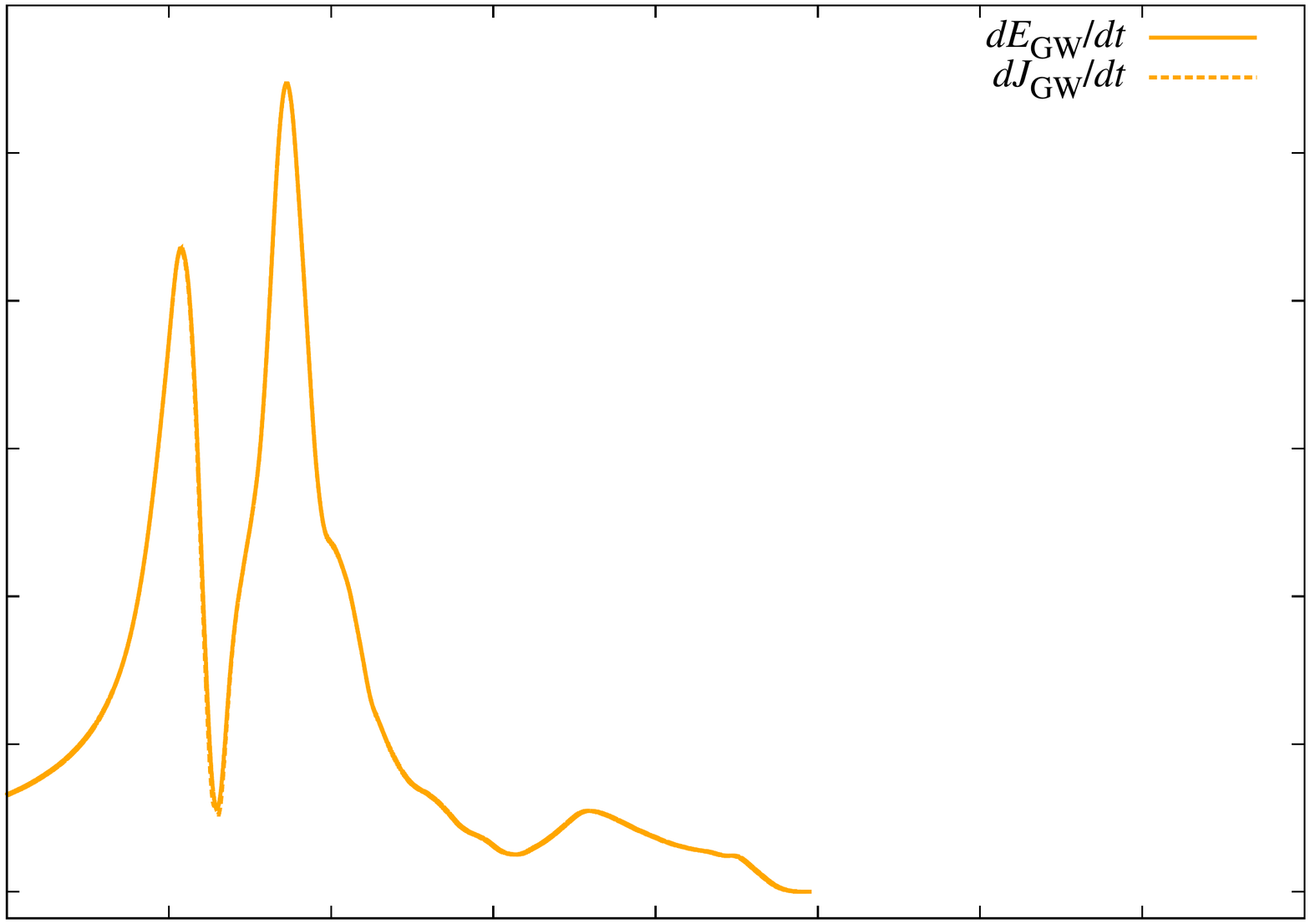}
\end{center}
\end{minipage}
\hspace{-13.35mm}
\begin{minipage}{0.27\hsize}
\begin{center}
\includegraphics[width=4.5cm,angle=0]{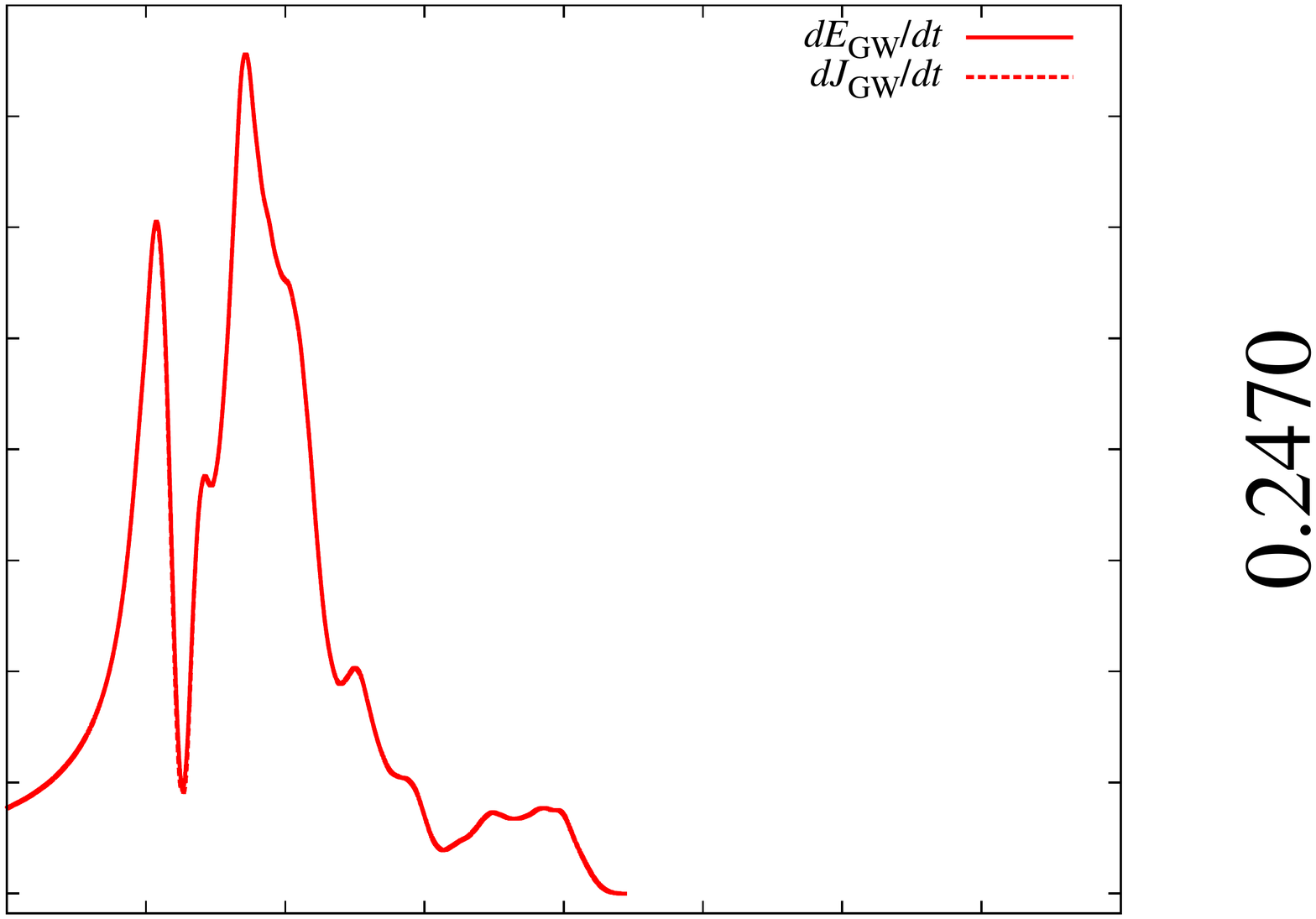}
\end{center}
\end{minipage}\\
\vspace{-9mm}
\hspace{-18.0mm}
\begin{minipage}{0.27\hsize}
\begin{center}
\includegraphics[width=4.5cm,angle=0]{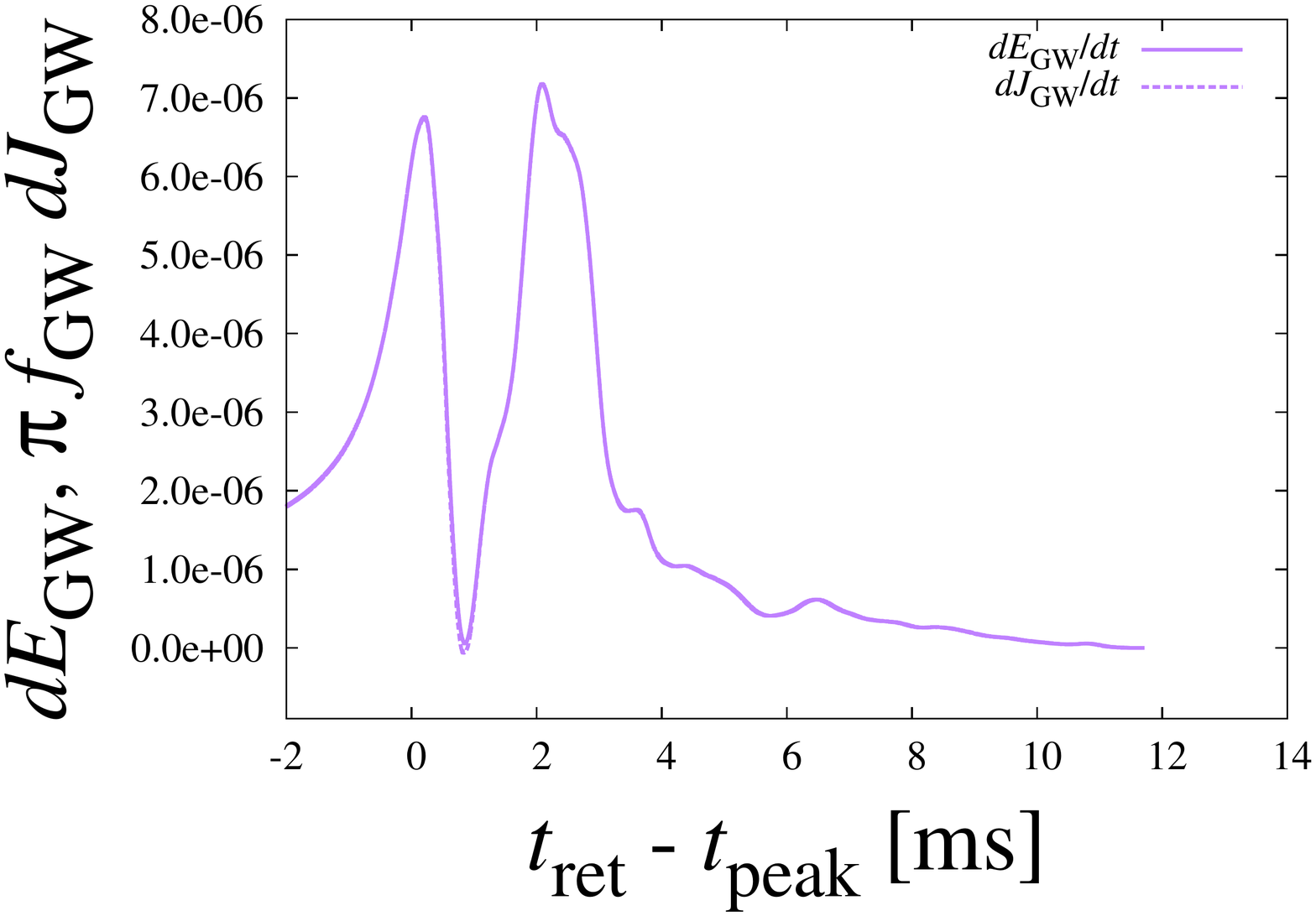}
\end{center}
\end{minipage}
\hspace{-13.35mm}
\begin{minipage}{0.27\hsize}
\begin{center}
\includegraphics[width=4.5cm,angle=0]{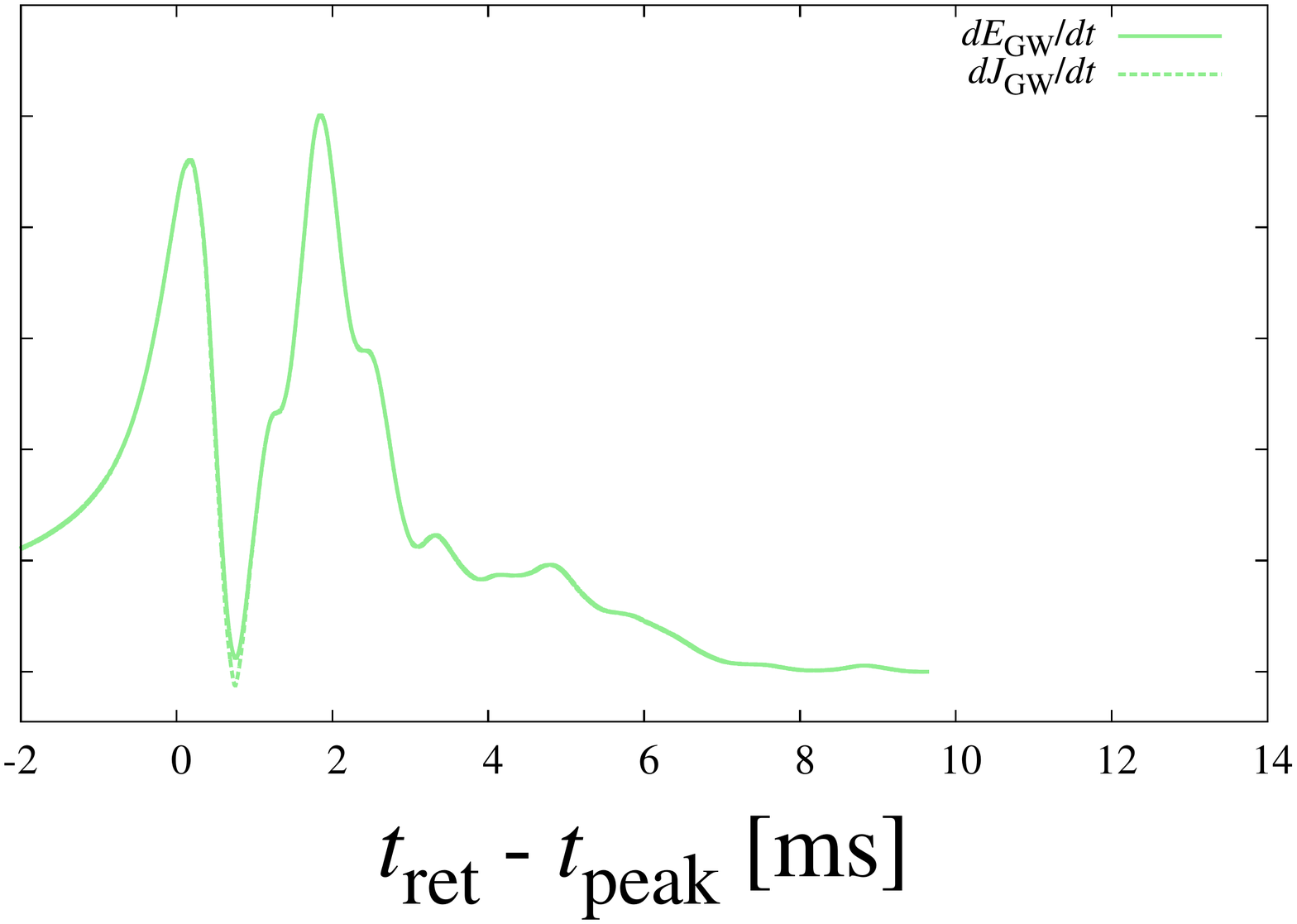}
\end{center}
\end{minipage}
\hspace{-13.35mm}
\begin{minipage}{0.27\hsize}
\begin{center}
\includegraphics[width=4.5cm,angle=0]{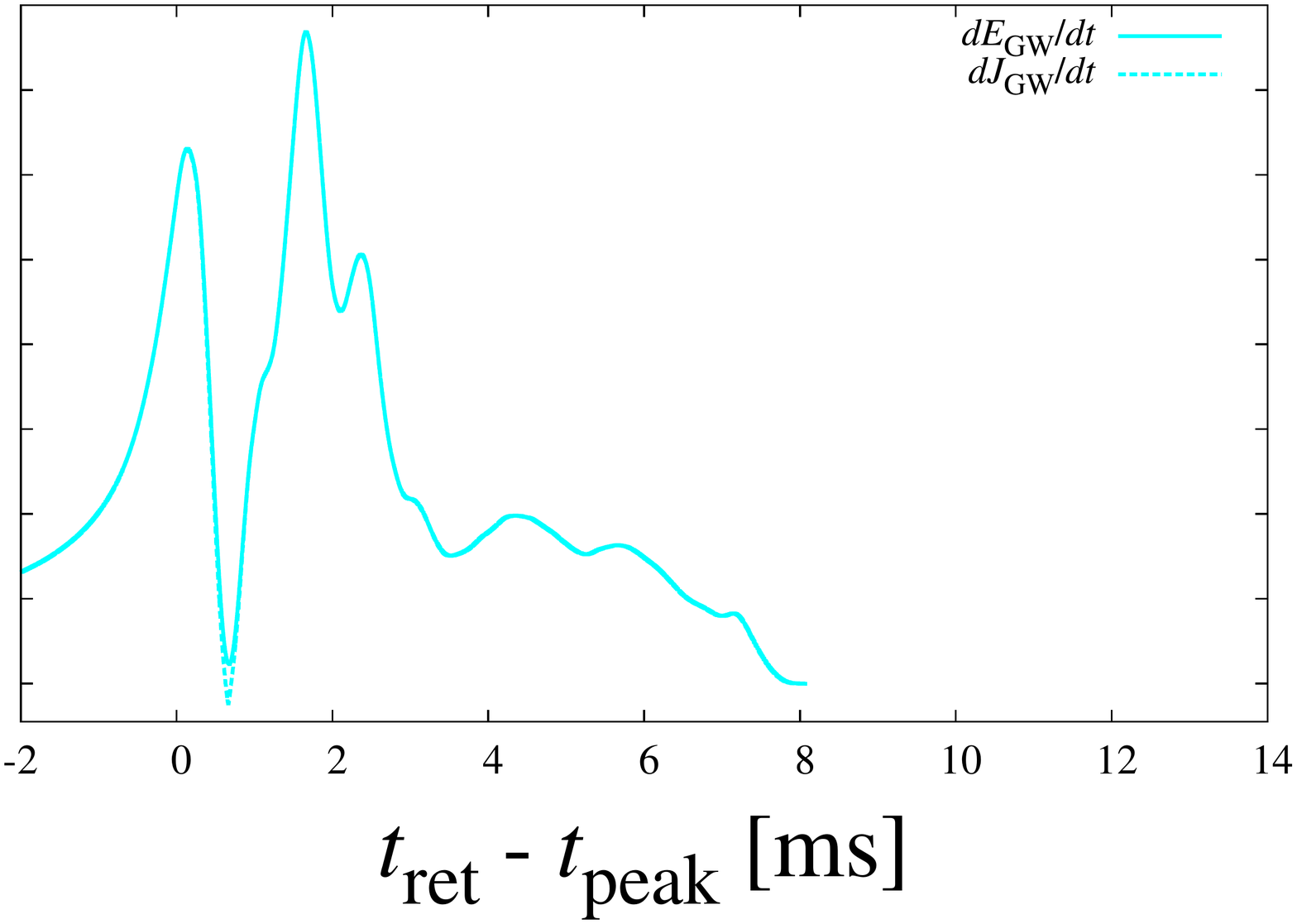}
\end{center}
\end{minipage}
\hspace{-13.35mm}
\begin{minipage}{0.27\hsize}
\begin{center}
\includegraphics[width=4.5cm,angle=0]{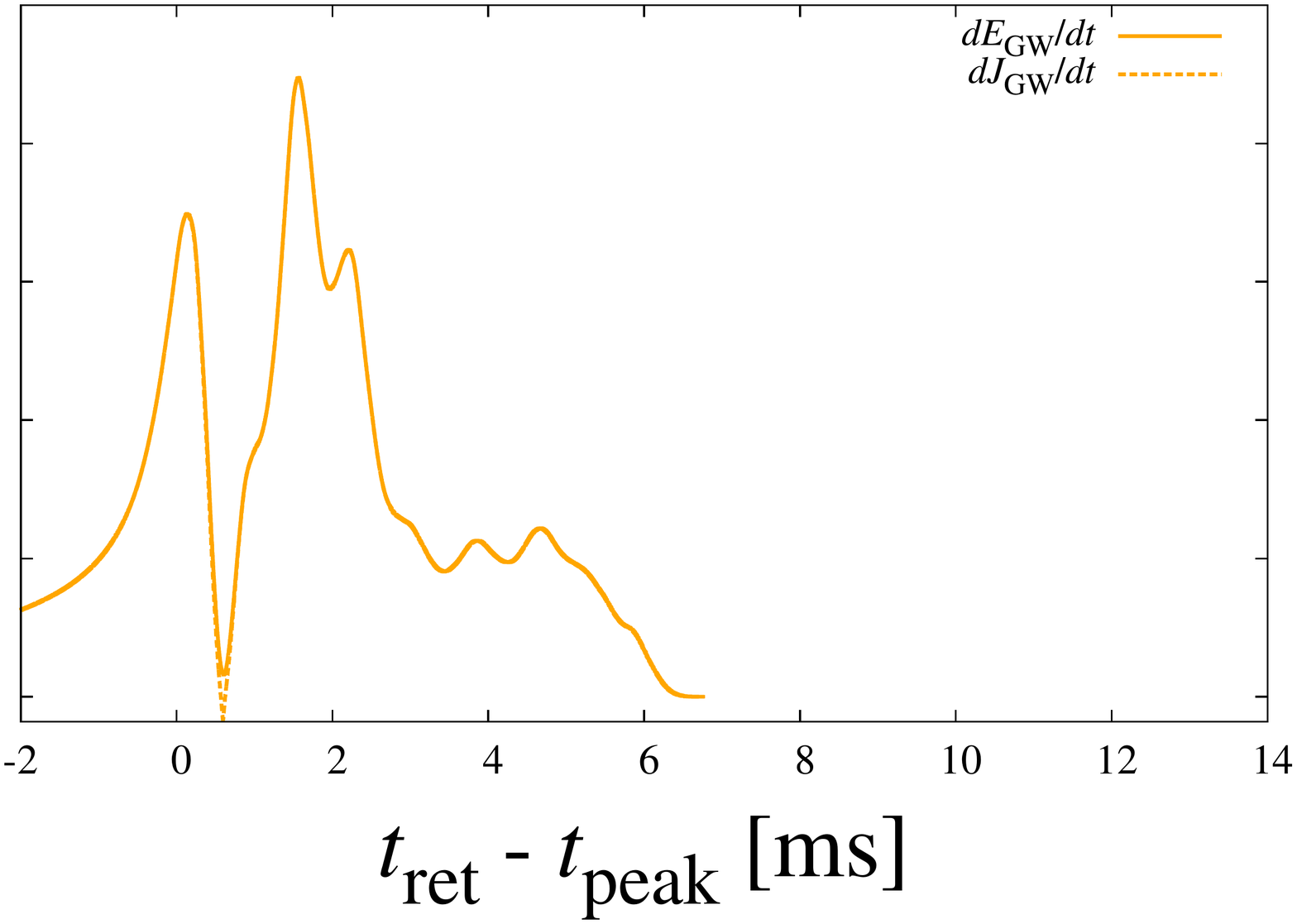}
\end{center}
\end{minipage}
\hspace{-13.35mm}
\begin{minipage}{0.27\hsize}
\begin{center}
\includegraphics[width=4.5cm,angle=0]{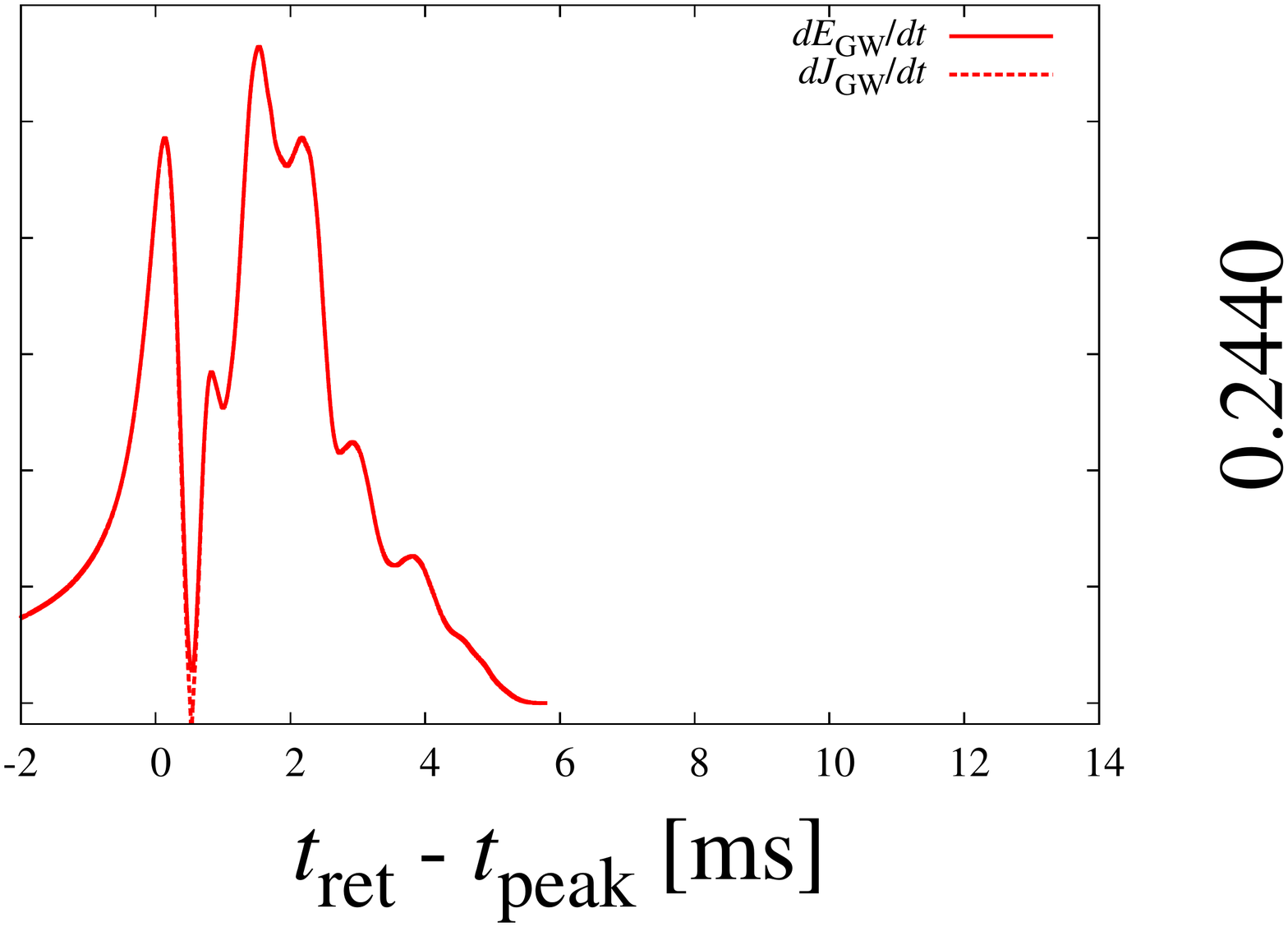}
\end{center}
\end{minipage}\\
\caption{\label{fig:dEdJb}
 The same as Fig.~\ref{fig:dEdJ}, but for ${\cal M}_c= 1.0882M_\odot$.
}
\end{figure*}

\begin{figure}[t]
\includegraphics[width=.9\linewidth]{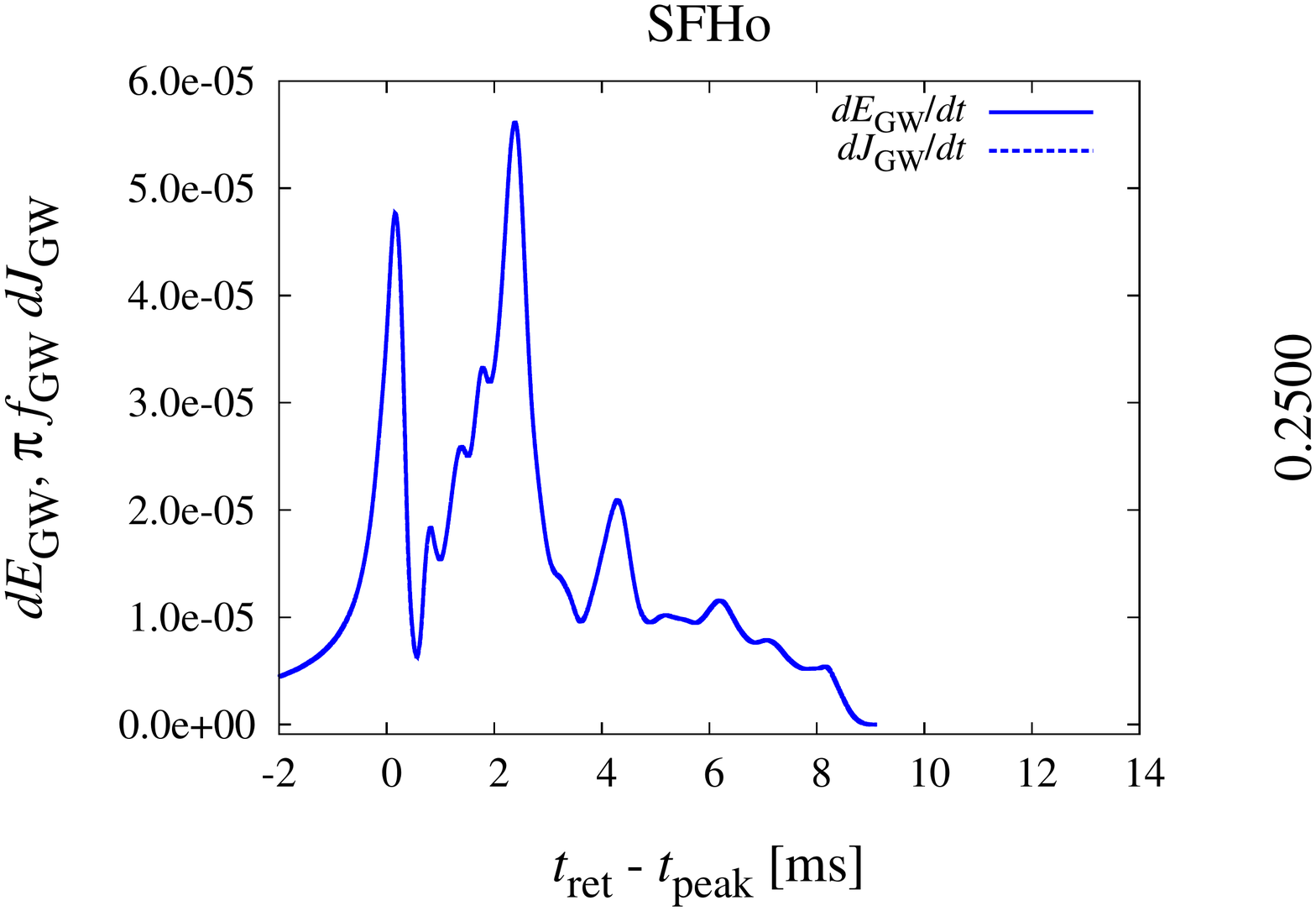}
\caption{\label{fig:dEdJ2}The same as Fig.~\ref{fig:dEdJ}, but for the SFHo (tabulated) EOS case.
}
\end{figure}

\begin{figure*}[t]
  	 \includegraphics[width=.5\linewidth]{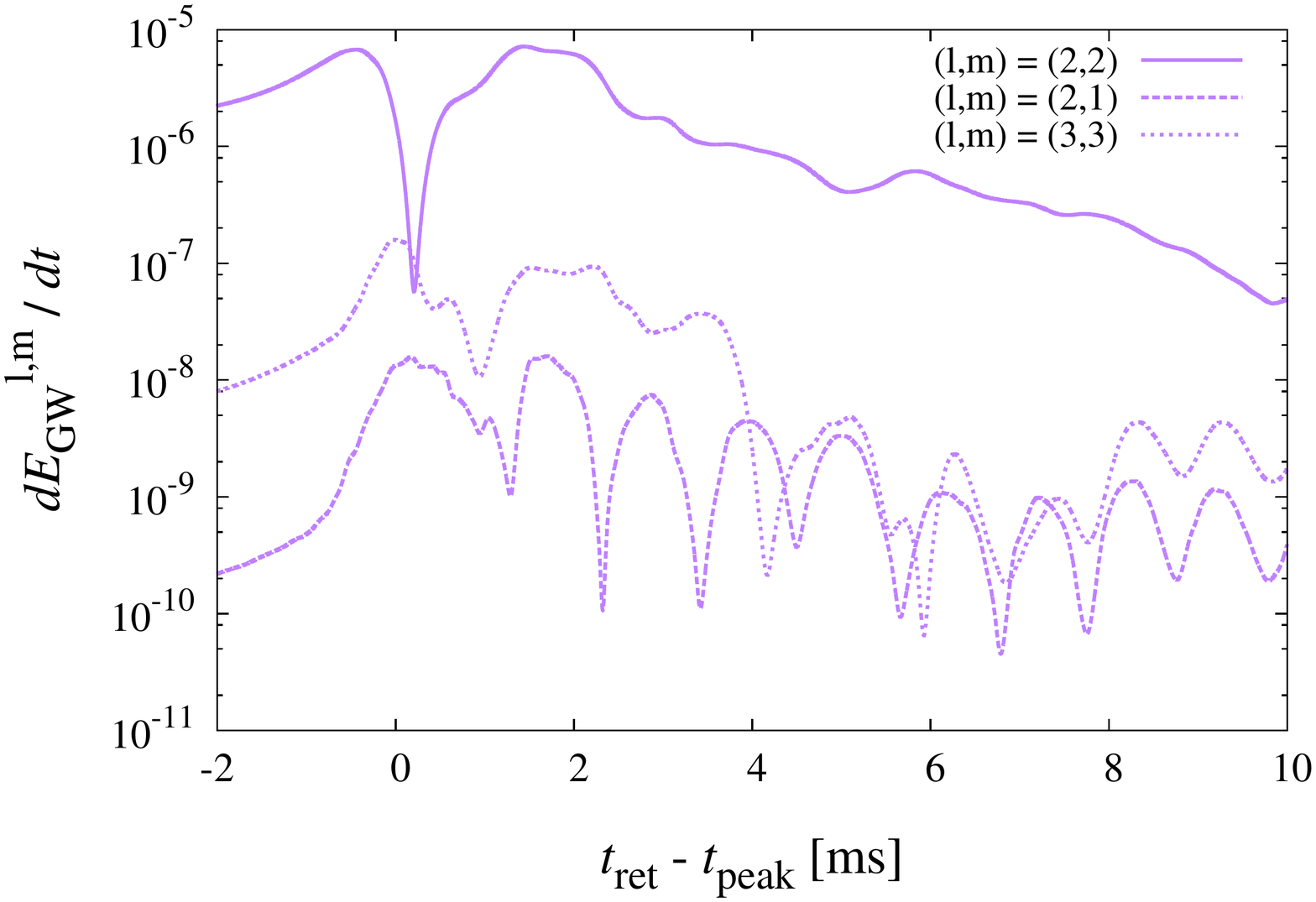}
 	 \includegraphics[width=.5\linewidth]{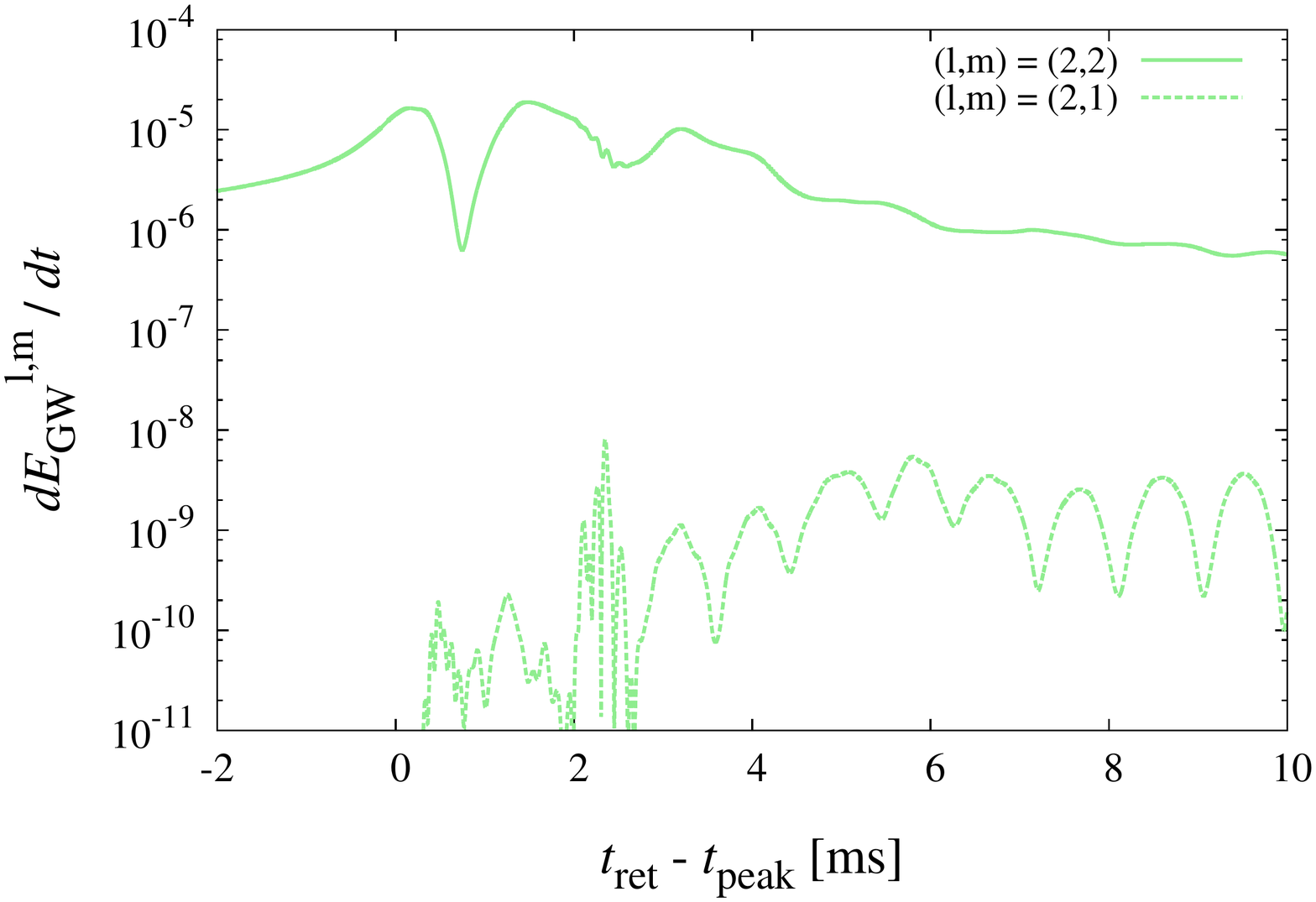}
         \caption{\label{fig:dEdJ3}
           (Top) Gravitational-wave energy flux~(\ref{eq:EGW}) for $(l,m)=(2,2)$, $(2,1)$, and $(3,3)$ modes for 15H107-146 with $N=182$. 
           (Bottom) The same as the top panel, but for $(l,m)=(2,2)$ and $(2,1)$ modes for 125H125-125 with $N=182$. 
}
\end{figure*}

As already mentioned, the post-merger gravitational wave signal is dominated by the f--mode oscillation with $(l,m)=(2,2)$ of the remnant massive NS~\cite{Bauswein:2011tp,Hotokezaka:2013iia}. Thus, it is natural to expect 
that a relation holds between the energy emission rate and angular momentum emission rate (\ref{eq:EGW})--(\ref{eq:JGW}) with
instantaneous gravitational-wave frequency~(\ref{eq:GWfreq}); 
\begin{align}
  \frac{dE_{\rm GW}^\text{post}}{dt} \approx \pi f_\text{GW} \frac{dJ_{\rm GW}^\text{post}}{dt}, \label{eq:dEdJ}
\end{align}
    where $dE_{\rm GW}^\text{post}/dt=\sum_{l,m}dE_{\rm GW}^{l,m}/dt$ and $dJ_{\rm GW}^\text{post}/dt=\sum_{l,m}dJ_{\rm GW}^{l,m}/dt$ for $t\ge t_\text{peak}$ in Eqs.~(\ref{eq:EGW}) and (\ref{eq:JGW}). 
    To investigate to what extent this relation is satisfied, we generate Figs.~\ref{fig:dEdJ}--\ref{fig:dEdJ2}. In these figures, the solid curve is the left hand side of Eq.~(\ref{eq:dEdJ}) and the dashed curve is the right hand side of Eq.~(\ref{eq:dEdJ}). 
    We find that they agree with each other with a relative error $\lesssim 8\%$ for any time. Because the emissivity reduces quickly to zero at $t_\text{ret}-t_\text{peak}\approx 0.5$ ms as shown in Figs.~\ref{fig:dEdJ}--\ref{fig:dEdJ2}, we estimate the error for $t_\text{ret}-t_\text{peak}\gtrsim 1$ ms.
    We also find that the time integrated values of Eq.~(\ref{eq:dEdJ}) agree with each other with a relative error $\lesssim 1\%$. This is also the case for the relation of $E_{\rm GW,p}\approx \pi f_2 J_{\rm GW,p}$.  

    We also confirm that a contribution from the one-arm spiral instability in the post-merger stage~\cite{Paschalidis:2015mla,Radice:2016gym} is negligible because the energy flux for $(l,m)=(2,1)$ mode is $\lesssim 0.5\%$ of that for $(l,m)=(2,2)$ mode even for the symmetric binary systems as shown in the bottom panel of Fig.~\ref{fig:dEdJ3}.
  Thus, we conclude that Eq.~(\ref{eq:dEdJ}) is well satisfied and confirm that the main gravitational-wave emission mechanism during the post-merger stage is the f--mode oscillation of the remnant massive NS, i.e, $f_\text{GW} \approx f_2$ (see also Figs.~\ref{fig:PSD}--\ref{fig:PSD2}). These findings encourage us to build a model for the post-merger gravitational-wave emission (see Ref.~\cite{Shibata:2019ctb}). 

In Table~\ref{tb:universal}, we summarize to what extent $E^\text{tot}_{\rm GW,i}/(m_0\eta)$--$\tilde{\Lambda}^{1/5}$ and $J_\text{rem}/(m_0^2\eta)$--$\tilde{\Lambda}^{1/5}$ relations of Eqs.~(\ref{eq:EGWi}) and (\ref{eq:Jrem2}) hold. 

\begin{table*}
\centering
\caption{Summary of the assessment of the universal relations for the non-spinning and non-magnetized binary systems. Neutrino radiation is not taken into account. We show the maximum relative errors produced by the original relation (upper row) and by the improved relation derived in this paper (lower row). For $f_1$, the error is unable to be estimated because of the absence of $f_1$ peak in the asymmetric binary systems. Therefore, we conclude there is no universal relation between $f_1$ and $\tilde{\Lambda}$.
 For $f_2$--$R_{1.6}$ relation, we do not propose an improved relation and sym. (asym.) in the parenthesis means the symmetric (asymmetric) binary.
  For $E^{2,2}_{\rm GW,p}$ and $J^{2,2}_{\rm GW,p}$, we do not propose an improved relation because uncertainties of the life time of the merger remnant NSs are large.   
  }
\begin{tabular}{ccccccccc}\hline\hline
 $m_0f_\text{peak}$--$\tilde{\Lambda}^{1/5}$ & $D h_\text{peak}/m_0$--$\tilde{\Lambda}^{1/5}$ & $f_1$--$\tilde{\Lambda}^{1/5}$ & $m_0 f_2$--$\tilde{\Lambda}^{1/5}$ & $f_2$--$R_{1.6}$ & $E^\text{tot}_{\rm GW,i}/(m_0\eta)$--$\tilde{\Lambda}^{1/5}$ & $J_\text{rem}/(m_0^2\eta)$--$\tilde{\Lambda}^{1/5}$\\
  \hline
  $\approx 17\%$ & N/A            & -- & $\approx 14\%$ & $\approx 6\%~\text{(sym.)}$ and $\approx 10\%~\text{(asym.)}$  & N/A           & N/A\\
  $\approx 3 \%$ & $\approx 4 \%$ & -- & $\approx 9 \%$ & --                                                             & $\approx 4\%$ & $\approx 3\%$\\
  \hline
\end{tabular}\label{tb:universal}
\end{table*}



\section{Summary} \label{sec:summary}

We performed long-term simulations for new 26 systems of the non-spinning BNS mergers in numerical relativity. To derive high-precision gravitational waveforms in a large parameter space, we systematically vary the EOSs of NS, the chirp mass, and the mass ratio.
To assess gravitational-wave phase error stemming from a finite grid resolution, we change the grid spacing by a factor of two for simulating each binary system. 

First, we found that the residual gravitational-wave phase error at the peak time of gravitational-wave amplitude is $\lesssim 0.5$ rad irrespective of the binary mass and NS EOS. By comparing the results for the piecewise polytropic and SFHo (tabulated) EOS systems, we also found that the interpolation of the thermodynamic quantities during the simulations could generate the phase error of
$\approx 0.2$--$0.3$ rad. However the gravitational-wave phase error for the SFHo (tabulated) EOS system still remains within the sub-radian accuracy level. 

Second, we validated our SACRA inspiral gravitational waveform template~\cite{Kawaguchi:2018gvj} by comparing with the high-precision gravitational waveforms derived in this paper. We found that for a variety of BNS the error in our inspiral waveform model is less than $0.1$ rad in the gravitational-wave phase and less than $20\%$ in the amplitude up to $f_\text{GW}=1000$ Hz.
This template can be used for a new gravitational wave data analysis for extracting tidal deformability from GW170817~\cite{Narikawa:2019xng} and for future event of BNS merger. 

Third, we assessed the universal relations between the gravitational-wave related quantities and the binary tidal deformability/NS radius proposed in the literature~\cite{Rezzolla:2016nxn,Read:2013zra,Zappa:2017xba,Bauswein:2011tp,Bauswein:2012ya,Bernuzzi:2014owa,Bernuzzi:2015rla}. We found that the gravitational-wave frequency at the peak time $f_\text{peak}$, the gravitational-wave amplitude at the peak time $h_\text{peak}$,
and the peak frequency $f_2$ associated with the f--mode oscillation of the remnant massive NS in the spectrum amplitude of post-merger gravitational waves depend strongly on the symmetric mass ratio and/or the grid resolution. This clearly illustrates that the universal relations proposed in the literature~\cite{Rezzolla:2016nxn,Read:2013zra,Zappa:2017xba,Bauswein:2011tp,Bauswein:2012ya,Bernuzzi:2014owa,Bernuzzi:2015rla} are not as universal as proposed. 

We proposed improved fitting formulae~(\ref{eq:fpeak}) for $m_0f_\text{peak}$--$\tilde{\Lambda}^{1/5}$, (\ref{eq:hpeak}) for $D h_\text{peak}/m_0$--$\tilde{\Lambda}^{1/5}$, and (\ref{eq:f2}) and for $m_0f_2$--$\tilde{\Lambda}^{1/5}$. However these fitting formulae may still suffer from systematics such as NS spin, NS magnetic fields, and the neutrino radiation, which are not taken into account in our simulations. In addition, the EOS of NS, in particular, for a high-density part of the NS, is still uncertain, and hence, the systematics due to this uncertainty should be kept in mind. 
We also note that we assessed the errors of these formulae only with our simulation data. A close comparison among the results of the independent BNS simulations with the existing numerical relativity codes is necessary to better understand the systematic error in these formulae. This should be done as a future project. We also found that $f_1$ frequency in the spectrum amplitude could be extracted only for the nearly symmetric binary systems. Unless we can determine the symmetric mass ratio accurately, using the universal relation for $f_1$ could lead to a misleading result in the gravitational-wave data analysis. 

Finally, we assessed the energy, $E_\text{GW}$, and angular momentum, $J_\text{GW}$, carried by gravitational waves in the inspiral and post-merger stages. As proposed in Ref.~\cite{Zappa:2017xba}, the correlation between $E_{\rm GW,i}^\text{tot}$ and the binary tidal deformability is tight and it does not depend significantly on the symmetric mass ratio. 
We found that the relation $E_\text{GW} \approx \pi f_2 J_\text{GW}$ is well satisfied in the post-merger gravitational wave signal irrespective of the binary mass and NS EOS because the signal from the remnant NSs is approximately monochromatically emitted by the f--mode oscillation. 
The angular momentum of the remnant massive NS, $J_\text{rem}$, correlates with the binary tidal deformability. This quantity is relevant to build a model of post-merger evolution of merger remnants~\cite{Shibata:2019ctb}.

\acknowledgments 

Numerical computation was performed on K computer at AICS
(project numbers hp160211, hp170230, hp170313, hp180179, hp190160), on Cray XC50 at cfca of
National Astronomical Observatory of Japan, Oakforest-PACS at Information
Technology Center of the University of Tokyo, and on Cray XC40 at Yukawa Institute for Theoretical Physics, Kyoto
University.  This work was supported by Grant-in-Aid for Scientific
Research (16H02183, 16H06342, 16H06341, 16K17706, 17H01131, 17H06361, 17H06363, 18H01213, 18H04595, 18H05236, 18K03642, 19H14720) of JSPS and by a post-K computer project (Priority issue No.~9) of Japanese MEXT. 
Our waveform data is publicly available on the web page.




\begin{thebibliography}{99}
\bibitem{footnote1}
  For the symmetric bianry, the tidal coupling constant $\kappa^T_2$ is equal to $3\tilde{\Lambda}/16$.
\end{thebibliography}
\end{document}